\documentclass[11pt,a4paper]{article}
\pdfoutput=1
\usepackage{amssymb,amsmath,amsfonts, mathtools, mathrsfs}
\usepackage[utf8]{inputenc} 
\usepackage[dvipsnames]{xcolor}

\newif\ifnatbibsort\natbibsorttrue

\relax

\ifnatbibsort\RequirePackage[numbers,sort&compress]{natbib}\else\RequirePackage[numbers,compress]{natbib}\fi
\RequirePackage[colorlinks=true
,urlcolor=blue
,anchorcolor=blue
,citecolor=blue
,filecolor=blue
,linkcolor=blue
,menucolor=blue
,pagecolor=blue
,linktocpage=true
,pdfproducer=medialab
,pdfa=true
]{hyperref}

\usepackage{graphicx}
\usepackage{caption}
\usepackage{tensor}
\usepackage{subfigure}
\usepackage{enumerate}
\usepackage{dsfont}
\usepackage{verbatim}
\usepackage{amsfonts,euscript,amssymb,stmaryrd,braket}
\usepackage{tikz}
\usetikzlibrary{arrows,decorations.markings,patterns}
\usepackage{slashed}

\def\clock{{\count0=\time
		\divide\count0 60
		\ifnum\count0<10 0\fi\the\count0
		\multiply\count0 -60 \advance\count0 \time
		:\ifnum\count0<10 0\fi \the\count0
}}
\newcommand{\timestamp}{{\small\vbox{\hbox{\tt\jobname.tex}
			\hbox{\the\day/\the\month/\the\year, \clock}}}}

\newcommand{\bea}{\begin{eqnarray}}
\newcommand{\eea}{\end{eqnarray}}

\newcommand{\be}{\begin{equation}}
\newcommand{\ee}{\end{equation}}

\makeatletter
\let\old@startsection=\@startsection
\let\oldl@section=\l@section
\renewcommand{\@startsection}[6]{\old@startsection{#1}{#2}{#3}{#4}{#5}{#6\mathversion{bold}}}
\renewcommand{\l@section}[2]{\oldl@section{\mathversion{bold}#1}{#2}}
\makeatother

\numberwithin{equation}{section}

\usepackage{color}

\begin{document}

\renewcommand{\thefootnote}{\arabic{footnote}}

	\overfullrule=0pt
	\parskip=2pt
	\parindent=12pt
	\headheight=0in \headsep=0in \topmargin=0in \oddsidemargin=0in

	\vspace{ -3cm} \thispagestyle{empty} \vspace{-1cm}
	\begin{flushright} 
		\footnotesize
		\textcolor{red}{\phantom{print-report}}
	\end{flushright}

\begin{center}
	\vspace{.5cm}

%
	
	{\Large\bf \mathversion{bold}
	Subsystem complexity after a global quantum quench \\
	}

	\vspace{0.8cm} {
		Giuseppe Di Giulio
		and 
		Erik Tonni
	}
	\vskip  0.7cm
	
	\small
	{\em
		SISSA and INFN Sezione di Trieste, via Bonomea 265, 34136, Trieste, Italy 
	}
	\normalsize
	
\end{center}

\vspace{0.3cm}
\begin{abstract} 
We study the temporal evolution of the circuit complexity for a subsystem 
in harmonic lattices after a global quantum quench of the mass parameter,
choosing the initial reduced density matrix as the reference state. 
Upper and lower bounds are derived
for the temporal evolution of the complexity for the entire system.
The subsystem complexity is evaluated by employing the 
Fisher information geometry for the covariance matrices.
We discuss numerical results for the temporal evolutions of the subsystem complexity 
for a block of consecutive sites in harmonic chains 
with either periodic or Dirichlet boundary conditions,
comparing them with the temporal evolutions of the entanglement entropy.
For infinite harmonic chains, 
the asymptotic value of the subsystem complexity is studied
through the generalised Gibbs ensemble.

\end{abstract}

\vspace{1cm}

\newpage

\tableofcontents


\section{Introduction}
\label{sec:intro}

The complexity of a quantum circuit is a quantity introduced in quantum information theory
\cite{Nielsen06,NielsenDowling06, DowlingNielsen08, Watrous2008quantum,  Aaronson:2016vto, aharonov1998quantum}
which has been studied also in the context of the holographic correspondence 
during the past few years
\cite{Susskind:2014rva,Susskind:2014jwa, Roberts:2014isa, Stanford:2014jda, Susskind:2014moa,
Alishahiha:2015rta, Brown:2015bva,Brown:2015lvg,
Barbon:2015ria,Carmi:2016wjl};
hence it provides an insightful way to explore a connection between
quantum information theory and quantum gravity.

A quantum circuit allows to construct a target state starting from a reference state through a sequence of gates.
The circuit complexity quantifies the difficulty to obtain the target state from the reference state 
by counting the minimum number of allowed gates that is necessary to construct the circuit in an optimal way. 
Besides the reference state, the target state and the set of allowed gates, 
the circuit complexity can 
depend also on the tolerance parameter for the target state. 
Many results have been obtained for the complexity of quantum circuits made by pure states
constructed through lattice models \cite{Jefferson:2017sdb,Chapman:2018hou,Guo:2018kzl,Hackl:2018ptj,Khan:2018rzm,Braccia:2019xxi,Chapman:2019clq,Doroudiani:2019llj,guo2020circuit} 
and in the gravitational side of the holographic correspondence.
Some proposals have been done also to study the circuit complexity in quantum fields theories 
\cite{Caputa:2017urj,Czech:2017ryf,Caputa:2017yrh,
Chapman:2017rqy, Bhattacharyya:2018wym,Caputa:2018kdj,Chapman:2018bqj,Camargo:2019isp,Ge:2019mjt,Sato:2019kik,Erdmenger:2020sup,Flory:2020eot,Flory:2020dja}.

Quantum quenches are insightful ways to explore
the dynamics of isolated quantum systems out of equilibrium 
(see \cite{Essler:2016ufo,Calabrese:2016xau} for recent reviews).
Given a quantum system prepared in the ground state $| \psi_0 \rangle$ of the hamiltonian $\widehat{H}_0$,
at $t=0$ a sudden change is performed  
such that the evolution Hamiltonian of the initial state $| \psi_0 \rangle$
becomes $\widehat{H} \neq \widehat{H}_0$.
Since $\widehat{H}$ and $\widehat{H}_0$ do not commute in general,
the unitary evolution 
$| \psi(t) \rangle =  e^{-\textrm{i} \widehat{H} t}  | \psi_0 \rangle$
for $t>0$ is highly non trivial. 
In the typical global quench, a parameter occurring in the Hamiltonian is suddenly changed 
from its value $\omega_0$ in $\widehat{H}_0$ to the value $\omega$ in $\widehat{H}$
%
\cite{Calabrese_2005,Calabrese:2006rx,Calabrese:2007rg,Essler:2016ufo}.
Insightful results have been obtained about
the asymptotic regime $t \to \infty$ of this unitary evolution by employing the 
generalised Gibbs ensemble (GGE)
(see the reviews \cite{Rigol_07,Ilievski_2015,Vidmar_2016}).

It is worth studying the circuit complexity 
with the target state given by the time-evolved pure state of certain unitary evolution
and the reference state by another pure state along the same evolution
\cite{Alves:2018qfv,Camargo:2018eof,Ali:2018fcz,Chapman:2018hou,Jiang:2018gft}.
In particular, considering a global quench protocol, 
we are interested in the optimal circuit and in the corresponding complexity 
where $| \psi(t) \rangle$ and $| \psi_0 \rangle$ 
are respectively the target and the reference states.
Within the gauge/gravity correspondence,
the temporal evolution of complexity for pure states has been explored in
\cite{Moosa:2017yvt,Chapman:2018dem,Chapman:2018lsv}.

Entanglement of spatial bipartitions
plays a crucial role both in quantum information theory and in quantum gravity,
hence it is a fundamental tool to understand the connections between them
(see \cite{Eisert:2008ur,Casini:2009sr,Calabrese:2009qy,Peschel_2009, 
Rangamani:2016dms,Headrick:2019eth,Tonni:2020bjq} for reviews).
The entanglement dynamics after global quantum quenches has been largely explored
by considering the temporal evolutions of various entanglement quantifiers.
The entanglement entropy has been mainly investigated 
through various methods 
\cite{Sotiriadis:2010si,Cotler:2016acd,ac-18-qp-quench,Calabrese:2016xau, 
Hubeny:2007xt,AbajoArrastia:2010yt},
but also other entanglement quantifiers
like the entanglement spectra 
\cite{Cardy:2016fqc,DiGiulio:2019lpb,Surace:2019mft},
the entanglement Hamiltonians 
\cite{Cardy:2016fqc,Wen:2018svb,DiGiulio:2019lpb},
the entanglement negativity \cite{Coser:2014gsa}
and the entanglement contours \cite{Chen_2014,Coser:2017dtb, DiGiulio:2019lpb} 
have been explored. 

In order to understand the relation between entanglement and complexity, 
it is useful to study the optimal circuits and the corresponding circuit complexity 
when both the reference and the target states are mixed states 
\cite{Caceres:2019pgf,DiGiulio:2020hlz,Ruan:2020vze,Camargo:2020yfv}.
The approach to the complexity of mixed states 
based on the purification complexity \cite{Agon:2018zso,Caceres:2019pgf,Camargo:2020yfv}
is general, but evaluating this quantity for large systems is technically complicated.
Some explicit results for large systems can be found 
by restricting to the simple case of bosonic Gaussian states
and by employing the methods of the information geometry 
\cite{Rao45,Atkinson81,Amari16book}.
In our analysis we adopt the approach to the complexity of mixed states 
based on the Fisher information geometry \cite{DiGiulio:2020hlz},
which allows to study large systems numerically.
The crucial assumption underlying this approach is that
all the states involved in the construction of the circuit are Gaussian.
We consider the important special case given by the subsystem complexity,
namely the circuit complexity corresponding to a circuit 
where both the reference and the target states 
are the reduced density matrices associated to a subsystem.

Within the gauge/gravity correspondence, 
 the subsystem complexity has been evaluated 
 both in static 
  \cite{Alishahiha:2015rta,Carmi:2016wjl,Abt:2018ywl,Agon:2018zso,Alishahiha:2018lfv,Auzzi:2019vyh} 
  and in time dependent gravitational backgrounds 
  \cite{Chen:2018mcc,Auzzi:2019mah,Ling:2019ien,Zhou:2019xzc}.
  In static backgrounds, it is given by the volume identified by 
  the minimal area hypersurface anchored to the boundary of the subsystem,
  whose area provides the holographic entanglement entropy
  \cite{Ryu:2006bv}
  (for static black holes, this hypersurface does not cross the horizon
  \cite{Headrick:2007km,Tonni:2010pv,Hubeny:2013gta}),
  while, in time dependent gravitational spacetimes,
  the extremal hypersurface occurring in the covariant proposal 
  for the holographic entanglement entropy \cite{Hubeny:2007xt} 
  must be employed.

In this manuscript we study 
the temporal evolution of the subsystem complexity after a global quantum quench
in harmonic lattices 
where the mass parameter is suddenly changed from $\omega_0$ to $\omega$.
Considering a ground state as the initial state,
the Gaussian nature of the state is preserved during the temporal evolution. 
In these bosonic systems, the reduced density matrices are characterised by
the corresponding reduced covariance matrices \cite{Weedbrook12b}.
By employing the approach to the complexity of bosonic mixed Gaussian states
based on the Fisher information geometry \cite{DiGiulio:2020hlz},
we evaluate numerically the subsystem complexity 
for one-dimensional harmonic lattices (i.e. harmonic chains)
and subsystems $A$ given by blocks of consecutive sites. 
We consider harmonic chains where either periodic boundary conditions (PBC) 
or Dirichlet boundary conditions (DBC) are imposed.
This allows to study the role of the zero mode.
The temporal evolution of the subsystem complexity 
at a generic time after the global quench w.r.t. the initial state
is compared with the 
temporal evolution of the 
corresponding  increment of the entanglement entropy.

This manuscript is organised as follows.
In Sec.\,\ref{sec:cov-mat} 
we introduce the main expressions to evaluate 
the circuit complexity after the global quench of the mass parameter 
through the covariance matrices of the reference and the target states
for harmonic lattices in a generic number of dimensions.
In the special case where the entire system is considered, 
these states are pure and bounds are obtained 
for the temporal evolution of the circuit complexity
w.r.t. the initial state.
In Sec.\,\ref{sec:purestates_HC_glob}
we specify this analysis to harmonic chains with either
PBC or DBC.
%
The main results of this manuscript are discussed in 
Sec.\,\ref{sec:mixedFinSize} and Sec.\,\ref{sec:GGE},
where the temporal evolution of the subsystem complexity
for a block of consecutive sites is investigated. 
In Sec.\,\ref{sec:mixedFinSize},
finite harmonic chains with either PBC or DBC are studied,
while in Sec.\,\ref{sec:GGE} we consider
infinite harmonic chains either on the line or on the semi-infinite line
with DBC at the origin.
In the cases of infinite chains, we employ known results about the GGE 
to determine
the asymptotic regime of the subsystem complexity.
In Sec.\,\ref{sec:conclusions} we draw some conclusions. 
Some technical details and supplementary results are discussed
in the appendices\;\ref{app:CMglobalquench}, 
\ref{app:unentangled}, \ref{app:large-N} and \ref{app:gge}.

\section{Complexity from the covariance matrix after the quench}
\label{sec:cov-mat}

In this section we discuss the expressions that allow to evaluate the 
temporal evolution of the circuit complexity based on the Fisher-Rao geometry 
for the harmonic lattices in a generic number of spatial dimensions
when both the reference and the target states are pure. 
Analytic expressions that bound this temporal evolution are also derived.

\subsection{Covariance matrix after the quench}
\label{sec-cov-mat-quench}

The Hamiltonian of the harmonic lattice made by $N$ sites with nearest neighbour spring-like interaction reads
\be
\label{HC ham}
\widehat{H} 
\,=\, 
\sum_{i=1}^{N} \left(
\frac{1}{2m}\,\hat{p}_i^2+\frac{m\omega^2}{2}\,\hat{q}_i^2 
\right) 
+
 \sum_{\langle i,j \rangle}\frac{\kappa}{2}(\hat{q}_{i} -\hat{q}_j)^2
 \,=\,
 \frac{1}{2}\, \hat{\boldsymbol{r}}^{\textrm t} H^{\textrm{\tiny phys}} \, \hat{\boldsymbol{r}}
\ee
where the position and the momentum operators  $\hat{q}_i$ and $\hat{p}_i$
are hermitean operators satisfying the canonical commutation relations
$[\hat{q}_i , \hat{q}_j]=[\hat{p}_i , \hat{p}_j] = 0$ 
and $[\hat{q}_i , \hat{p}_j]= \textrm{i} \delta_{i,j}$.
The matrix $H^{\textrm{\tiny phys}}$ in (\ref{HC ham}) has been defined
by collecting the position and the momentum operators into the vector
$\hat{\boldsymbol{r}} \equiv (\hat{q}_1 , \dots , \hat{q}_N, \hat{p}_1, \dots, \hat{p}_N)^{\textrm{t}}$.

In the Heisenberg picture, 
the unitary temporal evolution of the position and the momentum operators
$ \hat{q}_j(t)$ and $ \hat{p}_j(t)$ through the evolution Hamiltonian $\widehat{H}$ reads
\be
\label{heisemberg rps}
 \hat{q}_j(t) = e^{\textrm{i} \widehat{H} t}   \hat{q}_j(0) e^{-\textrm{i} \widehat{H} t} 
\qquad
 \hat{p}_j(t) = e^{\textrm{i} \widehat{H} t}   \hat{p}_j(0) e^{-\textrm{i} \widehat{H} t} \,.
\ee

In order to study the temporal evolution of the harmonic lattices 
after the global quantum quench of the mass parameter 
that we are considering,
we need to introduce
the $N \times N$ correlation matrices 
for operators (\ref{heisemberg rps})
whose elements read
\be
\label{time dep corrs}
\begin{array}{l}
Q_{i,j}(t) \equiv \langle \psi_0 |\, \hat{q}_i(t)  \,\hat{q}_j(t) \, | \psi_0 \rangle 
\\
\rule{0pt}{.6cm}
P_{i,j}(t) \equiv \langle \psi_0 | \,\hat{p}_i(t)  \,\hat{p}_j(t) \, | \psi_0 \rangle 
\\
\rule{0pt}{.6cm}
M_{i,j}(t) \equiv 
\textrm{Re} \big[ \langle \psi_0 | \,\hat{q}_i(t)  \,\hat{p}_j(t)  \,| \psi_0 \rangle \big]
\end{array}
\ee
where $| \psi_0 \rangle$ is the ground state of the Hamiltonian $\widehat{H}_0$,
defined by (\ref{HC ham}) with $\omega$ replaced by $\omega_0$.

At any time $t>0$ after the quench, the system is completely characterised by its covariance matrix $\gamma(t)$,
which is the following $2N \times 2N$ real, symmetric and positive definite matrix 
\be
\label{covariancematrix_t}
\gamma(t)
=\,
\bigg( 
\begin{array}{cc}
Q(t)  &  M(t) \\
M(t)^{\textrm{t}}  &  P(t)  \\
\end{array}   \bigg)
\ee
where the elements of the $N \times N$ block matrices are given by (\ref{time dep corrs}).
This covariance matrix has been already used to study
the entanglement dynamics e.g. in \cite{Sotiriadis:2010si,Coser:2014gsa,Cotler:2016acd}.

In the appendix\;\ref{subapp:covariancematrix} we discuss the fact that,
for the global quench we are exploring, 
the blocks of the covariance matrix (\ref{covariancematrix_t}) can be 
decomposed as
\be
\label{time dep corrs trans}
Q(t) =
\widetilde{V} \, \mathcal{Q}(t) \,\widetilde{V}^{\textrm{t}}
\,\,\qquad
\rule{0pt}{.6cm}
P(t) = \widetilde{V} \, \mathcal{P}(t)\, \widetilde{V}^{\textrm{t}}
\,\,\qquad
\rule{0pt}{.6cm}
M(t) = \widetilde{V}\, \mathcal{M}(t) \,\widetilde{V}^{\textrm{t}}
\ee
where $\widetilde{V}$ is a real orthogonal $N\times N$ matrix,
while $\mathcal{Q}(t)$, $\mathcal{P}(t)$ and $\mathcal{M}(t)$ are $N\times N$
diagonal matrices whose 
$k$-th element along the diagonal is
\cite{Calabrese:2007rg}
\be
\label{QPRmat t-dep-k}
\begin{array}{l}
\displaystyle
Q_{k}(t)
\equiv  \mathcal{Q}_{k,k}(t)=
\frac{1}{ 2m\Omega_k}
\left( \,\frac{\Omega_k}{\Omega_{0,k}} \, [ \cos(\Omega_k t) ]^2
+ \frac{\Omega_{0,k}}{\Omega_k} \, [\sin(\Omega_k t)]^2 \right)
\\
\displaystyle
\rule{0pt}{.9cm}
P_{k}(t)
\equiv  \mathcal{P}_{k,k}(t)=
 \frac{m\Omega_k}{2}
\left( \,\frac{\Omega_k}{\Omega_{0,k}}  \, [\sin(\Omega_k t)]^2
+ \frac{\Omega_{0,k}}{\Omega_k} \, [\cos(\Omega_k t)]^2 \right)
 \\
 \displaystyle
 \rule{0pt}{.9cm}
M_{k}(t)
\equiv \mathcal{M}_{k,k}(t)=
\frac{1}{2}\left(  \frac{\Omega_{0,k}}{\Omega_k} -\,\frac{\Omega_k}{\Omega_{0,k}} \right)
\sin(\Omega_k t) \cos(\Omega_k t) 
\end{array}
\ee
in terms of the dispersion relations  $\Omega_{0,k}$ and $\Omega_k$ of the Hamiltonians 
$\widehat{H}_0$ and $\widehat{H}$ respectively, 
which depend both on the dimensionality of the lattice and on the boundary conditions.

At $t=0$, the expressions in (\ref{QPRmat t-dep-k}) simplify respectively to
\be
\label{QPRmat tzero-dep-k}
Q_{k}(0) =  \frac{1}{2m\,\Omega_{0,k}} 
\;\;\qquad\;\;
P_{k}(0) =  \frac{m \,\Omega_{0,k}}{2} 
\;\;\qquad\;\;
M_{k}(0) = 0 \,.
\ee

From the above discussion, one realises that $\gamma(t)$ is a function of $t$ determined by the set of parameters given by 
$\{ m, \kappa, \omega, \omega_0\}$.

When the dispersion relation vanishes for certain value of $k$, e.g. $k=N$, 
the corresponding mode is a zero mode. 
The relations (\ref{QPRmat t-dep-k}) and (\ref{QPRmat tzero-dep-k}) 
are well defined when $\Omega_{0,k}$ does not vanish;
hence $\Omega_{k}$ can have a zero mode, while $\Omega_{0,k}$ cannot.
This highlights the asymmetric role of $\Omega_{0,k}$ and $\Omega_{k}$.

\subsection{Complexity for the system}
\label{sec-pure-states}

The circuit complexity is proportional to the length of the optimal quantum circuit
that creates a target state from a reference state.
In this manuscript we evaluate the complexity through the Fisher-Rao distance 
between two bosonic Gaussian states with vanishing first moments 
\cite{Weedbrook12b, Adesso14, Serafini17book}.
This approach allows to study also the circuits made by mixed states \cite{DiGiulio:2020hlz}.

Denoting by $\gamma_{\textrm{\tiny R}}$ and $\gamma_{\textrm{\tiny T}}$ 
the covariance matrices with vanishing first moments
of the reference and of the target state respectively,
the Fisher-Rao distance between them \cite{Atkinson81,Bhatia07book} 
provides the following definition of complexity 
\be
\label{c2 complexity}
\mathcal{C}
\,\equiv\,
\frac{1}{2\sqrt{2}}\;
\sqrt{\,
\textrm{Tr}\, \Big\{ \big[ \log \big( \gamma_{\textrm{\tiny T}} \,\gamma_{\textrm{\tiny R}}^{-1}  \big) \big]^2 \Big\}
}\;.
\ee
When both $\gamma_{\textrm{\tiny R}}$ and $\gamma_{\textrm{\tiny T}}$ characterise pure states,
this complexity corresponds to the one defined through
the $F_2$ cost function \cite{Chapman:2018hou}.

The analysis of the circuits made by bosonic Gaussian states based on the Fisher-Rao metric
provides also the optimal circuit between  $\gamma_{\textrm{\tiny R}}$ and $\gamma_{\textrm{\tiny T}}$.
It reads \cite{Bhatia07book}
\be
\label{optimal circuit}
G_s(\gamma_{\textrm{\tiny R}} \, , \gamma_{\textrm{\tiny T}})
\,\equiv \,
\gamma_{\textrm{\tiny R}}^{1/2} 
\Big(  \gamma_{\textrm{\tiny R}}^{- 1/2}  \,\gamma_{\textrm{\tiny T}} \,\gamma_{\textrm{\tiny R}}^{-1/2}  \Big)^s
\gamma_{\textrm{\tiny R}}^{1/2} 
\;\; \qquad \;\;
0 \leqslant s \leqslant 1
\ee
which gives $\gamma_{\textrm{\tiny R}}$ when $s=0$ 
and $\gamma_{\textrm{\tiny T}}$ when $s=1$.
The length of the optimal circuit (\ref{optimal circuit})
evaluated through the Fisher-Rao distance is proportional to the circuit complexity 
(\ref{c2 complexity}),
which has been explored both for pure states \cite{Chapman:2018hou} 
and for mixed states  \cite{DiGiulio:2020hlz}.

In this manuscript we are interested in the temporal evolution of the circuit complexity 
after a global quench.
In the following discussion and in Sec.\,\ref{sec:purestates_HC_glob}
we consider first the case where both the reference and the target states are pure states,
while in Sec.\,\ref{sec:mixedFinSize} and Sec.\,\ref{sec:GGE} 
we study the case where both the reference and the target states are mixed states.

Denoting by $t_{\textrm{\tiny R}}$ and $t_{\textrm{\tiny T}}$ the values of $t$
corresponding to the reference state and to the target state respectively, 
let us adopt the following notation
\be
\label{gamma-RT-time-def}
\gamma_{\textrm{\tiny R}} = \gamma(t_{\textrm{\tiny R}})
\;\;\qquad\;\;
\gamma_{\textrm{\tiny T}} = \gamma(t_{\textrm{\tiny T}})\,.
\ee
In the most general setup,
$\gamma_{\textrm{\tiny R}} $ is a function of $t_{\textrm{\tiny R}}$
characterised by the set of parameters 
$\{m_{\textrm{\tiny R}}, \kappa_{\textrm{\tiny R}}, \omega_{\textrm{\tiny R}}, \omega_{0,\textrm{\tiny R}}\}$,
while $\gamma_{\textrm{\tiny T}} $ is a function of $t_{\textrm{\tiny T}}$
parameterised by 
$\{m_{\textrm{\tiny T}}, \kappa_{\textrm{\tiny T}}, \omega_{\textrm{\tiny T}}, \omega_{0,\textrm{\tiny T}}\}$.
This means that the reference and target states are obtained as the time-evolved states at $t=t_\textrm{\tiny R} \geqslant 0$ and $t=t_\textrm{\tiny T} \geqslant t_\textrm{\tiny R}$
respectively, through two different global quenches
determined by 
 $\{\kappa_\textrm{\tiny R},m_\textrm{\tiny R},\omega_\textrm{\tiny R},\omega_{0,\textrm{\tiny R}}\}$
 and $\{\kappa_\textrm{\tiny T},m_\textrm{\tiny T},\omega_\textrm{\tiny T},\omega_{0,\textrm{\tiny T}}\}$
 respectively.

The covariance matrix (\ref{covariancematrix_t}) at a generic value of $t$ can be written as follows\footnote{We used that 
\be
\label{decomp_blockmat}
\bigg( \begin{array}{cc}
A  & B
\\
C  &  D
\end{array} \bigg)
=
\bigg( 
\begin{array}{cc}
S\,\mathcal{A}\, S^\dagger \,& S\,\mathcal{B}\, T^\dagger
\\
T\,\mathcal{C}\, S^\dagger \,&  T\,\mathcal{D}\, T^\dagger
\end{array}
\bigg)
=
\bigg(
\begin{array}{cc}
S & \boldsymbol{0}
\\
\boldsymbol{0} &  T
\end{array} \bigg)
\;
\bigg(\begin{array}{cc}
\mathcal{A}  \,& \mathcal{B}
\\
 \mathcal{C}  \,&  \mathcal{D}
\end{array}\bigg)
\;
\bigg(\begin{array}{cc}
S^\dagger \!& \boldsymbol{0}
\\
\boldsymbol{0} \!&  T^\dagger
\end{array} \bigg)
\ee
where $\mathcal{A}$, $\mathcal{B}$, $\mathcal{C}$ and $\mathcal{D}$ are diagonal matrices. 
}
\be
\label{VGammaV dec}
\gamma(t)=V^\textrm{t} \,\Gamma(t)\, V
\,\,\qquad\,\, V= \widetilde{V} \oplus \widetilde{V}
\ee
where $V$ is an orthogonal and symplectic matrix because $\widetilde{V}$ is orthogonal 
and the block decomposition of $\Gamma(t)$ reads
\be
\label{covariancematrix_diags_t}
\Gamma(t)
=
\bigg( 
\begin{array}{cc}
\mathcal{Q}(t)  & \!\,\, \mathcal{M}(t) 
\\
\mathcal{M}(t)  & \!\,\, \mathcal{P}(t)  
\end{array}  
\bigg)
\ee
in terms of the diagonal matrices whose elements have been defined in (\ref{QPRmat t-dep-k}).

Hereafter we enlighten the expressions by avoiding to indicate explicitly the dependence on $t$,
wherever this is possible. 
The inverse of (\ref{covariancematrix_diags_t})  is\footnote{The expression (\ref{Gamma-inv-diag}) is a special case of the following general formula
\be
\label{gamma-block-QPM}
\gamma
\equiv
\bigg(\begin{array}{cc}
A & B
\\
B^{\textrm t} &C
\end{array}\bigg)
\;\;\;\;\qquad\;\;\;\;
\gamma^{-1} 
\equiv
\bigg(\begin{array}{cc}
\mathfrak{A} & \mathfrak{B}
\\
\mathfrak{B}^{\textrm t} &\mathfrak{C}
\end{array}\bigg)
\;\;\qquad\;\;
\left\{
\begin{array}{l}
\mathfrak{A} \equiv  \big(A  - B\, C^{-1} B^{\textrm t}   \big)^{-1}
\\
\mathfrak{C} \equiv  \big(C  - B^{\textrm t}\, A^{-1} B   \big)^{-1}
\\
\mathfrak{B} \equiv  -  A^{-1} B \, \big(C  - B^{\textrm t}\, A^{-1} B   \big)^{-1}\,.
\end{array}
\right.
\ee
}
\be
\label{Gamma-inv-diag}
\Gamma^{-1}
=
\big(\mathcal{Q} \,\mathcal{P}  - \mathcal{M}^2\big)^{-1} 
\bigg(
\begin{array}{cc}
  \mathcal{P} \! & - \,\mathcal{M} 
\\
- \,\mathcal{M}  \! &   \mathcal{Q} 
\end{array}
\bigg)\,.
\ee

Since $\gamma$ in (\ref{covariancematrix_t}) describes a pure state, the condition 
$(\textrm{i} J \gamma)^2 = \tfrac{1}{4}\, \boldsymbol{1}$ holds;
hence the blocks $\mathcal{Q} $, $\mathcal{P} $ and $\mathcal{M} $ are not independent.
More explicitly, this constraint reads
\be
(\textrm{i} J \gamma)^2 
=
\bigg(\begin{array}{cc}
PQ -(M^{\textrm t})^2 \, & PM - M^{\textrm t}P
\\
QM^{\textrm t} - MQ \,& QP -M^2
\end{array}\bigg)
=\,
 V^\textrm{t}\,
 \bigg(\begin{array}{cc}
\mathcal{P} \mathcal{Q} - \mathcal{M}^2  \,& \mathcal{P} \mathcal{M} - \mathcal{M}^{\textrm t}\mathcal{P}
\\
\mathcal{Q}\mathcal{M}^{\textrm t} - \mathcal{M}\mathcal{Q} \,& \mathcal{Q}\,\mathcal{P} -\mathcal{M}^2
\end{array}\bigg)
\,  V
= 
\frac{1}{4}\, \boldsymbol{1}
\ee
which implies
\be
\label{purity-condition}
\mathcal{Q}\,\mathcal{P} -\mathcal{M}^2
=
\frac{1}{4}\, \boldsymbol{1}
\qquad \Longleftrightarrow \qquad
Q_k P_k -M_k^2 = \frac{1}{4}
\;\qquad\;
1\leqslant k \leqslant N\,.
\ee
This result allows to further simplify (\ref{Gamma-inv-diag}), which becomes
\be
\label{covariancematrix_diags_inv_t}
\Gamma^{-1} 
=\,
4\,
\bigg(
\begin{array}{cc}
  \mathcal{P} & - \mathcal{M} 
\\
- \mathcal{M}  &   \mathcal{Q} 
\end{array} 
\bigg) \,.
\ee

In this manuscript we restrict to cases where a symplectic matrix $V$ exists such that 
\be
\label{VV-diag-hyp}
\gamma_{\textrm{\tiny R}} = V^\textrm{t} \,\Gamma_{\textrm{\tiny R}}\, V
\;\;\qquad\;\;
\gamma_{\textrm{\tiny T}} = V^\textrm{t} \,\Gamma_{\textrm{\tiny T}}\, V
\ee
where both $\Gamma_{\textrm{\tiny R}}$ and $\Gamma_{\textrm{\tiny T}}$ have the form 
(\ref{covariancematrix_diags_t}), in terms of the corresponding diagonal matrices. 
When (\ref{VV-diag-hyp}) holds, the matrix occurring in the argument of the logarithm in (\ref{c2 complexity}) becomes
\be
\label{gamma-TR-global}
\gamma_{\textrm{\tiny T}}\, \gamma_{\textrm{\tiny R}}^{-1} 
=
V^\textrm{t} \,\Gamma_{\textrm{\tiny T}} \,\Gamma_{\textrm{\tiny R}}^{-1}\, V^{-\textrm{t}}
\ee
which tells us that the complexity (\ref{c2 complexity}) is provided by the eigenvalues of $\Gamma_{\textrm{\tiny T}} \,\Gamma_{\textrm{\tiny R}}^{-1}$.
Thus, the matrix $V$ does not influence the temporal evolution of the complexity 
after the global quench when both the reference and the target states are pure states.
Instead, they play a crucial role for
the temporal evolution of the subsystem complexity discussed in 
Sec.\,\ref{sec:mixedFinSize}
and Sec.\,\ref{sec:GGE}.

By using (\ref{covariancematrix_diags_t}) for $\Gamma_\textrm{\tiny T}$
and (\ref{covariancematrix_diags_inv_t}) for $\Gamma^{-1}_\textrm{\tiny R}$, 
we obtain the following block matrix 
\be
\label{relativeCM_diag}
\Gamma_\textrm{\tiny T} \,\Gamma^{-1}_\textrm{\tiny R}
=
4\,
\bigg( 
\begin{array}{cc}
\mathcal{P}_\textrm{\tiny R}\mathcal{Q}_\textrm{\tiny T}-\mathcal{M}_\textrm{\tiny R}\mathcal{M}_\textrm{\tiny T} 
& \;\;
\mathcal{Q}_\textrm{\tiny R}\mathcal{M}_\textrm{\tiny T}-\mathcal{M}_\textrm{\tiny R}\mathcal{Q}_\textrm{\tiny T} 
\\
 \mathcal{P}_\textrm{\tiny R}\mathcal{M}_\textrm{\tiny T}-\mathcal{M}_\textrm{\tiny R}\mathcal{P}_\textrm{\tiny T} 
 & \;\;
\mathcal{Q}_\textrm{\tiny R}\mathcal{P}_\textrm{\tiny T}-\mathcal{M}_\textrm{\tiny R}\mathcal{M}_\textrm{\tiny T} 
\end{array}   \bigg)
\ee
whose blocks are diagonal matrices.
By using also (\ref{purity-condition}), for 
the eigenvalues of (\ref{relativeCM_diag}) we find\footnote{Considering a $2N\times 2N$ matrix $M$ partitioned into four $N\times N$ blocks 
$\mathcal{A}$, $\mathcal{B}$, $\mathcal{C}$ and $\mathcal{D}$
which are diagonal matrices,
its eigenvalues equation can be written through the formula for the determinant of a block matrix, 
finding 
\be
\label{det_1}
M =
\bigg( \!
\begin{array}{cc}
\mathcal{A}  & \!\!\,\,\,\mathcal{B} \\
\mathcal{C} & \!\!\,\,\, \mathcal{D}  \\
\end{array}  \! \bigg)
\qquad
\det(M-\lambda\,\boldsymbol{1})
\,=\,
\det\!\big[ \mathcal{D}-\lambda\,\boldsymbol{1} \big]
\,
\det\!\big[\mathcal{A}-\lambda\,\boldsymbol{1}
-\mathcal{B}\,\mathcal{C}\,(\mathcal{D}-\lambda\boldsymbol{1})^{-1}\big]
= 0
\ee
where $\boldsymbol{1}$ is the identity matrix. 
Since the matrices in (\ref{det_1}) are diagonal, this equation becomes
$\prod_{k=1}^N[(d_k-\lambda)(a_k-\lambda)-b_k c_k]=0$;
hence the $2N$ eigenvalues of $M$  in (\ref{det_1}) are
\be
\label{eigenvalues block mat diag}
\lambda_{k}^{(\pm)}=\frac{a_k+d_k\pm\sqrt{(a_k-d_k)^2+4 b_k c_k}}{2}
\,\,\qquad\,\,
1 \leqslant k \leqslant N\,.
\ee
}
\bea
\label{eigenvalues relative CM}
g_{\textrm{\tiny TR},k}^{(\pm)}
&\equiv&
2
\bigg(P_{\textrm{\tiny R},k}Q_{\textrm{\tiny T},k}+Q_{\textrm{\tiny R},k}P_{\textrm{\tiny T},k}-2 M_{\textrm{\tiny R},k}M_{\textrm{\tiny T},k}
\\
&& \hspace{.6cm}
\pm \,
\sqrt{
\big(P_{\textrm{\tiny R},k}Q_{\textrm{\tiny T},k}-Q_{\textrm{\tiny R},k}P_{\textrm{\tiny T},k}\big)^2
+4\big(Q_{\textrm{\tiny R},k}M_{\textrm{\tiny T},k}-M_{\textrm{\tiny R},k}Q_{\textrm{\tiny T},k}\big)
\big(P_{\textrm{\tiny R},k}M_{\textrm{\tiny T},k}-M_{\textrm{\tiny R},k}P_{\textrm{\tiny T},k}\big)}
\;\bigg)
\nonumber
\eea
labelled by $1 \leqslant k \leqslant N$, which can be written as
\be
\label{gpm-from-C}
g_{\textrm{\tiny TR},k}^{(\pm)}
=
C_{\textrm{\tiny TR},k} \pm \sqrt{C_{\textrm{\tiny TR},k}^2 - 1}
\ee
where 
\be
\label{CTR generic}
C_{\textrm{\tiny TR},k} 
\equiv 
2\big( Q_{\textrm{\tiny T},k} \, P_{\textrm{\tiny R},k} + P_{\textrm{\tiny T},k} \, Q_{\textrm{\tiny R},k} - 2\, M_{\textrm{\tiny T},k} \, M_{\textrm{\tiny R},k} \big) 
\ee
in terms of the expressions in (\ref{QPRmat t-dep-k}) specialised to the reference and the target states.

From (\ref{gpm-from-C}) and (\ref{CTR generic}), one observes that
\be
\label{g-pm-product}
g_{\textrm{\tiny TR},k}^{(+)} \,g_{\textrm{\tiny TR},k}^{(-)}
 =
 16\big(Q_{\textrm{\tiny R},k}P_{\textrm{\tiny R},k}-M_{\textrm{\tiny R},k}^2\big)
 \big(Q_{\textrm{\tiny T},k}P_{\textrm{\tiny T},k}-M_{\textrm{\tiny T},k}^2\big)\,.
 \ee
By employing (\ref{purity-condition}) in this result, we find
$g_{\textrm{\tiny TR},k}^{(+)}= 1/ g_{\textrm{\tiny TR},k}^{(-)}$ for pure states,
for any $1 \leqslant k \leqslant N$.

From (\ref{relativeCM_diag}), (\ref{gpm-from-C}) and (\ref{g-pm-product}),
for the complexity (\ref{c2 complexity})
one obtains\footnote{The last step expression in (\ref{c2-log-lambda-arcosh}) is obtained through the identity 
$\log(x+\sqrt{x^2 -1}\,) = \textrm{arccosh}(x)$ for $x\geqslant 1$.}
\be
\label{c2-log-lambda-arcosh}
\mathcal{C} 
\,=\, 
\frac{1}{2}\,\sqrt{ \sum_{k=1}^N \!\big[ \log(g_{\textrm{\tiny TR},k}^{(+)})\big]^2}
\,=\,
\frac{1}{2}\,\sqrt{ 
 \sum_{k=1}^N\! \big[ \log(g_{\textrm{\tiny TR},k}^{(-)})\big]^2}
\,=\,
\frac{1}{2}\,\sqrt{  \sum_{k=1}^N\! \big[ \textrm{arccosh}(C_{\textrm{\tiny TR},k})\big]^2}\;.
\ee
In the most general setup described below (\ref{gamma-RT-time-def}),
the complexity can be found by writing (\ref{QPRmat t-dep-k}) 
for the reference and the target states first and then 
and plugging the results into (\ref{CTR generic})  and (\ref{c2-log-lambda-arcosh}).
The final result is a complicated expressions which can be seen as a function of $t_{\textrm{\tiny R}}$ and $t_{\textrm{\tiny T}}$
parameterised by  $\{\kappa_\textrm{\tiny R},m_\textrm{\tiny R},\omega_\textrm{\tiny R},\omega_{0,\textrm{\tiny R}}\}$
 and $\{\kappa_\textrm{\tiny T},m_\textrm{\tiny T},\omega_\textrm{\tiny T},\omega_{0,\textrm{\tiny T}}\}$.
We remark that (\ref{c2-log-lambda-arcosh}) can be employed when (\ref{VV-diag-hyp}) holds. 
Furthermore, we consider only cases where the matrix $V$ in (\ref{VV-diag-hyp}) 
depends on the geometric parameters of the system and of the subsystem
but it is independent of the physical parameters occurring in the Hamiltonians
(see Sec.\,\ref{subsec:initial HC}).

In the appendix\,\ref{subapp:squeezing}, 
the expression (\ref{c2-log-lambda-arcosh})
is obtained through the Williamson's decomposition 
\cite{Williamson36}
of the covariance matrices (\ref{gamma-RT-time-def}).

A remarkable simplification occurs 
when the reference and the target states are pure states along the time evolution of a given quench.
In this case, the parameters to fix in (\ref{QPRmat t-dep-k}) are 
$m_{\textrm{\tiny R}}=m_{\textrm{\tiny T}}=m$, 
$\kappa_{\textrm{\tiny R}}=\kappa_{\textrm{\tiny T}}=\kappa$, $\omega_{\textrm{\tiny R}}=\omega_{\textrm{\tiny T}}=\omega$, 
and $\omega_{0,\textrm{\tiny R}}=\omega_{0,\textrm{\tiny T}}=\omega_0$;
hence (\ref{CTR generic}) simplifies to
\be
\label{CTRomegaReqomegaT}
C_{\textrm{\tiny TR},k}
=
1+\frac{1}{2}
\Bigg(
\frac{\Omega_{k}^2 -\Omega^2_{0,k}
}{\Omega_{k}\, \Omega_{0,k}}\, 
\sin[\Omega_{k}(t_{\textrm{\tiny R}}- t_{\textrm{\tiny T}})]
\Bigg)^2
\ee
which must be plugged into (\ref{c2-log-lambda-arcosh}) to get the complexity of pure states after the global quench.
Notice that (\ref{CTRomegaReqomegaT}) is not invariant under the exchange $\Omega_{k} \leftrightarrow \Omega_{0,k}$ for a given $k$.
We remark that (\ref{CTRomegaReqomegaT})
and the corresponding complexity depend on $|t_{\textrm{\tiny R}}- t_{\textrm{\tiny T}}|$.
This is not the case for the most generic choice of the parameters.

\subsection{Complexity with respect to the initial state}
\label{sec:comp-initial-state}

A very natural choice for the reference state is the initial state $| \psi_0 \rangle$, 
which is a crucial ingredient of the quench protocol. 
This corresponds to choose $t_{\textrm{\tiny R}}=0$ in (\ref{gamma-RT-time-def}).
In this case, from (\ref{QPRmat tzero-dep-k}) and (\ref{purity-condition})
we have that $\mathcal{M}_{\textrm{\tiny R}}=\boldsymbol{0} $ and
$\mathcal{Q}_{\textrm{\tiny R}}\mathcal{P}_{\textrm{\tiny R}}=\frac{1}{4}\boldsymbol{1}$,
which allow to write (\ref{CTR generic}) as
\be
\label{C-TR-camargo}
C_{\textrm{\tiny TR},k} 
\,\equiv\,  
\frac{1}{2} \left( \frac{Q_{\textrm{\tiny T},k}}{Q_{\textrm{\tiny R},k}}  + \frac{P_{\textrm{\tiny T},k}}{P_{\textrm{\tiny R},k}} \right) .
\ee

Setting $m_{\textrm{\tiny R}}=m_{\textrm{\tiny T}}=m$ for simplicity 
and $t_{\textrm{\tiny T}}=t$ and $t_{\textrm{\tiny R}}=0$
in the most general setup described below (\ref{gamma-RT-time-def})
and then using (\ref{QPRmat t-dep-k}) and (\ref{QPRmat tzero-dep-k}), 
this expression becomes
\be
\label{complexity_equalm}
C_{\textrm{\tiny TR},k}
=
\frac{
\big(\Omega_{0,\textrm{\tiny T},k}^2+\Omega_{0,\textrm{\tiny R},k}^2\big)\,\Omega_{\textrm{\tiny T},k}^2 \,[\cos(\Omega_{\textrm{\tiny T},k}t)]^2
+
\big(
\Omega_{\textrm{\tiny T},k}^4+\Omega_{0,\textrm{\tiny T},k}^2\,\Omega_{0,\textrm{\tiny R},k}^2
\big)
[\sin(\Omega_{\textrm{\tiny T},k}t)]^2
}{
2\,\Omega_{\textrm{\tiny T},k}^2 \,\Omega_{0,\textrm{\tiny R},k}\, \Omega_{0,\textrm{\tiny T},k}}
\ee
in terms of the dispersion relations $\Omega_{0,\textrm{\tiny S},k}$ (with $\textrm{S}\in \{\textrm{R} , \textrm{T}\}$) 
before the quenches providing the reference and the target states
and of the dispersion relations $\Omega_{\textrm{\tiny T},k}$ after the quench
($\Omega_{\textrm{\tiny R},k}$ does not occur because $t_{\textrm{\tiny R}}=0$, hence (\ref{QPRmat tzero-dep-k}) must be employed).

The expression (\ref{C-TR-camargo})  is consistent with the result reported in \cite{Camargo:2018eof},
where the temporal evolution of the complexity of this free bosonic system has been also studied
through a different quench profile that does not include the quench protocol that we are considering. 
In many studies the reference state is the unentangled product state 
\cite{Jefferson:2017sdb,Chapman:2018hou,Guo:2018kzl,Alves:2018qfv}.
In appendix\;\ref{app:unentangled} we briefly discuss
the temporal evolution of the complexity
given by (\ref{c2-log-lambda-arcosh}) and  (\ref{complexity_equalm})
in the case where the initial state is the unentangled product state.

When the same quench is employed to construct the reference and the target states
$\Omega_{0,\textrm{\tiny R},k}=\Omega_{0,\textrm{\tiny T},k}=\Omega_{0,k} $ for any $k$
and (\ref{complexity_equalm}) simplifies.
This choice corresponds to evaluate 
the complexity between the initial state and the state at time $t$ after the quench. 
Specialising (\ref{complexity_equalm}) to this case and renaming $\Omega_{\textrm{\tiny T},k}\equiv\Omega_{k}$, we obtain 
\be
\label{CTR}
C_{\textrm{\tiny TR},k}
=
1+
\frac{1}{2}\Bigg(
\frac{\Omega_{k}^2 -\Omega^2_{0,k}
}{\Omega_{k} \,\Omega_{0,k}}\, \sin(\Omega_{k}t)\Bigg)^2
\ee
which coincides with (\ref{CTRomegaReqomegaT}) for $t_{\textrm{\tiny R}}=0$ and $t_{\textrm{\tiny T}}=t$, as expected.
Plugging (\ref{CTR}) into (\ref{c2-log-lambda-arcosh}) and 
using the identity $| \textrm{arccosh}(1+x^2/2)| = 2\,| \textrm{arcsinh}(x/2) | $,
one finds 
\be
\label{comp-pure-global-general}
\mathcal{C} 
\,=\,
\sqrt{\, \sum_{k=1}^N\! 
\left[ \textrm{arcsinh}\! 
\left( \,
 \frac{\Omega_{k}^2 - \Omega_{0,k}^2}{2\,\Omega_{k}\, \Omega_{0,k}} \,
 \sin (\Omega_{k} t ) 
 \right)
 \right]^2}\,.
\ee
In this expression the dispersion relations $\Omega_k$ and $\Omega_{0,k}$
(which depend on the number of spatial dimensions and on the boundary conditions of the lattice)
do not occur in a symmetric way.

We find it worth highlighting the contribution of the $N$-th mode by denoting
\be 
\label{zeromode-def}
c_0
\equiv \bigg[\,
\textrm{arcsinh}\!\,
\bigg( \,
 \frac{\Omega_{N}^2 - \Omega_{0,N}^2}{2\,\Omega_{N}\, \Omega_{0,N}} \,
 \sin (\Omega_{N} t ) 
 \bigg)\bigg]^2
\; \qquad\;
 \mathcal{C}_0^2
\equiv
\sum_{k=1}^{N-1}\! 
\left[ \textrm{arcsinh}\! 
\left( \,
 \frac{\Omega_{k}^2 - \Omega_{0,k}^2}{2\,\Omega_{k}\, \Omega_{0,k}} \,
 \sin (\Omega_{k} t ) 
 \right)
 \right]^2
\ee
which lead to write (\ref{comp-pure-global-general}) as 
\be
\label{eta-c0-decomposition}
\mathcal{C}^2 = \eta\,c_0 +  \mathcal{C}_0^2
\ee
where either $\eta =1$ or $\eta = 0$, 
depending on whether the $N$-th mode plays a particular role,
as one can read from the dispersion relation. 
This is the case e.g. for the zero mode in the harmonic lattices that are invariant under spatial translations,
which is briefly discussed also at the end of Sec.\,\ref{sec-cov-mat-quench};
hence hereafter we refer to $c_0$ as the zero mode contribution.
For instance, $\eta = 1$ in the harmonic chains with PBC,
while $\eta=0$ when DBC are imposed, 
as discussed later in Sec.\,\ref{subsec:initial HC}.
%
The result (\ref{comp-pure-global-general}), 
which can be applied for harmonic lattices in generic number of dimensions 
and for diverse boundary conditions, 
has been already reported in \cite{Ali:2018fcz} for harmonic chains with PBC.

It is interesting to determine the initial growth of the complexity by considering the series expansion of (\ref{comp-pure-global-general}) as $t \to 0$.
The function $\mathcal{C}^2 $ obtained from (\ref{comp-pure-global-general}) is an even function of $t$, 
hence its expansion for $t \to 0$ contains only even powers of $t$.
Since $\mathcal{C}|_{t=0} =0$, we have
\be
\label{comp-pure-initialgrowth}
\mathcal{C}^2 
=
b_1 \, t^2 +b_2 \,t^4+b_3 \,t^6+ O(t^8)
\quad \Longrightarrow \quad
\mathcal{C} 
=
\sqrt{b_1} \; t\,
\bigg(1
+ \frac{b_2}{2b_1}\, t^2 
+ \frac{4\,b_1 b_3-b_2^2}{8b_1^2}\, t^4
+O(t^6)\bigg)
\ee
where the coefficients $b_1$, $b_2$ and $b_3$ are
\be
\label{c_12_coeff_exp}
b_1
=
\frac{1}{4}\sum_{k=1}^N 
\left( \frac{ \Omega_{k}^2-  \Omega_{0,k}^2}{ \Omega_{0,k}} \right)^{2}
\;\;\qquad\;\;
b_2
=
-\frac{1}{48}\sum_{k=1}^N 
\left( \frac{ \Omega_{k}^4 -  \Omega_{0,k}^4}{ \Omega_{0,k}^2} \right)^{2}
\ee
and
\be
\label{c_3_coeff_exp}
b_3
=
\frac{1}{360}\sum_{k=1}^N  \frac{ (\Omega_{k}^4-  \Omega_{0,k}^4)^2 (\Omega_{k}^4+  \Omega_{0,k}^4-\Omega_{k}^2 \, \Omega_{0,k}^2)}{ \Omega_{0,k}^6}\,.
\ee
Since $b_1 >0$, the expansion (\ref{comp-pure-initialgrowth}) tells us that
the initial growth of the complexity (\ref{comp-pure-global-general}) is linear in $t$.

The temporal evolution of the circuit complexity for a bosonic system after a global quench 
has been studied also in \cite{Alves:2018qfv},
by employing a smooth quench and 
the unentangled product state 
as the reference state. 
This smooth quench
becomes the one that we are considering in the limit of sudden quench
but it is different from the quench considered in \cite{Camargo:2018eof}.
In appendix\;\ref{app:unentangled},
where the unentangled product state is considered as the initial state, 
we find a different result with respect to \cite{Alves:2018qfv}
because of the different sets of allowed gates.

\subsubsection{Bounds and the zero mode contribution}
\label{subsec:zeromodesboundsgeneral}

We find it worth studying some bounds for the complexity with respect to the initial state.
From (\ref{comp-pure-global-general}),
it is straightforward to observe that $\eta \,c_0  \leqslant  \mathcal{C}^2  \leqslant\widetilde{\mathcal{C}}^2$, 
where $c_0$ is the time dependent expression defined in (\ref{zeromode-def}) and 
\be
\label{upperbound-general-zeromode}
\widetilde{\mathcal{C}}^2
\,\equiv\,
\eta\,c_0+ \sum_{k=1}^{N-1}\! 
\left[ \textrm{arcsinh}\! 
\left( \,
 \frac{\Omega_{k}^2 - \Omega_{0,k}^2}{2\,\Omega_{k}\, \Omega_{0,k}} \, 
 \right)
 \right]^2
\ee
hence for the complexity (\ref{comp-pure-global-general}) we find
\be
\label{naive-bounds}
\sqrt{\eta\,c_0} \, \leqslant \, \mathcal{C} \, \leqslant \, \widetilde{\mathcal{C}}\,.
\ee

The zero mode contribution determines the behaviour of these bounds for large $t$.
\\
The occurrence of a zero mode in the dispersion relation $\Omega_k$
e.g. for $k=N$
means that $\Omega_N =0$.
In the absence of a zero mode,
 $\Omega_k$ is non vanishing for any value of $k$;
 hence $c_0$ and $ \widetilde{\mathcal{C}}$ are finite for any $t$
 and (\ref{naive-bounds}) tells us that
the complexity (\ref{comp-pure-global-general}) is always finite after the quench. 
Instead, when a zero mode for $k=N$ occurs, 
the time dependent zero mode contribution $c_0$ in (\ref{zeromode-def}) becomes
\be
c_0
=
\big[ \textrm{arcsinh} ( \Omega_{0,N} \,t/2 )\big]^2
\ee
which diverges at large $t$ because $\textrm{arcsinh}(x)\sim \log (2x)$ as $x\to+\infty$.
The terms labelled by $1\leqslant k \leqslant N-1$ in the sum in (\ref{upperbound-general-zeromode})
are bounded functions of $t$ because $\Omega_k$ is non vanishing. 
Thus, in the presence of a zero mode, the bounds (\ref{naive-bounds}) tell us that
the complexity for pure states in  (\ref{comp-pure-global-general}) diverges logarithmically when $t \to \infty$.

The bounds (\ref{naive-bounds}) can be significantly improved by employing the decomposition (\ref{eta-c0-decomposition}).
The following integral representation 
\be 
\label{int arcsinh}
\textrm{arcsinh}(x) \,=\int_0^1\frac{x}{\sqrt{1+x^2 s^2}}\,ds
\ee
leads to rewrite $ \mathcal{C}_0^2$ in (\ref{zeromode-def}) as 
\be 
\label{sum over modes int arcsinh}
 \mathcal{C}_0^2
\, =
  \sum_{k=1}^{N-1}
 \Bigg[\int_0^1\! \frac{1}{\sqrt{1+\tilde{x}_k^2  \sin^2 (\Omega_{k} t ) s^2}}\, ds \Bigg]^2
  \tilde{x}_k^2\,  \big[ \sin (\Omega_{k} t )\big]^2
  \;\;\qquad\;\;
   \tilde{x}_k\equiv
\frac{\Omega_k^2 - \Omega_{0,k}^2}{2\,\Omega_{k}\, \Omega_{0,k}}\,.
\ee
Then, by using (\ref{int arcsinh}), one observes that
\be
\frac{\textrm{arcsinh}(\tilde{x}_k)}{\tilde{x}_k} 
\,\leqslant 
\int_0^1\! \frac{1}{\sqrt{1+\tilde{x}_k^2  \sin^2 (\Omega_{k} t ) s^2}}\,ds
\, \leqslant \,1
\ee
which can be employed to bound (\ref{sum over modes int arcsinh}) as follows
\be
 \sum_{k=1}^{N-1}
  \big[ \textrm{arcsinh}(\tilde{x}_k) \,\sin(\Omega_{k} t ) \big]^2
\, \leqslant\,
 \mathcal{C}_0^2
\,\leqslant 
 \sum_{k=1}^{N-1}
  \tilde{x}_k^2 \, \big[\sin(\Omega_{k} t )\big]^2\,.
\ee

This result, combined with (\ref{zeromode-def}), provides the following bounds
 for the complexity (\ref{comp-pure-global-general})
\be
\label{new bounds_main}
\mathcal{C}^2_{\textrm{\tiny L}} \, \leqslant \,\mathcal{C}^2\,\leqslant \,\mathcal{C}^2_{\textrm{\tiny U}}
\ee 
where we have introduced 
\be
\label{new bounds C-L-U}
\mathcal{C}^2_{\textrm{\tiny B}} 
\,\equiv\, 
\eta \,c_0
+\! \sum_{k=1}^{N-1} f_{\textrm{\tiny B}}(\tilde{x}_k) \big[ \sin (\Omega_{k} t ) \big]^2
=
\Bigg(
\eta \,c_0
+
\frac{1}{2} \sum_{k=1}^{N-1} f_{\textrm{\tiny B}}(\tilde{x}_k)
\Bigg)
- 
\frac{1}{2} \sum_{k=1}^{N-1}
f_{\textrm{\tiny B}}(\tilde{x}_k) \cos(2\Omega_{k}\, t ) 
\ee
with $\textrm{B} \in \{ \textrm{L} , \textrm{U} \}$ and 
\be
\label{fLs-def}
f_{\textrm{\tiny L}}(x) = \big[\textrm{arcsinh}(x) \big]^2
\;\;\qquad\;\;
f_{\textrm{\tiny U}}(x) = x^2
\ee
in terms of $\tilde{x}_k$ defined in (\ref{sum over modes int arcsinh}), 
of the time dependent zero mode contribution $c_0$ introduced in (\ref{zeromode-def})
and of the parameter $\eta$, 
which is either $\eta=1$ or $\eta=0$, depending on whether the zero mode contribution occurs or not respectively.

The bounds (\ref{new bounds_main}) can be employed to improve 
the bounds reported in (\ref{naive-bounds}).
Indeed, in the presence of a zero mode,
$\mathcal{C}^2_{\textrm{\tiny L}}\geqslant c_0$ and therefore $\mathcal{C}^2_{\textrm{\tiny L}}$ provides a better lower bound than (\ref{zeromode-def}). 
Instead,
the relation between $\mathcal{C}^2_{\textrm{\tiny U}}$ in (\ref{new bounds C-L-U})
and $\widetilde{\mathcal{C}}^2$ in (\ref{upperbound-general-zeromode}) depends on the parameters; 
hence the optimal upper bound is given by 
$\min\!\big[\mathcal{C}_{\textrm{\tiny U}}(t)^2,\widetilde{\mathcal{C}}(t)^2\big]$.

\section{Complexity for harmonic chains}
\label{sec:purestates_HC_glob}

In this section we apply the results discussed in Sec.\,\ref{sec:cov-mat}
to the harmonic chains where either PBC or DBC are imposed. 
The numerical data reported in all the figures of the manuscript have been obtained 
by setting $\kappa=1$ and $m=1$.

\subsection{Complexity}
\label{subsec:initial HC}

The Hamiltonian of the harmonic chain made by $N$ oscillators
with the same frequency $\omega$, the same mass $m$ and coupled through the 
elastic constant $\kappa$ is (\ref{HC ham}) specialised to one spatial dimension,
i.e.
\be
\label{HC ham-1d}
\widehat{H} 
\,=\, 
\sum_{i=1}^{N} \left(\,
\frac{1}{2m}\,\hat{p}_i^2+\frac{m\omega^2}{2}\,\hat{q}_i^2 
+ \frac{\kappa}{2}(\hat{q}_{i} -\hat{q}_{i-1})^2
\right)
\,=\,
 \frac{1}{2}\, \hat{\boldsymbol{r}}^{\textrm t} H^{\textrm{\tiny phys}} \, \hat{\boldsymbol{r}}
\ee
where the vector
$\hat{\boldsymbol{r}} \equiv (\hat{q}_1 , \dots , \hat{q}_N, \hat{p}_1, \dots, \hat{p}_N)^{\textrm{t}}$
collects the position and momentum operators. 
Imposing PBC means that $\hat{q}_0=\hat{q}_N $, 
while DBC are satisfied when 
$\hat{q}_0=\hat{q}_N=0$ and $\hat{p}_N=0$.

When PBC hold, 
the orthogonal matrix $\widetilde{V}$ defined in (\ref{time dep corrs trans}),
when $N$ is even, is \cite{Serafini17book}
\be
\label{Vtilde-def-even}
\widetilde{V}_{i,k} \equiv
\left\{\begin{array}{ll}
\sqrt{2/N}\; \cos(2\pi \,i\,k/N)  \hspace{1cm}&   1\leqslant k < N/2
\\
\rule{0pt}{.5cm}
(-1)^i/\sqrt{N} &    k = N/2
\\
\rule{0pt}{.5cm}
\sqrt{2/N}\; \sin(2\pi\, i\,k/N)  &   N/2+1 \leqslant k < N-1
\\
\rule{0pt}{.5cm}
1/\sqrt{N} &    k = N
\end{array}
\right.
\ee
while, when $N$ is odd, it reads
\be
\label{Vtilde-def-odd}
\widetilde{V}_{i,k} \equiv
\left\{\begin{array}{ll}
\sqrt{2/N}\; \cos(2\pi\, i\,k/N)  &   1\leqslant k < (N-1)/2
\\
\rule{0pt}{.5cm}
\sqrt{2/N}\; \sin(2\pi \,i\,k/N)  \hspace{1cm}&   (N-1)/2+1 \leqslant k < N-1
\\
\rule{0pt}{.5cm}
1/\sqrt{N} &    k = N\,.
\end{array}
\right.
\ee
The dispersion relations of $\hat{H}_0$ and $\hat{H}$ for PBC are respectively 
\be
\label{dispersion relations}
\Omega_{0,k} = \sqrt{\omega_0^2 + \frac{4\kappa}{m} [ \sin(\pi k/N) ]^2}
\,\,\qquad\,\,
\Omega_k = \sqrt{\omega^2 + \frac{4\kappa}{m} [ \sin(\pi k/N)]^2}
\,\,\qquad\,\,
1 \leqslant k \leqslant N\,.
\ee

When DBC hold, only $N-1$ sites display some dynamics
because the ones labelled by $i=0$ and $i=N$ are fixed by the boundary conditions;
hence the vector $\hat{\boldsymbol{r}}$ contains $2(N-1)$ operators and,
correspondingly, the covariance matrix $\gamma(t)$ 
is the $(2N-2)\times (2N-2)$ symmetric matrix given by (\ref{covariancematrix_t}),
where $Q$, $P$ and $R$ are $(N-1)\times (N-1)$ matrices. 
For DBC and independently of the parity of $N$,
the matrix $\widetilde{V}$ defined in (\ref{time dep corrs trans}) becomes
\be
\label{Vtilde-HC-DBC}
\widetilde{V}_{i,k}=\sqrt{\frac{2}{N}}\, \sin(i\,k\,\pi/N)
\,\,\qquad\,\,
1 \leqslant i,k \leqslant N-1\,.
\ee
The dispersion relations of $\hat{H}_0$ and $\hat{H}$ for DBC read respectively
\be
\label{dispersion DBC}
\Omega_{0,k}=\sqrt{\omega_0^2+\frac{4\kappa}{m}\, [ \sin(\pi k/(2N))]^2}
\qquad
\Omega_k=\sqrt{\omega^2+\frac{4\kappa}{m}\,  [ \sin(\pi k/(2N))]^2}
\,\qquad\,
1 \leqslant k \leqslant N-1\,.
\ee

We remark that, both for PBC and DBC,
the matrix $V = \widetilde{V} \oplus \widetilde{V}$ defined in (\ref{VGammaV dec}) 
depends only on $N$;
hence the corresponding harmonic chains can be studied as special cases of the 
harmonic lattices considered in Sec.\,\ref{sec-pure-states}
because the condition (\ref{VV-diag-hyp}) is satisfied. 
Since $\eta=1$ for PBC and $\eta=0$ for DBC,
 the complexity (\ref{comp-pure-global-general}) for these harmonic chains becomes
\bea
\label{comp-pure-global-DBCPBC}
\mathcal{C}
&=&
\sqrt{ \sum_{k=1}^{N-1+\eta}\! 
\left[ \textrm{arcsinh}\! 
\left( \,
 \frac{\omega^2 - \omega_0^2}{2\,\Omega_{k}\, \Omega_{0,k}} \,
 \sin (\Omega_{k} t ) 
 \right)
 \right]^2}
 \\
 \rule{0pt}{1.1cm}
 &=&
 \sqrt{ \eta\left[ \textrm{arcsinh}\! 
\left( \,
 \frac{\omega^2 - \omega_0^2}{2\,\omega\, \omega_0} \,
 \sin (\omega t ) 
 \right)
 \right]^2+ \sum_{k=1}^{N-1}\! 
\left[ \textrm{arcsinh}\! 
\left( \,
 \frac{\omega^2 - \omega_0^2}{2\,\Omega_{k}\, \Omega_{0,k}} \,
 \sin (\Omega_{k} t ) 
 \right)
 \right]^2}
 \nonumber
\eea
where the dispersion relations $\Omega_{0,k}$ and $\Omega_{k}$
are given by (\ref{dispersion relations}) for PBC
and by (\ref{dispersion DBC}) for DBC.

When PBC are imposed,
the first term under the square root in the last expression of (\ref{comp-pure-global-DBCPBC}) 
comes from the zero mode $k=N$ and it does not occur for DBC.
This crucial difference between the two models leads to different 
qualitative behaviours for the complexity.

The dispersion relations of the harmonic chain with PBC given in (\ref{dispersion relations})  
are invariant under the exchange $k \leftrightarrow N-k$.
This symmetry leads to an expression for the complexity which is simpler to evaluate numerically. 
Indeed, by introducing 
\be
\label{low-bound}
c_{0}
\equiv 
\left[ \textrm{arcsinh}\! 
\left( \,
 \frac{\omega^2 - \omega_0^2}{2\,\omega\, \omega_{0}} \,
 \sin (\omega  t ) 
 \right)
 \right]^2
\ee
and
\be
\label{c-Nover2-def}
c_{N/2} \equiv
\left\{ \begin{array}{l l}
\displaystyle 
\left[ \textrm{arcsinh}\! 
\left( \,
 \frac{\omega^2 - \omega_0^2}{2\,\Omega_{N/2}\, \Omega_{0,N/2}} \,
 \sin (\Omega_{N/2}\, t ) 
 \right)
 \right]^2
\hspace{1cm}&
\textrm{even $N$}
\\
\rule{0pt}{.7cm}
\;0
&
\textrm{odd $N$}
\end{array}
\right.
\ee
one observes that (\ref{comp-pure-global-DBCPBC}) for PBC can be written as
\be
\label{comp-pure-global-zm}
\mathcal{C} 
\,=\,
\sqrt{\;
c_{0}
+
2 \sum_{k=1}^{ \lfloor \frac{N-1}{2} \rfloor}\! 
\left[ \textrm{arcsinh}\! 
\left( \,
 \frac{\omega^2 - \omega_0^2}{2\,\Omega_{k}\, \Omega_{0,k}} \,
 \sin (\Omega_{k} t ) 
 \right)
 \right]^2
 + c_{N/2}
 }
 \ee
where $\lfloor x \rfloor$ denotes the integer part of $x$.
Notice that $c_{N/2}$ in (\ref{c-Nover2-def}),  as function of $t$,
is bounded by a constant.

We find it worth considering the small quench regime, 
defined by setting $\omega_0=\omega+\delta\omega$ 
and taking $|\delta\omega| \ll 1$ 
in (\ref{comp-pure-global-DBCPBC}).
As $\delta\omega \to 0$, the leading term of the expansion reads
\be
\label{comp-pure-global-smallquench}
\mathcal{C}
\,=\,
\omega\,\delta\omega\;
\sqrt{\eta\;\frac{[\sin(\omega t)]^2}{\omega^4}+\sum_{k=1}^{N-1}\frac{[\sin(\Omega_k t)]^2}{\Omega_k^4}}+O\big(\delta\omega^2\big)
\ee
This result simplifies to
$\mathcal{C} = \eta \,\delta \omega \, t + O(\delta\omega^2)$ 
when $\omega \to 0$;
which tells us that the $O(\delta\omega)$ term does not occur in this limit 
when DBC hold.

\subsection{Critical evolution}
\label{subsec:criticalevolution}

An important case that we find worth emphasising is the global quench 
where the evolution Hamiltonian is gapless, i.e. when $\omega=0$.

When PBC are imposed, 
by specialising (\ref{CTR}) and (\ref{dispersion relations}) to $\omega=0$, we obtain
\be
\label{CTRk_HC massless}
C_{\textrm{\tiny TR},k}
\,=\,
1 +
\frac{\omega_0^4\, \big[ \sin\!\big(2\sqrt{\kappa/m}\; t \, \sin(\pi k/N)\big) \big]^2}{
8\, (\kappa/m)\, [\sin(\pi k/N)]^2 \,\big( \omega_0^2+4 (\kappa/m) [\sin(\pi k/N)]^2\big)}
\ee
which satisfies the following bounds
\be
1 \,<\, 
C_{\textrm{\tiny TR},k} 
\,<\,
1 +\frac{\omega_0^4}{8\, (\kappa/m)\, [\sin(\pi k/N)]^2 \,
\big( \omega_0^2+4 (\kappa/m) [\sin(\pi k/N)]^2\big)}
\;\qquad\;
1\leqslant k \leqslant N-1\,.
\ee
For $k=N$, the expression (\ref{CTRk_HC massless}) simplifies to
$C_{\textrm{\tiny TR},N}=1+\tfrac{\omega_0^2}{2}\, t^2$,
which diverges as $t \to \infty$.

Instead, when DBC hold and therefore the zero mode does not occur,
by using (\ref{dispersion DBC}) and (\ref{CTR}) with $\omega=0$, 
we obtain 
\be
\label{CTRk_HC massless DBC}
C_{\textrm{\tiny TR},k}
\,=\,
1 +\frac{\omega_0^4\, \big[ \sin\!\big(2\sqrt{\kappa/m}\,t \,\sin(\pi k/(2N))\big) \big]^2}{
8\, (\kappa/m) [\sin(\pi k/(2N))]^2 \,\big( \omega_0^2+4 \kappa/m [\sin(\pi k/(2N))]^2\big)}
\ee
which is finite when $t \to \infty$,  for any allowed value of $k$.

Plugging the expressions discussed above for $C_{\textrm{\tiny TR},k}$ into (\ref{c2-log-lambda-arcosh}),
we find that, when the evolution Hamiltonian is critical,
the complexity of the pure state at time $t$ with respect to the initial state can be written 
by highlighting the zero mode contribution as follows
\be
\label{C-pure-both-eta}
\mathcal{C}^2 
\,=\, 
 \frac{\eta}{4}\left[ \,
\log\!\left(1+\frac{(\omega_0 \,t)^2}{2} + \frac{\omega_0 \,t}{2} \,\sqrt{(\omega_0 \,t)^2+4}\,\right) 
\right]^2
+
\frac{1}{4} \sum_{k=1}^{N-1}\! \big[ \textrm{arccosh}\big(C_{\textrm{\tiny TR},k}\big)\big]^2
\ee
where either $\eta = 1$ for PBC or $\eta = 0$ for DBC
(see the text above (\ref{comp-pure-global-DBCPBC})) 
and $C_{\textrm{\tiny TR},k}$ is given by 
(\ref{CTRk_HC massless}) for PBC and by (\ref{CTRk_HC massless DBC}) for DBC.
In particular, (\ref{C-pure-both-eta}) tells us that, for PBC and finite $N$,
the complexity diverges logarithmically as $t  \to \infty$ because of the zero mode contribution.
Instead, for DBC (i.e. $\eta=0$) and finite $N$, all the terms in (\ref{C-pure-both-eta}) are finite
as $t\to  \infty$.

\begin{figure}[t!]
\subfigure
{\hspace{-1.55cm}
\includegraphics[width=.57\textwidth]{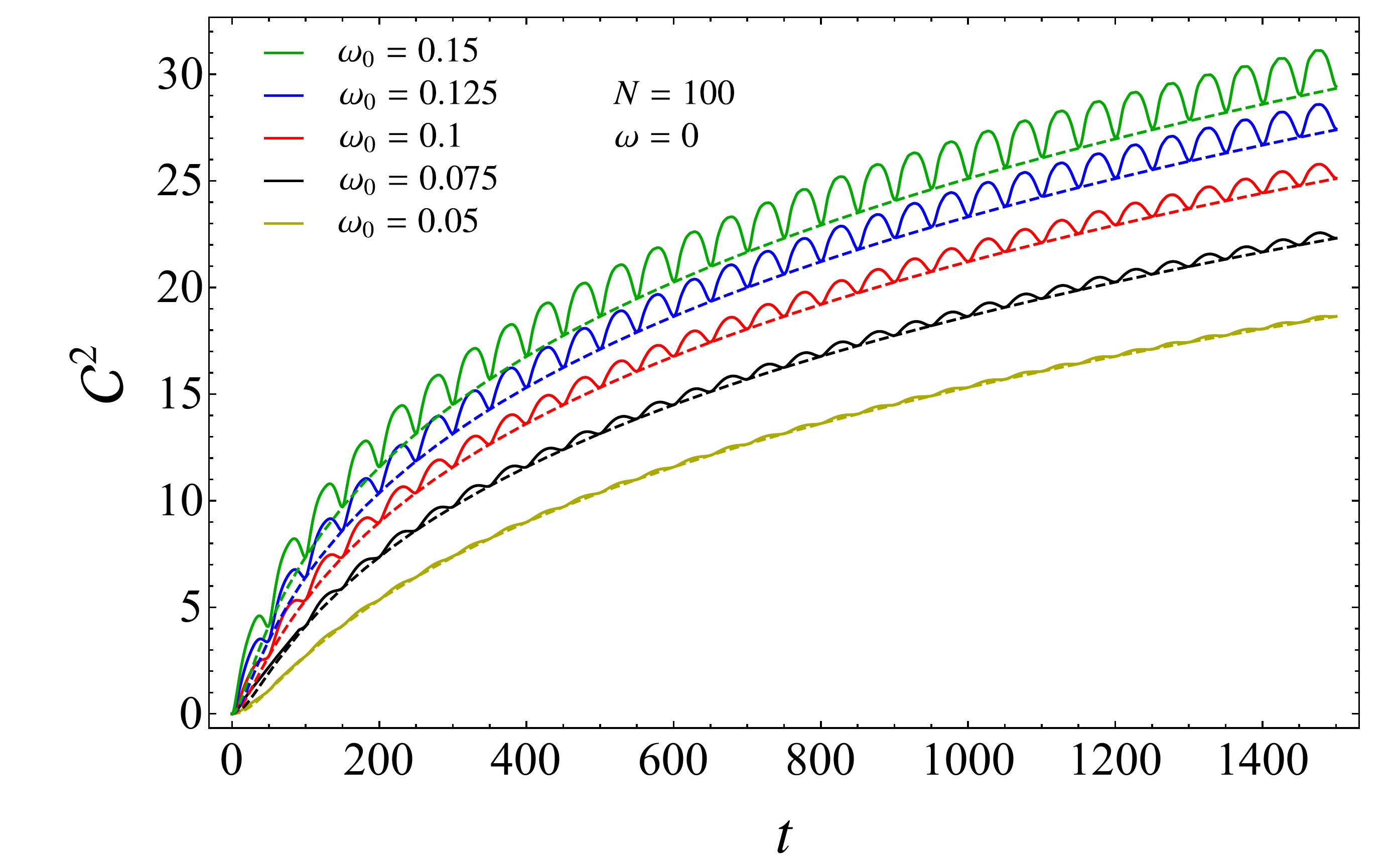}}
\subfigure
{
\hspace{-0.35cm}\includegraphics[width=.57\textwidth]{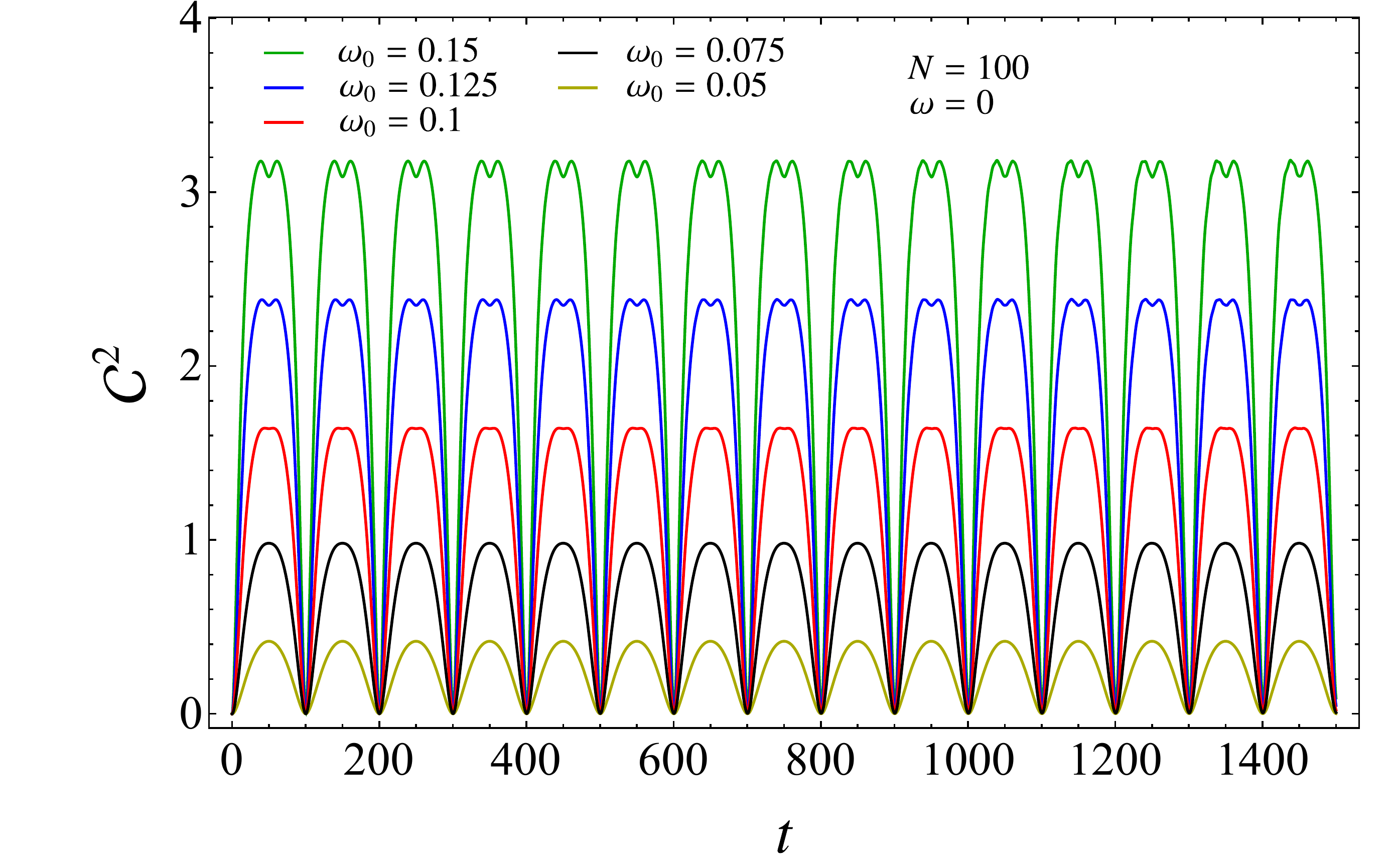}}
\subfigure
{
\hspace{-1.55cm}\includegraphics[width=.57\textwidth]{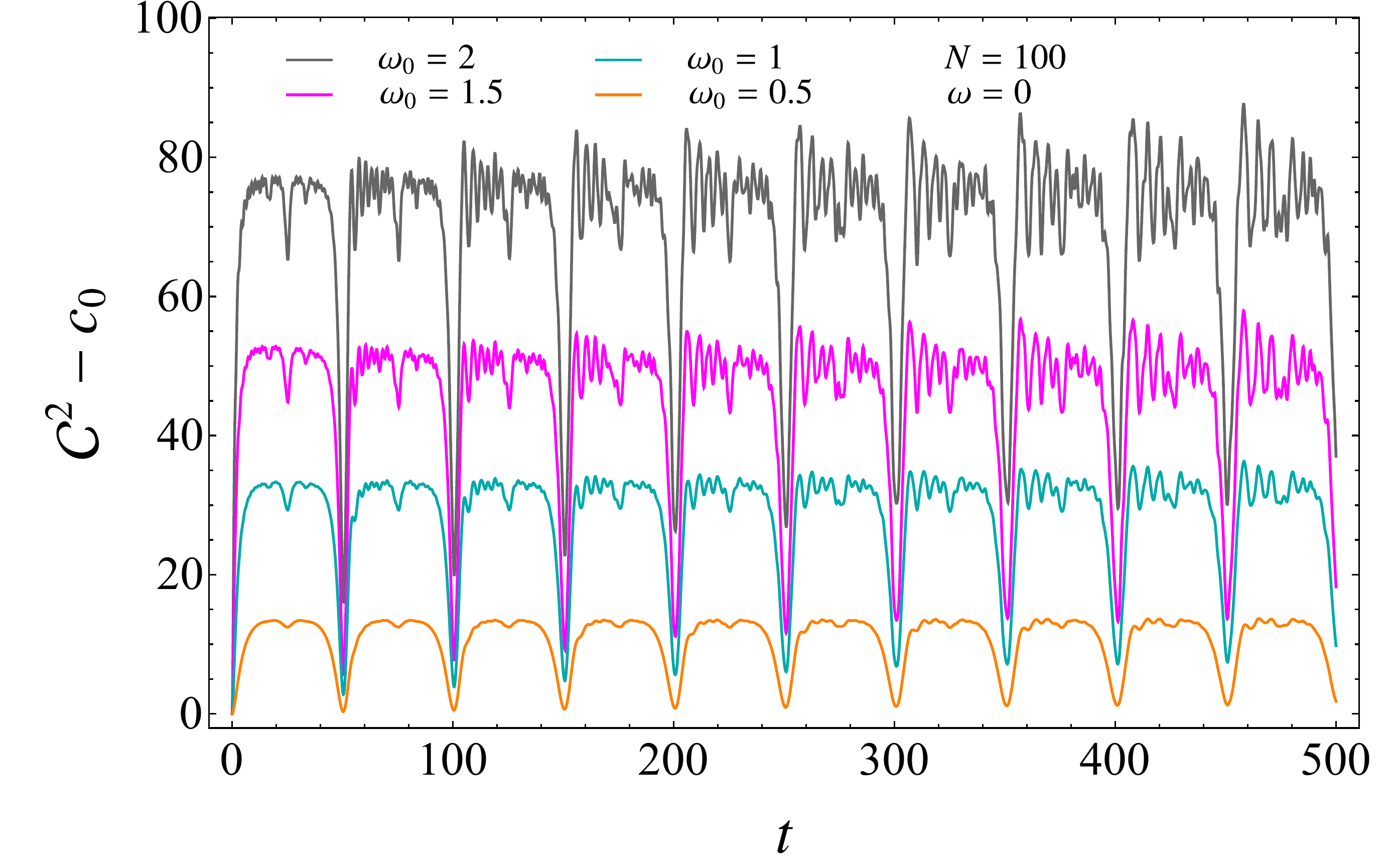}}
\subfigure
{\hspace{-0.35cm}
\includegraphics[width=.57\textwidth]{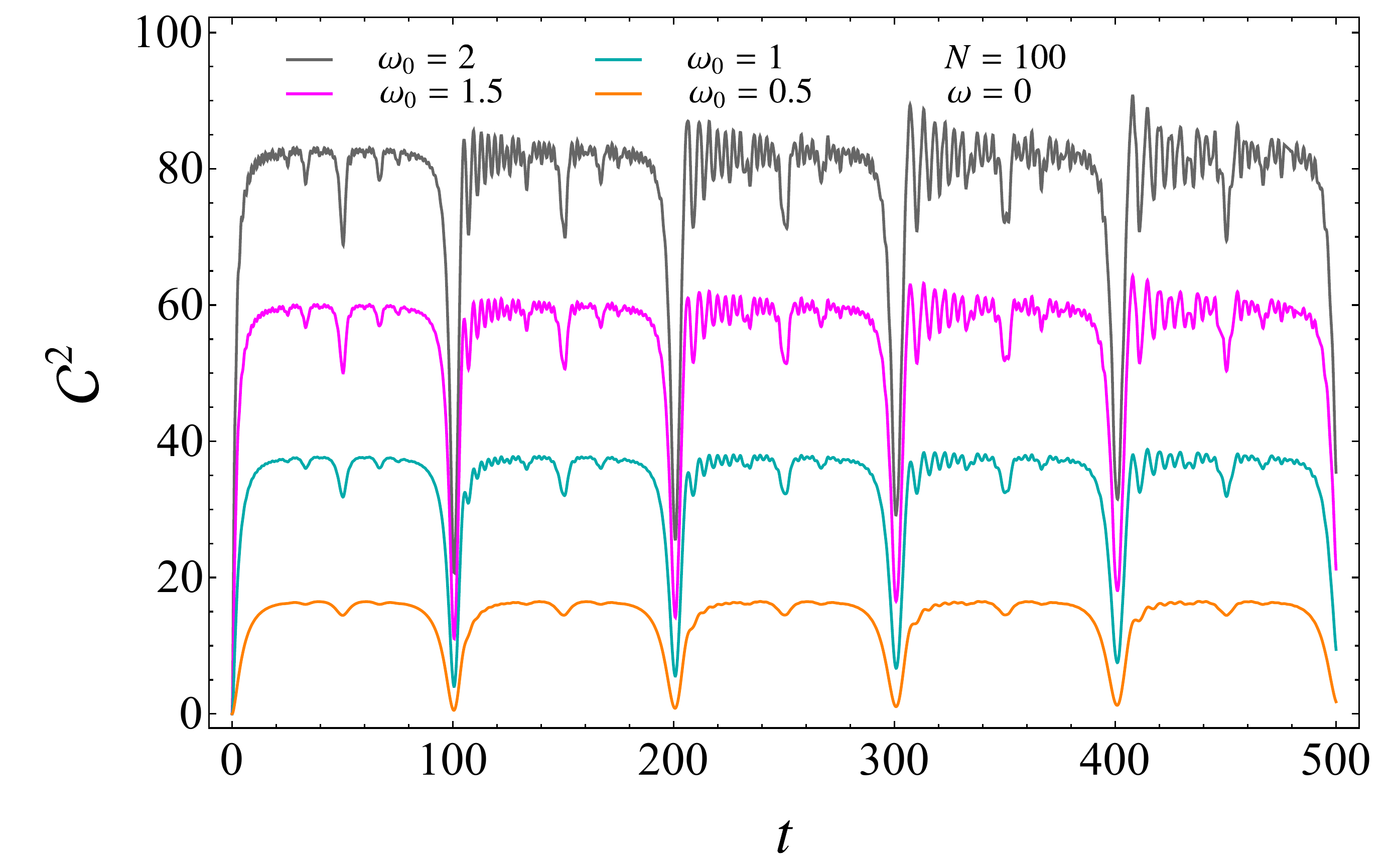}}
\caption{Temporal evolution of the complexity 
after the global quench  w.r.t. the initial state at $t=0$
for harmonic chains 
with either PBC (left panels) or DBC (right panels)
made by $N=100$ sites.
The solid lines correspond to the complexity (\ref{C-pure-both-eta}).
In the top left panel, the dashed lines show the zero mode term $c_0$
(i.e. the expression multiplyed by $\eta$ in (\ref{C-pure-both-eta}),
which has been subtracted to obtain the bottom left panel),
with the same colour code for the corresponding value of $\omega_0$.
}
\vspace{0.4cm}
\label{fig:PureStateCritical}
\end{figure}

In Fig.\,\ref{fig:PureStateCritical} we show the temporal evolution of the complexity
(\ref{C-pure-both-eta}) for various $\omega_0$'s,
when either PBC (left panels) or DBC (right panels) are imposed. 
Since $N$ is finite, the revivals 
already studied in the temporal evolutions of other quantities \cite{Cardy:2014rqa}
are observed also in the temporal evolution of the complexity,
with a period given by $N/2$ for PBC and by $N$ for DBC.
The most important qualitative difference between PBC and DBC is
the overall growth observed for PBC, which does not occur for DBC. 
This growth is due to the zero mode contribution occurring in 
the complexity (\ref{C-pure-both-eta}) for PBC.
Indeed, when the corresponding term 
is subtracted, as done in the 
bottom left panel of Fig.\,\ref{fig:PureStateCritical}, 
the resulting curve is similar to the temporal evolution of the complexity when DBC hold. 

Finally, let us remark that
the effect of the decoherence as $t$ increases is more evident for higher values of $\omega_0$.
For PBC this is observed once the zero mode contribution has been subtracted
(see the bottom left panel of Fig.\,\ref{fig:PureStateCritical}).

In \cite{Chapman:2018hou} the temporal evolution of the complexity 
of a thermofield double state is considered
by taking the unentangled product state as the reference state
(in this setup, the choice $\omega =0$ is not allowed).
Despite this temporal evolution is different from the one investigated in this manuscript, 
it also exhibits an overall logarithmic growth due to the zero mode contribution.

\subsection{Bounds}
\label{subsec:bounds}

\begin{figure}[t!]
\subfigure
{
\hspace{3.0cm}\includegraphics[width=.57\textwidth]{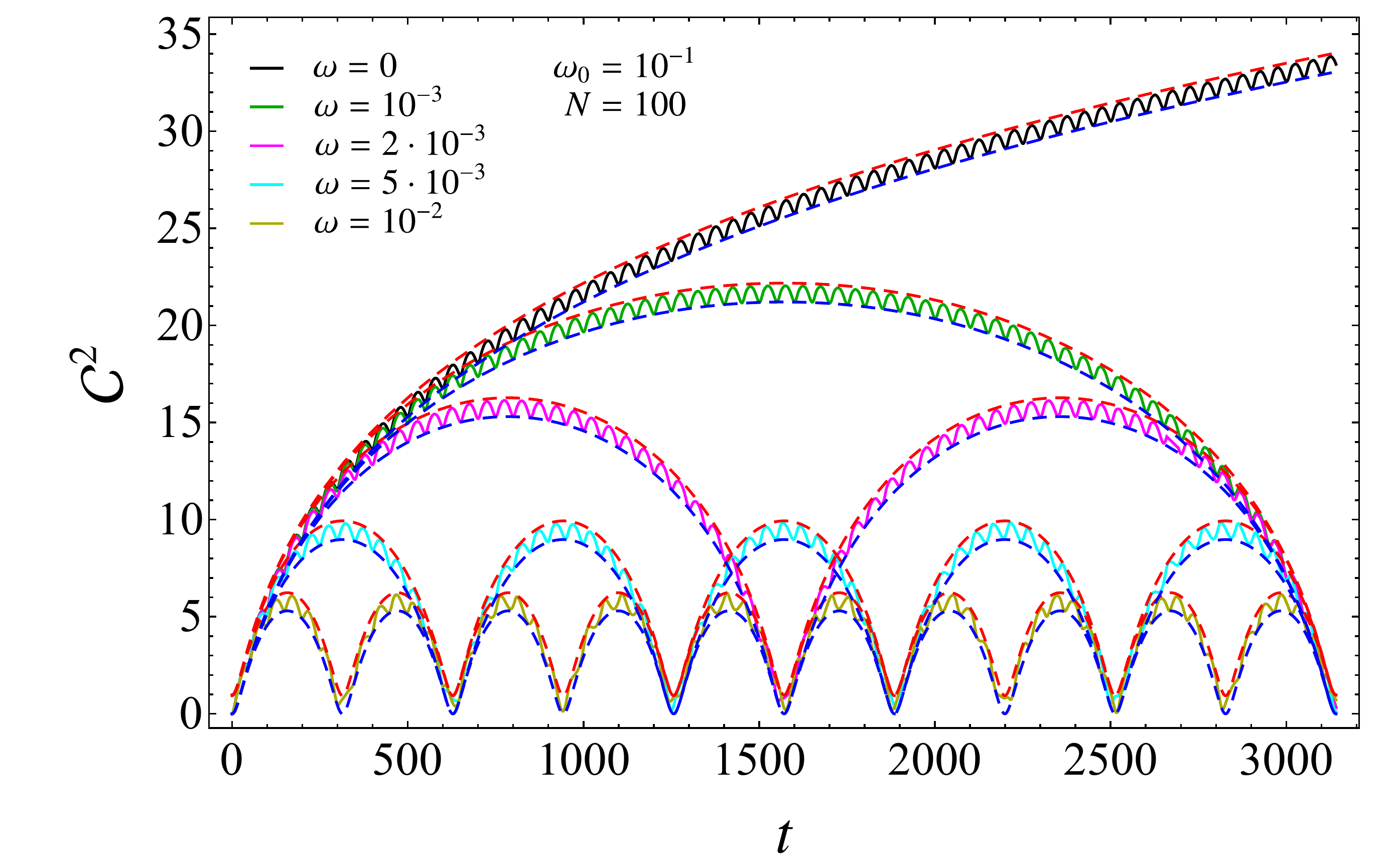}}
\\
\subfigure
{
\hspace{-1.55cm}\includegraphics[width=.57\textwidth]{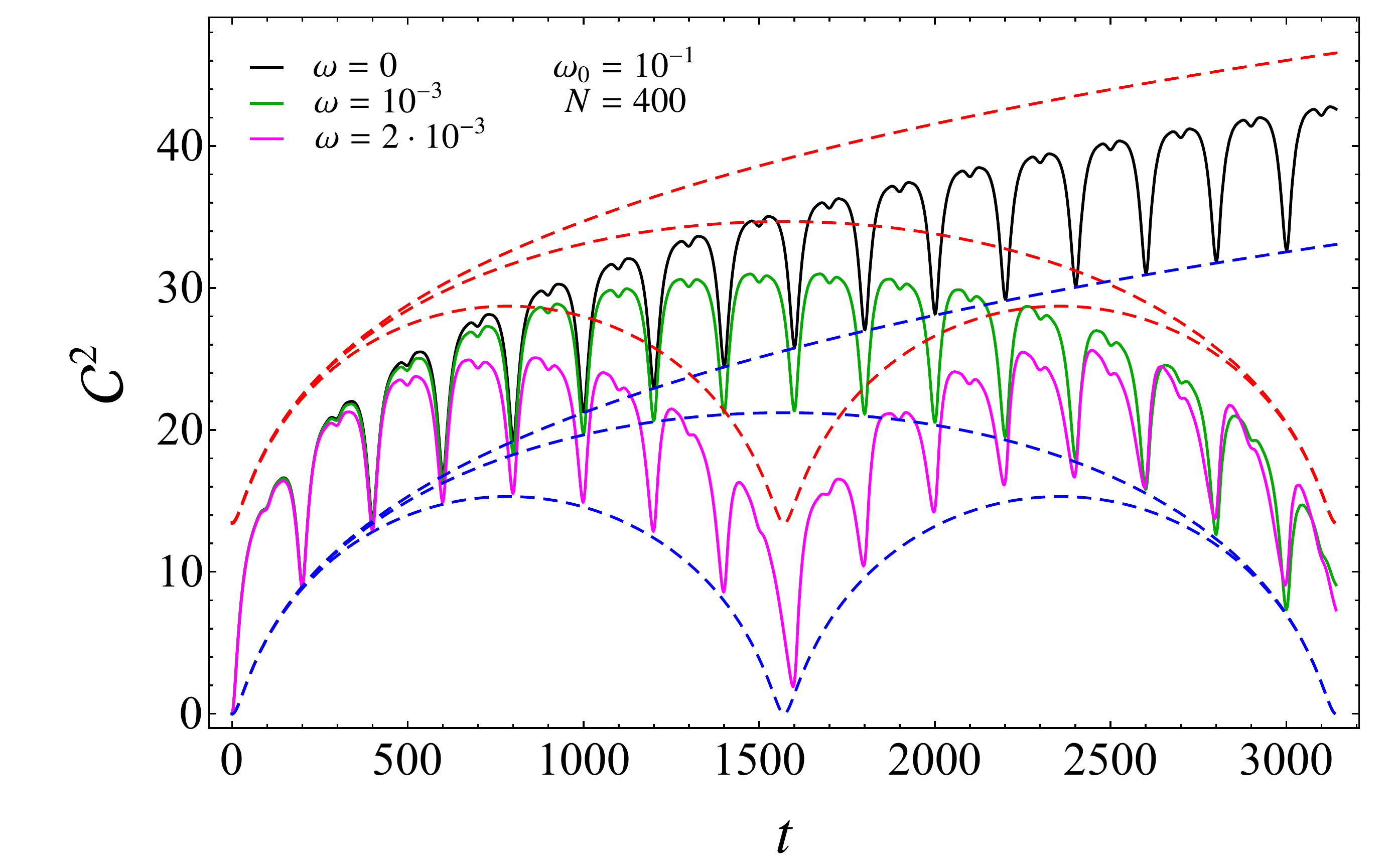}}
\subfigure
{\hspace{-0.45cm}
\includegraphics[width=.57\textwidth]{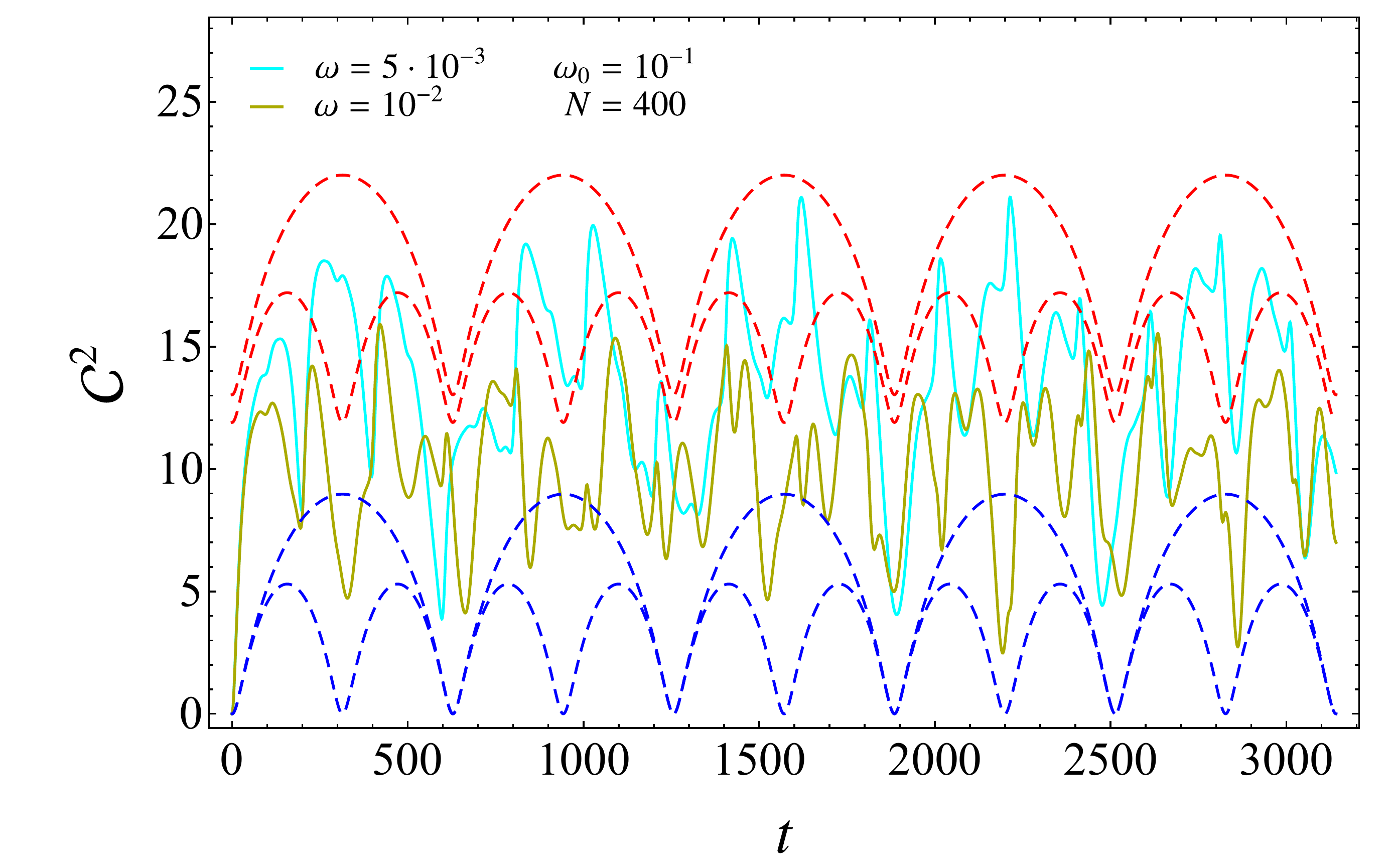}}
\caption{
Temporal evolution of the complexity (\ref{comp-pure-global-DBCPBC}) (solid lines)
and of the corresponding bounds in (\ref{naive-bounds}) for PBC.
%
The blue and red dashed lines show the lower and the upper bounds, 
from  (\ref{zeromode-def}) and (\ref{upperbound-general-zeromode}) respectively.
} 
\vspace{0.4cm}
\label{fig:BoundNaive}
\end{figure}

It is instructive to discuss further the bounds for the complexity 
introduced in Sec.\,\ref{subsec:zeromodesboundsgeneral}
in the special cases of the harmonic chains  with either PBC or DBC.

In Fig.\,\ref{fig:BoundNaive} we show the complexity (\ref{comp-pure-global-DBCPBC}) 
and the corresponding bounds (\ref{naive-bounds}) for harmonic chains with PBC.
In this case the zero mode term influences the bounds in a crucial way. 
In Fig.\,\ref{fig:BoundNaive},
the bounds (\ref{naive-bounds})
correspond to the red and blue dashed lines,
while in the top left panel of Fig.\,\ref{fig:PureStateCritical}, where $\omega =0$,
the lower bound in (\ref{naive-bounds}) is shown through the dashed curves.

In the temporal evolutions of the complexity for PBC 
displayed in the top panel of Fig.\,\ref{fig:BoundNaive},
we can identify two periods approximatively given by $\pi /\omega$ and $N/2$.
Considering also the bottom panels of Fig.\,\ref{fig:BoundNaive},
the revivals observed 
for the critical evolution in Fig.\,\ref{fig:PureStateCritical} for PBC and $\omega =0$
occur also when $\omega >0$ whenever $\frac{\pi}{\omega}\gg \frac{N}{2}$.
The bottom panels in Fig.\,\ref{fig:BoundNaive} highlight that
the revivals are not observed when $\omega$ is large enough with respect to $1/N$.

For PBC, by comparing the top panel with the  bottom ones in Fig.\,\ref{fig:BoundNaive},
which differ for the size $N$ of the chain,
we notice  that  the bounds (\ref{naive-bounds}) are very efficient when 
$\frac{\pi}{\omega}\gg \frac{N}{2}$, while they become not useful away from this regime. 
In our numerical investigations we have also observed that the bounds (\ref{naive-bounds})
are not useful when $\omega>\omega_0$.

When DBC hold, 
the lower bound in (\ref{naive-bounds}) is trivial
and the upper bound is a constant.

\begin{figure}[t!]
\subfigure
{\hspace{-1.55cm}
\includegraphics[width=.57\textwidth]{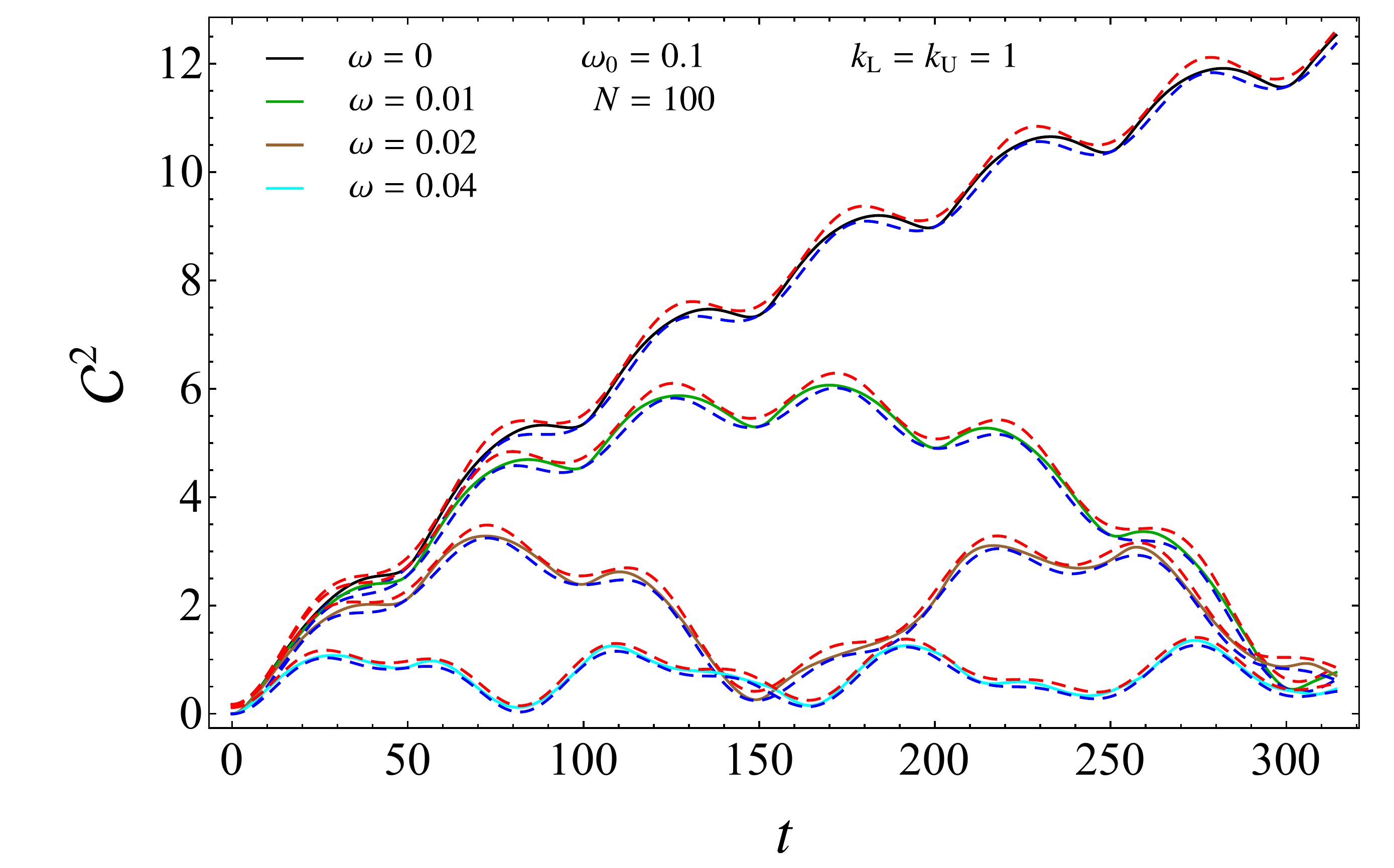}}
\subfigure
{
\hspace{-.45cm}\includegraphics[width=.57\textwidth]{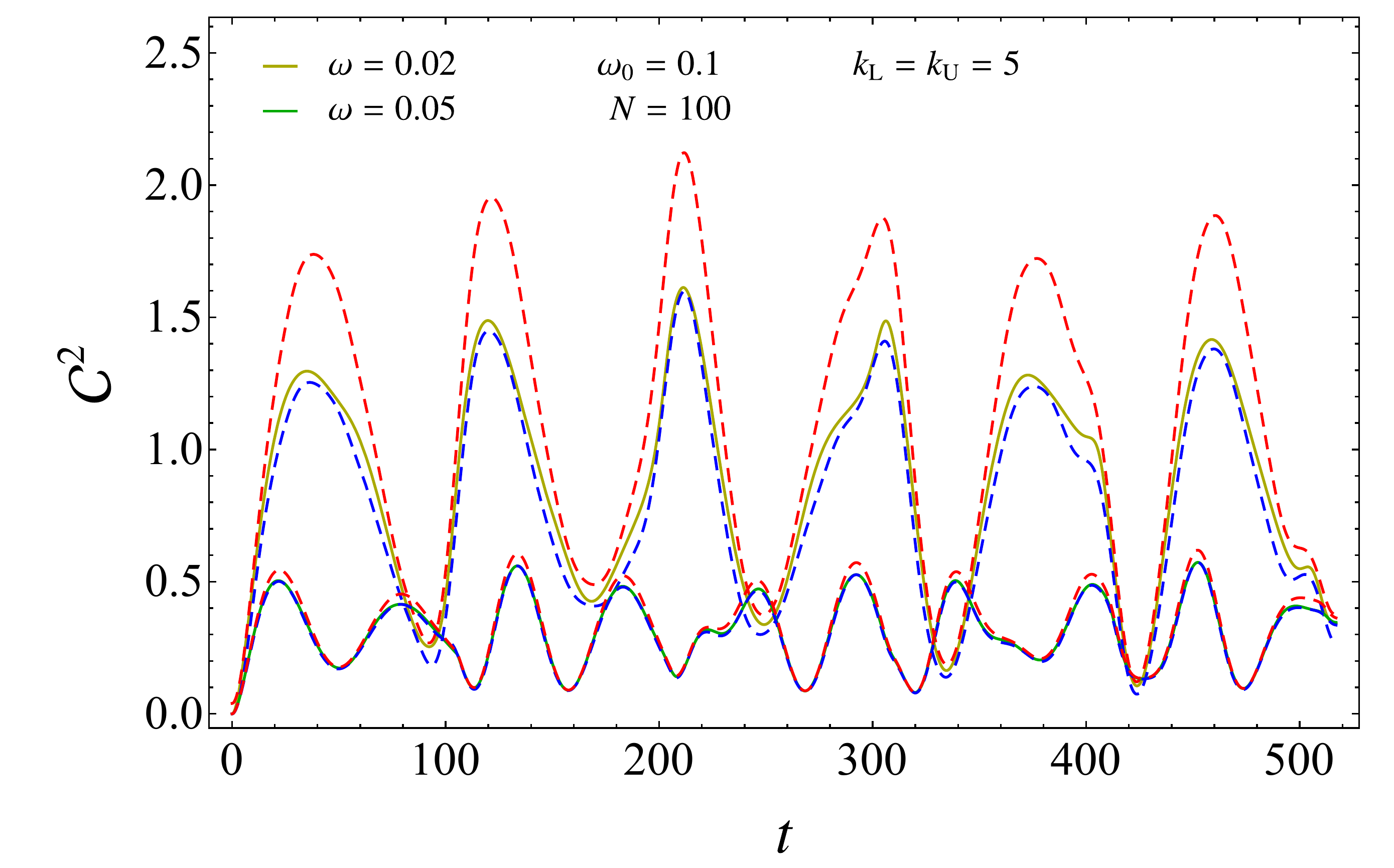}}
\subfigure
{
\hspace{-1.55cm}\includegraphics[width=.57\textwidth]{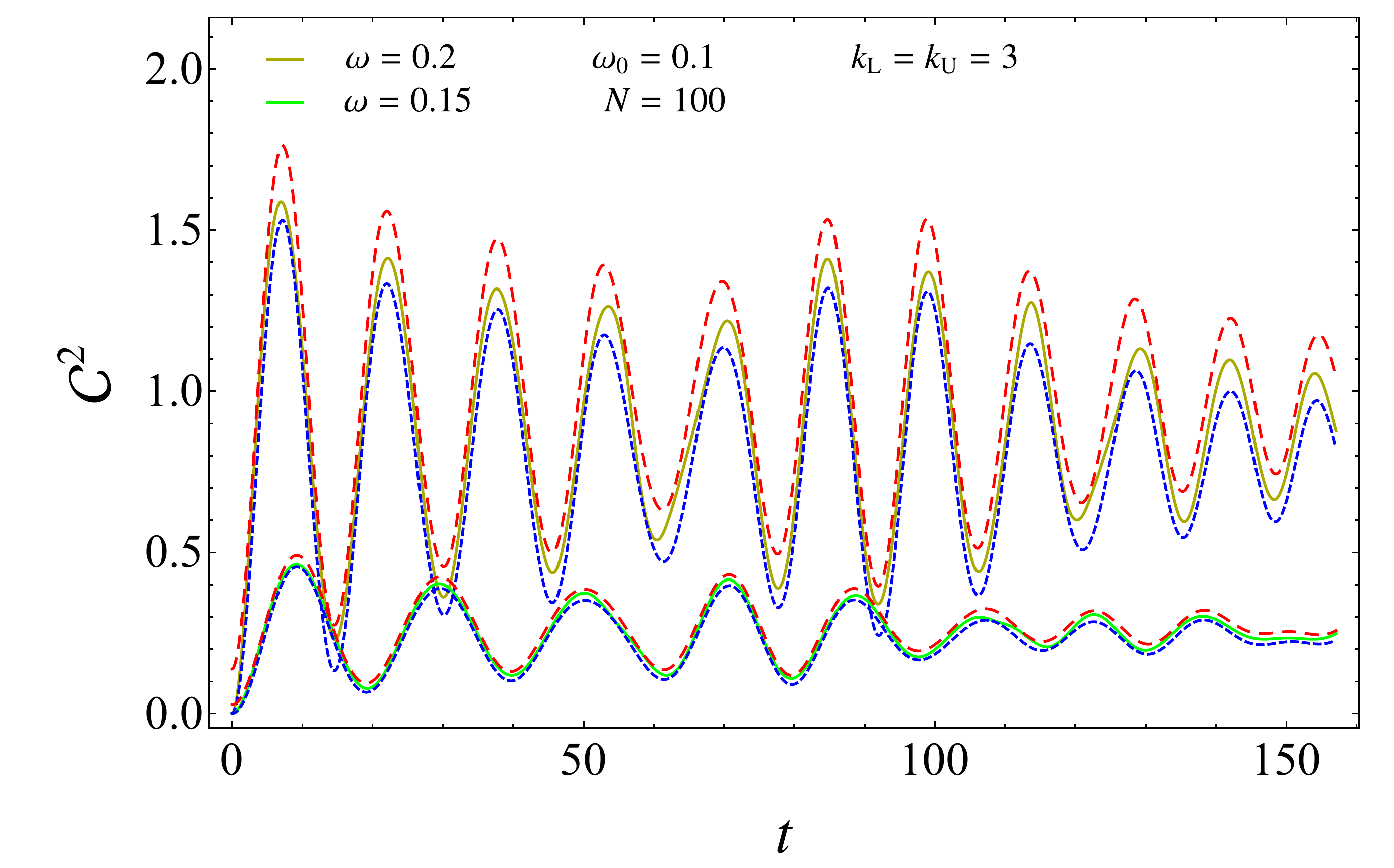}}
\subfigure
{\hspace{-.45cm}
\includegraphics[width=.57\textwidth]{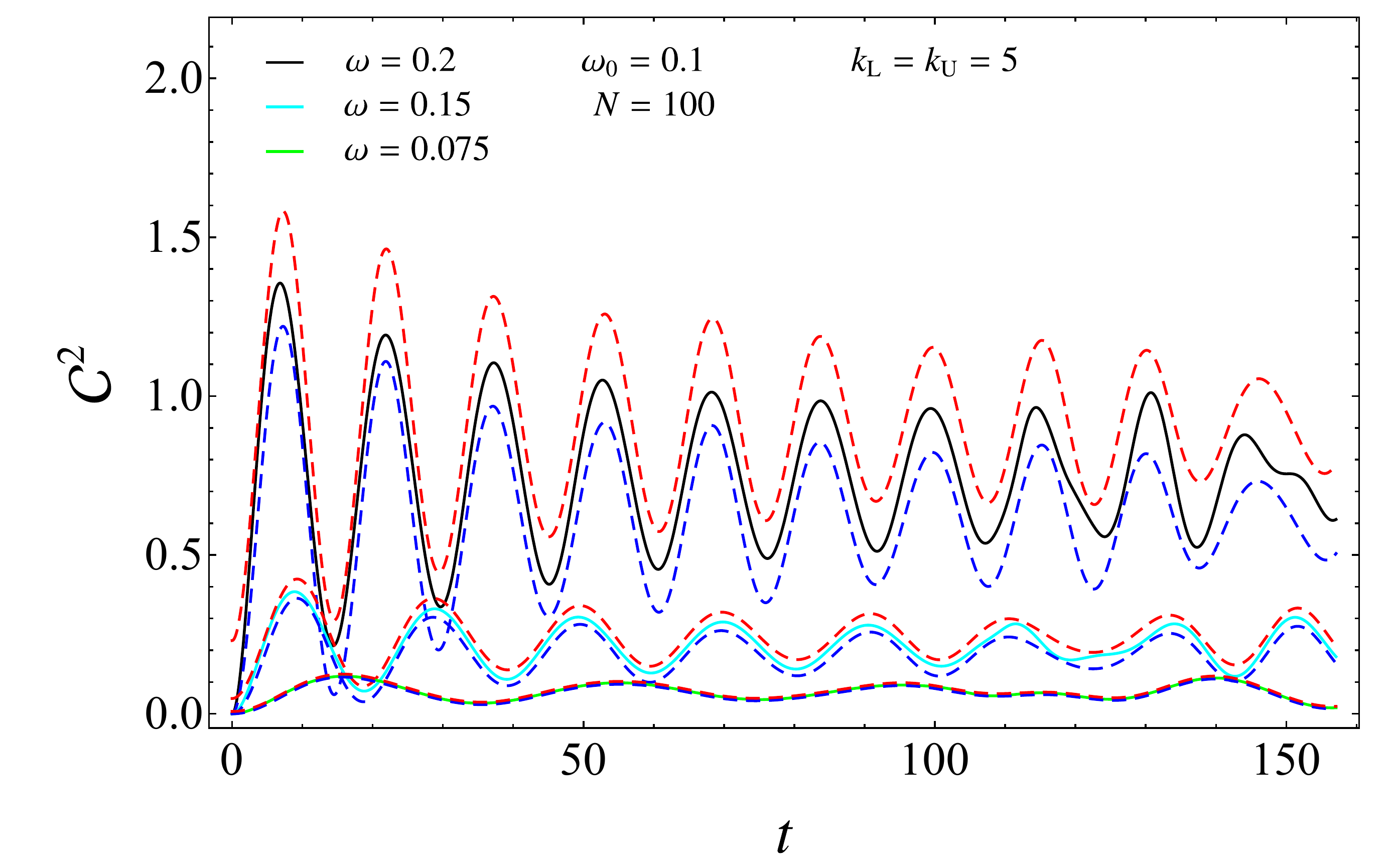}}
\caption{
Temporal evolution of the complexity (\ref{comp-pure-global-DBCPBC}) (solid lines)
and of the corresponding bounds in (\ref{new bounds_main_weaker}) 
for harmonic chains with either PBC (left panels) or DBC (right panels)
and $N=100$.
The blue (red) dashed lines correspond to the lower (upper) bound
(see (\ref{new upper bound relaxed}) and (\ref{new lower bound relaxed}) for the left panels
and (\ref{new bounds relaxed DBC}) for the right panels).
In all the panels $\omega_0=0.1$.
}
\vspace{0.4cm}
\label{fig:BoundsImproved}
\end{figure}

The bounds (\ref{new bounds_main})
can be written explicitly for the harmonic chains that we are considering 
by setting either $\eta=1$ or $\eta=0$
and employing either (\ref{dispersion relations}) or (\ref{dispersion DBC}) for the dispersion relations
when either PBC or DBC respectively are imposed.
The resulting expressions for these bounds require to sum either $N$ or $N-1$ terms
and we can obtain less constraining but still insightful bounds 
by keeping only few terms in these sums, i.e.
\be
\label{new bounds_main_weaker}
\mathcal{C}^2_{\textrm{\tiny L}, k_{\textrm{\tiny L}}}
\, \leqslant \,\mathcal{C}^2\,\leqslant \,
\mathcal{C}^2_{\textrm{\tiny U}, k_{\textrm{\tiny U}}}
\ee 
where $k_{\textrm{\tiny L}}$ and $k_{\textrm{\tiny U}}$ are independent parameters
related to the number of terms in the sum kept to define the corresponding bound.
Since the explicit expressions of the dispersion relations are important to write explicitly 
the bounds in (\ref{new bounds_main_weaker}), the cases of PBC and DBC must be studied separately.

Considering PBC first, 
one observes that the corresponding $f_{\textrm{\tiny L}}(\tilde{x}_k)$ as function of $k$ 
(that can be constructed from (\ref{fLs-def}), (\ref{sum over modes int arcsinh}) and (\ref{dispersion relations}))
is large when $k\simeq 1$ and $k\simeq N-1$,
while it becomes negligible in the middle of the interval $[1, N-1]$. 
This leads to sum just over $k=1,\dots,k_{\textrm{\tiny L}}$ 
and $k=N-k_{\textrm{\tiny L}},\dots,N-1$, for some $k_{\textrm{\tiny L}}$. 
Thus, by employing also the symmetry $ k \leftrightarrow N-k$ of the dispersion relations (\ref{dispersion relations}),
the lower bound in (\ref{new bounds_main_weaker}) for PBC reads
\be
\label{new lower bound relaxed}
\mathcal{C}^2_{\textrm{\tiny L}, k_{\textrm{\tiny L}}}
= c_0
+
2\sum_{k=1}^{k_{\textrm{\tiny L}}}
f_{\textrm{\tiny L}}(\tilde{x}_k)\, [ \sin(\Omega_{k} t )]^2\,.
\ee
The upper bound $\mathcal{C}^2_{\textrm{\tiny U}, k_{\textrm{\tiny U}}}$ can be found through
similar considerations applied to the function $f_{\textrm{\tiny U}}(\tilde{x}_k)$ introduced in 
(\ref{fLs-def}).
This leads to sum the terms whose $k$ is close to the boundary of $[1, N-1]$ keeping their dependence on $t$
and to set $\sin^2(\Omega_k t)=1$ in the remaining ones, which must not be discarded. 
The resulting bound is 
\be
\label{new upper bound relaxed}
\mathcal{C}^2_{\textrm{\tiny U}, k_{\textrm{\tiny U}}}
\,=\, 
c_0
+
2\sum_{k=1}^{k_{\textrm{\tiny U}}}
f_{\textrm{\tiny U}}(\tilde{x}_k)\, [ \sin(\Omega_{k} t )]^2
+\!\!
\sum_{k=k_{\textrm{\tiny U}}+1}^{\lfloor \frac{N-1}{2} \rfloor}\!\!
 2\,f_{\textrm{\tiny U}}(\tilde{x}_k)
 +  
  f_{\textrm{\tiny U}}(\tilde{x}_{N/2})\,
 \big| \cos(\pi N/2) \big|\,.
\ee
The bounds (\ref{new bounds_main}) are recovered when $k_{\textrm{\tiny L}}=k_{\textrm{\tiny U}}=\lfloor \frac{N-1}{2} \rfloor$, 
by setting to zero the second sum in the r.h.s. of (\ref{new upper bound relaxed}) 
and by restoring the time dependence in the term having $k=N/2$ when $N$ is even, 
both in (\ref{new lower bound relaxed}) and (\ref{new upper bound relaxed}).

When DBC are imposed, a similar analysis can be carried out, 
with the crucial difference that the symmetry $k\leftrightarrow N-k$ 
in the dispersion relations (\ref{dispersion DBC}) does not occur in this case. 
Setting $\eta=0$ and employing the dispersion relations (\ref{dispersion DBC}),
one obtains (\ref{new bounds_main_weaker}) with 
\be
\label{new bounds relaxed DBC}
\mathcal{C}^2_{\textrm{\tiny L}, k_{\textrm{\tiny L}}}
\equiv
\sum_{k=1}^{k_{\textrm{\tiny L}}}
 f_{\textrm{\tiny L}}(\tilde{x}_k) \, [\sin (\Omega_{k} t )]^2
\;\;\qquad\;\;
\mathcal{C}^2_{\textrm{\tiny U}, k_{\textrm{\tiny U}}}
\equiv
 \sum_{k=1}^{k_{\textrm{\tiny U}}}
 f_{\textrm{\tiny U}}(\tilde{x}_k) \, [\sin (\Omega_{k} t )]^2
+\!\!
\sum_{k=k_{\textrm{\tiny U}}+1}^{N-1}\!\!
 f_{\textrm{\tiny U}}(\tilde{x}_k)
\ee
where $1\leqslant k_{\textrm{\tiny L}},  k_{\textrm{\tiny U}} \leqslant N-1$.
In order to recover (\ref{new bounds_main})  from (\ref{new bounds_main_weaker}),
we have to choose $k_{\textrm{\tiny L}}=k_{\textrm{\tiny U}}=N-1$
and set to zero the last sum in the second expression of (\ref{new bounds relaxed DBC}).

By construction, we have 
$\mathcal{C}^2_{\textrm{\tiny L}, k_{\textrm{\tiny L}}} \leqslant 
\mathcal{C}^2_{\textrm{\tiny L}}$
and
$\mathcal{C}^2_{\textrm{\tiny U}, k_{\textrm{\tiny U}}} \geqslant 
\mathcal{C}^2_{\textrm{\tiny U}}$,
but  $\mathcal{C}^2_{\textrm{\tiny L}, k_{\textrm{\tiny L}}}$ and $\mathcal{C}^2_{\textrm{\tiny U}, k_{\textrm{\tiny U}}}$ 
contain less terms than  $\mathcal{C}^2_{\textrm{\tiny L}}$ and $\mathcal{C}^2_{\textrm{\tiny U}}$ respectively, 
hence they are easier to evaluate and to study analytically. 
For both PBC and DBC, considering either the lower or the upper bound in (\ref{new bounds_main_weaker}),
it improves as either $k_{\textrm{\tiny L}}$ or $ k_{\textrm{\tiny U}}$ respectively increases.

In Fig.\,\ref{fig:BoundsImproved} we show the bounds (\ref{new bounds_main_weaker})
when either PBC (left panels) or DBC (right panels) are imposed
and  small values of $k_{\textrm{\tiny L}}$ and $ k_{\textrm{\tiny U}}$
are considered. 
For given values of $k_{\textrm{\tiny L}}$ and $ k_{\textrm{\tiny U}}$,
the agreement between the bounds and the exact curve improves as $|\omega_0-\omega|$ decreases. 
Notice that higher values of $k_{\textrm{\tiny L}}$ and $ k_{\textrm{\tiny U}}$ 
are needed for DBC 
to reach an agreement with the exact curve comparable with the one obtained for PBC.

\subsection{Large $N$}
\label{subsec:TD limit HC}

It is important to study approximate expressions for the temporal evolution of the complexity
when large values of $N$ are considered.

\begin{figure}[htbp!]
\vspace{-.8cm}
\subfigure
{\hspace{-1.2cm}
\includegraphics[width=.54\textwidth]{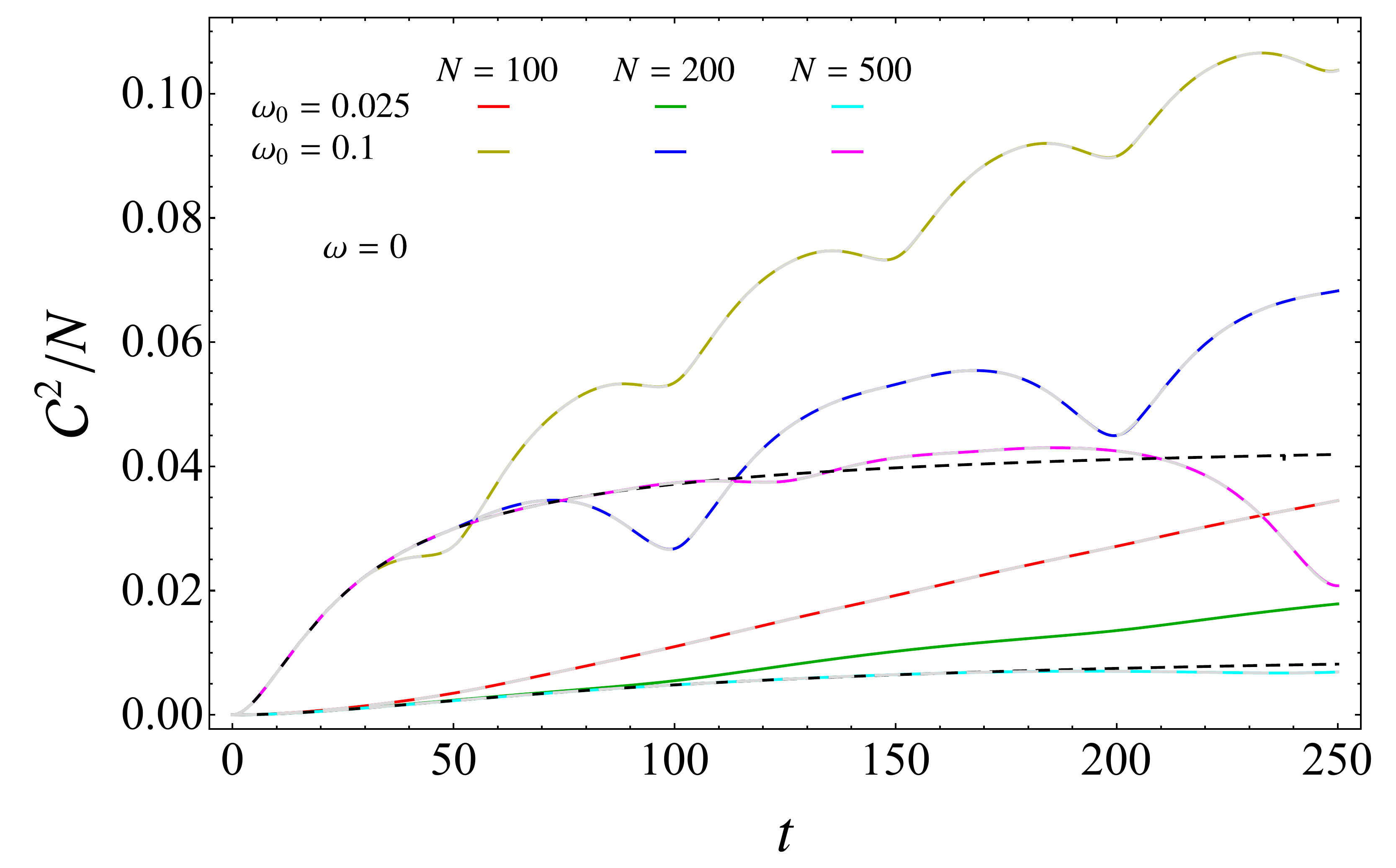}}
\vspace{-.25cm}
\subfigure
{
\hspace{.0cm}\includegraphics[width=.54\textwidth]{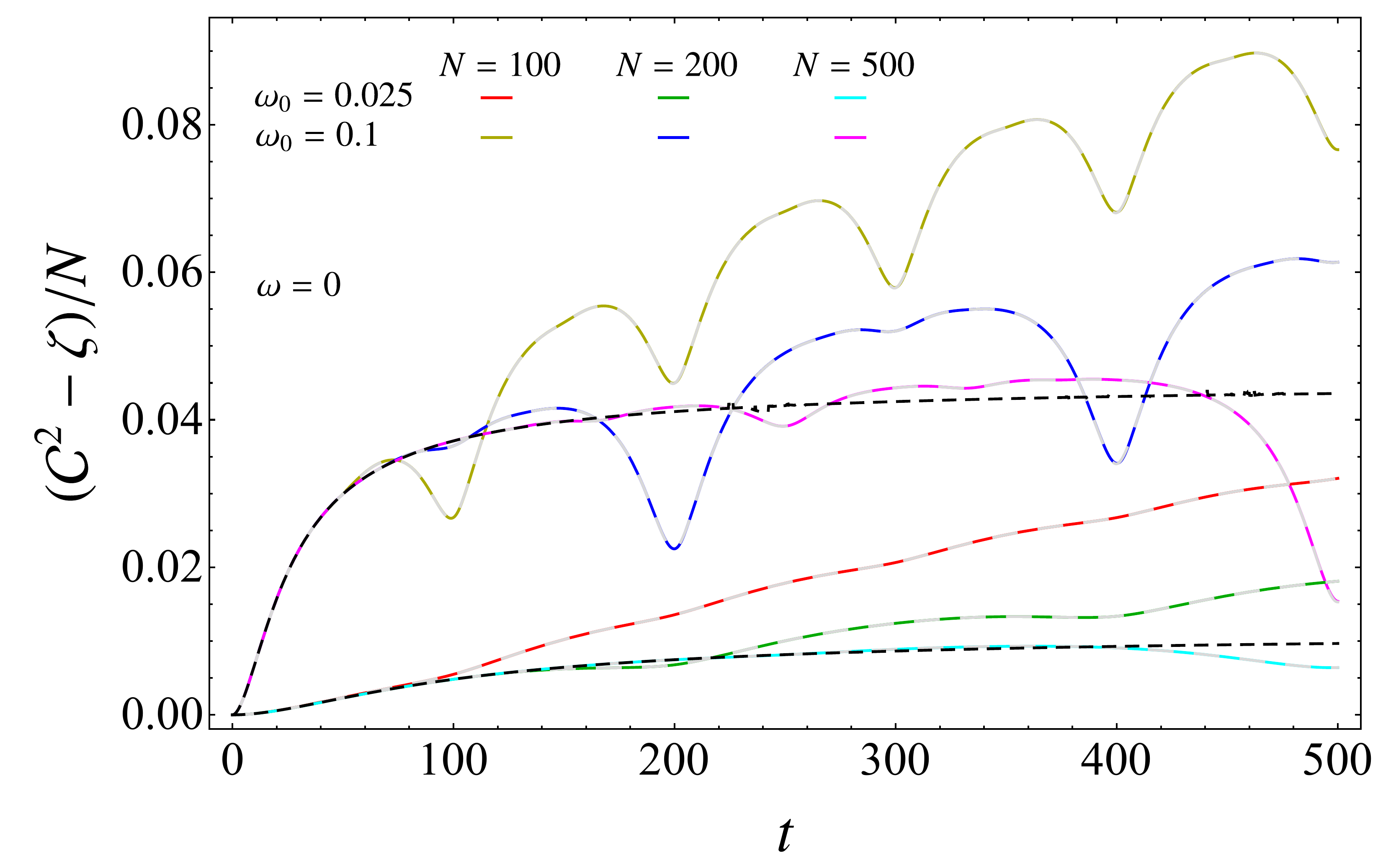}}
\subfigure
{
\hspace{-1.2cm}\includegraphics[width=.54\textwidth]{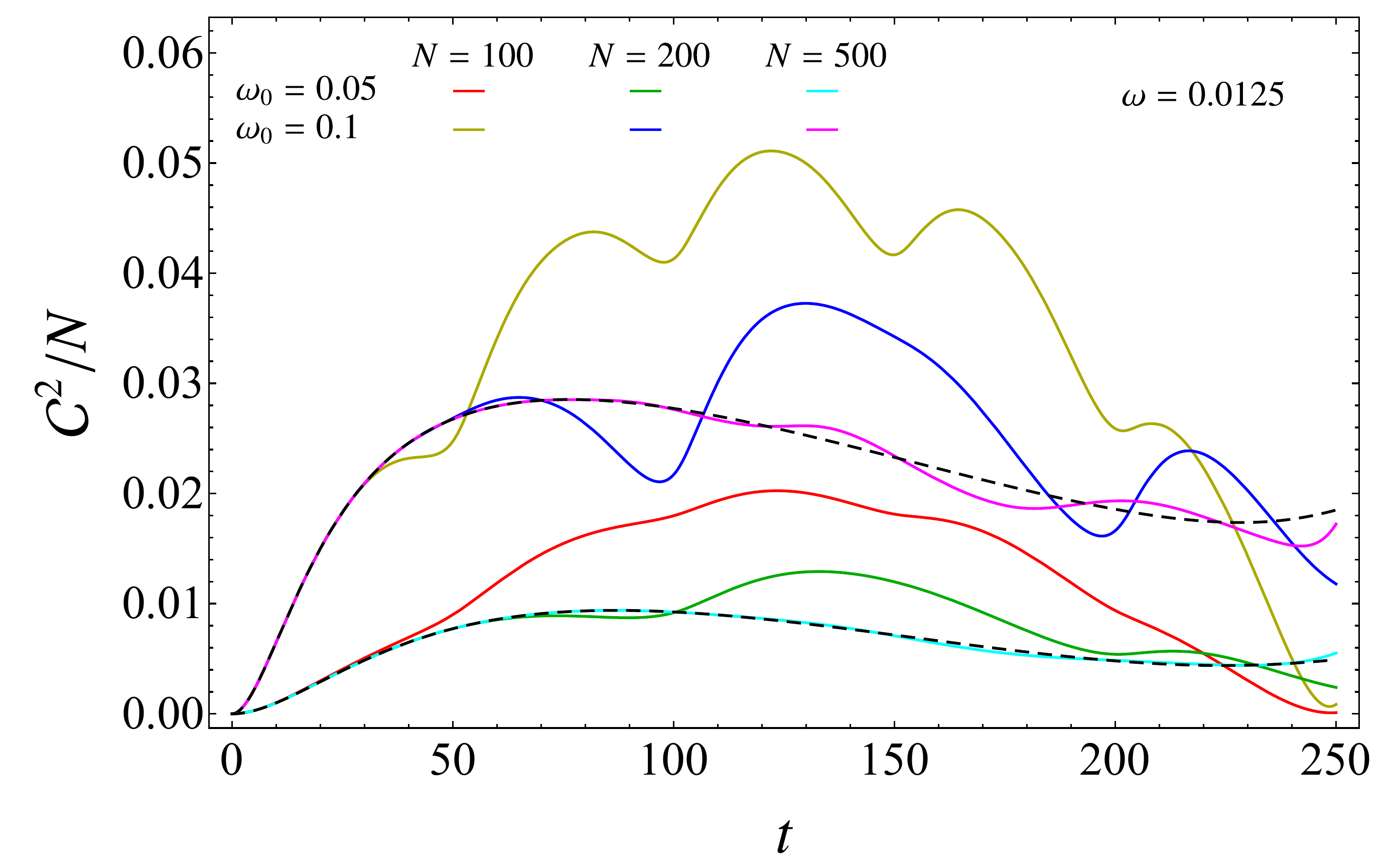}}
\vspace{-.25cm}
\subfigure
{\hspace{.0cm}
\includegraphics[width=.54\textwidth]{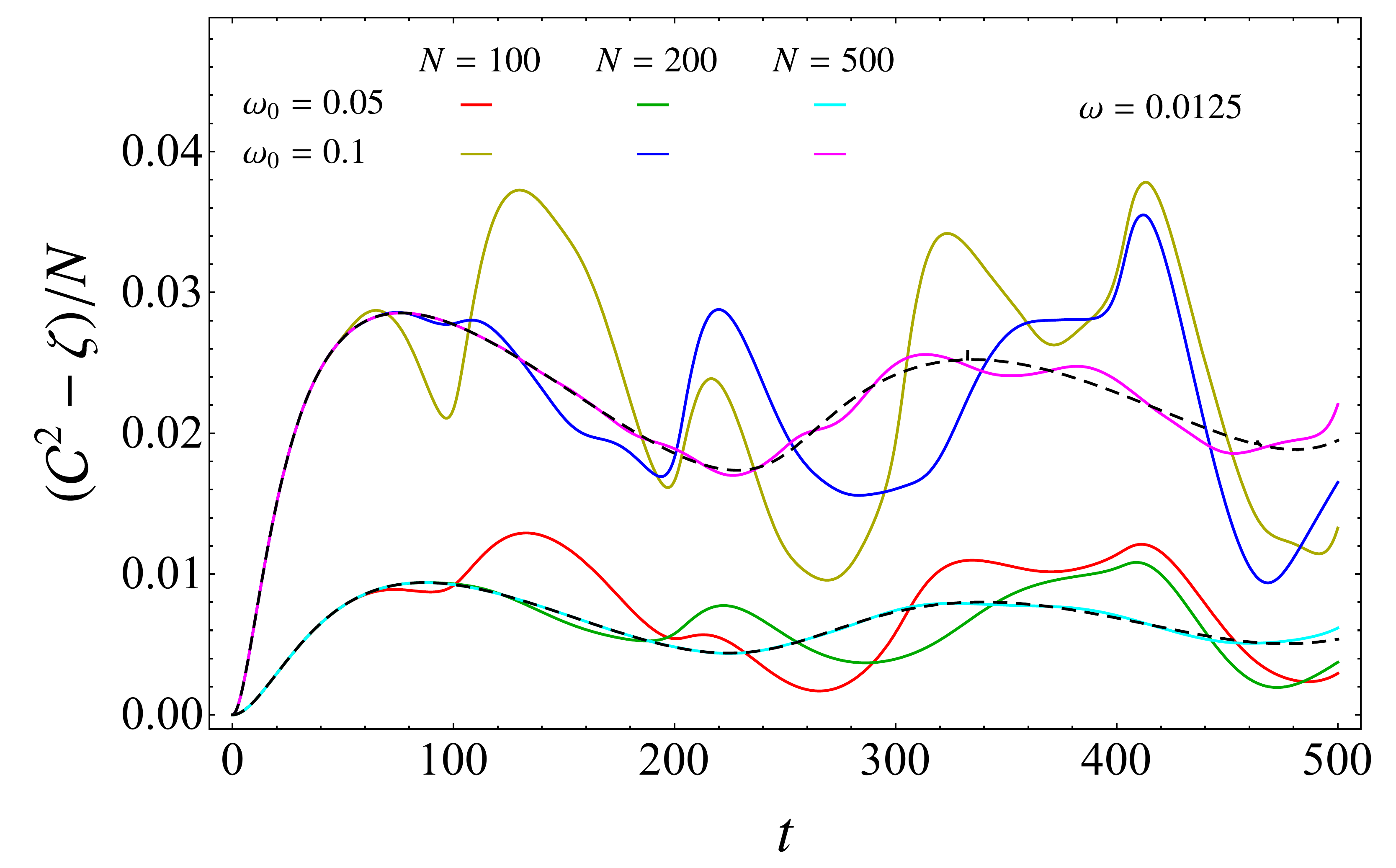}}
\subfigure
{
\hspace{-1.2cm}\includegraphics[width=.54\textwidth]{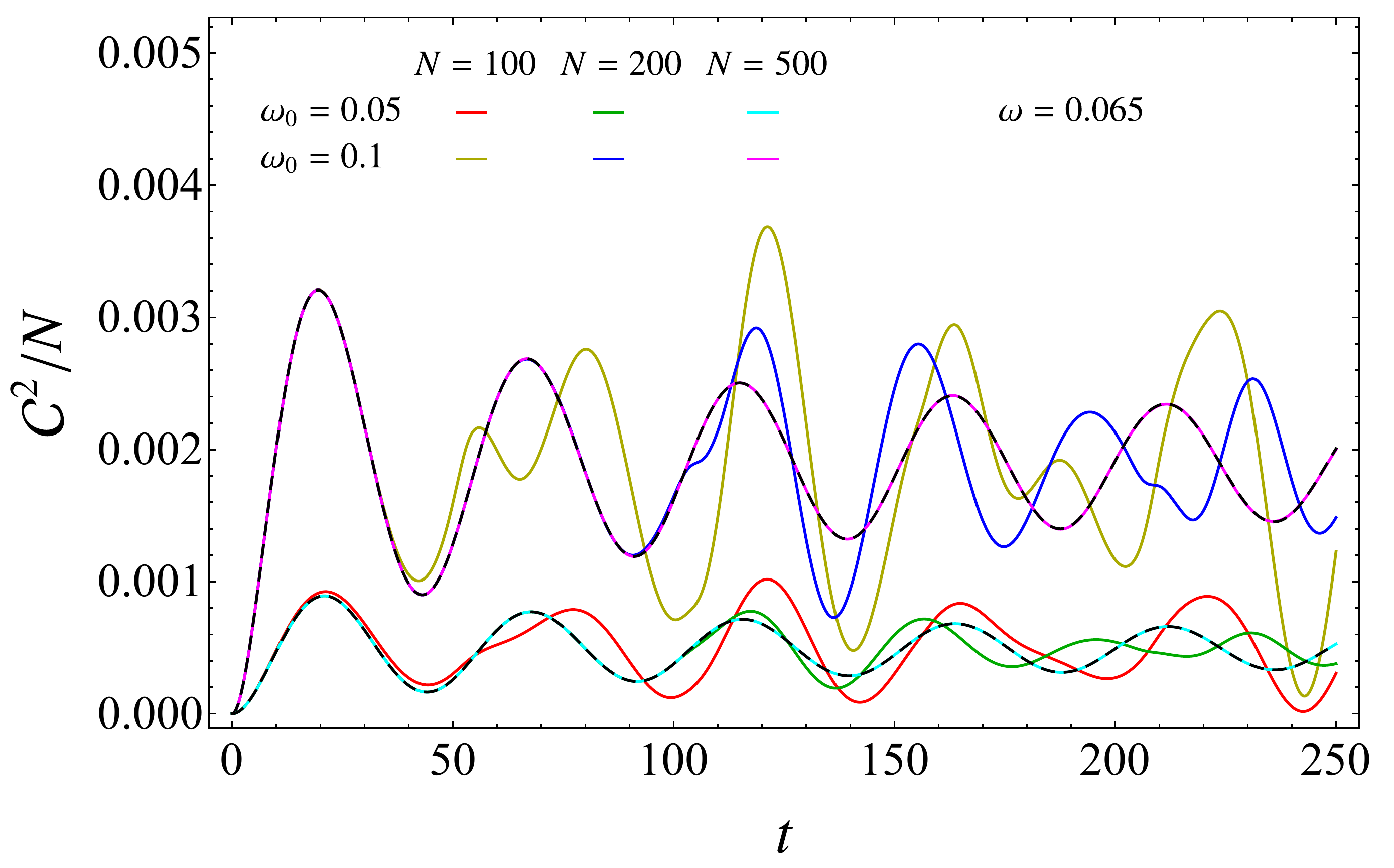}}
\vspace{-.25cm}
\subfigure
{\hspace{.0cm}
\includegraphics[width=.54\textwidth]{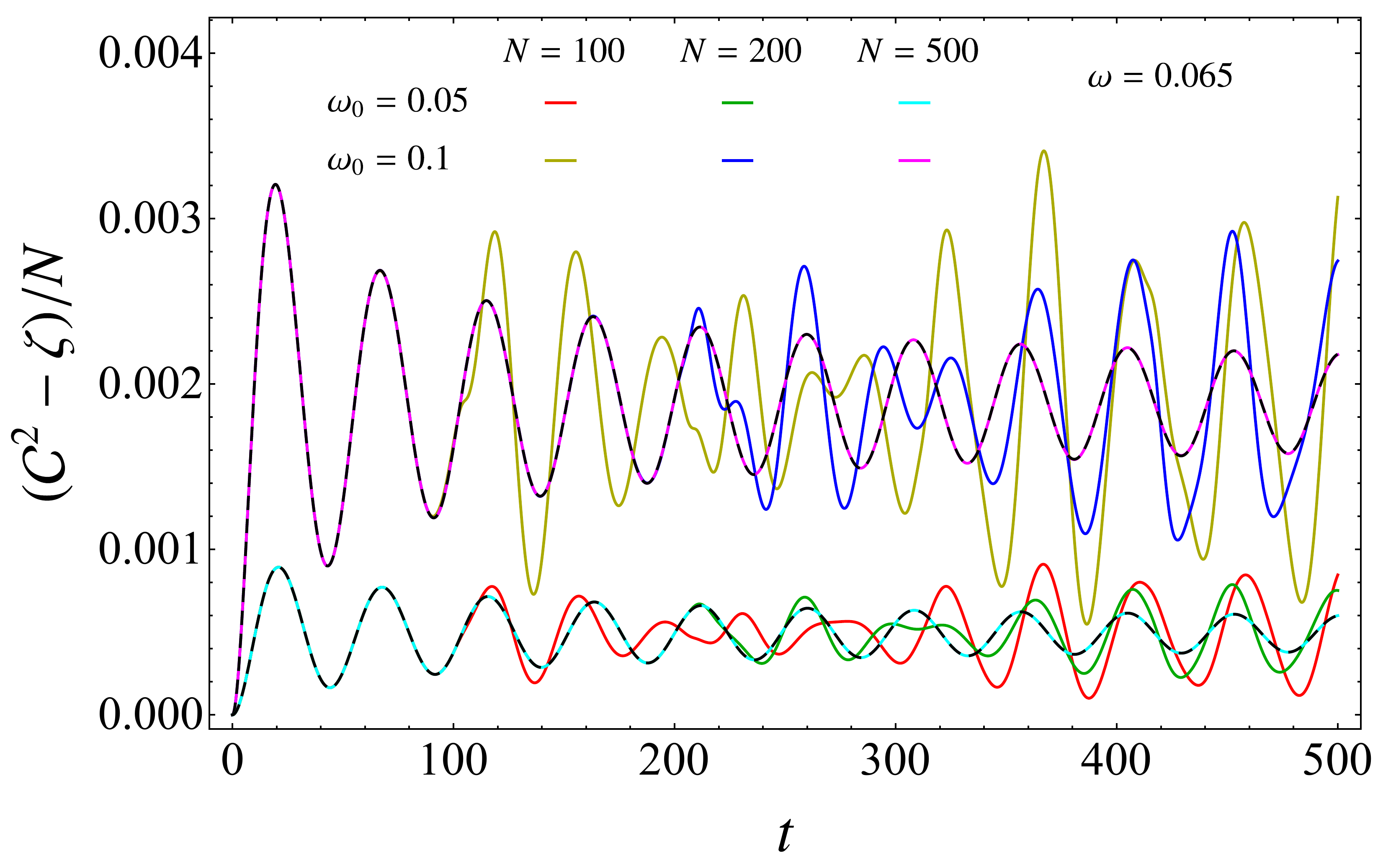}}
\subfigure
{
\hspace{-1.2cm}\includegraphics[width=.54\textwidth]{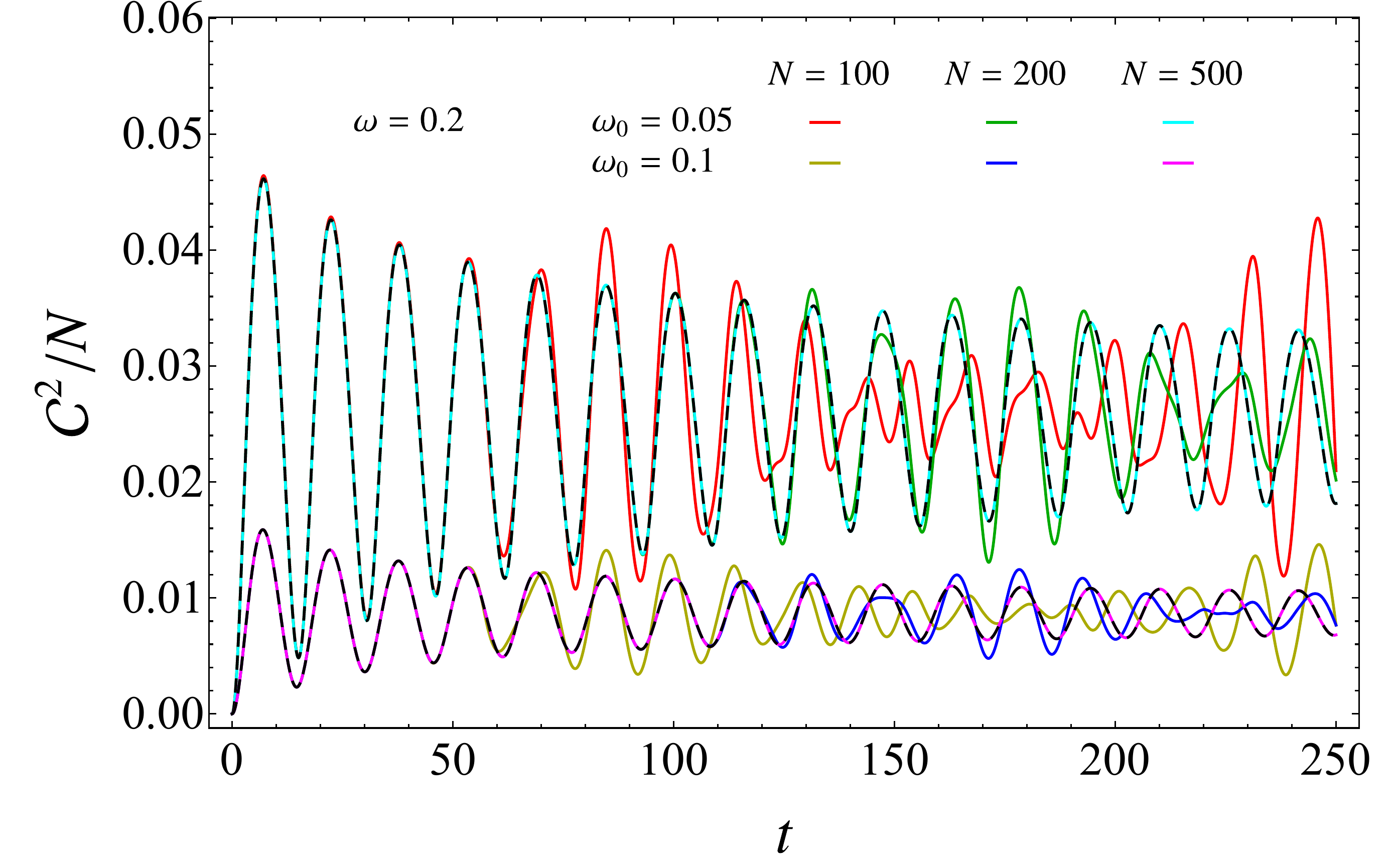}}
\vspace{-.25cm}
\subfigure
{\hspace{.0cm}
\includegraphics[width=.54\textwidth]{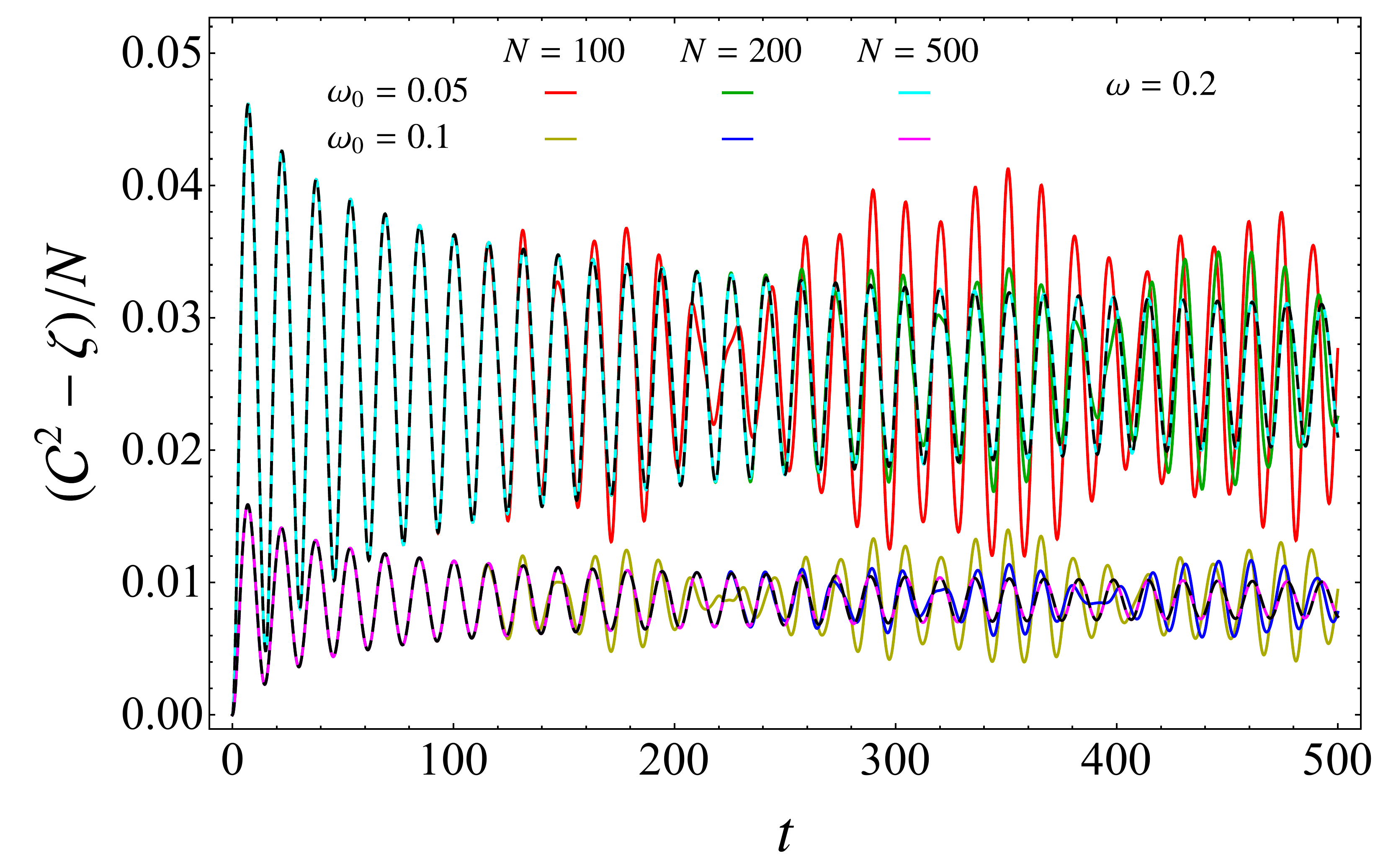}}
\caption{Temporal evolutions of the complexity for harmonic chains 
with either PBC (left panels) or DBC (right panels).
The solid lines show $\mathcal{C}^2/N$ for PBC
and $(\mathcal{C}^2-\zeta)/N$ for DBC,
with $\mathcal{C}$ given by (\ref{comp-pure-global-DBCPBC})  
and $\zeta$ by (\ref{corr_Csq Dir}). 
The dashed black lines represent $\mathcal{C}_{\textrm{\tiny TD}}^2/N$,
from (\ref{comp TD limit}).
}
\label{fig:TDvsfinite}
\end{figure}

In our numerical analysis, we noticed that, 
for finite but large enough values of $N\gtrsim 10$
the complexity (\ref{comp-pure-global-DBCPBC})
is well described by a function of $\omega N$, $\omega_0 N$ and $t/N$.
This function, which depends on whether PBC or DBC are imposed,
can be written by introducing the approximation $\sin(x) \simeq x$ into the dispersion relations 
and keeping only the leading term (see appendix \ref{subapp:smallkapprox} for a more detailed discussion).
For PBC we find
\be
\label{smallkapprox_PBC_main}
\mathcal{C}_{\textrm{\tiny approx}}
\,=\,
\sqrt{c_{0}(t) 
+
2 \!\! \sum_{k=1}^{\lfloor \frac{N-1}{2} \rfloor }\! 
\left[ \textrm{arcsinh}\! 
\left( \,
 \frac{(\omega N)^2 - (\omega_0 N)^2}{2\,\widetilde{\Omega}^{\textrm{\tiny (P)}}_{k}\, \widetilde{\Omega}^{\textrm{\tiny (P)}}_{0,k}} \,
 \sin \! \big(\widetilde{\Omega}^{\textrm{\tiny (P)}}_{k} t / N \big) 
 \right)
 \right]^2
 }
\ee
where $c_{0}(t)$ is (\ref{low-bound});
while for DBC we get
\be
\label{smallkapprox_DBC_main}
\mathcal{C}_{\textrm{\tiny approx}}
=
\sqrt{\,\sum_{k=1}^{N-1}\left[ \textrm{arcsinh}\! 
\left( \,
 \frac{(\omega N)^2 - (\omega_0 N)^2}{2\,\widetilde{\Omega}^{\textrm{\tiny (D)}}_k \widetilde{\Omega}^{\textrm{\tiny (D)}}_{0,k}} \,
 \sin \! \big(\widetilde{\Omega}^{\textrm{\tiny (D)}}_{k} t / N \big) 
 \right)
 \right]^2}
\ee
where
\be
\label{smallkapprox_dispersion}
\widetilde{\Omega}^{\textrm{\tiny (P)}}_k
=\sqrt{(\omega N)^2+\frac{4\pi^2 \kappa}{m}\, k^2}
\,\,\qquad\,\,
\widetilde{\Omega}^{\textrm{\tiny (D)}}_k
=
\sqrt{(\omega N)^2+\frac{\pi^2 \kappa}{m}\, k^2}
\ee
while $\widetilde{\Omega}^{\textrm{\tiny (P)}}_{0,k}$ 
and $\widetilde{\Omega}^{\textrm{\tiny (D)}}_{0,k}$ are obtained by replacing $\omega$ with $\omega_0$
in these expressions.
Notice that both (\ref{smallkapprox_PBC_main}) and (\ref{smallkapprox_DBC_main}) 
depend on $\omega N$, $\omega_0 N$ and $t/N$. 
These approximate expressions have been used to plot the dashed light grey curves
in the top panels of Fig.\,\ref{fig:TDvsfinite}, 
which nicely agree with the corresponding solid coloured curves.


The thermodynamic limit $N\to\infty$ of the complexity can be studied
through the standard procedure. 
Introducing $\theta \equiv \pi k/N$ 
and substituting $\sum_k\to \frac{N}{\pi}\int_{0}^\pi d\theta$ 
in (\ref{comp-pure-global-DBCPBC}), at the leading order we find
\be
\label{comp TD limit}
\mathcal{C}_{\textrm{\tiny TD}}
=
\sqrt{\frac{N}{\pi}}\;
\sqrt{\int_{0}^\pi\left[ \textrm{arcsinh}\! 
\left( \,
 \frac{\omega^2 - \omega_0^2}{2\,\Omega_{\theta}\, \Omega_{0,\theta}} \,
 \sin (\Omega_{\theta} \,t ) 
 \right)
 \right]^2 \! d\theta}
\ee
where the dispersion relations for PBC and DBC become respectively
\be
\label{dispersion relations TD}
\Omega_{0,\theta} = \sqrt{\omega_0^2 + \frac{4\kappa}{m}  \,(\sin\theta)^2 }
\,\,\qquad\,\,
\Omega_\theta = \sqrt{\omega^2 + \frac{4\kappa}{m} \,( \sin \theta )^2}
\ee
and 
\be
\label{dispersion DBC TD}
\Omega_{0,\theta}=\sqrt{\omega_0^2+\frac{4\kappa}{m} \,[\sin(\theta/2)]^2}
\,\,\qquad\,\, 
\Omega_\theta=\sqrt{\omega^2+\frac{4\kappa}{m} \,[\sin(\theta/2)]^2}\;.
\ee

Notice that, for PBC, the zero mode does not contribute because $c_0/N \to 0$ as $N \to \infty$.
When DBC hold, by using the dispersion relations (\ref{dispersion DBC TD}),
changing of variable $\tilde{\theta}=\theta/2$ in (\ref{comp TD limit})
and exploiting the symmetry of the function $\sin (x)$ in the interval $[0,\pi]$,
one finds that (\ref{comp TD limit}) with (\ref{dispersion relations TD}) holds for both PBC and DBC.
Thus, the leading order of this limit is independent of the boundary conditions. 
This means that the complexity does not distinguish the boundary conditions
in this regime.
Indeed, in the left and right panels of Fig.\,\ref{fig:TDvsfinite}, 
the same function (just described) has been used to plot the dashed black curves.

%

The boundary conditions become crucial in the subleading term of the expansion of 
(\ref{comp-pure-global-DBCPBC})  as  $N\to\infty$,
which can be studied through the Euler-Maclaurin formula \cite{Kac02bookCalculus}.
The details of this analysis are discussed in appendix \ref{app:EulerMcLaurin}
and  the final result is 
\be
\label{EmcL maintext}
 \mathcal{C}^2-\mathcal{C}_{\textrm{\tiny TD}}^2
 =
 R_{1,\infty}^{\textrm{\tiny (B)}}
 \;\;\qquad\;\;
 \textrm{B} \in \big\{  \textrm{P},  \textrm{D} \big\}
  \;\;\qquad\;\;
  R_{1,\infty}^{\textrm{\tiny (B)}} 
  =
  \left\{\begin{array}{ll}
  R_{1,\infty}^{\textrm{\tiny (P)}} & \textrm{PBC}
  \\
  \rule{0pt}{.7cm}
  \displaystyle
  R_{1,\infty}^{\textrm{\tiny (D)}} 
  =
  \frac{R_{1,\infty}^{\textrm{\tiny (P)}}}{2} + \zeta  \hspace{.5cm}  & \textrm{DBC}
  \end{array}\right.
\ee
where $R_{1,\infty}^{\textrm{\tiny (P)}}$ and $\zeta$  are the time-dependent functions 
given in (\ref{bound rest_ p0PBC v2}) and in (\ref{corr_Csq Dir}) respectively. 
Numerical checks for these results are shown in Fig.\,\ref{fig:TDvsfinite}.
In the top panels of this figure we have displayed also $\mathcal{C}_{\textrm{\tiny approx}}/N$ 
from (\ref{smallkapprox_PBC_main}) (left panel)
and $(\mathcal{C}_{\textrm{\tiny approx}}-\zeta)/N$ from (\ref{smallkapprox_DBC_main})
(right panel) through dashed light grey  lines.

In the continuum limit,
$N\to\infty$ and the lattice spacing $a \equiv \sqrt{m/\kappa} \to 0$
is vanishing while $N a\equiv \ell$ is kept fixed. 
In this limit,
the expression (\ref{comp-pure-global-DBCPBC}) for the complexity
(which holds for both PBC and DBC) becomes
\be 
\label{comp HC cont}
\mathcal{C}_{\textrm{\tiny cont}}
=
\sqrt{\frac{\ell}{2\pi}}\,
\sqrt{\int_{-\infty}^{\infty}
 \left[ \textrm{arcsinh}\! 
\left( \,
 \frac{\omega^2 - \omega_0^2}{2\,\Omega_p\, \Omega_{0,p}} \,
 \sin (\Omega_p t ) 
 \right)
 \right]^2\! dp}
\ee
(see appendix \ref{app:continuum} for a detailed discussion)
where
\be
\label{dispersion continuum HC}
\Omega_{0,p}=\sqrt{\omega_0^2 + p^2}
\,\,\qquad\,\,
\Omega_p=\sqrt{\omega^2 + p^2}\,.
\ee
Since $\Omega_p\simeq p$ when $p\gg\omega$, 
the vanishing of the integrand in (\ref{comp HC cont})
as $p\to\pm\infty$ is such that the complexity is UV finite.
We remark that, instead, when the reference state is the unentangled product state, 
the continuum limit of the complexity is UV divergent,
as discussed in appendix \ref{app:continuum};
hence a UV cutoff in the integration domain over $p$ must be introduced.

\subsection{Initial growth}

It is worth discussing the initial growth of the complexity for the harmonic chains 
that we are considering. 
Since the complexity (\ref{comp-pure-global-DBCPBC}) is a special case of (\ref{comp-pure-global-general}), 
its expansion as $ t \to 0$ can be found by specialising the expansion (\ref{comp-pure-initialgrowth}) 
and its coefficients (\ref{c_12_coeff_exp}) and (\ref{c_3_coeff_exp}) to the harmonic chains with either PBC or DBC.
For the sake of simplicity, in the following we discuss only the leading term
(i.e. only the coefficient $b_1$ in (\ref{c_12_coeff_exp})), 
which provides the linear growth,
but a similar analysis can be applied straightforwardly 
to the coefficients of the higher order terms in the $t \to 0$ expansion.

For the harmonic chains with either PBC or DBC, the linear growth 
in (\ref{comp-pure-initialgrowth}) becomes
\be
\label{initial growth HC 1d}
\mathcal{C}
\,=\,
\frac{|\omega^2-\omega^2_0|}{2}\,
\Bigg( \sum_{k=1}^{N-1+\eta}  \! \Omega_{0,k}^{-2} \Bigg)^{1/2}
t
\, +O(t^3)
\ee
where $\eta=1$ and (\ref{dispersion relations}) must be used for PBC,
while $\eta=0$ and (\ref{dispersion DBC}) must be employed for DBC.
We remark that the slope of the initial linear growth in (\ref{initial growth HC 1d}) is proportional to $|\omega-\omega_0|$.

\begin{figure}[t!]
\subfigure
{\hspace{-1.5cm}
\includegraphics[width=.57\textwidth]{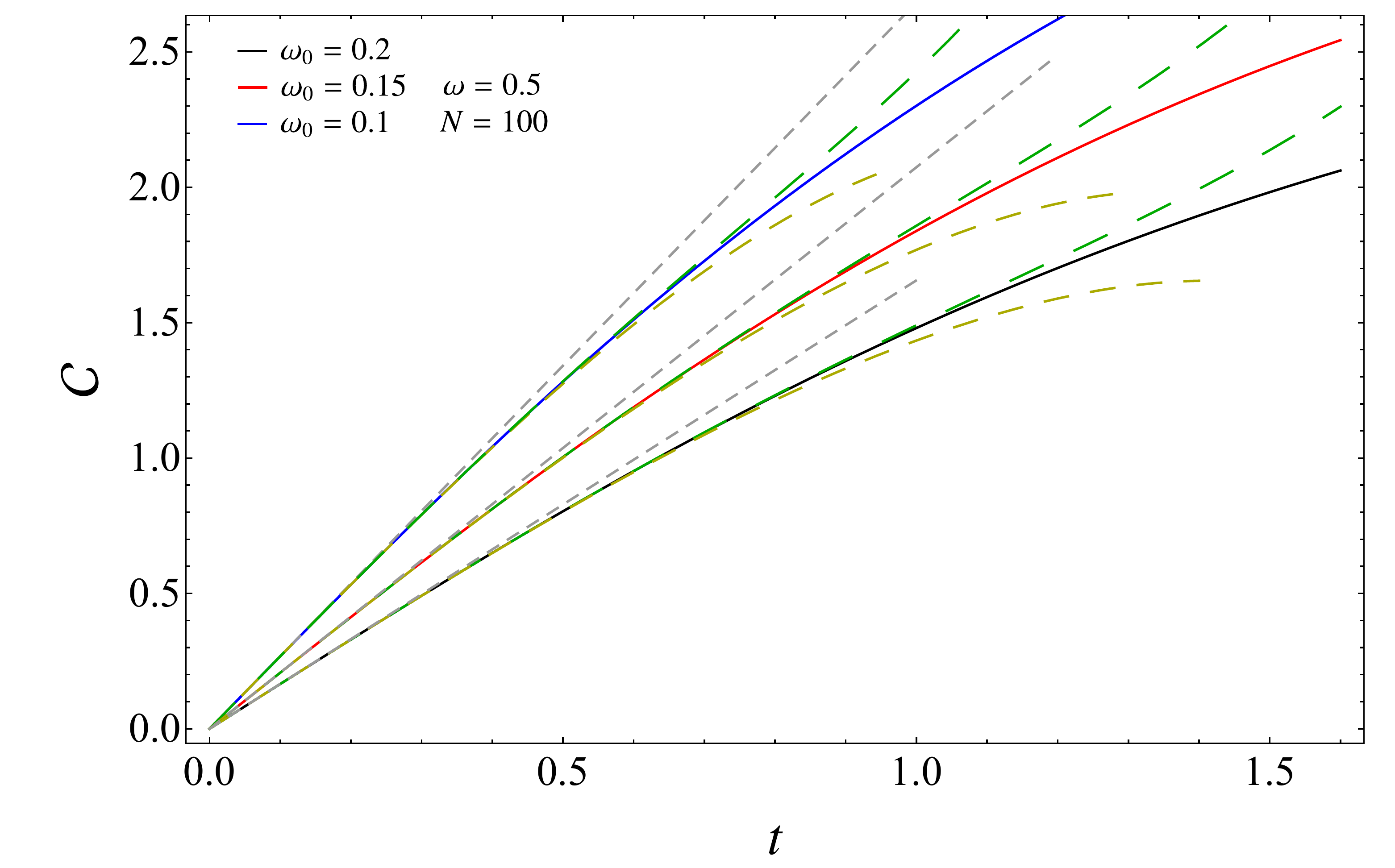}}
\subfigure
{
\hspace{-.45cm}\includegraphics[width=.57\textwidth]{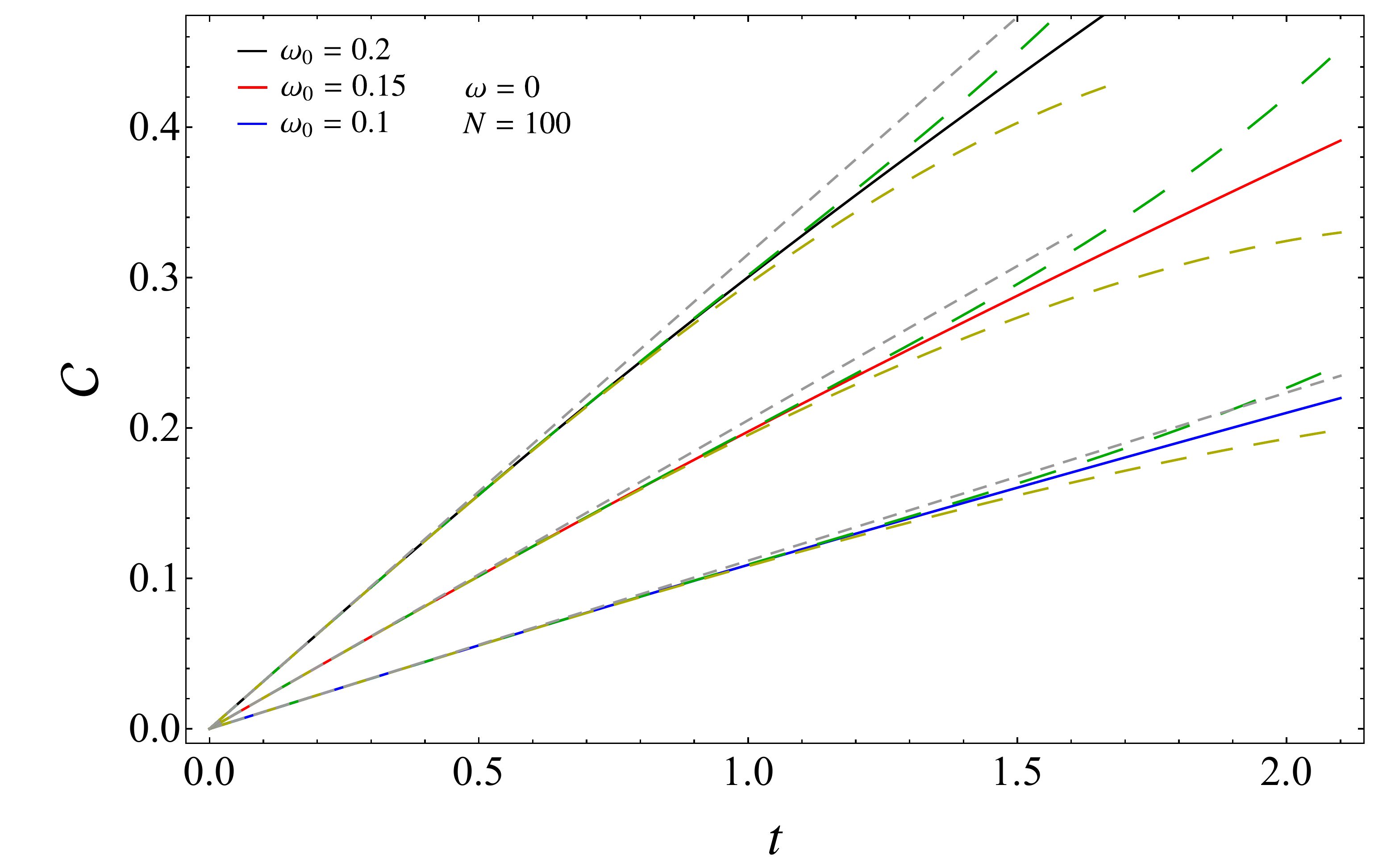}}
\caption{Initial growth of the complexity 
for harmonic chains with PBC and $N=100$.
The evolution Hamiltonian is either massive (left panel) or massless (right panel),
for three values $\omega_0$.
The solid lines show the complexity (\ref{comp-pure-global-DBCPBC})
(with (\ref{dispersion relations}))
and the dashed lines represent its expansion (\ref{comp-pure-initialgrowth})
up to the $O(t)$ (grey),  $O(t^3)$ (yellow) and $O(t^5)$ (green) term included. 
} 
\vspace{0.1cm}
\label{fig:PureStateGlobalSmallTimes}
\end{figure}

In Fig.\,\ref{fig:PureStateGlobalSmallTimes}, we consider
the initial growth of the complexity (\ref{comp-pure-global-DBCPBC}) when PBC are imposed,
comparing the exact curve against its expansion (\ref{comp-pure-initialgrowth}).
The corresponding analysis for DBC provides curves that are very similar to the ones displayed in Fig.\,\ref{fig:PureStateGlobalSmallTimes};
hence it has not been reported in this manuscript.

Let us conclude our discussion about the temporal evolution of the complexity of pure states 
with a brief qualitative comparison between 
the results discussed above and 
the corresponding ones for the temporal evolution of the holographic complexity
\cite{Stanford:2014jda,Susskind:2014jwa,Brown:2015bva,Brown:2015lvg,Roberts:2014isa,Moosa:2017yvt,Chapman:2018dem,Chapman:2018lsv}.

The Vaidya spacetimes are the typical backgrounds employed as the gravitational duals 
of global quantum quenches in the conformal field theory on their boundary.
They describe the formation of a black hole through the collapse of a matter shell. 
In Vaidya spacetimes, the temporal evolution of the holographic entanglement entropy
has been largely studied \cite{Hubeny:2007xt,AbajoArrastia:2010yt,
Balasubramanian:2011ur,
Balasubramanian:2011at,Allais:2011ys,Callan:2012ip,Hubeny:2013hz,Hubeny:2013dea}
and
the temporal evolutions of the holographic complexity
for the entire spatial section of the conformal field theory on the boundary
has been investigated in \cite{Moosa:2017yvt,Chapman:2018dem,Chapman:2018lsv, Fan:2018xwf}.
Considering the temporal evolution of the rate $\frac{d\mathcal{C}}{dt}$ 
allows to avoid the problem of choosing the reference state,
which deserves further clarifications for the holographic complexity,
even for static gravitational backgrounds. 
The analysis of $\frac{d\mathcal{C}}{dt}$ in Vaidya spacetimes,
both for the CV and for the CA prescriptions, 
shows that these temporal evolutions are linear in time both 
at very early and at late time \cite{Moosa:2017yvt, Chapman:2018dem}.
While also the initial growth of the complexity 
that we have explored is linear (see (\ref{initial growth HC 1d})),
the late time growth is at most logarithmic. 
This disagreement, which deserves further analysis,
has been discussed in \cite{Chapman:2018hou}.

We find it worth observing also that 
the coefficient of the initial growth (\ref{initial growth HC 1d}) is proportional to $|\omega^2 - \omega_0^2|$
and that the corresponding coefficient for the holographic complexity 
is proportional to the mass of the 
final black hole \cite{Chapman:2018dem, Moosa:2017yvt}.

\section{Subsystem complexity in finite harmonic chains}
\label{sec:mixedFinSize}

In this section we study the temporal evolution of the subsystem complexity after a global quench.
The reference and the target states are the reduced density matrices associated to a given subsystem. 
We focus on the simple cases where the subsystem $A$ 
is a block made by consecutive sites in harmonic chains 
with either PBC or DBC.

\subsection{Subsystem complexity}

In the harmonic lattices that we are considering, 
the reduced density matrix associated to $A$ 
characterises a Gaussian state
which can be described equivalently through 
its reduced covariance matrix $\gamma_A$.
This matrix is constructed by considering 
the reduced correlation matrices  $Q_A$, $P_A$ and $M_A$,
whose elements are respectively given  by
$(Q_A)_{i,j}=\langle \psi_0 |\, \hat{q}_i(t)  \,\hat{q}_j(t) \, | \psi_0 \rangle $,
$(P_A)_{i,j}=\langle \psi_0 |\, \hat{p}_i(t)  \,\hat{p}_j(t) \, | \psi_0 \rangle $
and 
$(M_A)_{i,j}=\textrm{Re}\big[\langle \psi_0 |\, \hat{q}_i(t)  \,\hat{p}_j(t) \, | \psi_0 \rangle \big]$
with $i,j\in A$,
which depend also on the time after the global quench.
These matrices provide the following block decomposition of the reduced covariance matrix
\be
\label{reduced CM}
\gamma_A(t)
=\,
\bigg( 
\begin{array}{cc}
Q_A(t)  \,& M_A(t) \\
M_A(t)^{\textrm{t}}  \,& P_A(t)  \\
\end{array}   \bigg)\,.
\ee

For the harmonic chains with either PBC or DBC 
introduced in Sec.\,\ref{sec:purestates_HC_glob}
and $A$ made by $L$ consecutive sites,
$Q_A$ and $P_A$ are $L \times L$ symmetric matrices
and $\gamma_A$ is a real, symmetric and positive definite $2L \times 2L$ matrix.

Adapting the analysis made in Sec.\,\ref{sec:purestates_HC_glob} for pure states
to the mixed states described by the reduced covariance matrices $\gamma_A(t)$,
we have that
the reference state is given by the reduced density matrix for the interval $A$
at time $t_\textrm{\tiny R} \geqslant 0$ 
obtained through the quench protocol characterised by
$\big(\kappa_\textrm{\tiny R},m_\textrm{\tiny R},\omega_\textrm{\tiny R},\omega_{0,\textrm{\tiny R}}\big)$
and the target state by the reduced density matrix for the same interval
at time $t_\textrm{\tiny T} \geqslant  t_\textrm{\tiny R}$ 
constructed through the quench protocol described by
$\big(\kappa_\textrm{\tiny T},m_\textrm{\tiny T},\omega_\textrm{\tiny T},\omega_{0,\textrm{\tiny T}}\big)$.
The corresponding reduced covariance matrices 
are denoted by $ \gamma_{\textrm{\tiny R},A} (t_\textrm{\tiny R}) $
and $ \gamma_{\textrm{\tiny T},A} (t_\textrm{\tiny T}) $ respectively. 
These reduced covariance matrices 
are decomposed in terms of the
correlation matrices of the subsystem like in (\ref{reduced CM}).

The approach to the circuit complexity of mixed states based on the Fisher information geometry \cite{DiGiulio:2020hlz} 
allows to construct the optimal circuit 
between $ \gamma_{\textrm{\tiny R},A} (t_\textrm{\tiny R}) $ and $ \gamma_{\textrm{\tiny T},A} (t_\textrm{\tiny T}) $.
The covariance matrices along this optimal circuit are
\be
\label{optimal circuit rdm}
G_{s}( \gamma_{\textrm{\tiny R},A}(t_\textrm{\tiny R}) \, , \gamma_{\textrm{\tiny T},A}(t_\textrm{\tiny T}))
\,\equiv \,
\gamma_{\textrm{\tiny R},A}(t_\textrm{\tiny R})^{1/2} 
\Big(  \gamma_{\textrm{\tiny R},A}(t_\textrm{\tiny R})^{- 1/2} \,
\gamma_{\textrm{\tiny T},A}(t_\textrm{\tiny T}) \,
\gamma_{\textrm{\tiny R},A}(t_\textrm{\tiny R})^{-1/2} \Big)^s
\gamma_{\textrm{\tiny R},A}(t_\textrm{\tiny R})^{1/2}
\ee
where $0 \leqslant s \leqslant 1$ parameterises the optimal circuit.
The length of this optimal circuit is proportional to its complexity 
\be
\label{c2-complexity-rdm}
\mathcal{C}_A
\,=\,
\frac{1}{2\sqrt{2}}\;
\sqrt{\,
\textrm{Tr} \,\Big\{ \big[
\log \!\big( \gamma_{\textrm{\tiny T},A}(t_\textrm{\tiny T}) \, \gamma_{\textrm{\tiny R},A}(t_\textrm{\tiny R})^{-1} \big)
\big]^2  \Big\}}\;.
\ee

Considering harmonic chains made by $N$ sites where PBC are imposed,
by using (\ref{time dep corrs trans}), (\ref{QPRmat t-dep-k})
and either (\ref{Vtilde-def-even}) or (\ref{Vtilde-def-odd}), 
one obtains the elements of the correlation matrices whose reduction to $A$ provides 
(\ref{reduced CM}).
They read
\bea
\label{QPRmat t-dep 1d}
Q_{i,j}(t) 
&=&
\frac{1}{N} \sum_{k=1}^N Q_{k}(t) \,\cos\!\big[(i-j)\,2\pi k/N\big] 
\nonumber
\\
\rule{0pt}{.8cm}
P_{i,j}(t) 
&=&
\frac{1}{N}\sum_{k=1}^N P_{k}(t)\, \cos\!\big[(i-j)\,2\pi k/N\big] 
 \\
 \rule{0pt}{.8cm}
M_{i,j}(t) 
&=&
\frac{1}{N}\,\sum_{k=1}^N M_{k}(t) \,\cos\!\big[(i-j)\,2\pi k/N\big] 
\nonumber
\eea
where $1 \leqslant i,j \leqslant N$;
while for DBC, by using (\ref{Vtilde-HC-DBC}), one obtains the following correlators
\bea
\label{QPRmat t-dep 1d DBC}
Q_{i,j}(t) 
&=&
 \frac{2}{N}\sum_{k=1}^N Q_{k}(t) \,\sin\!\big(i\,\pi k/N\big)\,\sin\!\big(j\,\pi k/N\big)
 \nonumber
\\
\rule{0pt}{.8cm}
P_{i,j}(t) 
&=&
\frac{2}{N}\sum_{k=1}^N P_{k}(t) \,\sin\!\big(i\,\pi k/N\big)\,\sin\!\big(j\,\pi k/N\big) 
 \\
 \rule{0pt}{.8cm}
M_{i,j}(t) 
&=&
\frac{2}{N}\,\sum_{k=1}^N M_{k}(t)\,\sin\!\big(i\,\pi k/N\big)\,\sin\!\big(j\,\pi k/N\big)
\nonumber
\eea
where $1 \leqslant i,j \leqslant N-1$.
In these correlators, the functions  $Q_{k}(t)$, $P_{k}(t)$ and $M_{k}(t)$ 
are given by (\ref{QPRmat t-dep-k}), with either (\ref{dispersion relations}) for PBC or (\ref{dispersion DBC}) for DBC.

The reduced covariance matrices 
$ \gamma_{\textrm{\tiny R},A} (t_\textrm{\tiny R}) $ 
and $ \gamma_{\textrm{\tiny T},A} (t_\textrm{\tiny T}) $
for the block $A$ providing the optimal circuit (\ref{optimal circuit rdm}) 
and its complexity (\ref{c2-complexity-rdm}) are constructed as in (\ref{reduced CM}),
through the reduced correlation matrices $Q_A$, $P_A$ and $M_A$,
obtained by restricting to $i,j \in A$ the indices of the correlation matrices whose elements 
are given in (\ref{QPRmat t-dep 1d}) and (\ref{QPRmat t-dep 1d DBC}).

We remark that the matrix $\widetilde{V}$ in (\ref{time dep corrs trans})
(given in (\ref{Vtilde-def-even}) or (\ref{Vtilde-def-odd}) for PBC and in (\ref{Vtilde-HC-DBC}) for DBC)
is crucial to write (\ref{QPRmat t-dep 1d}) and (\ref{QPRmat t-dep 1d DBC});
hence it enters in a highly non-trivial way in the evaluation of the subsystem complexity. 
Instead, it does not affect the complexity for the entire system, 
where both the reference and the target states are pure states, as remarked below (\ref{gamma-TR-global}).

\subsection{Numerical results}
\label{sec:subsysteem-num-res}

Considering the global quench that we are exploring,
in the following we discuss some numerical results 
for the temporal evolution of the subsystem complexity 
of a block $A$ made by $L$ consecutive sites
in harmonic chains made by $N$ sites, 
where either PBC or DBC are imposed.  
We focus on the simplest setup where the reference state is the initial state (hence $t_\textrm{\tiny R}=0$)
and the target state corresponds to a generic value of $t_\textrm{\tiny T}=t  \geqslant 0$ after the quench. 
The remaining parameters are fixed to
$\omega_{0,\textrm{\tiny R}}=\omega_{0,\textrm{\tiny T}}\equiv \omega_{0}$, $\omega_{\textrm{\tiny R}}=\omega_{\textrm{\tiny T}}\equiv \omega$, $\kappa_{\textrm{\tiny R}}=\kappa_{\textrm{\tiny T}}=1$ and $m_{\textrm{\tiny R}}=m_{\textrm{\tiny T}}=1$.
In the case of DBC, we consider both $A$ adjacent to the boundary and separated from it.

In this setup, the subsystem complexity (\ref{c2-complexity-rdm}) can be written as
\be
\label{c2-complexity-rdm-our-case}
\mathcal{C}_A
\,=\,
\frac{1}{2\sqrt{2}}\;
\sqrt{\,
\textrm{Tr} \,\Big\{ \big[
\log \!\big( \gamma_{A}(t) \, \gamma_{A}(0)^{-1} \big)
\big]^2  \Big\}}\;.
\ee

It is natural to introduce also the entanglement entropy $S_A(t)$ and its initial value $S_A(0)$,
which lead to define the increment of the entanglement entropy w.r.t. its initial value, i.e.
\be
\label{Delta-S_A-def}
\Delta S_A \equiv S_A(t) - S_A(0) 
\ee
where $S_A(t)$ and $S_A(0)$ can be evaluated from the symplectic spectrum of 
$\gamma_{A}(t)$ and of $\gamma_{A}(0)$ respectively in the standard way 
\cite{Peschel_2009,Eisert:2008ur,Weedbrook12b,Peschel03,Audenaert:2002xfl,Plenio:2004he,Cramer:2005mx,Rajabpour_14,Cotler:2016acd,Hackl:2017ndi}.

\begin{figure}[t!]
\subfigure
{\hspace{-1.6cm}
\includegraphics[width=.58\textwidth]{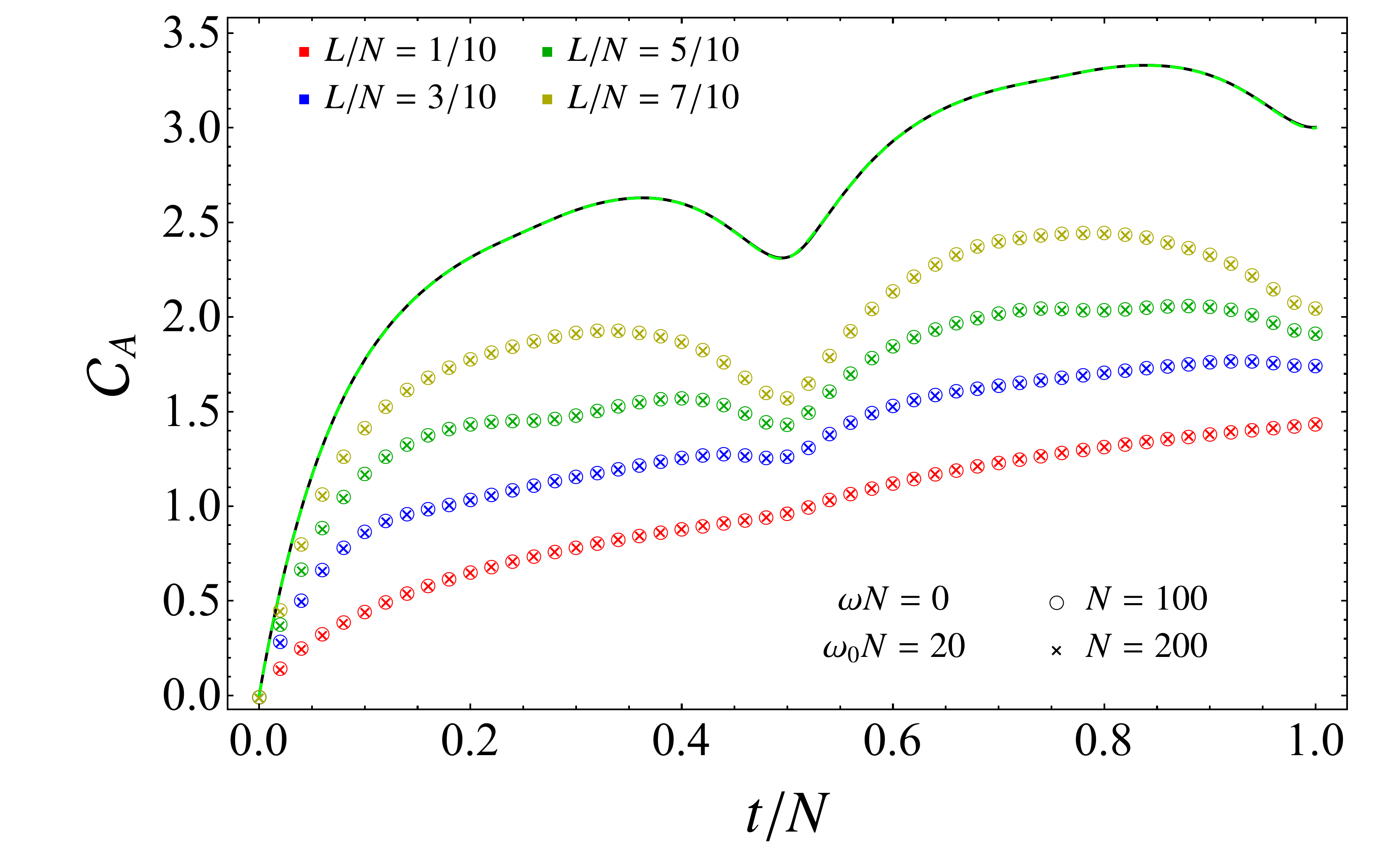}}
\subfigure
{
\hspace{-.7cm}\includegraphics[width=.58\textwidth]{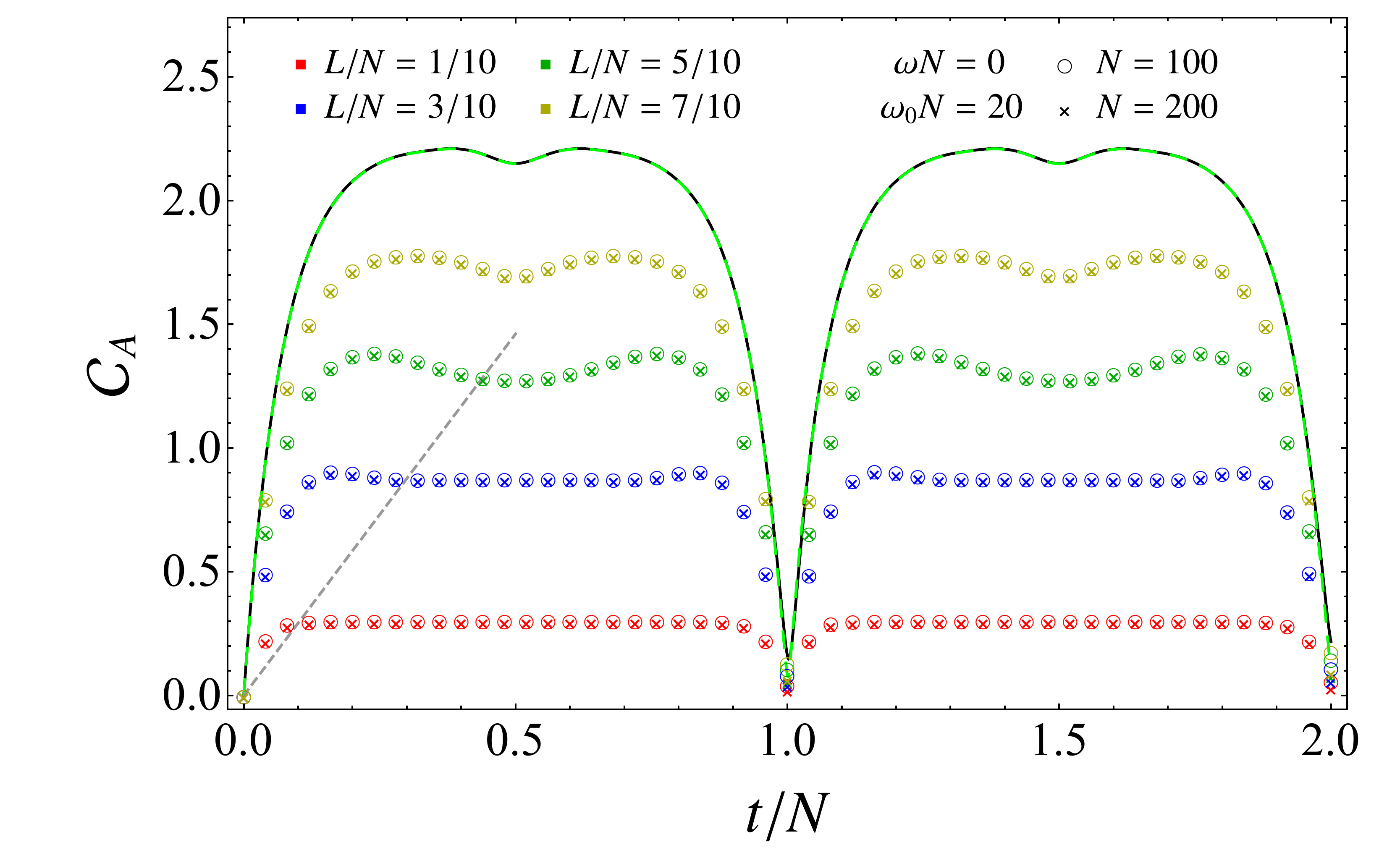}}
\caption{
Temporal evolution of $\mathcal{C}_A$ in (\ref{c2-complexity-rdm-our-case})
after the global quench with gapless evolution Hamiltonian and $\omega_0 N =20$,
for a block $A$ made by $L$ consecutive sites in harmonic chains
with  either PBC (left panels) or DBC (right panels) made by $N$ sites
(in the latter case $A$ is adjacent to a boundary). 
When $L=N$, the complexity (\ref{comp-pure-global-DBCPBC}) is shown 
for $N=100$ (solid black lines) and $N=200$ (dashed green lines).
}
\vspace{0.4cm}
\label{fig:MixedStateGlobalMasslessEvolutionDimensionlessNoentropy}
\end{figure}

\begin{figure}[htbp!]
\vspace{-1cm}
\subfigure
{
\hspace{-1.6cm}\includegraphics[width=.58\textwidth]{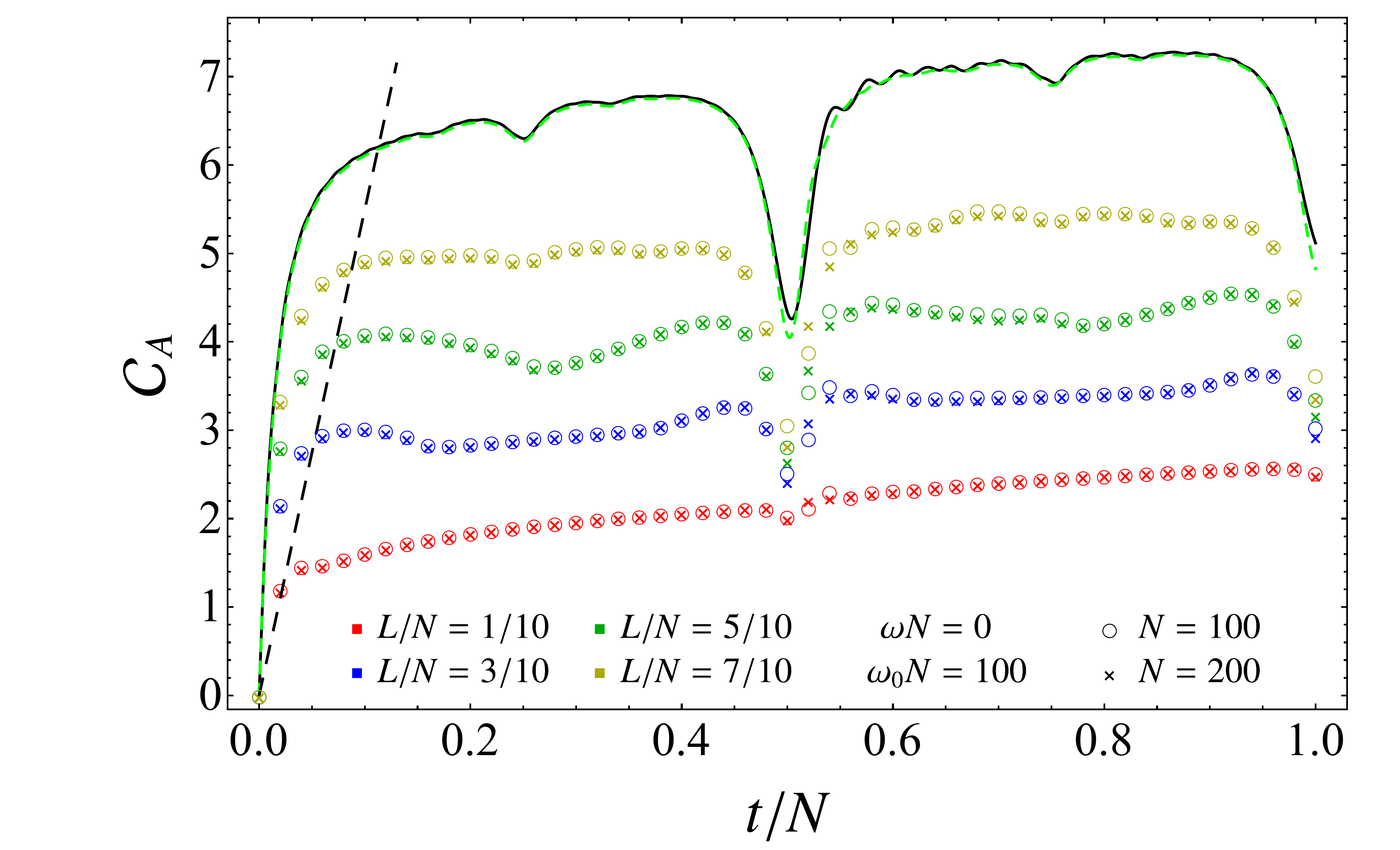}}
\subfigure
{\hspace{-.7cm}
\includegraphics[width=.58\textwidth]{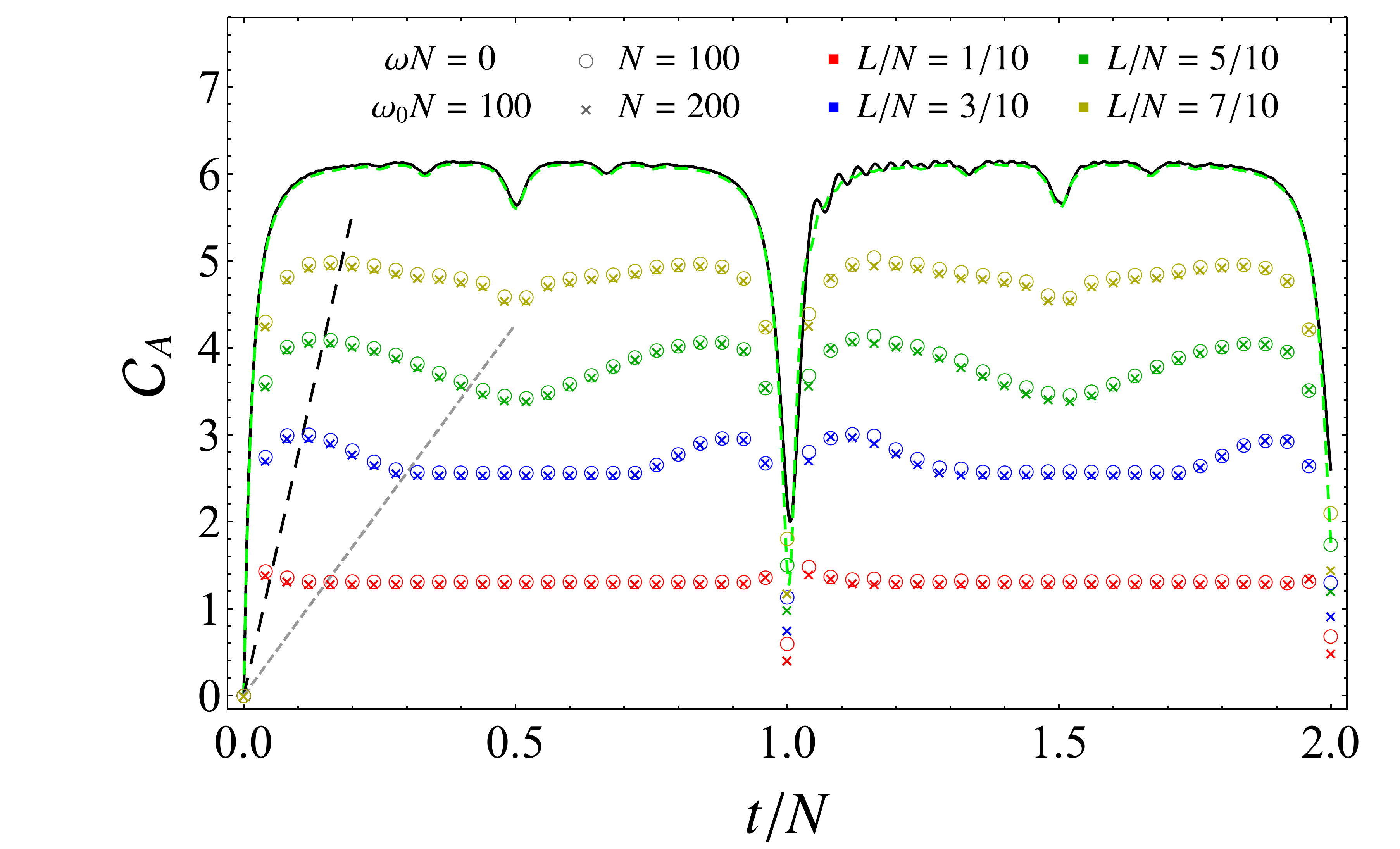}}
\subfigure
{
\hspace{-1.6cm}\includegraphics[width=.58\textwidth]{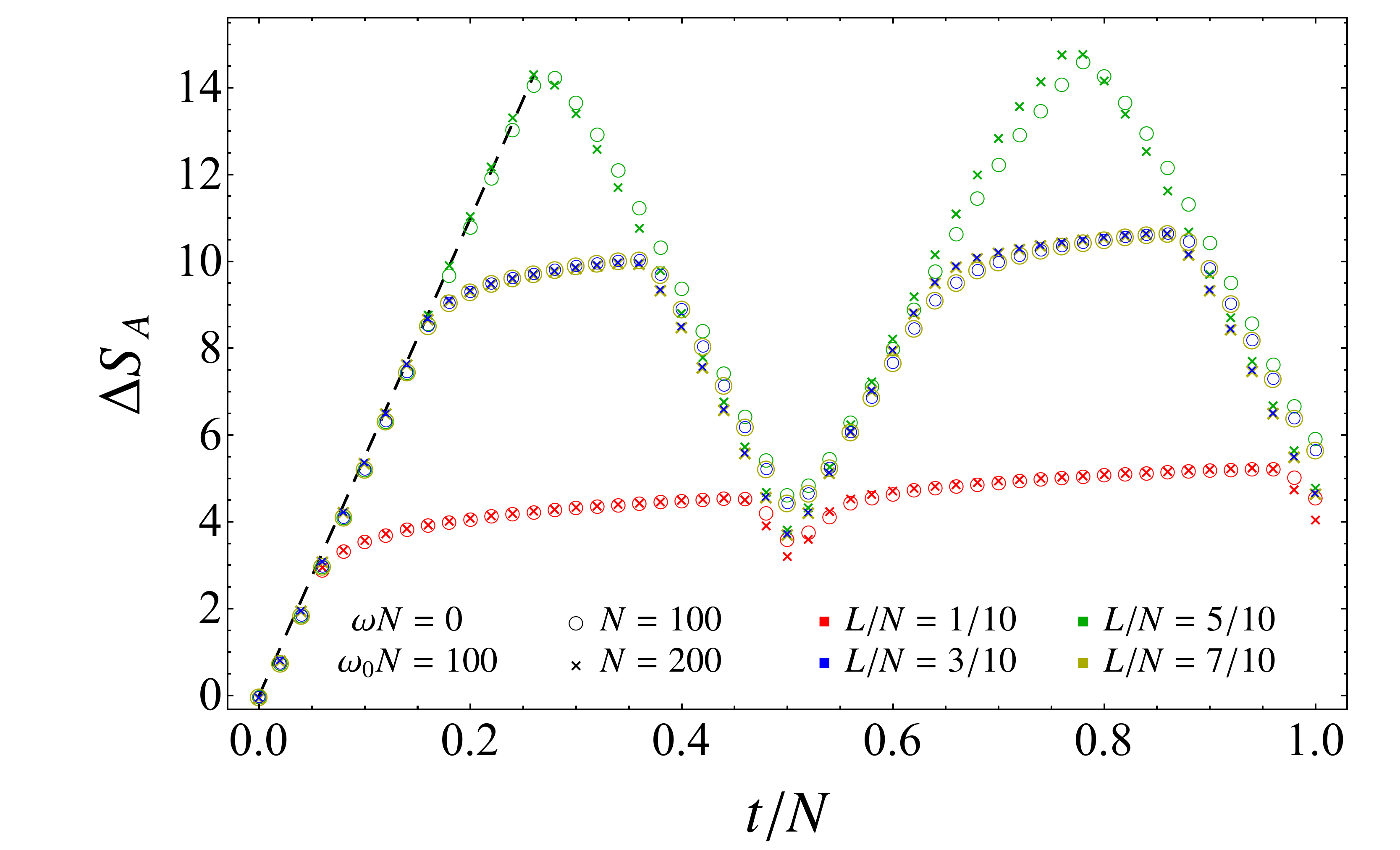}}
\subfigure
{\hspace{-.7cm}
\includegraphics[width=.58\textwidth]{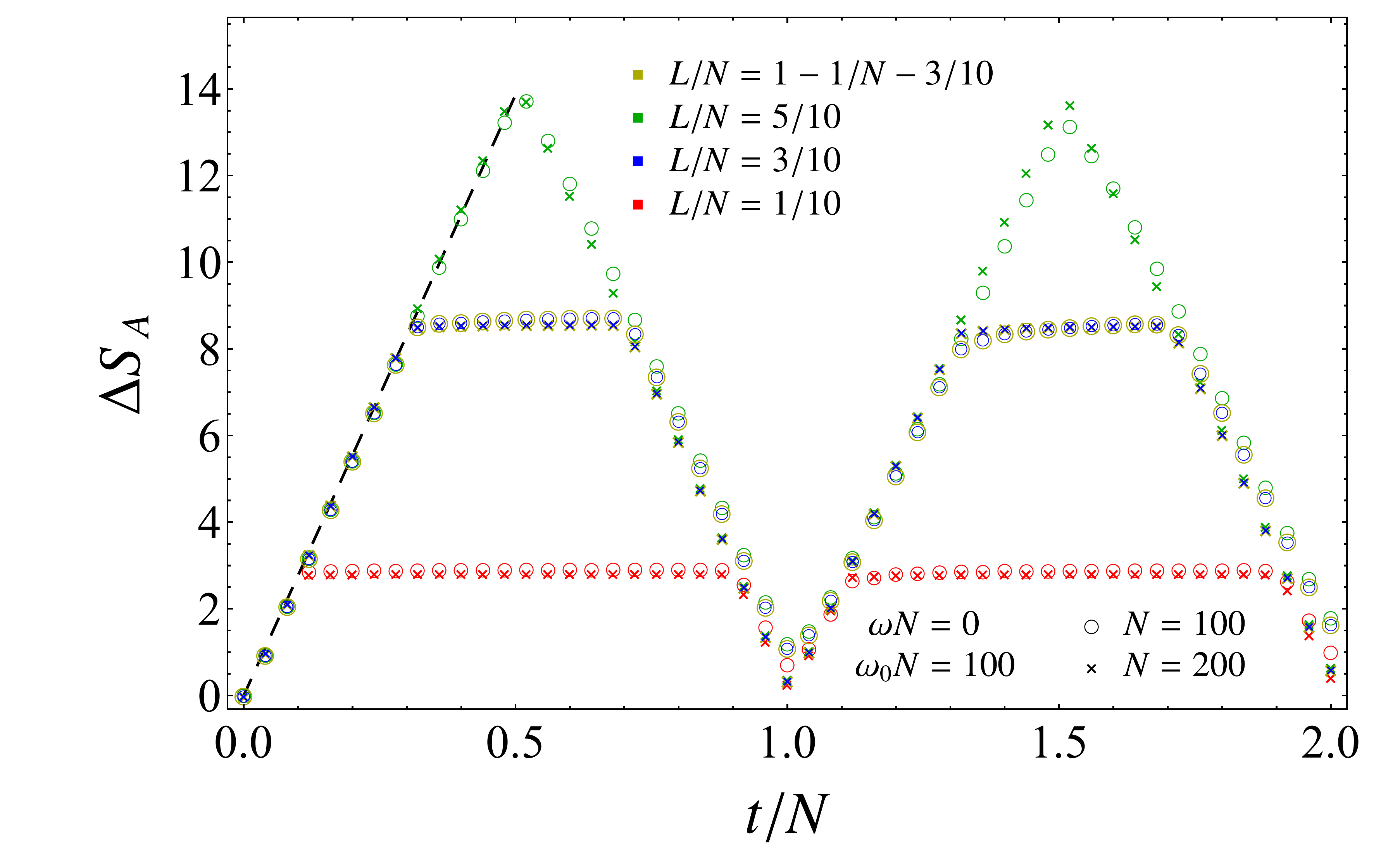}}
\subfigure
{
\hspace{-1.6cm}\includegraphics[width=.58\textwidth]{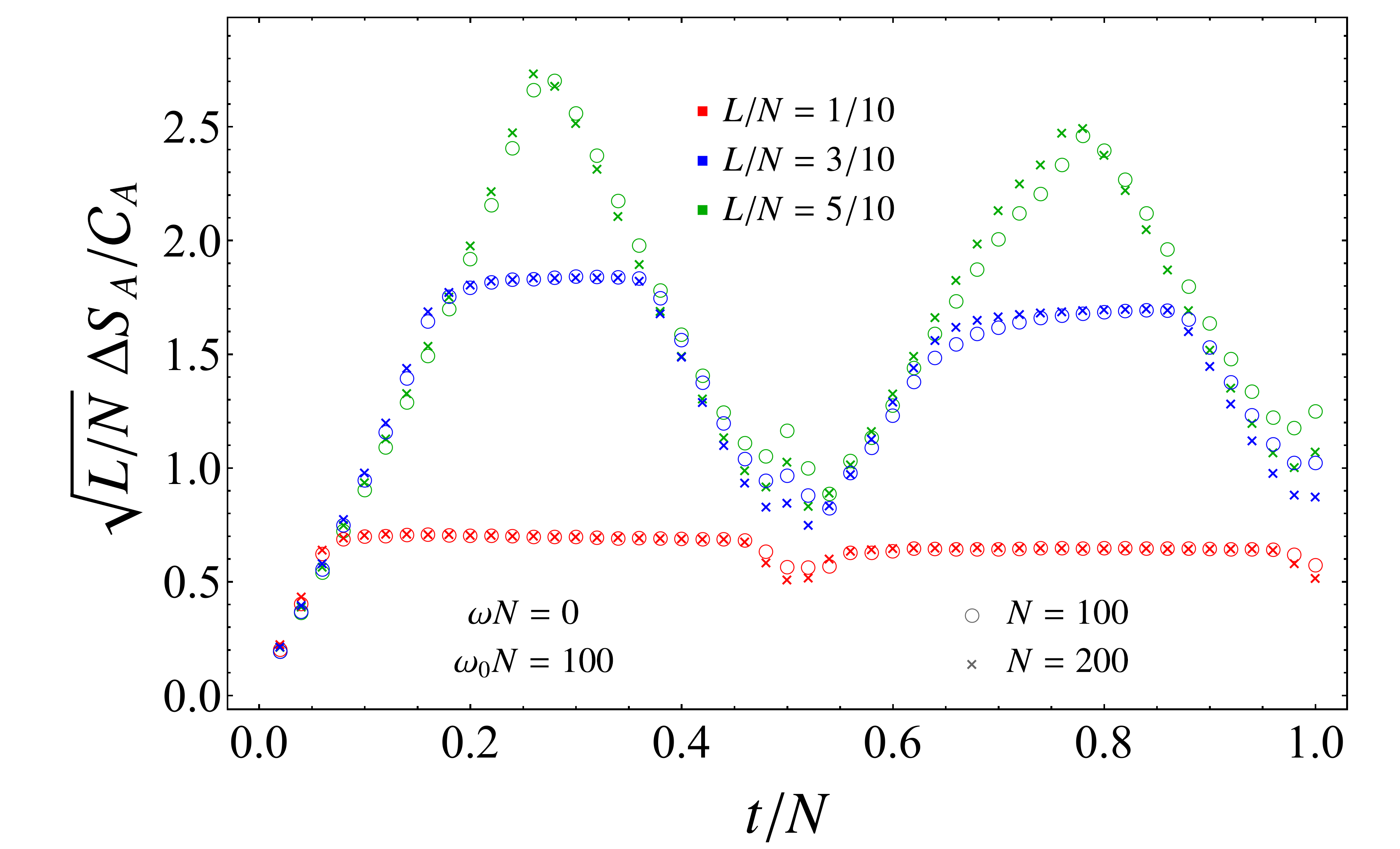}}
\subfigure
{\hspace{-.7cm}
\includegraphics[width=.58\textwidth]{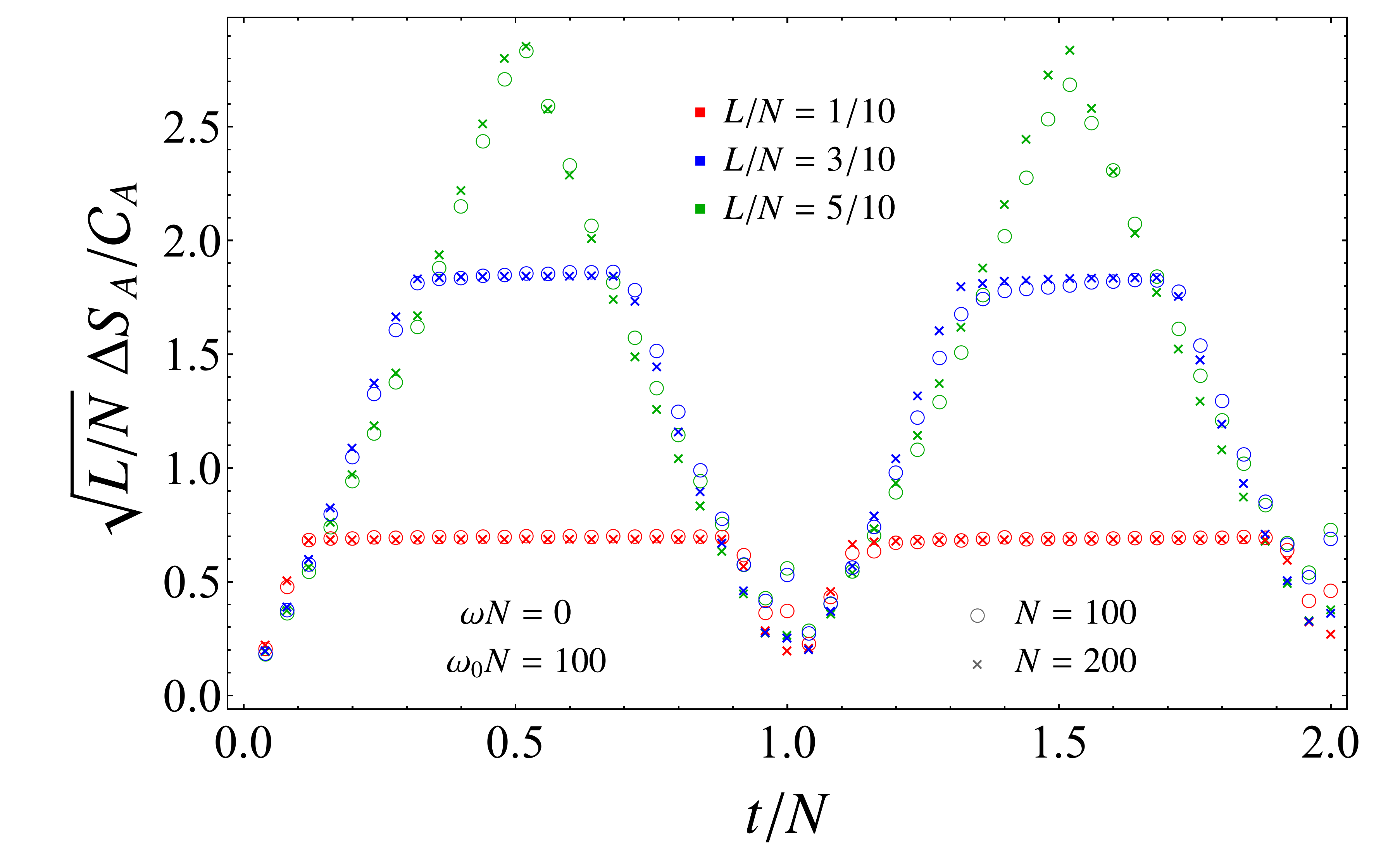}}
\caption{Temporal evolution 
of $\mathcal{C}_A$ in (\ref{c2-complexity-rdm-our-case}) (top panels), 
of $\Delta S_A$ in (\ref{Delta-S_A-def}) (middle panels) 
and of $\sqrt{L/N}\,\Delta S_A/\mathcal{C}_A $ (bottom panels)
after the global quench with gapless evolution Hamiltonian and $\omega_0 N =100$,
for a block $A$ made by $L$ consecutive sites in a harmonic chains
with  either PBC (left panels) or DBC (right panels) made by $N$ sites
(in the latter case $A$ is adjacent to a boundary).
When $L=N$ the complexity (\ref{comp-pure-global-DBCPBC}) is shown 
for $N=100$ (solid black lines) and $N=200$ (dashed green lines).
}
\vspace{0.4cm}
\label{fig:MixedStateGlobalMasslessEvolutionDimensionless}
\end{figure}

\begin{figure}[htbp!]
\vspace{-1cm}
\subfigure
{\hspace{-1.6cm}
\includegraphics[width=.58\textwidth]{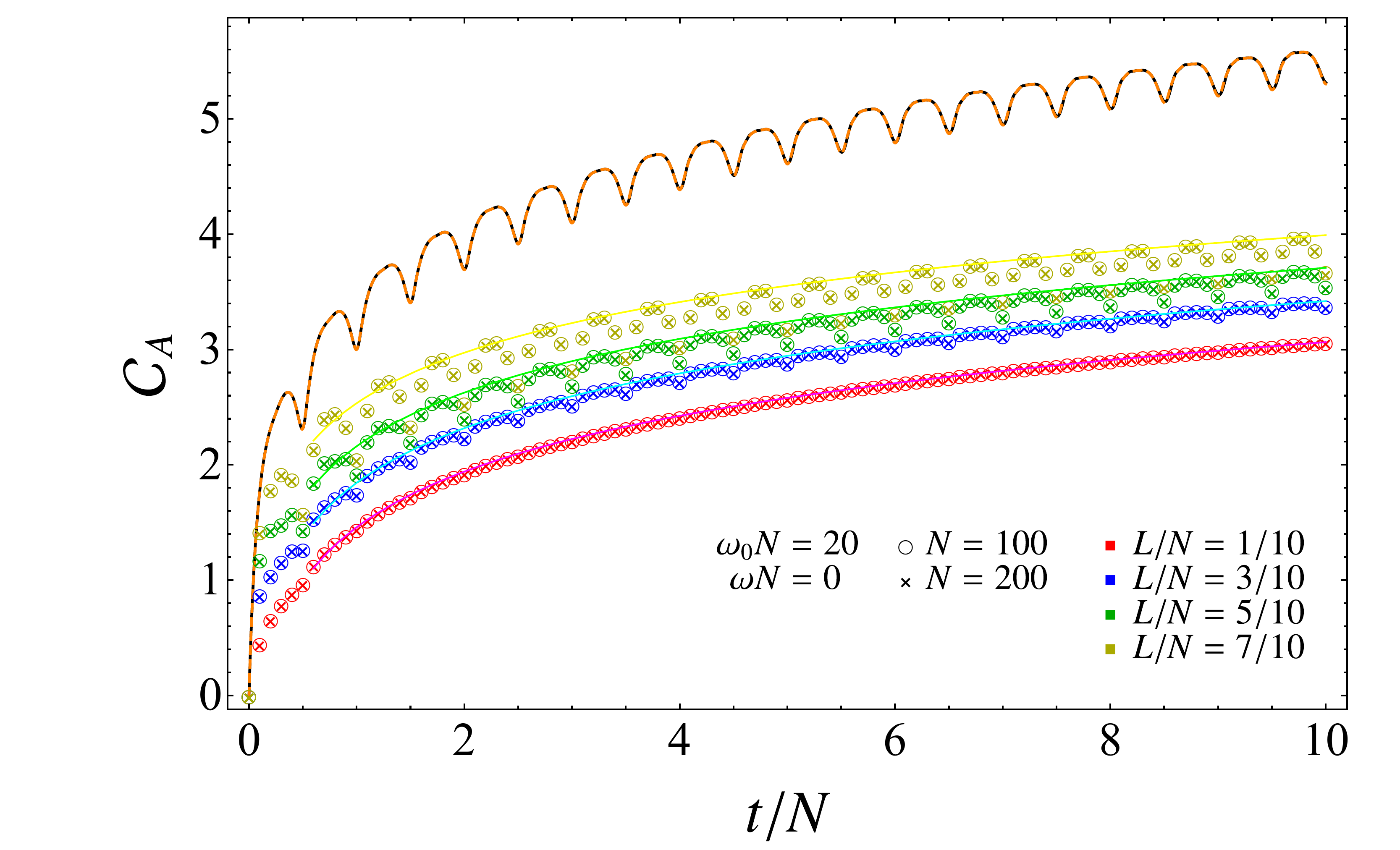}}
\subfigure
{
\hspace{-.7cm}\includegraphics[width=.58\textwidth]{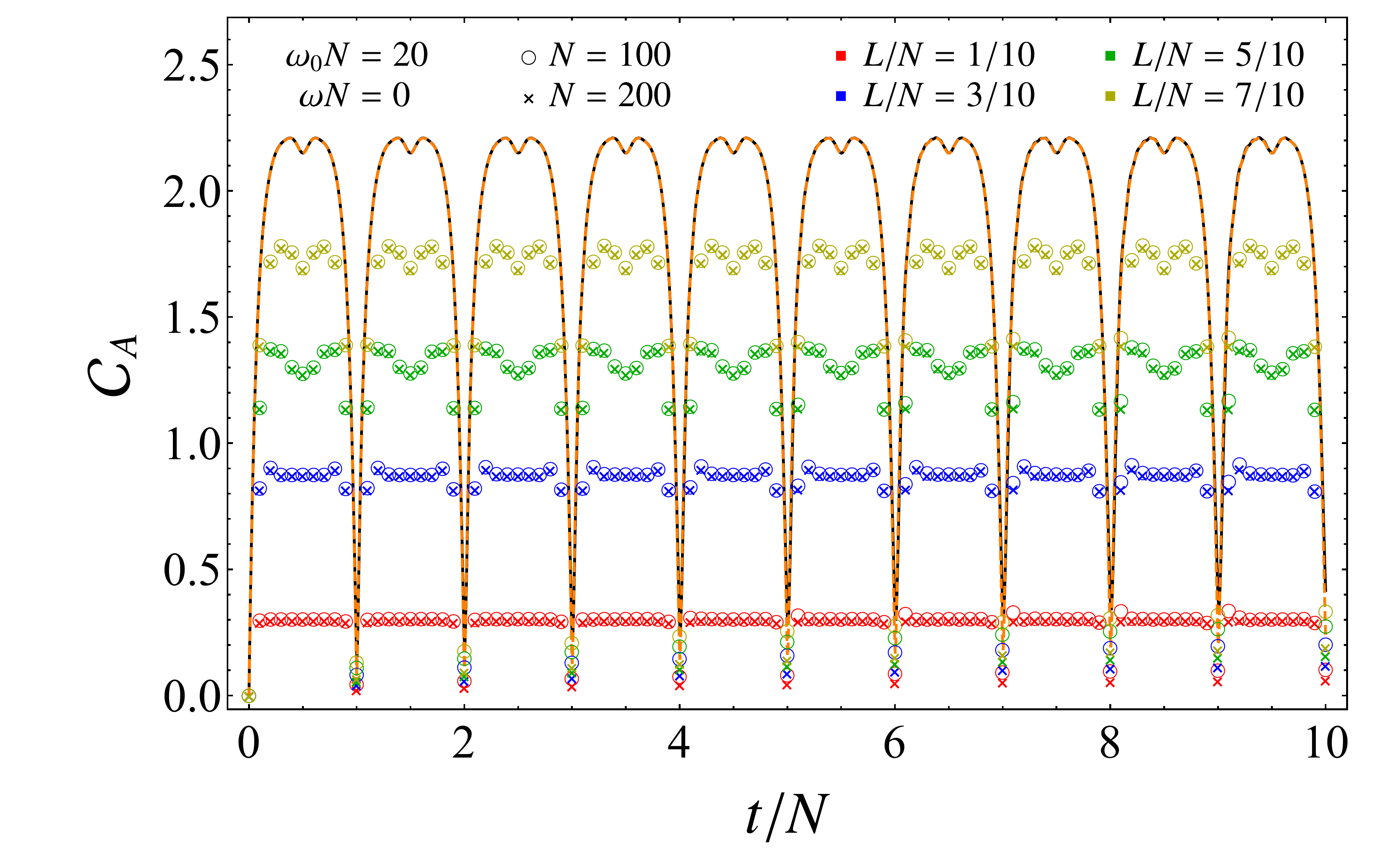}}
\subfigure
{
\hspace{-1.6cm}\includegraphics[width=.58\textwidth]{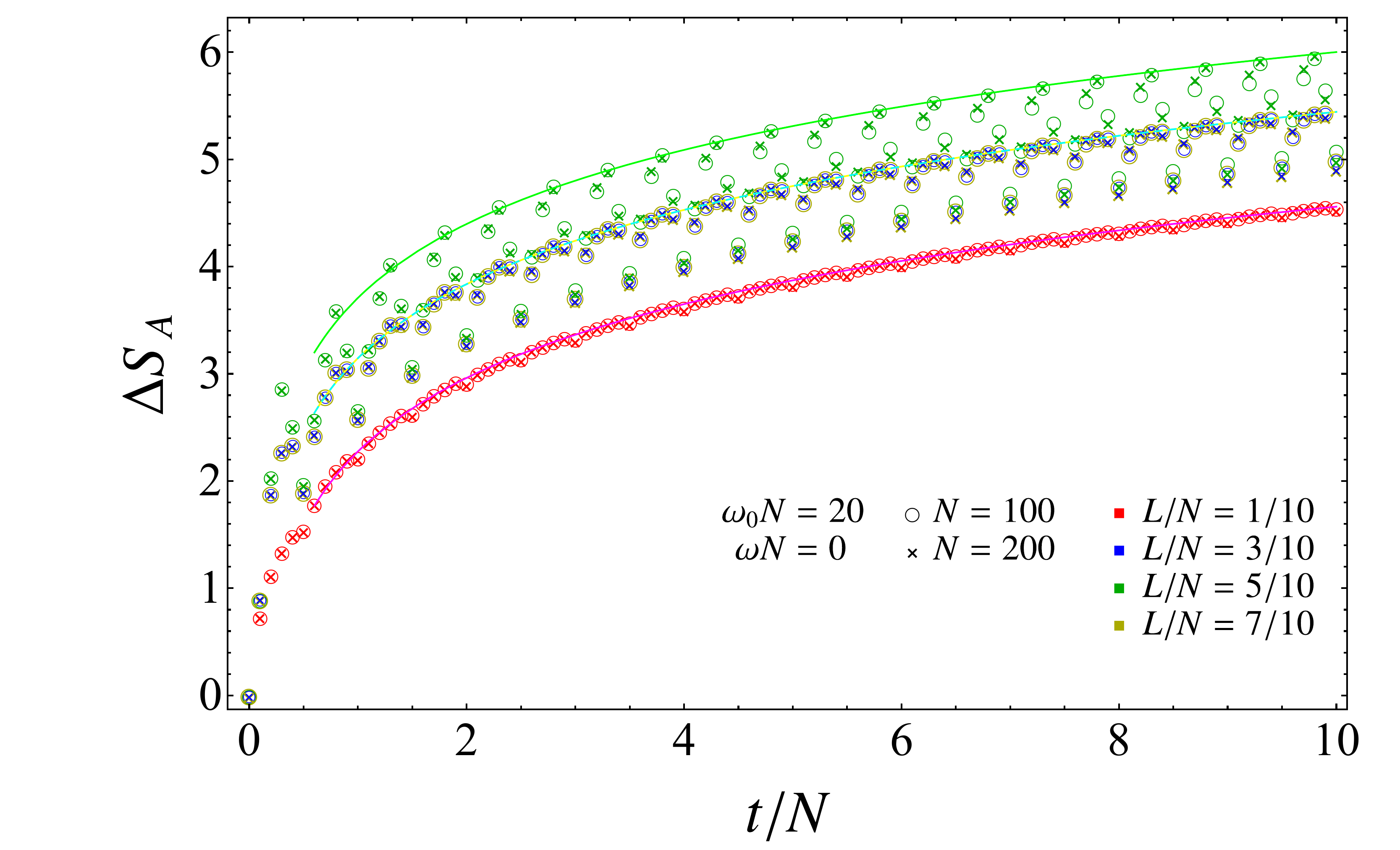}}
\subfigure
{\hspace{-.7cm}
\includegraphics[width=.58\textwidth]{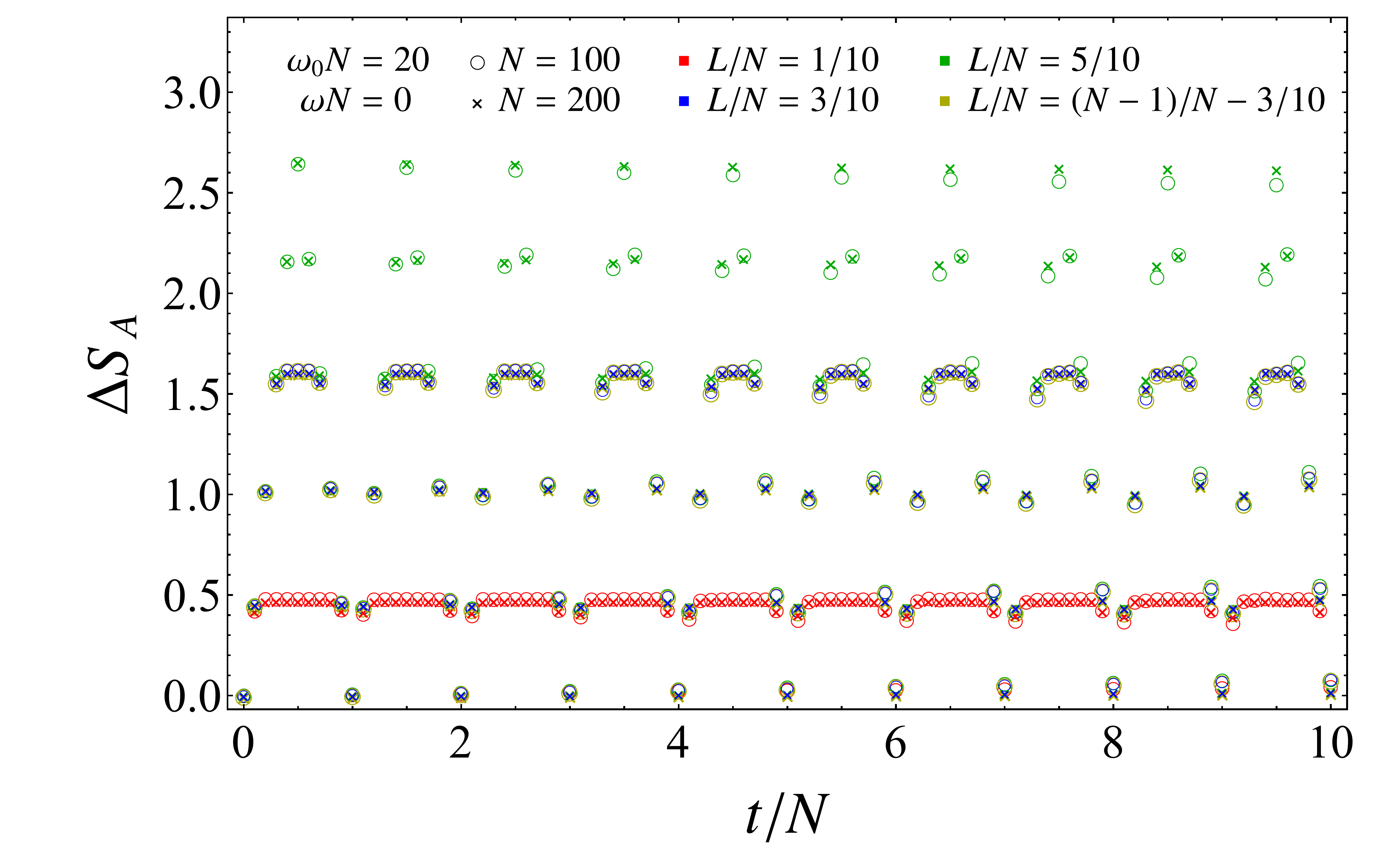}}
\subfigure
{
\hspace{-1.6cm}\includegraphics[width=.58\textwidth]{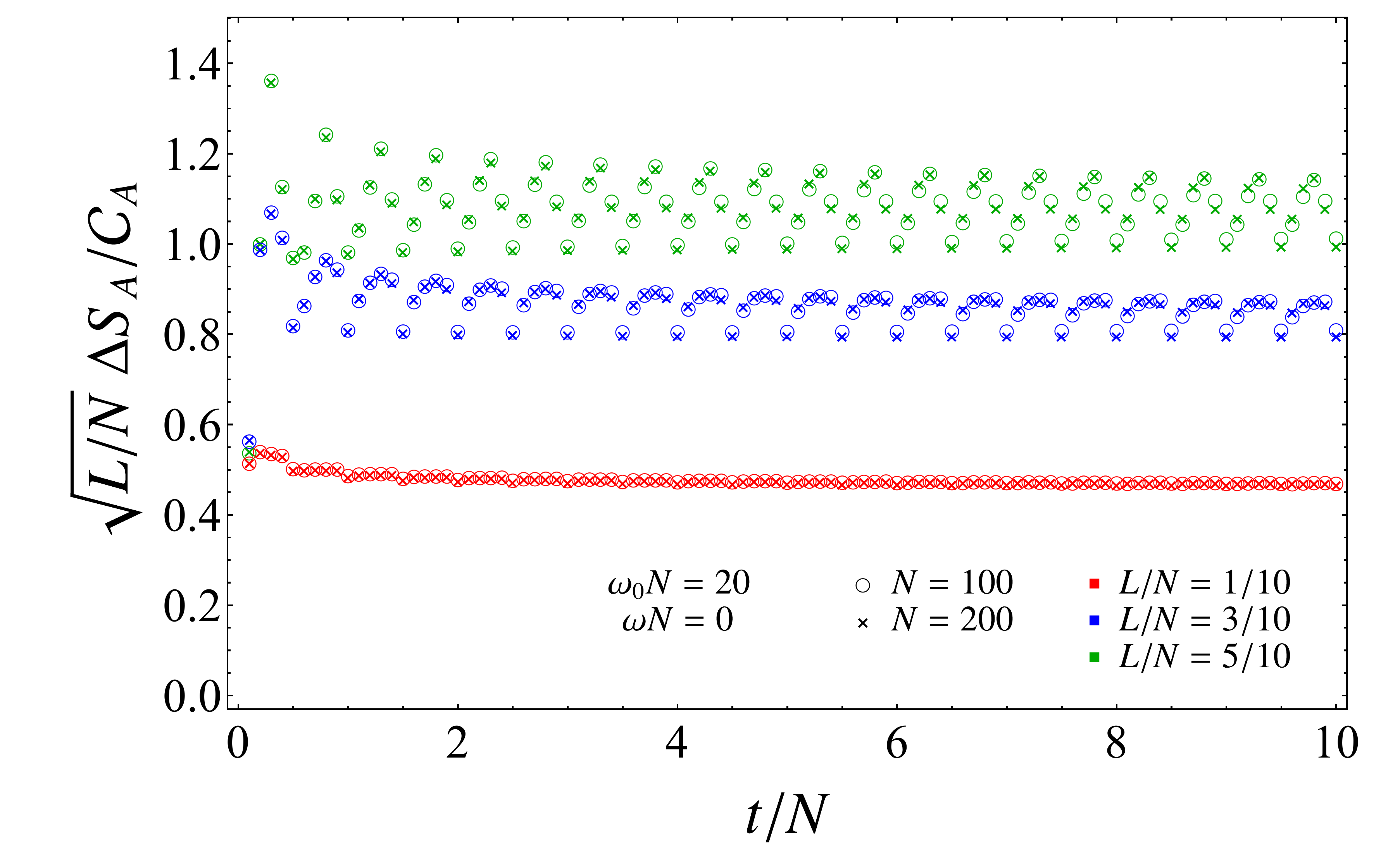}}
\subfigure
{\hspace{-.7cm}
\includegraphics[width=.58\textwidth]{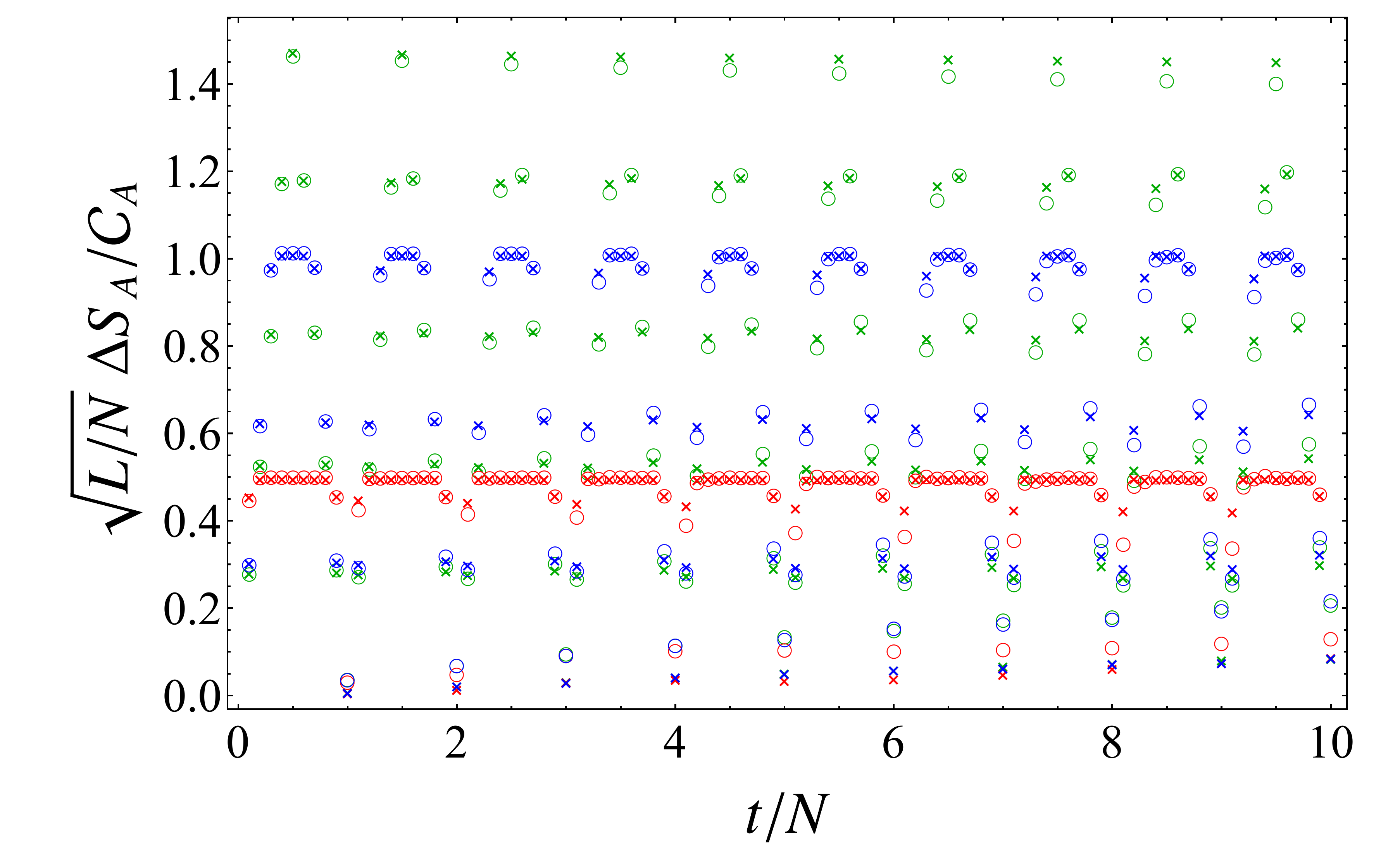}}
\caption{Temporal evolution 
of $\mathcal{C}_A$ in (\ref{c2-complexity-rdm-our-case}) (top panels), 
of $\Delta S_A$ in (\ref{Delta-S_A-def}) (middle panels) 
and of $\sqrt{L/N}\,\Delta S_A/\mathcal{C}_A$ (bottom panels)
after the global quench with a gapless evolution Hamiltonian and $\omega_0 N =20$,
for harmonic chains with either PBC (left panels) or DBC (right panels),
in the same setups of 
Fig.\,\ref{fig:MixedStateGlobalMasslessEvolutionDimensionlessNoentropy}
are considered. 
}
\vspace{.4cm}
\label{fig:CompvsEntStateGlobalMasslessEvolutionDimensionless}
\end{figure}

\begin{figure}[t!]
\vspace{.1cm}
\subfigure
{\hspace{-1.6cm}
\includegraphics[width=.58\textwidth]{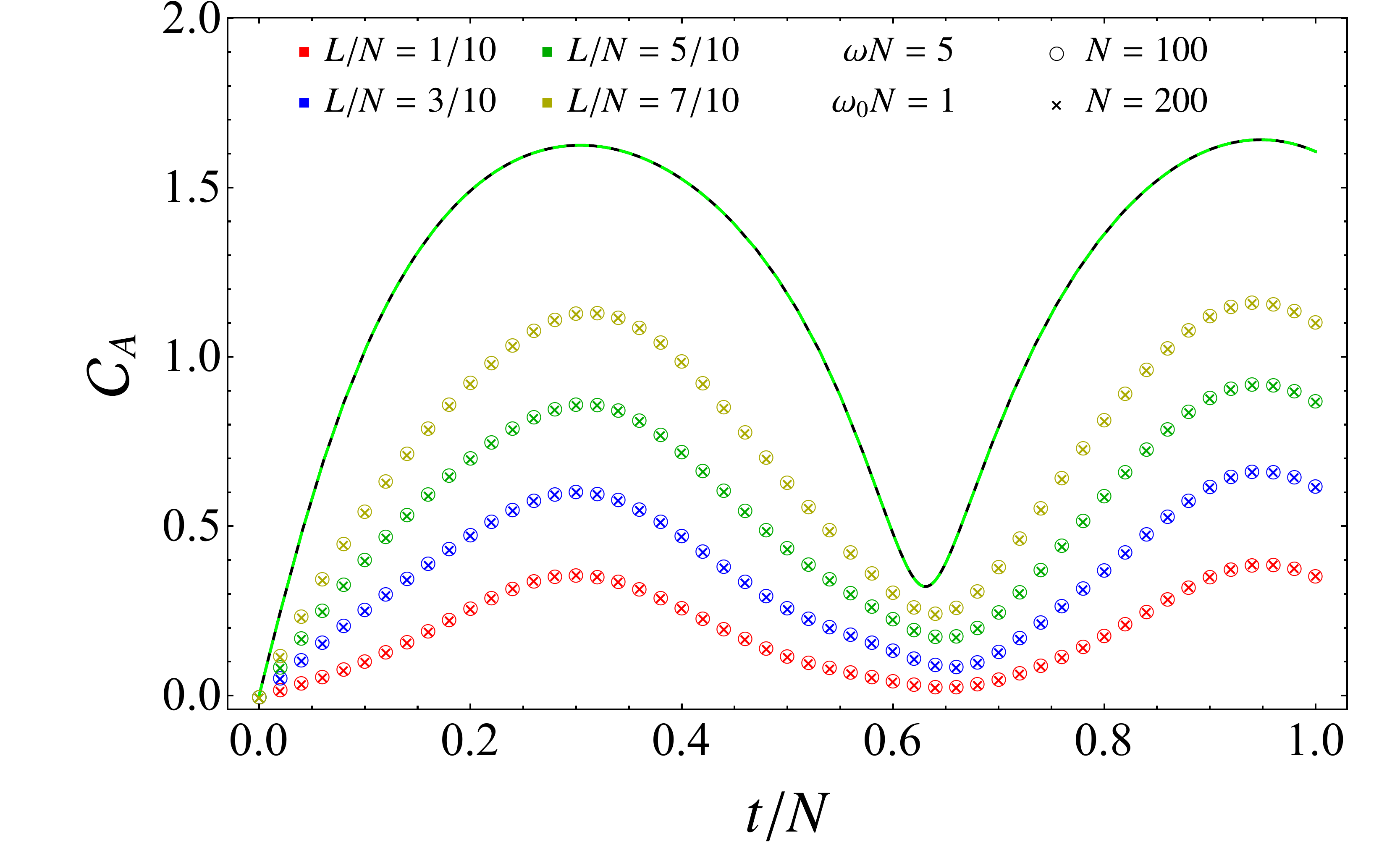}}
\subfigure
{
\hspace{-.7cm}\includegraphics[width=.58\textwidth]{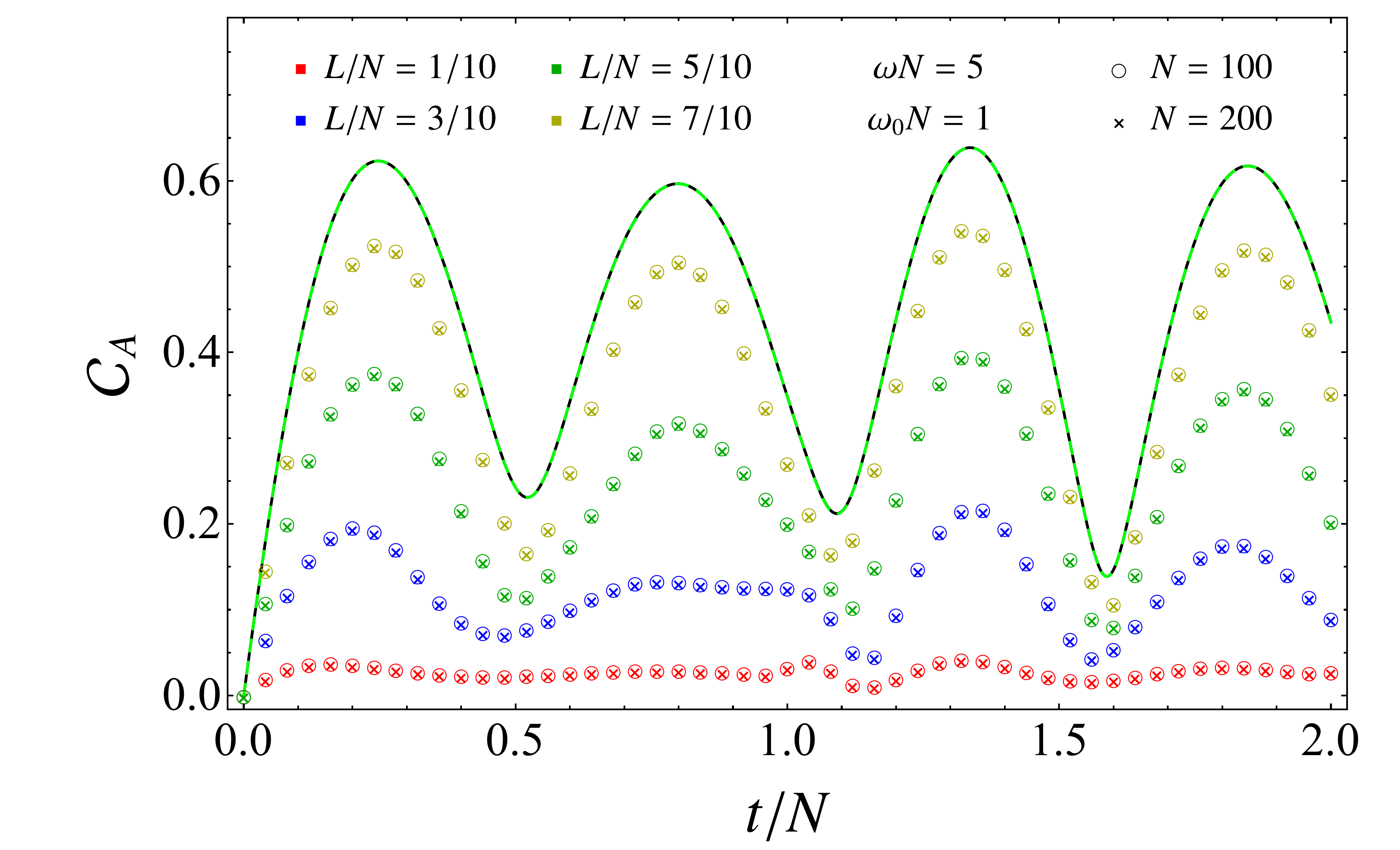}}
\caption{
Temporal evolution of $\mathcal{C}_A$ in (\ref{c2-complexity-rdm-our-case}) 
after the global quench with a gapped evolution Hamiltonian
for a block $A$ made by $L$ consecutive sites in harmonic chains
with  either PBC (left panels) or DBC (right panels) made by $N$ sites
(in the latter case $A$ is adjacent to a boundary of the segment). 
When $L=N$ the complexity (\ref{comp-pure-global-DBCPBC}) is shown 
for $N=100$ (solid black lines) and $N=200$ (dashed green lines).
}
\vspace{0.4cm}
\label{fig:MixedStateGlobalMassiveEvolutionDimensionlessNoEntropy}
\end{figure}

\begin{figure}[t!]
\vspace{.1cm}
\subfigure
{
\hspace{-1.6cm}\includegraphics[width=.58\textwidth]{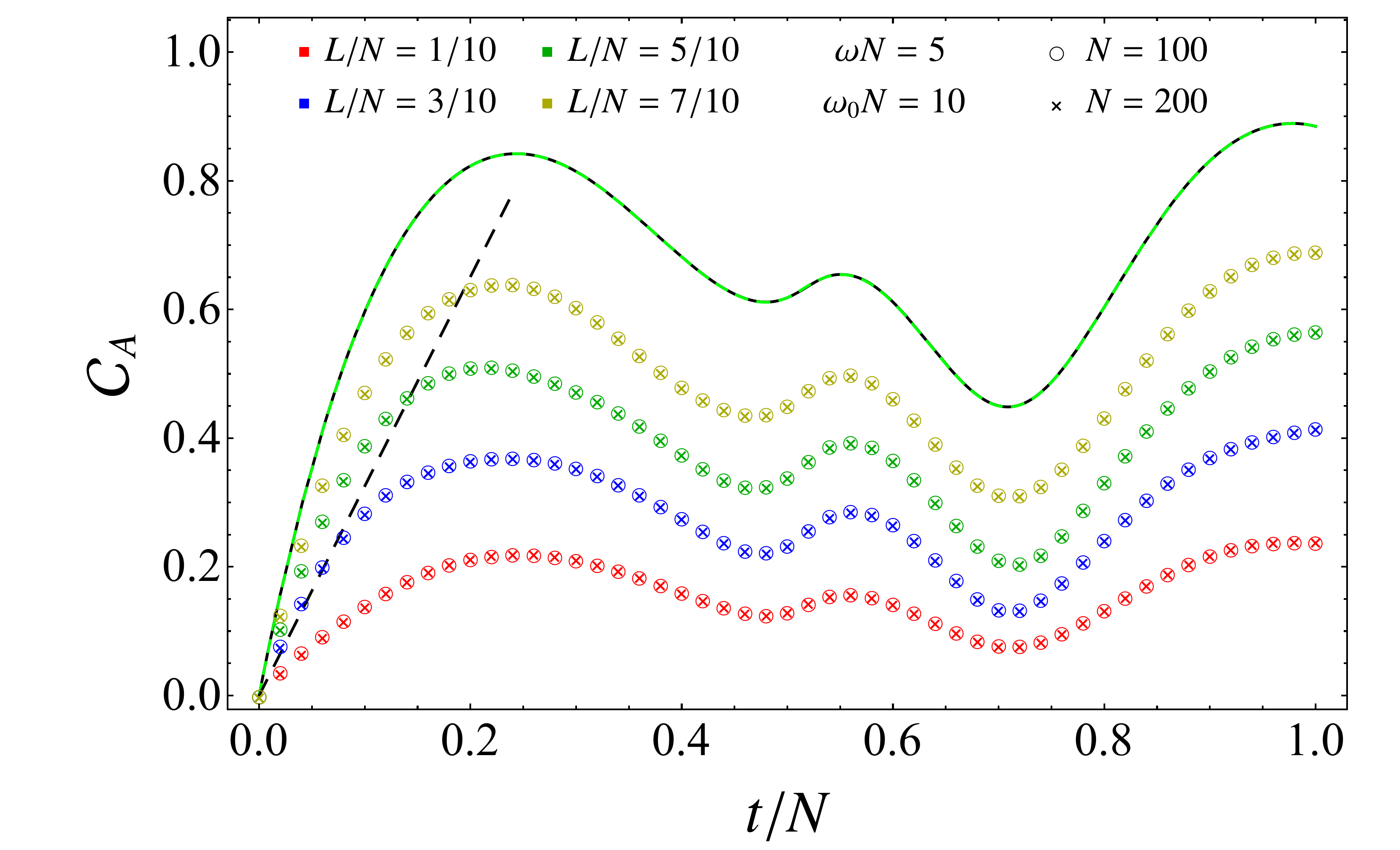}}
\subfigure
{\hspace{-.7cm}
\includegraphics[width=.58\textwidth]{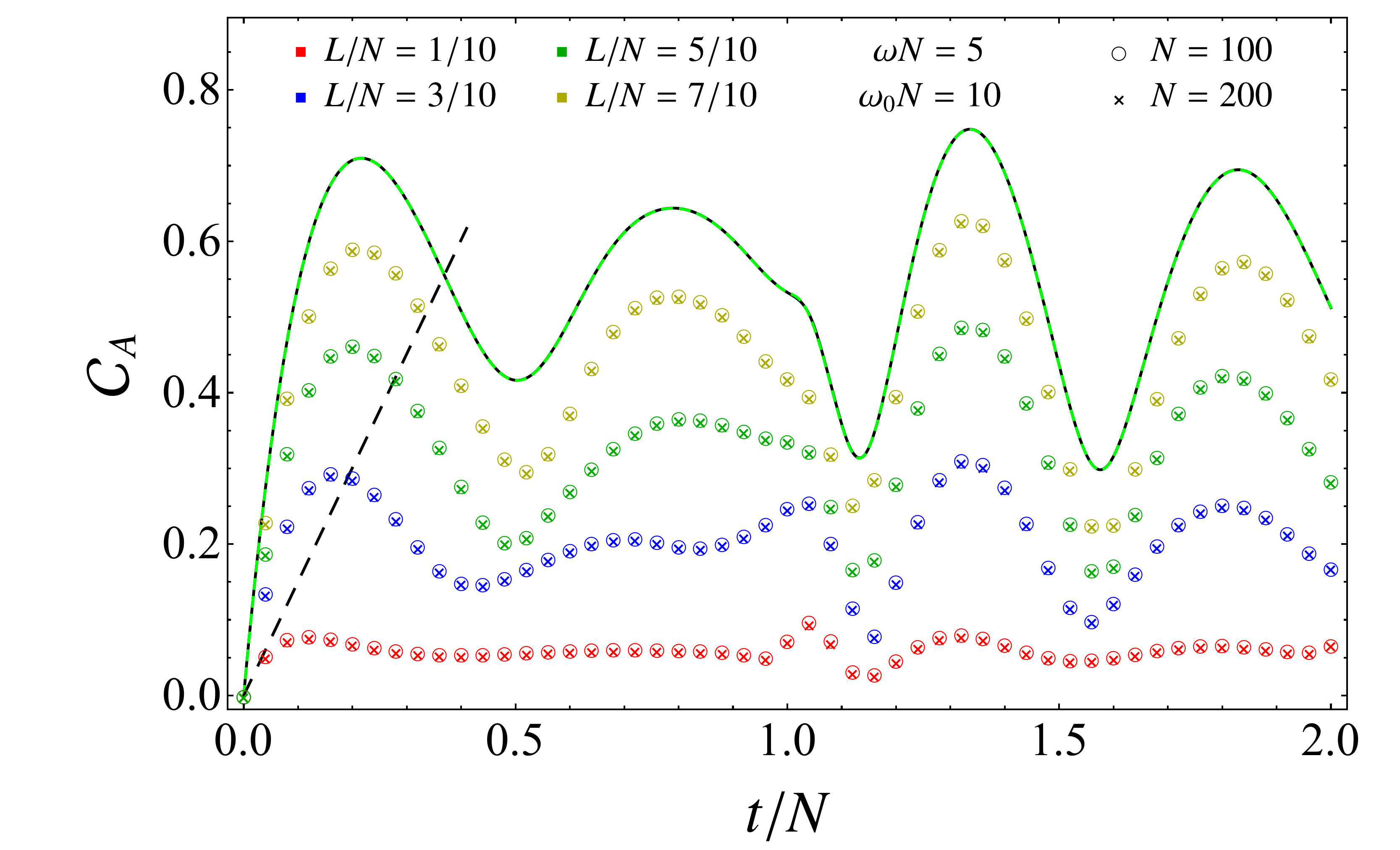}}
\subfigure
{
\hspace{-1.6cm}\includegraphics[width=.58\textwidth]{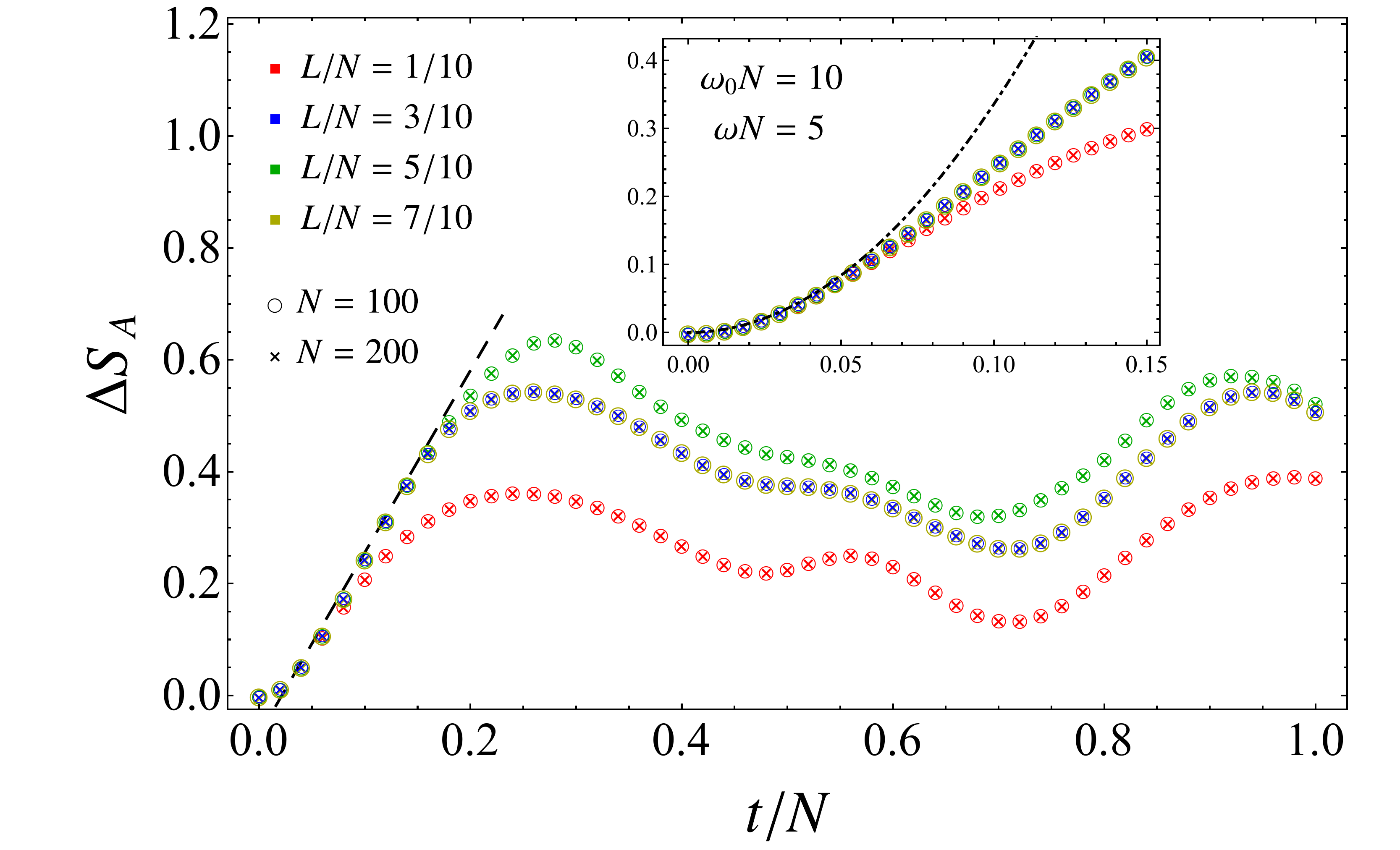}}
\subfigure
{\hspace{-.7cm}
\includegraphics[width=.58\textwidth]{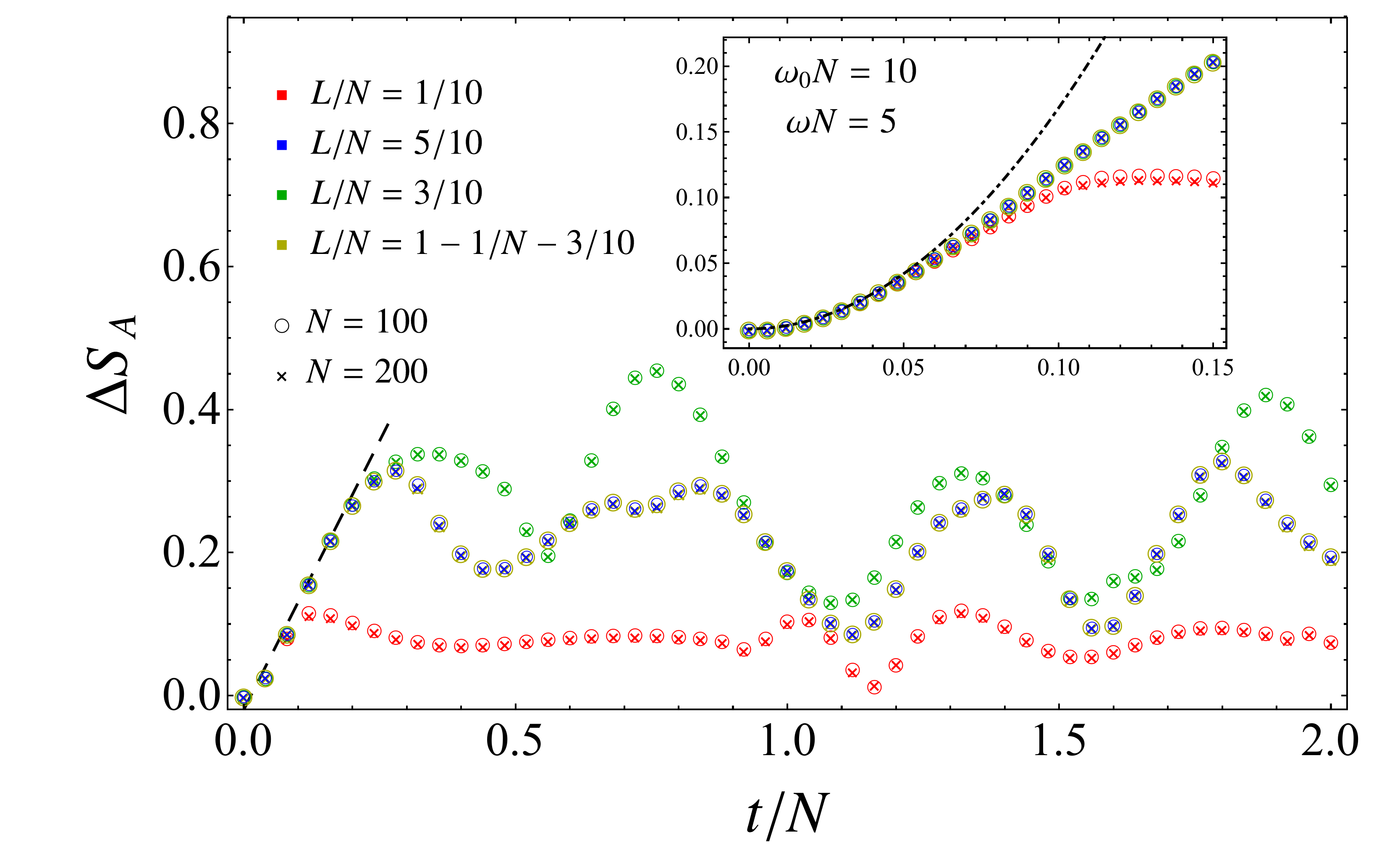}}
\subfigure
{\hspace{-1.6cm}
\includegraphics[width=.58\textwidth]{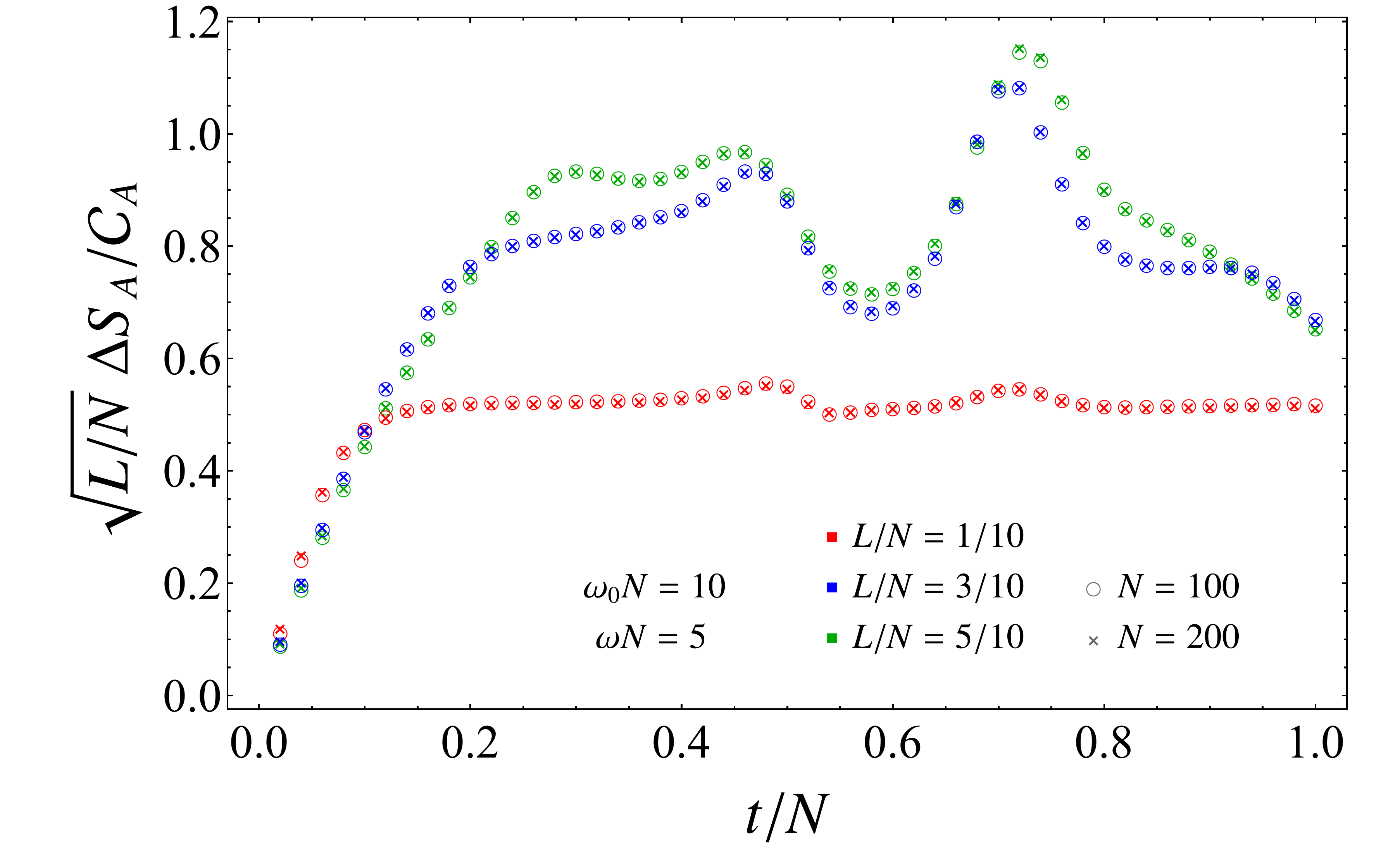}}
\subfigure
{
\hspace{-.7cm}\includegraphics[width=.58\textwidth]{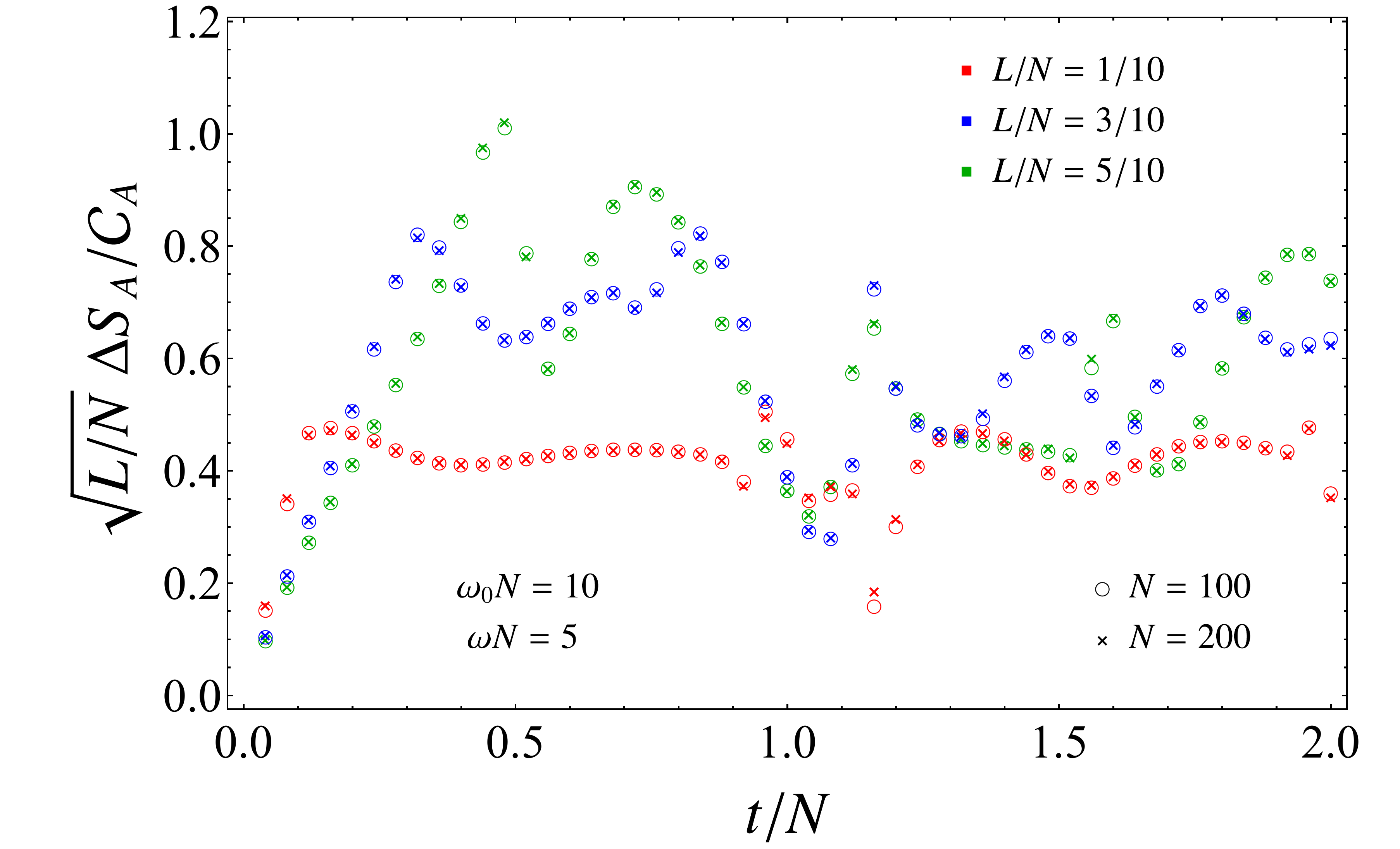}}
\caption{
Temporal evolution after the global quench with a gapped evolution Hamiltonian
of $\mathcal{C}_A$ in (\ref{c2-complexity-rdm-our-case}) 
(top panels), of $\Delta S_A$ in (\ref{Delta-S_A-def}) (middle panels) 
and of $ \sqrt{L/N}\,\Delta S_A/\mathcal{C}_A $ (bottom panels)
for a block $A$ made by $L$ consecutive sites in harmonic chains 
with  either PBC (left panels) or DBC (right panels) made by $N$ sites
(in the latter case $A$ is adjacent to a boundary of the segment).
When $L=N$, the complexity (\ref{comp-pure-global-DBCPBC}) is shown 
for $N=100$ (solid black lines) and $N=200$ (dashed green lines).
}
\vspace{0.4cm}
\label{fig:MixedStateGlobalMassiveEvolutionDimensionless}
\end{figure}

\begin{figure}[htbp!]
\vspace{-1cm}
\subfigure
{\hspace{-1.2cm}
\includegraphics[width=.54\textwidth]{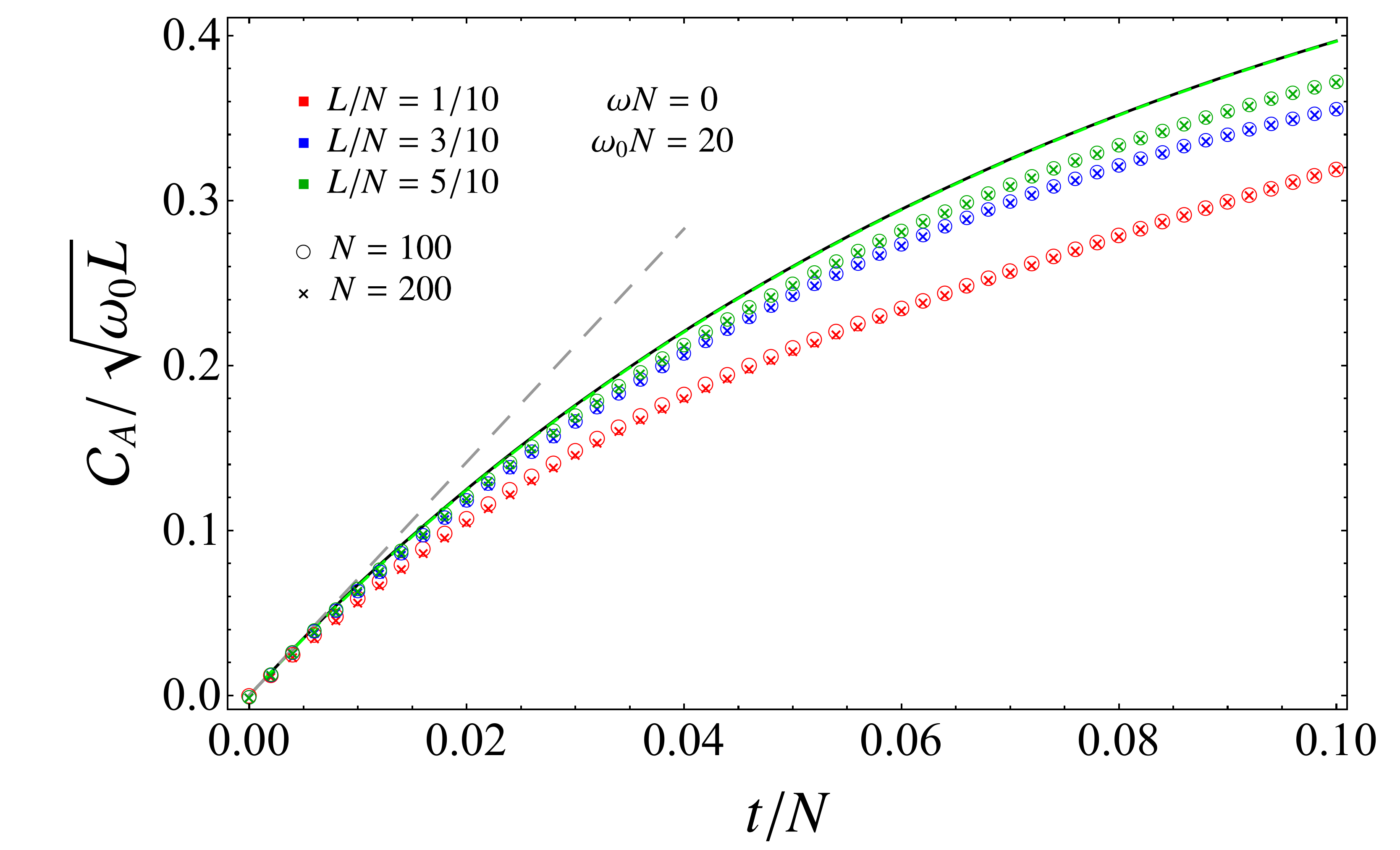}}
\vspace{-.25cm}
\subfigure
{
\hspace{0.cm}\includegraphics[width=.54\textwidth]{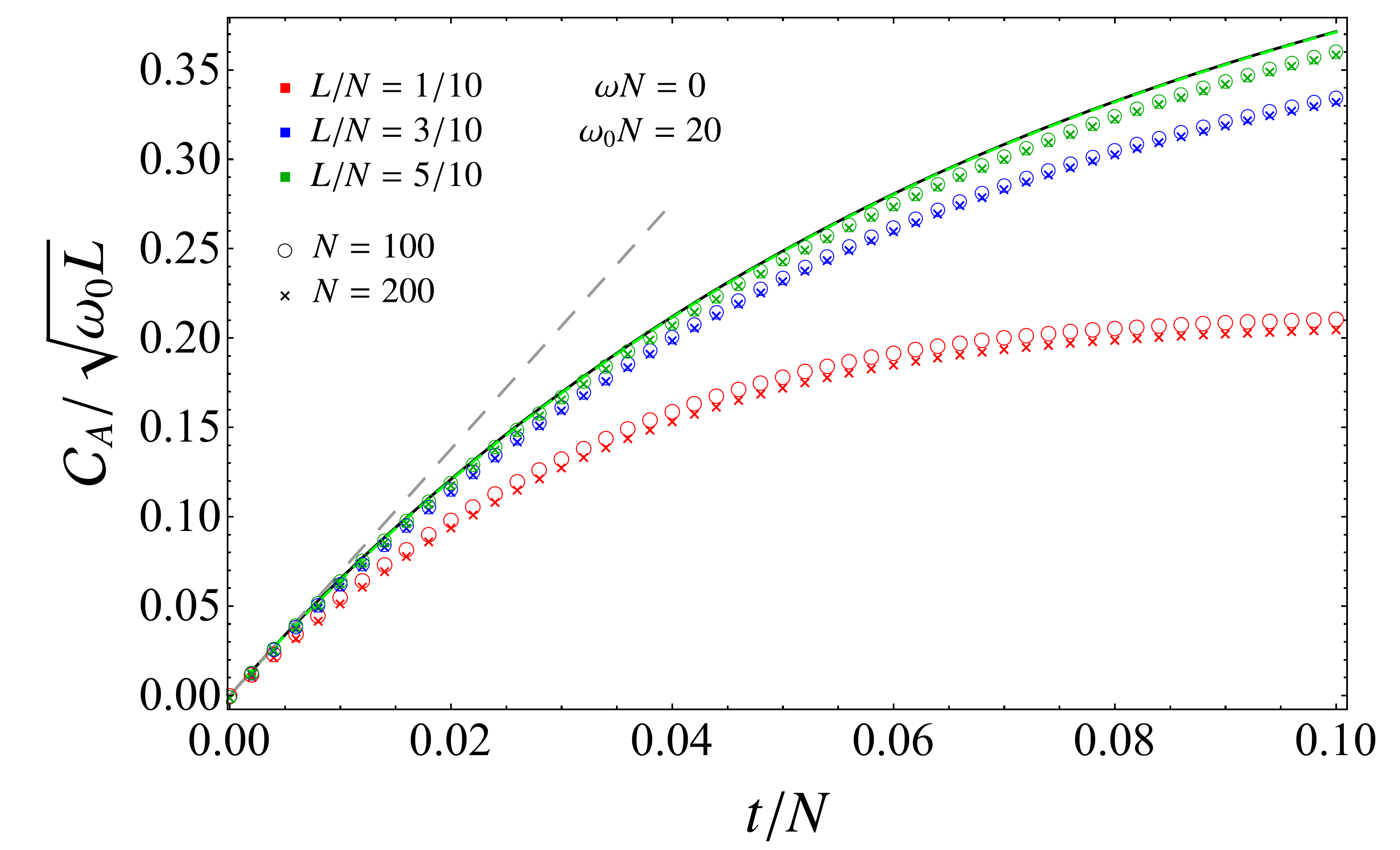}}
\subfigure
{
\hspace{-1.2cm}\includegraphics[width=.54\textwidth]{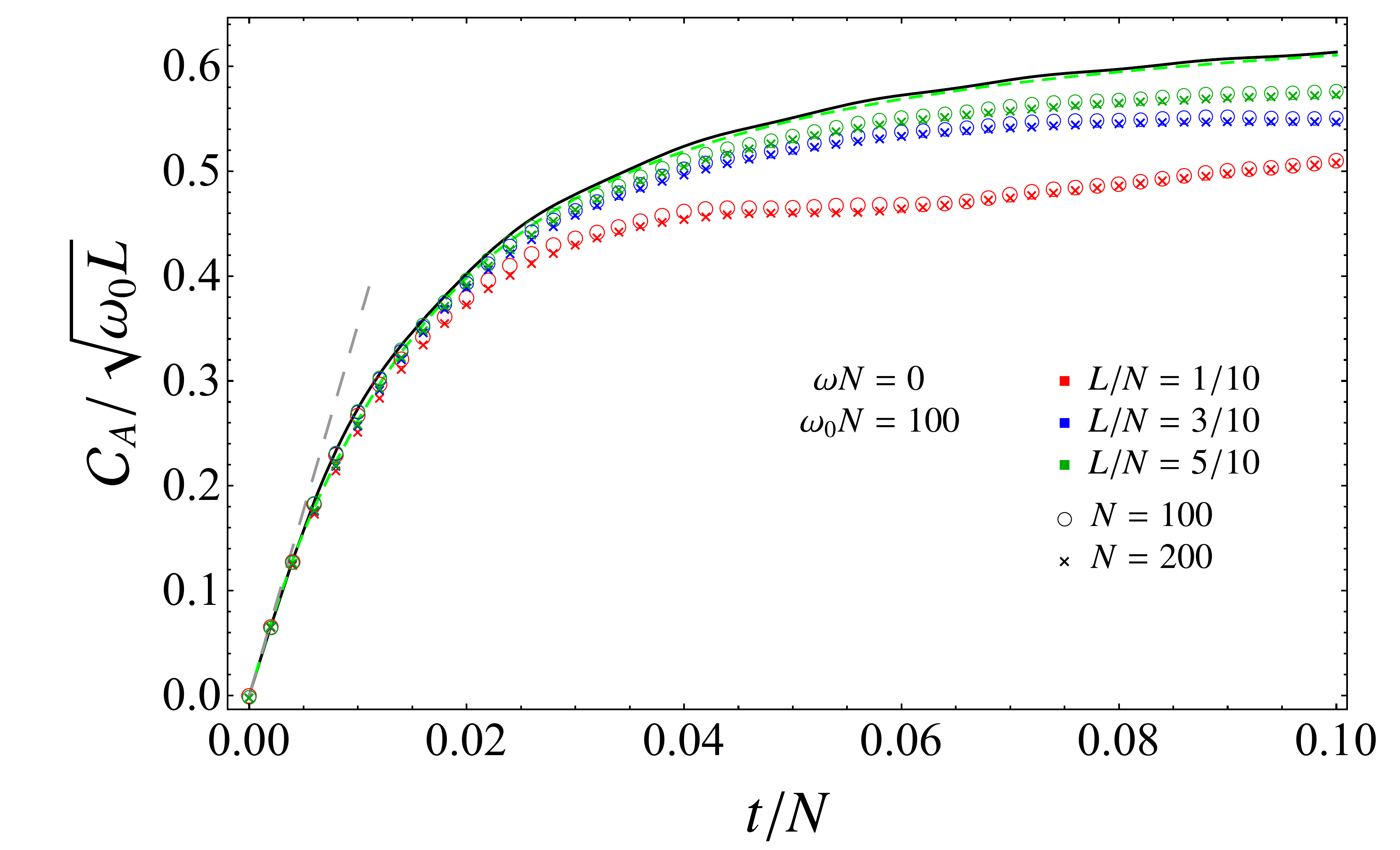}}
\vspace{-.25cm}
\subfigure
{\hspace{0.cm}
\includegraphics[width=.54\textwidth]{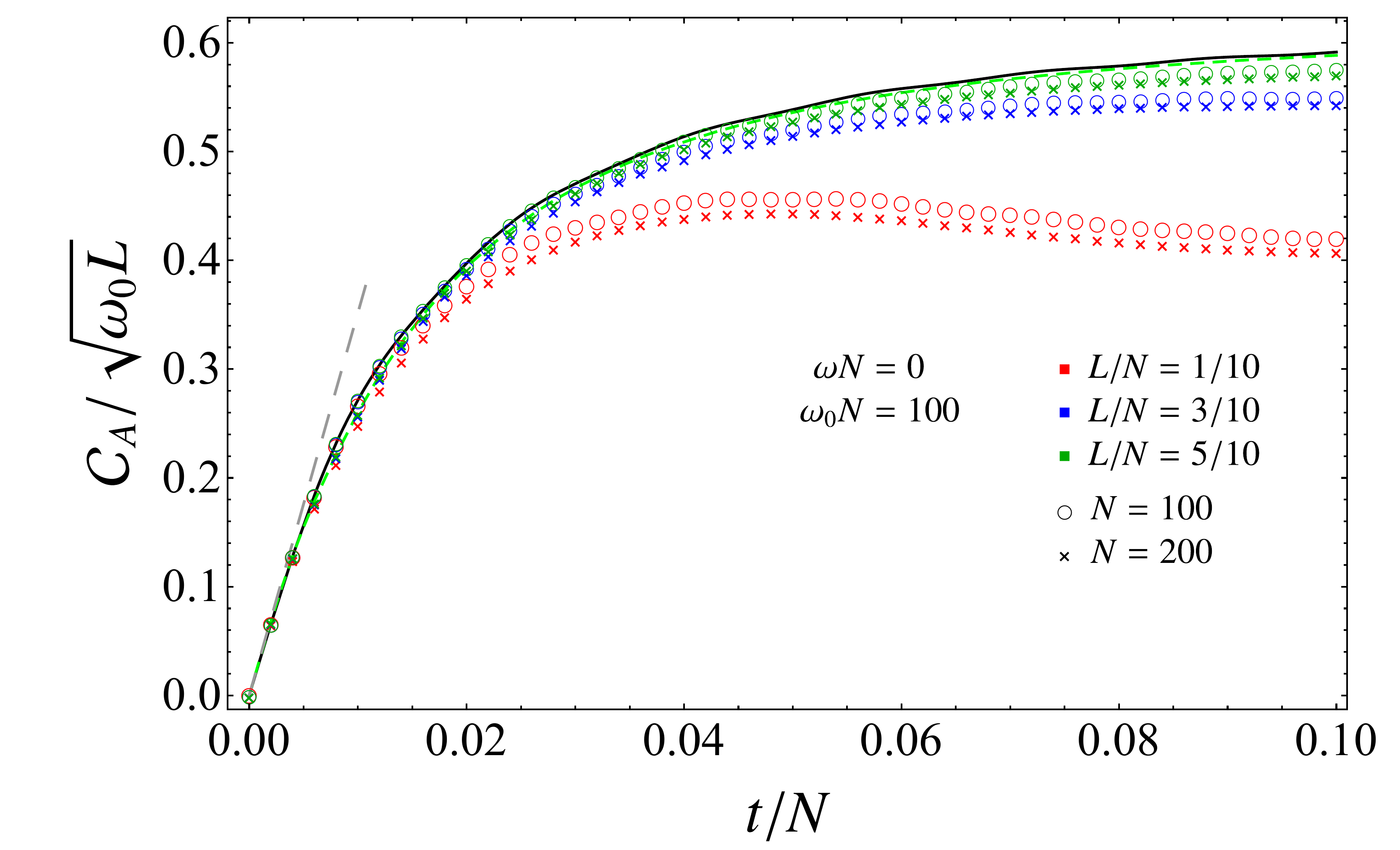}}
\subfigure
{
\hspace{-1.2cm}\includegraphics[width=.54\textwidth]{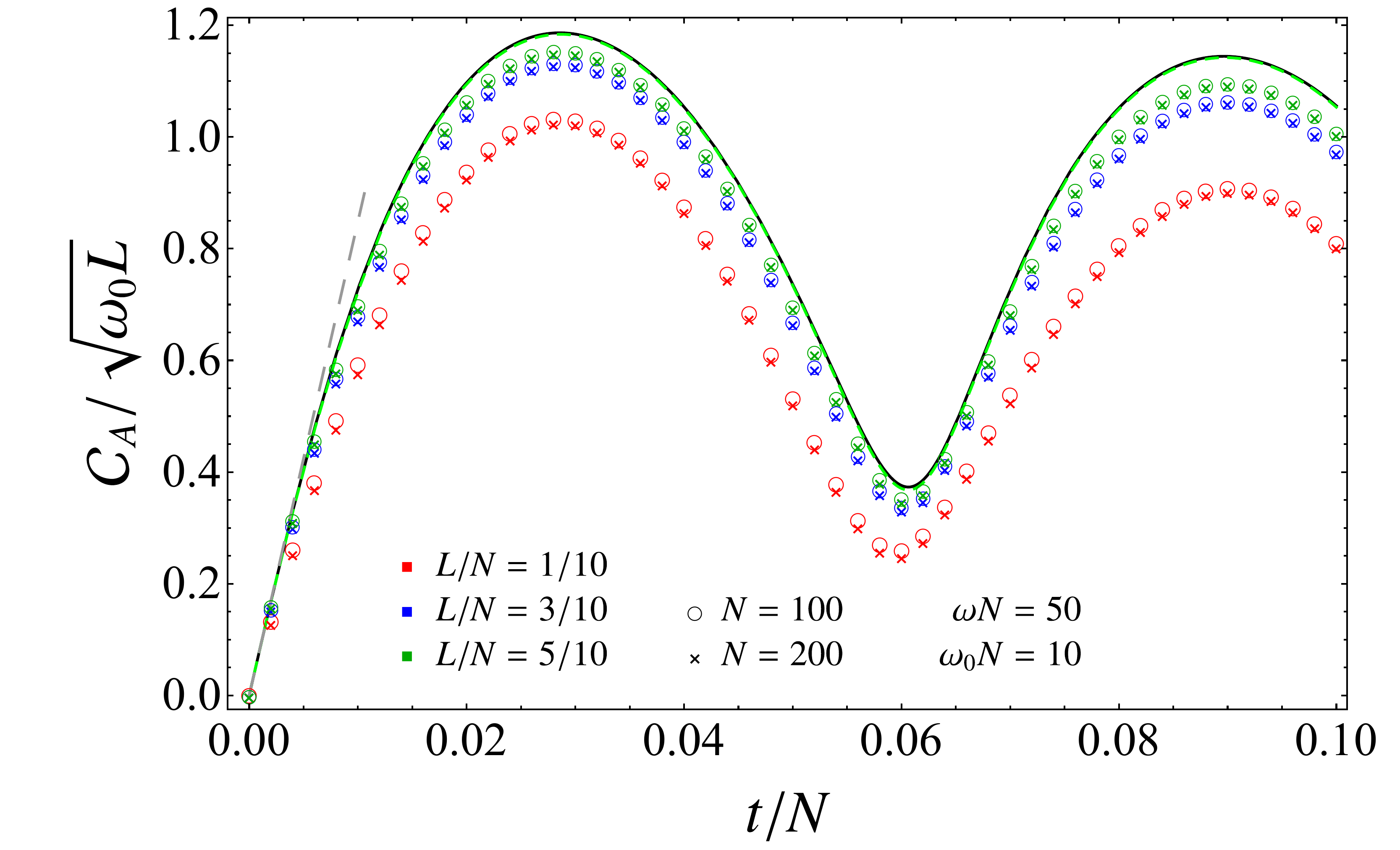}}
\vspace{-.25cm}
\subfigure
{\hspace{0.cm}
\includegraphics[width=.54\textwidth]{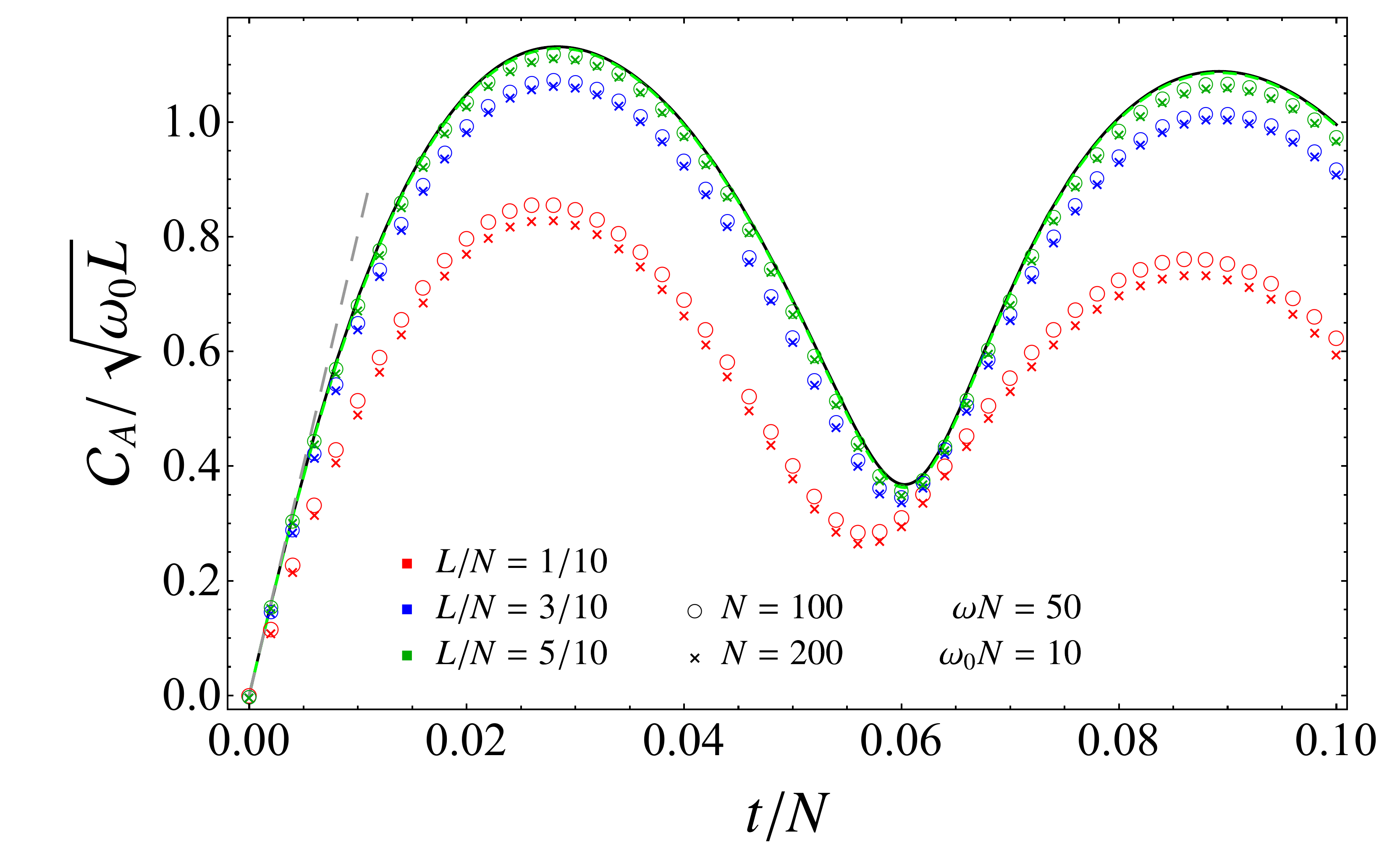}}
\subfigure
{
\hspace{-1.2cm}\includegraphics[width=.54\textwidth]{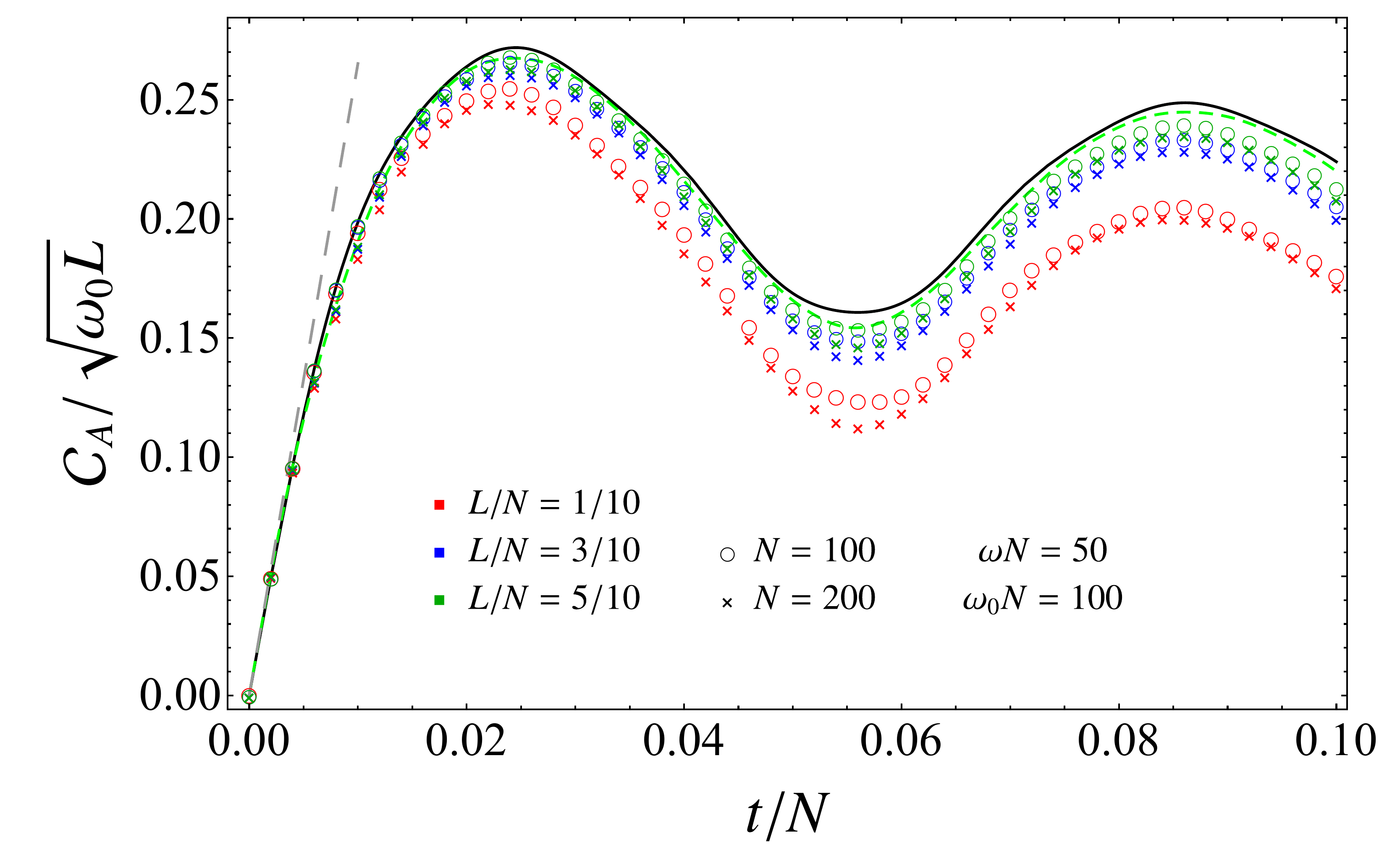}}
\subfigure
{\hspace{0.cm}
\includegraphics[width=.54\textwidth]{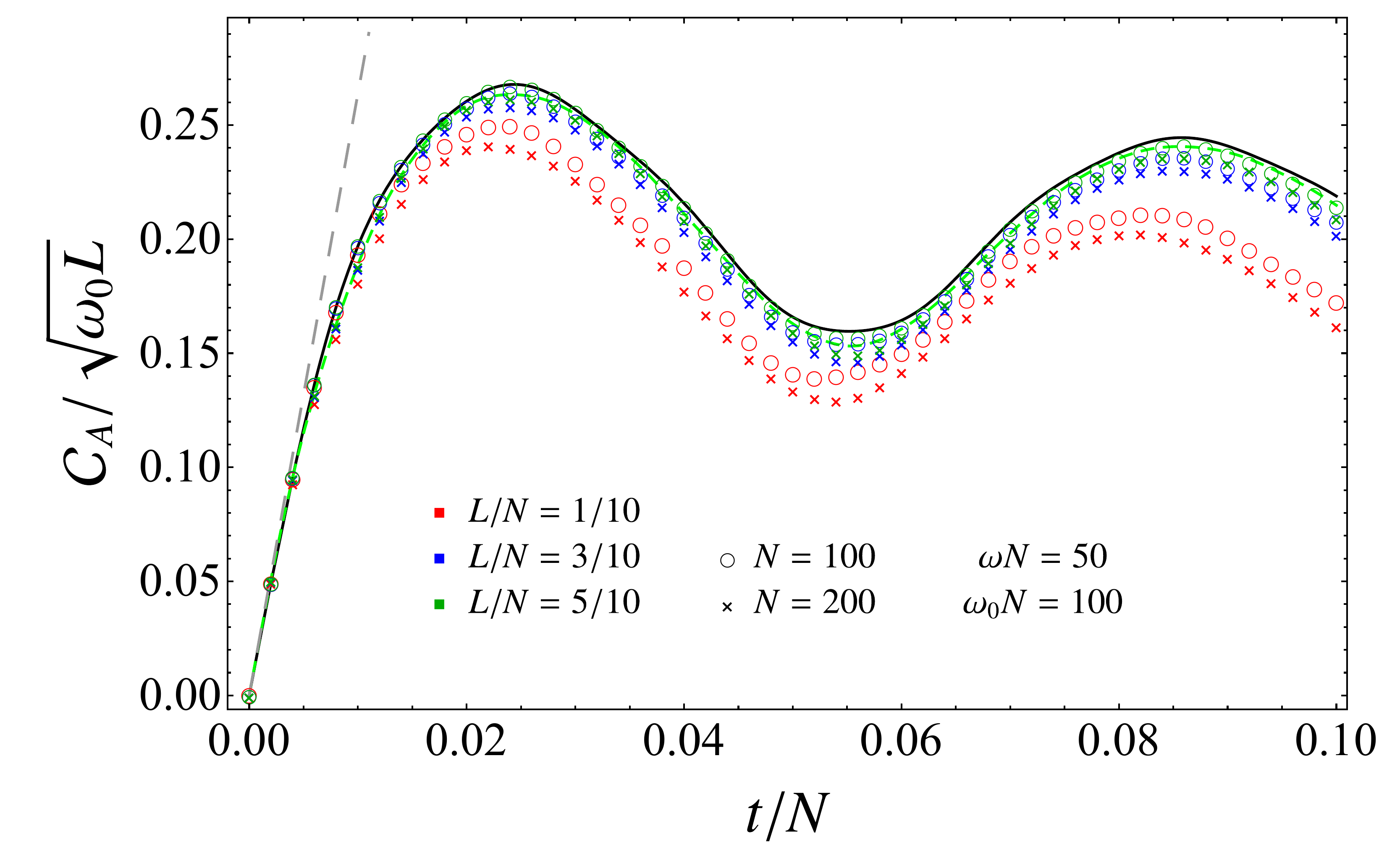}}
\vspace{-.5cm}
\caption{Initial growth of $\mathcal{C}_A$ in (\ref{c2-complexity-rdm-our-case})
for a block $A$ made by $L$ consecutive sites 
in harmonic chains with 
either PBC (left panels) or DBC (right panels) made by $N$ sites
(in the latter case $A$ is adjacent to a boundary).
When $L=N$, the complexity (\ref{comp-pure-global-DBCPBC}) is shown 
for $N=100$ (solid black lines) and $N=200$ (dashed green lines).
}
\label{fig:MixedStateGlobalSmallTimesDimensionless}
\end{figure}

\begin{figure}[t!]
\subfigure
{\hspace{-1.6cm}
\includegraphics[width=.58\textwidth]{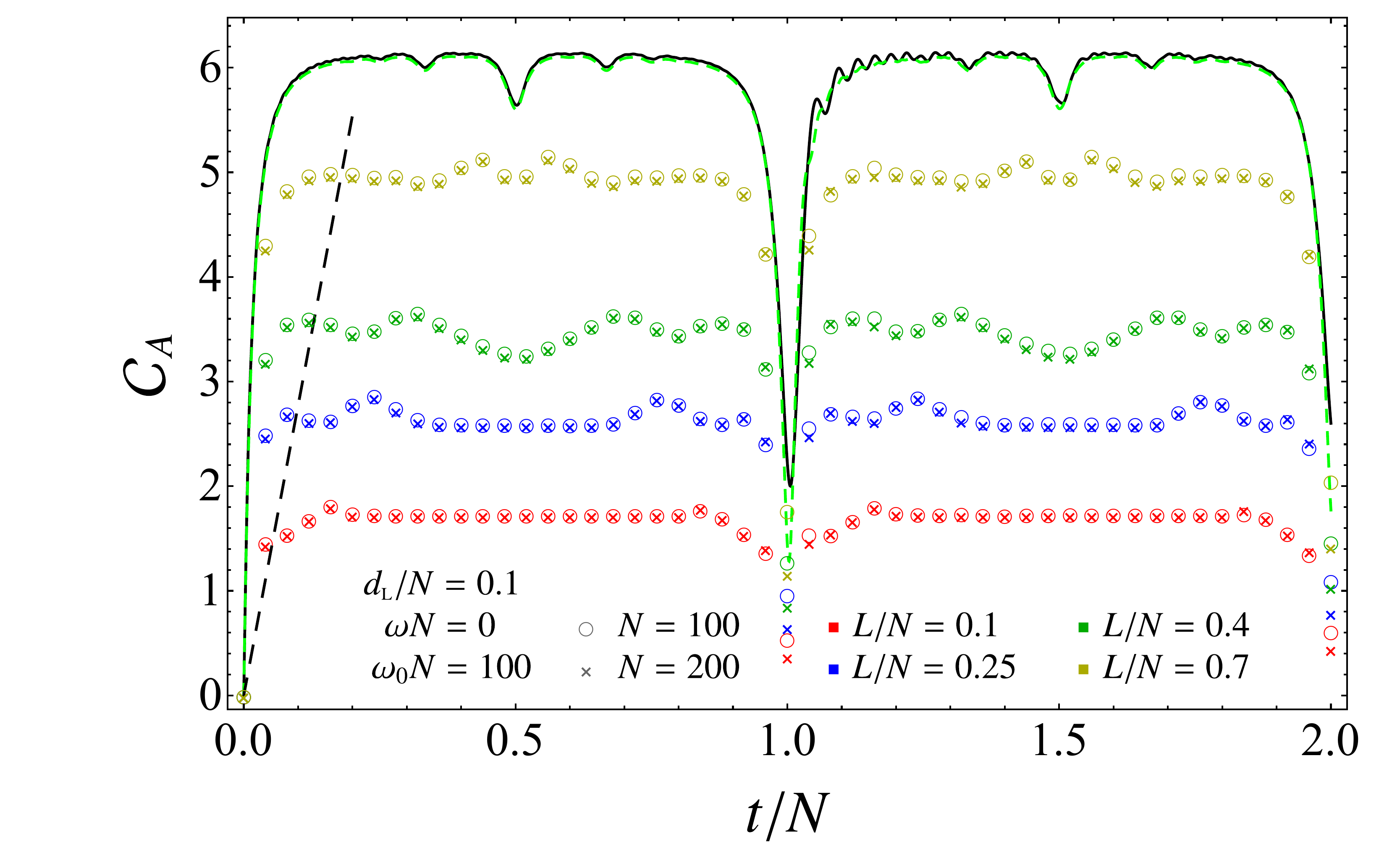}}
\subfigure
{
\hspace{-.7cm}\includegraphics[width=.58\textwidth]{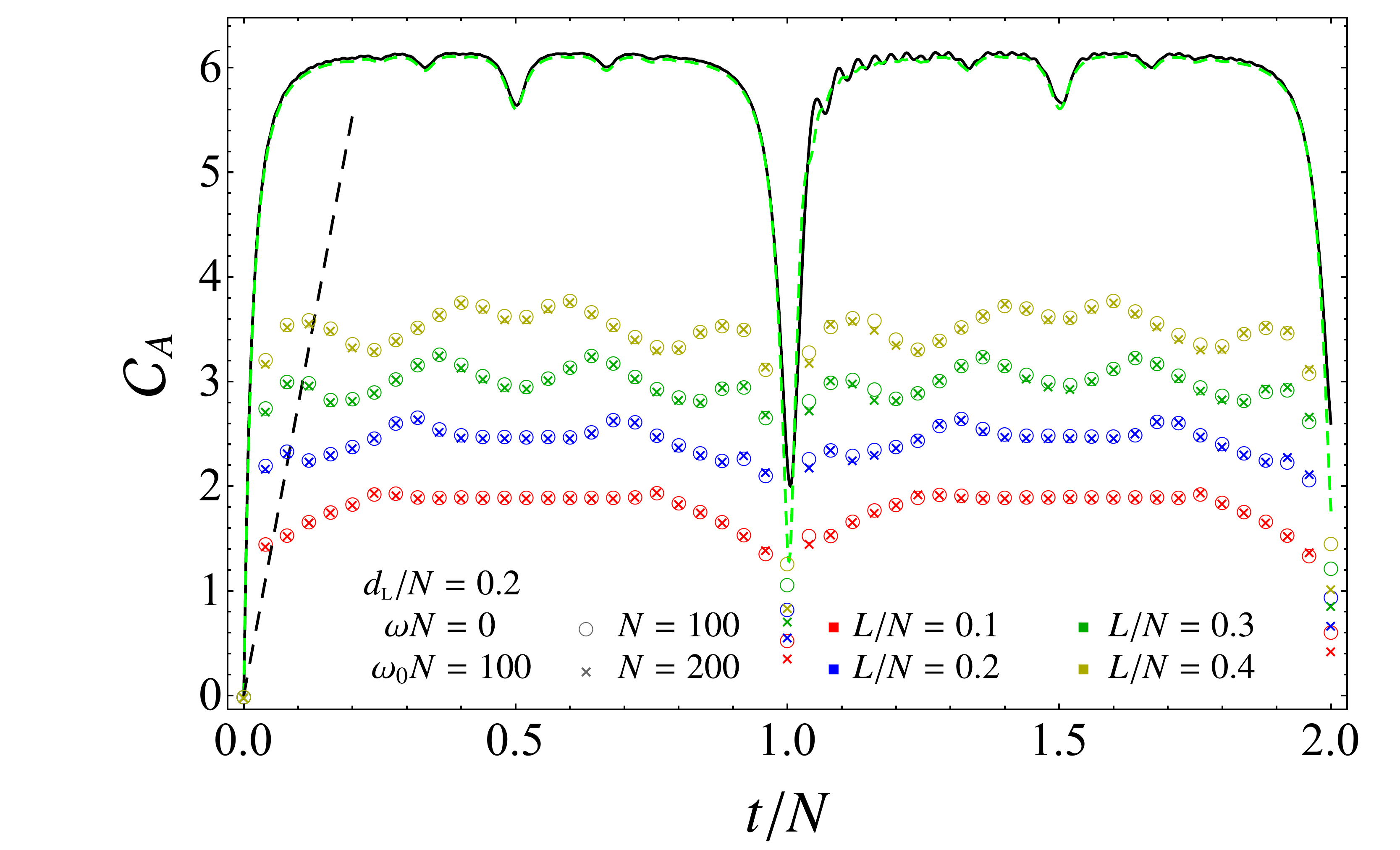}}
\subfigure
{\hspace{-1.6cm}
\includegraphics[width=.58\textwidth]{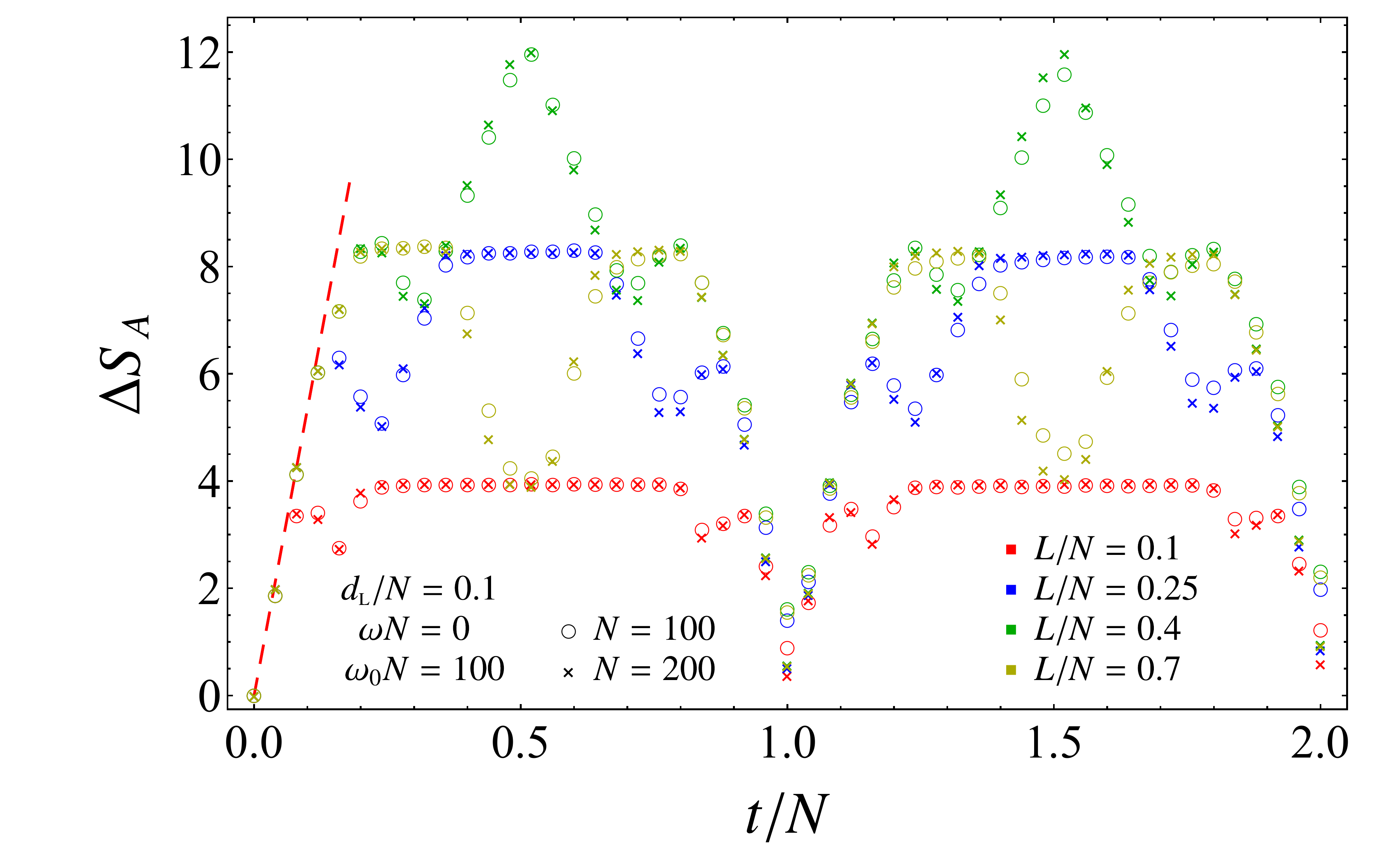}}
\subfigure
{
\hspace{-.7cm}\includegraphics[width=.58\textwidth]{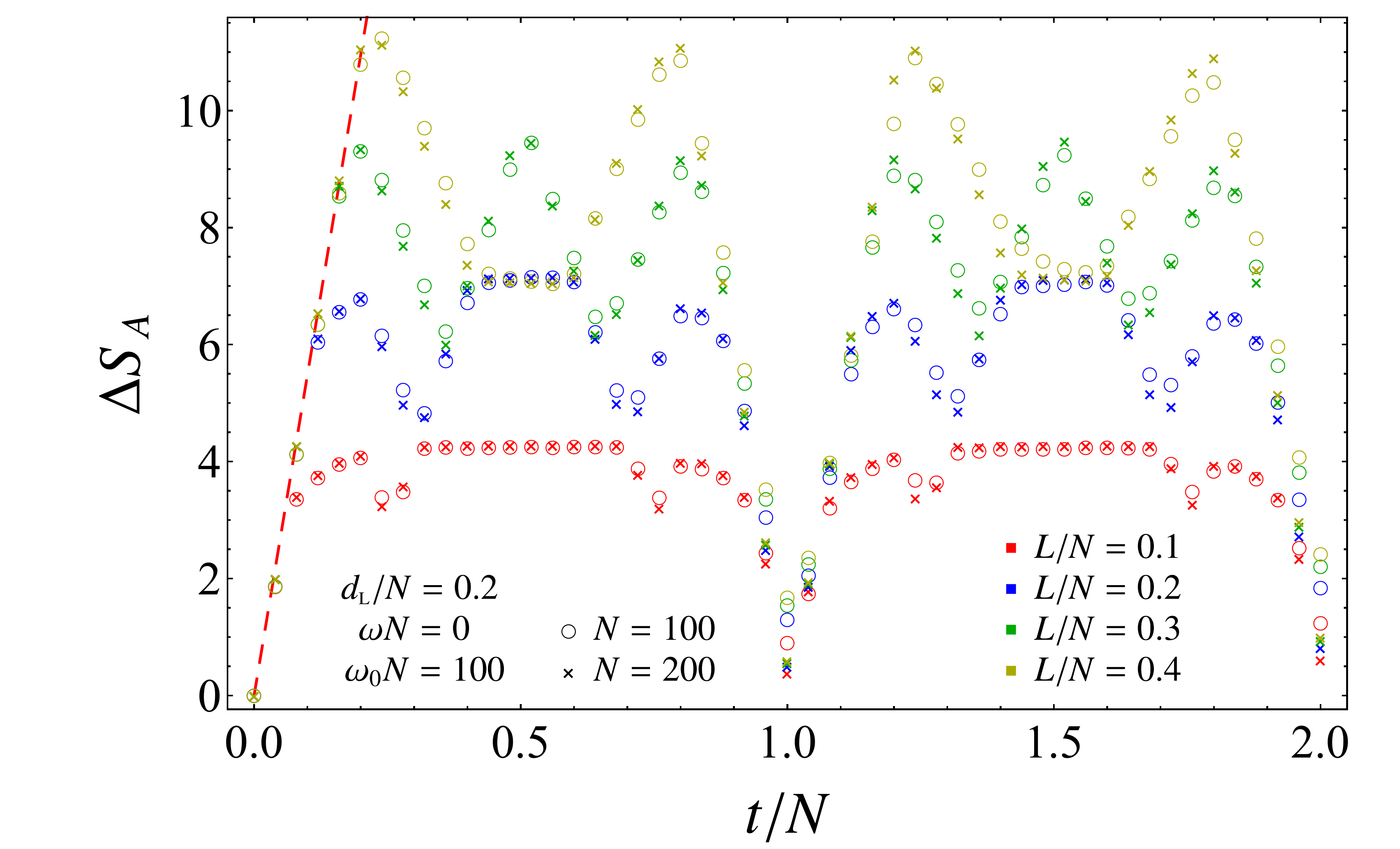}}
\caption{Temporal evolution after the global quench with a gapless evolution Hamiltonian
of $\mathcal{C}_A$ in (\ref{c2-complexity-rdm-our-case}) (top panels) 
and of $\Delta S_A$ in (\ref{Delta-S_A-def}) (bottom panels) 
for a block $A$ made by $L$ consecutive sites 
and separated by $d_{\textrm{\tiny L}}$ sites from the left boundary
of harmonic chains with DBC made by $N$ sites.
When $L=N$, the complexity (\ref{comp-pure-global-DBCPBC}) is shown 
for $N=100$ (solid black lines) and $N=200$ (dashed green  lines).
}
\vspace{0.4cm}
\label{fig:MixedStateGlobalMasslessDBCIntDet}
\end{figure}

In all the figures discussed in this section
we show the temporal evolutions of the subsystem complexity $\mathcal{C}_A$ in (\ref{c2-complexity-rdm-our-case})
or of the increment $\Delta S_A$ of the entanglement entropy in (\ref{Delta-S_A-def}) 
after the global quench.
In particular, we show numerical results corresponding to $N=100$ and $N=200$,
finding nice collapses of the data 
when $L/N$, $\omega_0 N$ and $\omega N$ are kept fixed, independently of the boundary conditions. 
The data reported in all the left panels have been obtained in harmonic chains with PBC,
whose dispersion relations  are (\ref{dispersion relations}),
while the ones in all the right panels correspond to a block adjacent to a boundary
of harmonic chains where DBC are imposed,
whose dispersion relations  are (\ref{dispersion DBC}),
if not otherwise indicated
(like in Fig.\ref{fig:MixedStateGlobalMasslessDBCIntDet}).
The evolution Hamiltonian is gapless 
in Fig.\,\ref{fig:MixedStateGlobalMasslessEvolutionDimensionlessNoentropy},
Fig.\,\ref{fig:MixedStateGlobalMasslessEvolutionDimensionless}, 
Fig.\,\ref{fig:CompvsEntStateGlobalMasslessEvolutionDimensionless} 
and Fig.\,\ref{fig:MixedStateGlobalMasslessDBCIntDet},
while it is gapped in Fig.\,\ref{fig:MixedStateGlobalMassiveEvolutionDimensionlessNoEntropy} and Fig.\,\ref{fig:MixedStateGlobalMassiveEvolutionDimensionless}, 
with $\omega N =5$.
In Fig.\,\ref{fig:MixedStateGlobalSmallTimesDimensionless}, where the initial growth is explored, 
both gapless and gapped evolution Hamiltonians have been employed. 
When $L=N$, the complexity (\ref{comp-pure-global-DBCPBC}) for pure states 
has been evaluated
with either $N=100$ (black solid lines) or $N=200$ (dashed green lines).

In Fig.\,\ref{fig:MixedStateGlobalMasslessEvolutionDimensionlessNoentropy},
Fig.\,\ref{fig:MixedStateGlobalMasslessEvolutionDimensionless}
and Fig.\,\ref{fig:CompvsEntStateGlobalMasslessEvolutionDimensionless} 
all the data have been obtained with $\omega N =0$
and either $\omega_0 N =20$ (Fig.\,\ref{fig:MixedStateGlobalMasslessEvolutionDimensionlessNoentropy}
and Fig.\,\ref{fig:CompvsEntStateGlobalMasslessEvolutionDimensionless})
or $\omega_0 N =100$ 
(Fig.\,\ref{fig:MixedStateGlobalMasslessEvolutionDimensionless}).
Revivals are observed and the different cycles correspond 
to $p < 2t/N < p+1$ for PBC and to $p < t/N < p+1$ for DBC, 
where $p$ is a non-negative integer.

The qualitative behaviour 
of the temporal evolution of the subsystem complexity
crucially depends on the boundary conditions of the harmonic chain. 
For DBC, 
considering the data having $L/N < 1/2$ when $t/N < 1/2$,
we can identify three regimes:
an initial growth until a local maximum is reached,
a decrease and then a thermalisation regime after certain value of $t/N$,
where the subsystem complexity remains constant.
For PBC and $L/N < 1/2$,
the latter regime is not observed and $\mathcal{C}_A$ keeps growing.
Comparing the right panel in Fig.\,\ref{fig:MixedStateGlobalMasslessEvolutionDimensionlessNoentropy} with the top right panel in 
Fig.\,\ref{fig:MixedStateGlobalMasslessEvolutionDimensionless},
one realises that, for DBC,
the height of the plateaux increases as either $L/N$ or $\omega_0 N$ increases, as expected.
The absence of thermalisation regimes for PBC could be related 
to the occurrence of the zero mode,
as suggested by the fact that, for pure states, 
the zero mode contribution provides the logarithmic growth 
of the complexity (\ref{C-pure-both-eta}).
However, we are not able to identify explicitly the zero mode contribution in the subsystem complexity,
hence we cannot subtract it as done 
in the bottom left panel of Fig.\,\ref{fig:PureStateCritical}
for the temporal evolution of the complexity of pure states.

For DBC, the plateau in the thermalisation regime is not observed when $L/N \geqslant 1/2$
and, considering the interval $t/N\in[\nu,\nu+1]$ with $\nu=\{0,1\}$, 
it approximately begins at  $t - \nu N \simeq L$ and ends at $t - \nu N \simeq N-L+1$.
The straight dashed grey lines approximatively indicate 
the beginning of the plateaux for different $L/N < 1/2$
(in particular, they are obtained by joining the origin with the point of the curve made by the blue data points at $t/N =0.3$).

We remark that the temporal evolution of $\mathcal{C}_A$ in infinite chains
is made by the three regimes mentioned above 
(see Fig.\,\ref{fig:MixedStateGlobalMassiveEvolutionTDDetDimensionless}, 
Fig.\,\ref{fig:RatioGlobalMassiveEvolutionTDDetDimensionless} 
and Fig.\,\ref{fig:MixedStateGlobalMassiveEvolutionTDLargeomega0DetDimensionless}),
as largely discussed in Sec.\,\ref{sec:GGE}.

Comparing the temporal evolutions of $\mathcal{C}_A$ and $\Delta S_A$ 
for the same quench protocol and the same subsystem in Fig.\,\ref{fig:MixedStateGlobalMasslessEvolutionDimensionless},
we observe that the initial growth of $\mathcal{C}_A$ in the first revival 
is faster than the linear initial growth of $\Delta S_A$,
as highlighted by the straight dashed black lines
in Fig.\,\ref{fig:MixedStateGlobalMasslessEvolutionDimensionless}.
Within the first revival, we do not observe a 
long range of $t/N$ where the evolution of $\mathcal{C}_A$ is linear. 
Nonetheless, the straight line characterising the initial growth of 
$\Delta S_A$ intersects the first local maximum corresponding to the end of the
initial growth of $\mathcal{C}_A$ when $L/N < 1/2$.
Considering the data points for $L/N < 1/2$ 
and the initial regime of $t/N$ corresponding to half of the first revival, 
we notice that, 
while the temporal evolution of $\Delta S_A$ 
displays a linear growth followed by a saturation regime,
the temporal evolution of $\mathcal{C}_A$ 
is characterised by the three regimes described above.
The saturation regimes of $\mathcal{C}_A$ and $\Delta S_A$ 
are qualitatively very similar 
and begin approximatively at the same value of $t/N$.
Notice that the amplitude of the decrease of $\mathcal{C}_A$
at the end of the first revival is smaller than the one of $\Delta S_A$.
%

The temporal evolutions of $\mathcal{C}_A$ and $\Delta S_A$ 
can be compared for $L/N \leqslant  1/2$.
Indeed, for a bipartite system in a pure state
the entanglement entropy of a subsystem is equal to
the entanglement entropy of the complementary subsystem.
This property, which does not hold for $\mathcal{C}_A$,
implies the overlap between the data for $\Delta S_A$
corresponding to $L/N =3/10$ and to $L/N =7/10$.
Furthermore, $\Delta S_A = 0$ identically when $L = N$.

In the bottom panels of Fig.\,\ref{fig:MixedStateGlobalMasslessEvolutionDimensionless}
we have reported the temporal evolutions of the ratio $\Delta S_A / \mathcal{C}_A$
for the data reported in the other panels of the figure.
The curves of $\Delta S_A / \mathcal{C}_A$ corresponding to
PBC (left panel) and DBC (right panel) are very similar.
For instance, the curves for $\sqrt{L/N}\,\Delta S_A / \mathcal{C}_A$ 
have the same initial growth for different values of $L/N$.
However, we remark that a mild logarithmic decrease
occurs in the thermalisation regime for PBC.

In Fig.\,\ref{fig:CompvsEntStateGlobalMasslessEvolutionDimensionless} 
the range $0 \leqslant t/N \leqslant 10$ is considered, which is 
made by 20 revivals for PBC and by 10 cycles for DBC.
The temporal evolutions of $\mathcal{C}_A$ in the top panels 
show that, up to oscillations due to the revivals,  
after the initial growth $\mathcal{C}_A$ keeps growing logarithmically for PBC
(the solid coloured lines in the top left panel are two-parameter fits 
through the function $a + b \log(t/N)$ of the corresponding data),
while it remains constant for DBC.
This feature is observed also in the corresponding 
temporal evolutions of $\Delta S_A$ 
(middle panels of Fig.\,\ref{fig:CompvsEntStateGlobalMasslessEvolutionDimensionless}).
These two logarithmic growths for PBC are very similar,
as shown by the temporal evolution of $\Delta S_A/\mathcal{C}_A$ 
displayed in the bottom left panel of
Fig.\,\ref{fig:CompvsEntStateGlobalMasslessEvolutionDimensionless}.

In Fig.\,\ref{fig:MixedStateGlobalMassiveEvolutionDimensionlessNoEntropy}
and Fig.\,\ref{fig:MixedStateGlobalMassiveEvolutionDimensionless}
we show some temporal evolutions of $\mathcal{C}_A$ 
when the evolution Hamiltonian is massive
($\omega_0 < \omega$ in Fig.\,\ref{fig:MixedStateGlobalMassiveEvolutionDimensionlessNoEntropy}
and $\omega_0 > \omega$ in Fig.\,\ref{fig:MixedStateGlobalMassiveEvolutionDimensionless},
with $\omega N = 5$ in both the figures).
In these temporal evolutions 
one observes that the local extrema of the curves for $\mathcal{C}_A$
having different $L/N$
roughly occur at the same values of $t/N$.
It is insightful to compare these temporal evolutions with the corresponding ones
characterised by $\omega =0$ in Fig.\,\ref{fig:MixedStateGlobalMasslessEvolutionDimensionlessNoentropy}
and Fig.\,\ref{fig:MixedStateGlobalMasslessEvolutionDimensionless}.
For PBC,  the underlying growth observed when $\omega =0$  does not occur if $\omega >0$.
For DBC, the plateaux observed in the saturation regime when $\omega =0$ 
are replaced by oscillatory behaviours if $\omega > 0$.

In Fig.\,\ref{fig:MixedStateGlobalMassiveEvolutionDimensionless}, 
we report the temporal evolutions of $\mathcal{C}_A$,
of $\Delta S_A$ and of $ \Delta S_A/\mathcal{C}_A$ for the same global quench. 
The evolutions of $\mathcal{C}_A$ and of $\Delta S_A$ are qualitatively similar when $L/N < 1/2$.
An important difference is the initial growth at very small values of $t/N$:
for $\mathcal{C}_A$ is linear 
(see also Fig.\,\ref{fig:MixedStateGlobalSmallTimesDimensionless} and the corresponding discussion),
while for $\Delta S_A$ is quadratic,
as highlighted in the insets of the middle panels
(the coefficient of this quadratic growth for PBC is twice the one obtained for DBC)
and also observed in \cite{Hubeny:2013hz,Liu:2013qca,Rajabpour_14,Unanyan_2014}.
Comparing the bottom panels of Fig.\,\ref{fig:MixedStateGlobalMassiveEvolutionDimensionless}
against the bottom panels of Fig.\,\ref{fig:MixedStateGlobalMasslessEvolutionDimensionless},
one notices that the similarity observed for PBC and DBC when $\omega =0$
does not occur when $\omega \neq 0$.
It is important to perform a systematic analysis considering many 
other values of $\omega N$ and $\omega_0 N$, 
in order to understand the effect of a gapped evolution Hamiltonian 
in the temporal evolution of $\mathcal{C}_A$.

In Fig.\,\ref{fig:MixedStateGlobalSmallTimesDimensionless} 
we consider the initial regime of the temporal evolution of $\mathcal{C}_A$ w.r.t. the initial state
for various choices of $\omega_0 N$ and $\omega N$
(in particular, $\omega =0$ in the first and in the second lines of panels, 
while $\omega >0$ in the third and in the fourth ones).
Very early values of $t$ are considered 
with respect to the ones explored in the previous figures.
In this regime, data collapses are observed for different values of $L/N$  
when $\mathcal{C}_A /\sqrt{\omega_0 L}$ is reported as function of  $t/N$.
In the special case of $L=N$, the complexity of pure states (\ref{comp-pure-global-DBCPBC})
discussed in Sec.\,\ref{sec:purestates_HC_glob} is recovered, 
as shown in Fig.\,\ref{fig:MixedStateGlobalSmallTimesDimensionless} 
by the black solid lines ($N=100$) and by the green dashed lines ($N=200$).

Each panel on the left in Fig.\,\ref{fig:MixedStateGlobalSmallTimesDimensionless}
is characterised by the same $\omega_0 N$ and $\omega N$
of the corresponding one on the right. 
From their comparison, one realises that the qualitative behaviour of the initial growth
at very early times is not influenced by the choice of the boundary conditions. 
Moreover, the linear growth of $\mathcal{C}_A / \sqrt{\omega_0 L}$
is independent of $L/N$ for very small values of $t/N$;
hence the slope of the initial growth can be found by considering the case $L=N$
(discussed in Sec.\,\ref{sec:purestates_HC_glob})
and the approximation described 
in Sec.\,\ref{subsec:TD limit HC} and in appendix\,\ref{subapp:smallkapprox}.
Combining these observations
with (\ref{compDBCapproximinitialt}) and (\ref{compPBCapproximinitialt}),
we obtain the initial linear growth $a_{\textrm{\tiny (B)}} \, t/N + \dots$
where the dots represent higher order in $t/N$ and the slope depends on the boundary conditions 
labelled by $\textrm{B} \in \{ \textrm{P}, \textrm{D}\}$
as follows
\bea
\label{lineargrowth_app_PBC_text}
a_{\textrm{\tiny (P)}} 
&=&
\frac{\big|(\omega N)^2 - (\omega_0 N)^2\big|}{2\omega_0 N}\;
\sqrt{\sqrt{\frac{m}{4\kappa}}\;\omega_0 N\coth \! \bigg(\sqrt{\frac{m}{4\kappa}}\;\omega_0 N\bigg)}
\\
\rule{0pt}{1.1cm}
\label{lineargrowth_app_DBC_text}
a_{\textrm{\tiny (D)}} 
&=&
\frac{\big|(\omega N)^2 - (\omega_0 N)^2\big|}{2\sqrt{2}\omega_0 N}\;
\sqrt{\sqrt{\frac{m}{\kappa}}\;\omega_0 N\coth \! \bigg(\sqrt{\frac{m}{\kappa}}\;\omega_0 N\bigg)-1}\;.
\eea
The grey dashed lines in Fig.\,\ref{fig:MixedStateGlobalSmallTimesDimensionless}
represent $a_{\textrm{\tiny (P)}} \, t/N$ (left panels)
and $a_{\textrm{\tiny (D)}} \, t/N$ (right panels).

Since for DBC and $\omega =0$ 
the temporal evolution of $\mathcal{C}_A$ 
displays a thermalisation regime 
after the initial growth and the subsequent decrease 
when the block $A$ with $L/N < 1/2$ is adjacent to a boundary,
we find it worth investigating also the case where $A$
is separated from the boundary. 
Denoting by $d_{\textrm{\tiny L}}$ the number of sites 
separating $A$ from the left boundary of the chain
(hence $d_{\textrm{\tiny R}}=N-L-d_{\textrm{\tiny L}}$ sites occur between $A$
and the right boundary),
$\mathcal{C}_A$ must be invariant under a spatial reflection w.r.t. the center of the chain,
i.e. when $d_{\textrm{\tiny L}}$ and $d_{\textrm{\tiny R}}$
are replaced  by 
$d_{\textrm{\tiny R}} -1$ and $d_{\textrm{\tiny L}} +1$ respectively.

In Fig.\,\ref{fig:MixedStateGlobalMasslessDBCIntDet} we show 
the temporal evolutions of $\mathcal{C}_A$ and of $\Delta S_A$
for this bipartition of the segment
when the evolution Hamiltonian is gapless and $\omega_0 N=100$,
for four different values of $L/N$ and fixed values of $d_{\textrm{\tiny L}}/N$ 
given by $d_{\textrm{\tiny L}}/N=0.1$ (left panels) 
or $d_{\textrm{\tiny L}}/N=0.2$ (right panels).
Once these parameters have been chosen,
the data points corresponding to $N=100$ and $N=200$ nicely collapse on the same curve.
%

When $d_{\textrm{\tiny L}}\neq 0$, 
a thermalisation regime where both the curves of 
$\mathcal{C}_A$  and $\Delta S_A$ are constant
occurs if $b/N<1/2$, with $b=\textrm{min}[d_{\textrm{\tiny L}}+L,d_{\textrm{\tiny R}}+L-1]$
(see the red and blue curves in Fig.\,\ref{fig:MixedStateGlobalMasslessDBCIntDet}).
The plateau is observed approximatively for $t/N\in [b/N,1-b/N]$
and its height depends on $\omega_0 L$, on $L/N$ and also on $d_{\textrm{\tiny L}}/N$.
A remarkable feature of the temporal evolution of $\mathcal{C}_A$ 
when $d_{\textrm{\tiny L}} > 0$ is the occurrence of two local maxima for $t/N < 1/2$,
while only one maximum is observed when $d_{\textrm{\tiny L}} = 0$  for $t/N < 1/2$
(see the top panels in Fig.\,\ref{fig:MixedStateGlobalMasslessEvolutionDimensionless}).
For any given value of $L/N \leqslant 1/2$ in the top panels of Fig.\,\ref{fig:MixedStateGlobalMasslessDBCIntDet},
the subsystem complexity grows 
in the temporal regime between the two local maxima, for $t/N < 1/2$.
The occurrence of two local maxima in the temporal evolution of $\mathcal{C}_A$ 
when $A$ is separated from the boundary is observed also when $N \to \infty$.
This is shown in Fig.\,\ref{fig:MixedStateGlobalMassiveEvolutionTDDetDimensionless}
and Fig.\,\ref{fig:MixedStateGlobalMassiveEvolutionTDLargeomega0DetDimensionless},
where we also highlight the logarithmic nature of the growth of $\mathcal{C}_A$ 
in the temporal regime between the two local maxima,
which can be compared with a logarithmic growth occurring in $\Delta S_A$
(see e.g. Fig.\,\ref{fig:RatioGlobalMassiveEvolutionTDDetDimensionless}).

Comparing each top panel with the corresponding bottom panel in 
Fig.\,\ref{fig:MixedStateGlobalMasslessDBCIntDet}, we observe that 
the black dashed straight line 
(it is the same in the two top panels)
captures the first local maximum of $\mathcal{C}_A$.
The slope of this line 
is twice the slope of the red dashed straight line in the bottom panels,
which identifies the initial linear growth of $\Delta S_A$.

\newpage
\section{Subsystem complexity and the generalised Gibbs ensemble}
\label{sec:GGE}

In this section we consider infinite harmonic chains, 
either on the infinite line or on the semi-infinite line with DBC at the origin, 
and discuss that the asymptotic value of $\mathcal{C}_A$
for a block made by consecutive sites can be found through the 
generalised Gibbs ensemble (GGE).

\subsection{Complexity of the GGE}

An isolated system prepared in a pure state and then 
suddenly driven out of equilibrium through a global quench does not relax. 
Instead, relaxation occurs for a subsystem 
\cite{Barthel08,Cramer:2008zz,Cramer_2010} 
(see also the review \cite{Essler:2016ufo} and the references therein).

Consider a spatial bipartition of a generic harmonic chain
given by a finite subsystem $A$ and its complement. 
Denoting by $\hat{\rho}(t)$ the density matrix of the entire system 
and by $\hat{\rho}_A(t)$ the reduced density matrix of $A$,
a quantum system relaxes locally to a stationary state if
the limit $\lim_{t\to\infty} \lim_{N\to\infty} \hat{\rho}_A(t) \equiv \hat{\rho}_A(t=\infty)$ 
exists for any $A$,
where $N$ is the number of sites in the harmonic chain.
This stationary state is described by the time independent density matrix 
$\hat{\rho}_{\textrm{\tiny E}}$ describing a statistical ensemble
if $\lim_{N\to\infty} \hat{\rho}_{\textrm{\tiny E},A}=\hat{\rho}_A(t=\infty) $,  for any  $A$,
where $\hat{\rho}_{\textrm{\tiny E},A}$ is obtained by tracing $\hat{\rho}_{\textrm{\tiny E}}$
over the degrees of freedom of the complement of $A$.
For the global quench of the mass parameter that we are investigating
in infinite harmonic chains,
the stationary state is described by a GGE
\cite{Rigol_07,Sotiriadis:2014uza,Ilievski_2015,Ilievski:2016fdy}
(see the review \cite{Vidmar_2016} for an extensive list of references).

In terms of the creation and annihilation operators (\ref{b_operators def}),
the evolution Hamiltonian reads
\be
\label{Hphys-op-Omega GGE}
\widehat{H}
=
\sum_{k=1}^N 
\Omega_{k} \! \left( \hat{\mathfrak{b}}_k^\dagger\, \hat{\mathfrak{b}}_k +\frac{1}{2}\, \right) .
\ee
The GGE that provides the stationary state reads \cite{Calabrese:2007rg}
\be
\label{rho_gge}
\hat{\rho}_{\textrm{\tiny GGE}}
=
\frac{e^{-\sum_{k=1}^N \lambda_k \hat{\mathfrak{b}}_k^\dagger \hat{\mathfrak{b}}_k}}{\mathcal{Z}_{\textrm{\tiny GGE}}}
\;\;\qquad\;\;
\mathcal{Z}_{\textrm{\tiny GGE}}
=
\textrm{Tr} (\hat{\rho}_{\textrm{\tiny GGE}})
=
\prod_{k=1}^N \frac{1}{1-e^{-\lambda_k}}
\ee
where $\hat{\rho}_{\textrm{\tiny GGE}}$ is normalised through the condition $\textrm{Tr} (\hat{\rho}_{\textrm{\tiny GGE}})  =1$.
The conservation of the number operators $\hat{\mathfrak{b}}_k^\dagger \hat{\mathfrak{b}}_k$ 
tells us that the relation between their expectation values and $\lambda_k$ reads \cite{Calabrese:2007rg}
\be
\label{nk lambdak}
n_k
\,\equiv\,
\textrm{Tr}\big(\hat{\mathfrak{b}}_k^\dagger \,\hat{\mathfrak{b}}_k \, \hat{\rho}_{\textrm{\tiny GGE}}\big) =\frac{1}{e^{\lambda_k}-1}
\,=\,
\langle \psi_0 |\,  \hat{\mathfrak{b}}_k^\dagger \,\hat{\mathfrak{b}}_k \,  | \psi_0 \rangle
\ee
which is strictly positive because $\lambda_k>0 $ for any value of $k$.

Since the GGE in (\ref{rho_gge}) is a bosonic Gaussian state, it is characterised by its covariance matrix 
$\gamma_{\textrm{\tiny GGE}}$,
which can be decomposed as follows
\be
\label{gamma-GGE-matrix}
\gamma_{\textrm{\tiny GGE}}
=
\,
\bigg( 
\begin{array}{cc}
Q_{\textrm{\tiny GGE}}  &  \, M_{\textrm{\tiny GGE}} 
\\
M_{\textrm{\tiny GGE}}^{\textrm{t}}  & \, P_{\textrm{\tiny GGE}}  
\end{array}  \bigg)
\ee
where (see the appendix\;\ref{app-sec-GGE-corr})
\be
(Q_{\textrm{\tiny GGE}})_{i,j}=\textrm{Tr}\big( \hat{q}_i  \,\hat{q}_j \,\hat{\rho}_{\textrm{\tiny GGE}} \big) 
\qquad
(P_{\textrm{\tiny GGE}})_{i,j}=\textrm{Tr}\big( \hat{p}_i  \,\hat{p}_j \, \hat{\rho}_{\textrm{\tiny GGE}} \big)
\qquad
(M_{\textrm{\tiny GGE}})_{i,j}=\textrm{Re}\big[\textrm{Tr}\big( \hat{q}_i  \,\hat{p}_j \,\hat{\rho}_{\textrm{\tiny GGE}} \big)\big] \,.
\ee

By adapting the computation reported in appendix \ref{subapp:covariancematrix} to this case, 
the operators $\hat{\boldsymbol{\mathfrak{q}}}$ and $\hat{\boldsymbol{\mathfrak{p}}}$ 
can be introduced as in (\ref{Williamson-basis-Hphys})
and for (\ref{gamma-GGE-matrix}) one finds
\bea
\label{QGGE}
Q_{\textrm{\tiny GGE}}
&=&
\widetilde{V} \, S^{-1}_{\textrm{\tiny phys}} \,
\textrm{Tr}\big( \hat{\boldsymbol{\mathfrak{q}}}  \,\hat{\boldsymbol{\mathfrak{q}}}^{\textrm{t}}\hat{\rho}_{\textrm{\tiny GGE}} \big) 
\, S^{-1}_{\textrm{\tiny phys}} \, \widetilde{V}^{\textrm{t}}
\equiv
\widetilde{V} \, \mathcal{Q}\textrm{\tiny GGE}  \, \widetilde{V}^{\textrm{t}}
\\
\rule{0pt}{.5cm}
\label{PGGE}
P_{\textrm{\tiny GGE}}
&=&
\widetilde{V} \, S_{\textrm{\tiny phys}} \,
\textrm{Tr}\big( \hat{\boldsymbol{\mathfrak{p}}}  \,\hat{\boldsymbol{\mathfrak{p}}}^{\textrm{t}}\hat{\rho}_{\textrm{\tiny GGE}} \big)
\, S_{\textrm{\tiny phys}} \, \widetilde{V}^{\textrm{t}}
\,\equiv\,
\widetilde{V} \, \mathcal{P}\textrm{\tiny GGE} \,\widetilde{V}^{\textrm{t}}
\\
\rule{0pt}{.5cm}
\label{MGGE}
M_{\textrm{\tiny GGE}}
&=&
\widetilde{V} \, S^{-1}_{\textrm{\tiny phys}} \,
\textrm{Re}\big[\textrm{Tr}\big( \hat{\boldsymbol{\mathfrak{q}}}  \,\hat{\boldsymbol{\mathfrak{p}}}^{\textrm{t}}\hat{\rho}_{\textrm{\tiny GGE}} \big)\big]
\, S_{\textrm{\tiny phys}} \,\widetilde{V}^{\textrm{t}}
\,\equiv\,
\widetilde{V} \,\mathcal{M}\textrm{\tiny GGE} \, \widetilde{V}^{\textrm{t}}\,.
\eea
Then, expressing $\hat{\boldsymbol{\mathfrak{q}}}$ and $\hat{\boldsymbol{\mathfrak{p}}}$ in terms of $\hat{\boldsymbol{\mathfrak{b}}}$ 
and $\hat{\boldsymbol{\mathfrak{b}}}^\dagger$ defined in (\ref{b_operators def}), 
exploiting the fact that the two points correlators vanish
when the indices of the annihilation and creation operators are different,
using (\ref{nk lambdak}) and 
$\textrm{Tr}\big(\hat{\mathfrak{b}}_k^\dagger \hat{\mathfrak{b}}_k^\dagger\, \hat{\rho}_{\textrm{\tiny GGE}}\big)
=\textrm{Tr}\big(\hat{\mathfrak{b}}_k \hat{\mathfrak{b}}_k\hat{\rho}_{\textrm{\tiny GGE}}\big)=0$, 
we find
$M\textrm{\tiny GGE}=\mathcal{M}\textrm{\tiny GGE}=\mathbf{0}$
and
\be
\label{Q-P-GGEdiag}
\mathcal{Q}_{\textrm{\tiny GGE}}
\,\equiv\,
\textrm{diag}\,\Big\{ Q_{\textrm{\tiny GGE},k} \, ; 1\leqslant k \leqslant N\Big\}
\;\;\qquad\;\;
\mathcal{P}_{\textrm{\tiny GGE}}
\,\equiv\,
\textrm{diag}\,\Big\{P_{\textrm{\tiny GGE},k} \, ; 1\leqslant k \leqslant N\Big\}
\ee
where
\be
Q_{\textrm{\tiny GGE},k} \equiv \frac{1+2n_k}{2m\,\Omega_k}
\;\;\qquad\;\;
P_{\textrm{\tiny GGE},k} \equiv \frac{m\,\Omega_k}{2}\, (1+2n_k)\,.
\ee
Thus, the covariance matrix (\ref{gamma-GGE-matrix}) simplifies to $\gamma_\textrm{\tiny GGE}=Q_\textrm{\tiny GGE}\oplus P_\textrm{\tiny GGE}$, where $Q\textrm{\tiny GGE}$ and $P\textrm{\tiny GGE}$ 
are given by (\ref{QGGE}) and (\ref{PGGE}).

We find it worth writing the Williamson's decomposition of $\gamma_\textrm{\tiny GGE}$, namely
\be
\label{Williamson-GGE}
\gamma_\textrm{\tiny GGE}
\,=\,
W^{\textrm{t}}_\textrm{\tiny GGE} \, \mathcal{D}_\textrm{\tiny GGE} \, W_\textrm{\tiny GGE}
\,\,\qquad\,\,
W_\textrm{\tiny GGE}
=
\mathcal{X}_\textrm{\tiny GGE} \, V^\textrm{t}
\ee
where the symplectic spectrum is given by 
\be
\label{sympspecGGE}
\mathcal{D}_\textrm{\tiny GGE}
=
\frac{1}{2} \, \boldsymbol{1}
+ 
\textrm{diag}\,\Big\{n_1,\dots,n_N,n_1,\dots,n_N\Big\}
\ee
and, like for the $2N \times 2N$ symplectic matrix $W_\textrm{\tiny GGE}$, 
we have $V=\widetilde{V}\oplus \widetilde{V}$
and that the diagonal matrix $\mathcal{X}_\textrm{\tiny GGE}=\mathcal{X}_\textrm{\tiny phys}^{-1}$ 
is the inverse of $\mathcal{X}_\textrm{\tiny phys}$ defined in (\ref{Chiphys-Wdec}).
We emphasise that $\gamma_{\textrm{\tiny GGE}}$  does not describe a pure state.
Indeed, since $n_k\geqslant 0$ for any $k$, from (\ref{sympspecGGE}) 
we have that the symplectic eigenvalues of $\gamma_{\textrm{\tiny GGE}}$
are greater than $1/2$, as expected for a mixed bosonic Gaussian state.

For the global quench in the harmonic chains that we considering, 
$n_k$ in (\ref{nk lambdak})
can be computed from the expectation value of $\hat{\mathfrak{b}}_k^\dagger \hat{\mathfrak{b}}_k$ on the initial state obtaining \cite{Calabrese:2007rg}
\be
\label{nk omegak}
n_k=
\frac{1}{4}\bigg(\frac{\Omega_k}{\Omega_{0,k}}+\frac{\Omega_{0,k}}{\Omega_k}\bigg)-\frac{1}{2}
\ee
where $\Omega_{0,k}$ and $\Omega_k$ are the dispersion relations of the Hamiltonian defining the initial state 
and of the evolution Hamiltonian respectively. 
Notice that (\ref{nk omegak}) is symmetric under the exchange $\Omega_k \leftrightarrow \Omega_{0,k}\,$.
We recall that the boundary conditions defining the harmonic chain influence
both the dispersion relations and the matrix $V$.

By introducing  the reduced covariance matrix $\gamma_{\textrm{\tiny GGE},A}$ for $A$,
obtained from (\ref{gamma-GGE-matrix}) in the usual way, 
the entanglement entropy 
\be
\label{EE-initialvsGGE}
S_{\textrm{\tiny GGE},A}\equiv-\textrm{Tr}(\hat{\rho}_{\textrm{\tiny GGE},A}\log \hat{\rho}_{\textrm{\tiny GGE},A})
\ee
can be evaluated from the symplectic spectrum of $\gamma_{\textrm{\tiny GGE},A}$ 
through standard methods \cite{Peschel03,Peschel_2009}.

The asymptotic value of the increment of the entanglement entropy $\Delta S_A$
when $t \to \infty$ can be computed as follows
\be
\label{stationary entropy density}
\lim_{L\to\infty}\frac{\lim_{t\to\infty}\lim_{N\to\infty}\Delta S_{A}}{L}
=
\lim_{L\to\infty}\frac{\lim_{N\to\infty}S_{\textrm{\tiny GGE},A}}{L}
=
\lim_{N\to\infty}\frac{S_{\textrm{\tiny GGE}}}{N}
\ee
where the order of the limits is important 
and in the last step we used that $S_{\textrm{\tiny GGE}}$ is an extensive quantity
(see the review \cite{Calabrese:2020wfx} and the references therein).

For the global quench in the harmonic chains that we are considering, 
the asymptotic value (\ref{stationary entropy density}) for the entanglement entropy reads \cite{ac-18-qp-quench}
\bea
\label{SGGE}
\lim_{N\to\infty}\frac{S_{\textrm{\tiny GGE}}}{N}
&=&
\int_{0}^\pi \Big[(n_\theta+1)\log(n_\theta+1)-n_\theta\log n_\theta\Big]\, \frac{d\theta}{\pi}
\\
\rule{0pt}{.8cm}
&=&
\label{SGGE dispersion}
\int_{0}^\pi \Bigg\{
\bigg[\frac{1}{4}\bigg(\frac{\Omega_\theta}{\Omega_{0,\theta}}+\frac{\Omega_{0,\theta}}{\Omega_\theta}\bigg)+\frac{1}{2}\bigg]\log\!\bigg[\frac{1}{4}\bigg(\frac{\Omega_\theta}{\Omega_{0,\theta}}+\frac{\Omega_{0,\theta}}{\Omega_\theta}\bigg)+\frac{1}{2}\bigg]
\\
\rule{0pt}{.7cm}
&& \hspace{1cm}
-\,
\bigg[\frac{1}{4}\bigg(\frac{\Omega_\theta}{\Omega_{0,\theta}}+\frac{\Omega_{0,\theta}}{\Omega_\theta}\bigg)-\frac{1}{2}\bigg]
\log\!\bigg[\frac{1}{4}\bigg(\frac{\Omega_\theta}{\Omega_{0,\theta}}+\frac{\Omega_{0,\theta}}{\Omega_\theta}\bigg)-\frac{1}{2}\bigg]
\Bigg\}\, \frac{d\theta}{\pi}
\nonumber
\eea
in terms of $n_\theta$ given in (\ref{nk omegak}),
where the dispersion relations to employ
are (\ref{dispersion relations TD}) for PBC and 
(\ref{dispersion DBC TD}) for DBC. 
A straightforward change of integration variable
leads to the same expression for both the boundary conditions,
as already noticed for (\ref{comp TD limit}).
Let us remark that (\ref{SGGE}) is finite  
for any choice of the parameter (including $\omega =0$),
both for PBC and DBC. 
It is also symmetric under the exchange $\Omega_{\theta}\leftrightarrow \Omega_{0,\theta}$;
hence under $\omega \leftrightarrow \omega_0$ as well.


We study the circuit complexity to construct the GGE (which is a mixed state)
starting from the (pure) initial state at $t=0$,
by employing the approach based on the Fisher information geometry \cite{DiGiulio:2020hlz}.
The optimal circuit to get $\gamma_{\textrm{\tiny GGE}}$ 
from the initial covariance matrix $\gamma(0)$ at $t=0$ reads
\cite{Bhatia07book,DiGiulio:2020hlz} 
\be
\label{optimal circuit GGE full}
G_s(\gamma(0) \, , \gamma_{\textrm{\tiny GGE}})
\,\equiv \,
\gamma(0)^{1/2} 
\Big(  \gamma(0)^{- 1/2}  \,\gamma_{\textrm{\tiny GGE}} \,\gamma(0)^{-1/2}  \Big)^s
\gamma(0)^{1/2} 
\ee
where $0 \leqslant s \leqslant 1$ parameterises the covariance matrix along the circuit.
The length of the optimal circuit (\ref{optimal circuit GGE full}) provides the circuit complexity
\be 
\label{comp-initialvsGGE-full}
\mathcal{C}_{\textrm{\tiny GGE}}
\,=\,
\frac{1}{2\sqrt{2}}\;
\sqrt{
\textrm{Tr} \,\Big\{ \big[
\log \!\big( \gamma_{\textrm{\tiny GGE}} \; \gamma(0)^{-1} \big)
\big]^2  \Big\}}\;.
\ee

Since $\mathcal{M}_\textrm{\tiny GGE}=\mathcal{M}(0)=\boldsymbol{0}$,
from (\ref{time dep corrs trans}), (\ref{QGGE}) and (\ref{PGGE}) we obtain
\be
\gamma_{\textrm{\tiny GGE}}
=
V
\big[ \mathcal{Q}_\textrm{\tiny GGE}\oplus \mathcal{P}_\textrm{\tiny GGE}\big]
V^{\textrm{t}} 
\;\;\qquad\;\;
\gamma(0)
=
V\big[ \mathcal{Q}(0)\oplus \mathcal{P}(0) \big]V^{\textrm{t}} \,.
\ee
Then, by exploiting (\ref{Q-P-GGEdiag}), (\ref{QPRmat tzero-dep-k}) 
and the fact that the matrix $V$ is the same for both 
$\gamma_{\textrm{\tiny GGE}} $ and $\gamma(0)$,
we find that the complexity (\ref{comp-initialvsGGE-full}) reads
\bea
\label{comp full GGE v0}
\mathcal{C}_{\textrm{\tiny GGE}}
&=&
\frac{1}{2\sqrt{2}} \;
\sqrt{\,\sum_{k=1}^N\bigg\{
\bigg[\log\bigg(\frac{Q_{\textrm{\tiny GGE},k}}{Q_{k}(0)}\bigg)\bigg]^2+\bigg[\log\bigg(\frac{P_{\textrm{\tiny GGE},k}}{P_{k}(0)}\bigg)\bigg]^2
\Bigg\}}
\\
\rule{0pt}{1cm}
&=&
\frac{1}{2\sqrt{2}}\;
\sqrt{\,\sum_{k=1}^N\Bigg\{
\bigg[\log\bigg(\frac{\Omega_{0,k}}{\Omega_k}\,(1+2n_k)\bigg)\bigg]^2
+\bigg[\log\bigg(\frac{\Omega_{k}}{\Omega_{0,k}}\,(1+2n_k)\bigg)\bigg]^2
\Bigg\}}\;.
\eea
By using (\ref{nk omegak}), this expression becomes
\be
\label{comp full GGE}
\mathcal{C}_{\textrm{\tiny GGE}}
\,=\,
\frac{1}{2\sqrt{2}}\;
\sqrt{\,\sum_{k=1}^N\Bigg\{
\bigg[\log\bigg(\frac{\Omega_{0,k}^2}{2\,\Omega_k^2}+\frac{1}{2}\bigg)\bigg]^2+\bigg[\log\bigg(\frac{\Omega_{k}^2}{2\,\Omega_{0,k}^2}+\frac{1}{2}\bigg)\bigg]^2
\Bigg\}}
\ee
which is symmetric under the exchange $\Omega_{k}\leftrightarrow \Omega_{0,k}$,
hence under $\omega \leftrightarrow \omega_0$ as well.

The leading order of this expression as $N \to \infty$ is given by
\be
\label{comp full GGE TD}
\mathcal{C}_{\textrm{\tiny GGE}}
=
\frac{\sqrt{N}}{2\,\sqrt{2\pi}} \;
\sqrt{\,\int_{0}^\pi
\left\{\bigg[\log\bigg(\frac{\Omega_{0,\theta}^2}{2\,\Omega_\theta^2}+\frac{1}{2}\bigg)\bigg]^2+\bigg[\log\bigg(\frac{\Omega_{\theta}^2}{2\,\Omega_{0,\theta}^2}+\frac{1}{2}\bigg)\bigg]^2\right\}
\,d\theta
}
\ee
where $\Omega_{0,\theta}$ and $\Omega_\theta$ are thermodynamic limits of the 
dispersion relations associated to the Hamiltonians before and after the quench respectively. 
By repeating the argument reported below  (\ref{comp TD limit}),
one finds that (\ref{comp full GGE TD}) with (\ref{dispersion relations TD}) 
can be employed for both PBC and DBC.
Moreover, the resulting expression for $\mathcal{C}_{\textrm{\tiny GGE}}/\sqrt{N}$ 
is finite for any choice of the parameters (including for $\omega=0$).

\begin{figure}[t!]
\subfigure
{\hspace{-1.6cm}
\includegraphics[width=.58\textwidth]{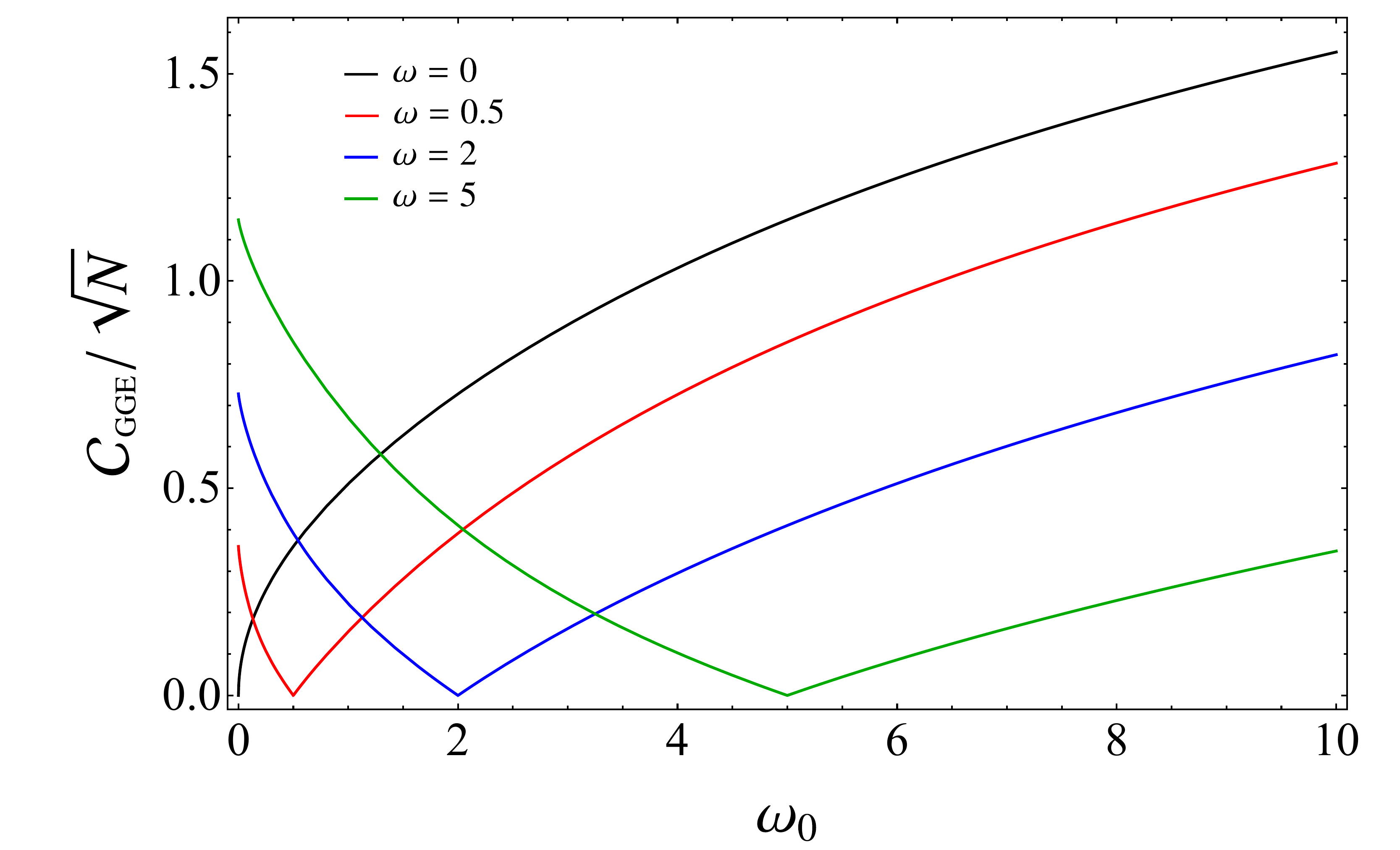}}
\subfigure
{
\hspace{-.7cm}\includegraphics[width=.58\textwidth]{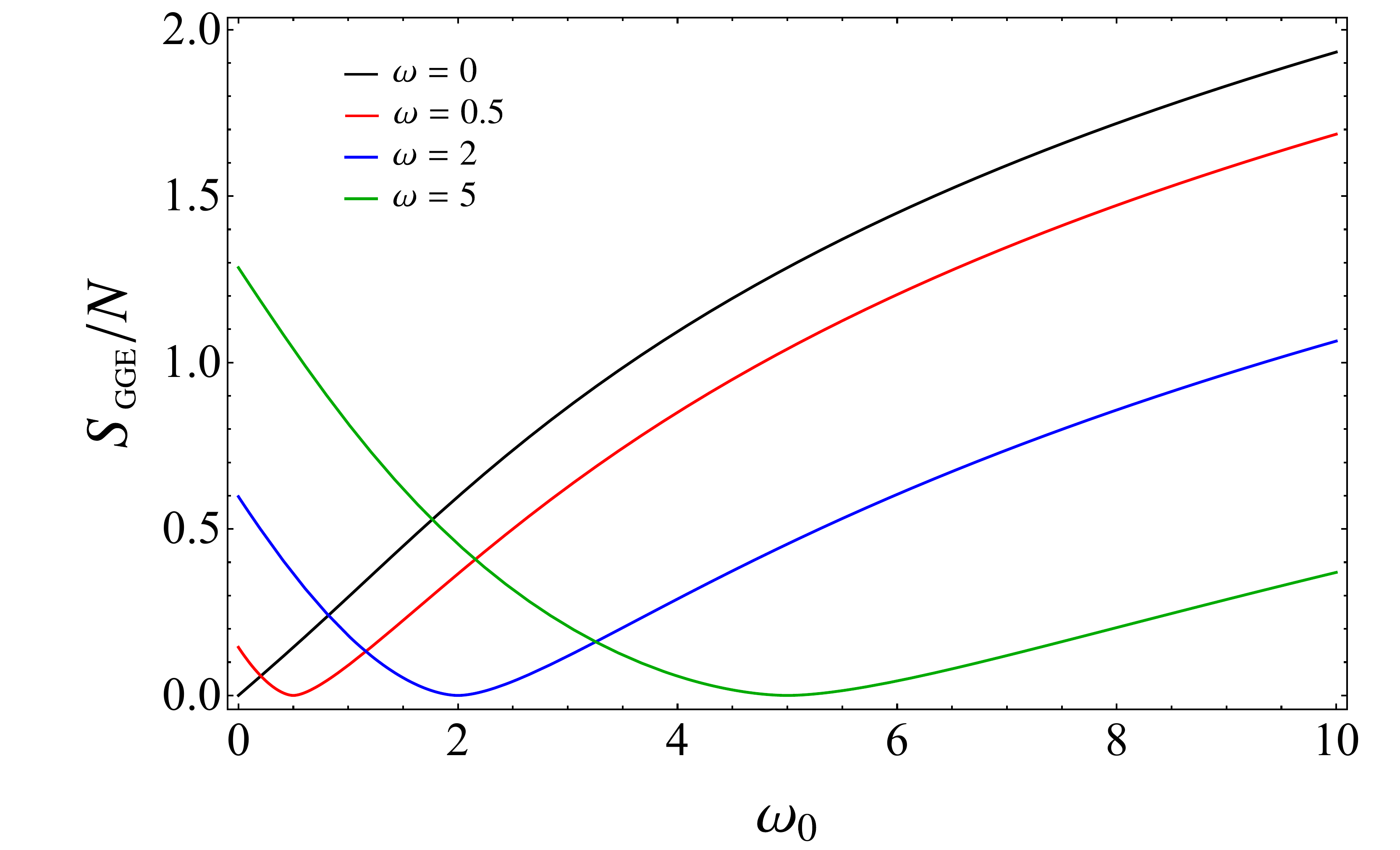}}
\caption{Asymptotic value of $\mathcal{C}_{\textrm{\tiny GGE}}/\sqrt{N}$ from (\ref{comp full GGE TD}) (left panel) 
and of $S_{\textrm{\tiny GGE}}/N$ from (\ref{SGGE}) (right panel) as functions of $\omega_0$, for some values of $\omega$.
}
\vspace{0.4cm}
\label{fig:CGGEanalytic}
\end{figure} 

In Fig.\,\ref{fig:CGGEanalytic} we show $\mathcal{C}_{\textrm{\tiny GGE}}/\sqrt{N}$ from (\ref{comp full GGE TD}) 
and $S_{\textrm{\tiny GGE}}/N$ from (\ref{SGGE}) as functions of $\omega_0$, for some values of $\omega$.  
The resulting curves are qualitatively similar.
At $\omega_0 = \omega$ they both vanish,
but $\mathcal{C}_{\textrm{\tiny GGE}}/\sqrt{N}$ is singular at this point, 
while $S_{\textrm{\tiny GGE}}/N$ is smooth.

The reduced covariance matrix $\gamma_{\textrm{\tiny GGE},A}$ associated to any finite subsystem $A$
is obtained by selecting the rows and the columns in (\ref{gamma-GGE-matrix}) corresponding to $A$.
The results of \cite{DiGiulio:2020hlz} can be applied again 
to write the optimal circuit  that provides $\gamma_{\textrm{\tiny GGE},A}$  
from the initial mixed state characterised by the reduced covariance matrix 
$\gamma_{A}(0)$ at $t=0$, 
obtained from $\gamma(0)$ through the usual reduction procedure. 
This optimal circuit reads
\be
\label{optimal circuit GGE}
G_s(\gamma_A(0) \, , \gamma_{\textrm{\tiny GGE},A})
\,\equiv \,
\gamma_A(0)^{1/2} 
\Big(  \gamma_A(0)^{- 1/2}  \,\gamma_{\textrm{\tiny GGE},A} \,\gamma_A(0)^{-1/2}  \Big)^s
\gamma_A(0)^{1/2} 
\ee
where $0 \leqslant s \leqslant 1$ parametrises the covariance matrix along the optimal circuit.
Its length corresponds to the subsystem complexity of the GGE w.r.t. the initial state
\be 
\label{comp-initialvsGGE}
\mathcal{C}_{\textrm{\tiny GGE},A}
\,=\,
\frac{1}{2\sqrt{2}}\;
\sqrt{
\textrm{Tr} \,\Big\{ \big[
\log \!\big( \gamma_{\textrm{\tiny GGE},A} \, \gamma_{A}(0)^{-1} \big)
\big]^2  \Big\}}\;.
\ee

Since the harmonic chain relaxes locally to the GGE after the quantum quench,
for the subsystem complexity of any finite subsystem $A$ we expect
\be
\label{stationary complexity}
\lim_{t\to\infty} \, \lim_{N\to\infty}\mathcal{C}_A=\lim_{N\to\infty}\mathcal{C}_{\textrm{\tiny GGE},A}
\ee
which is confirmed by 
the numerical results in 
Fig.\,\ref{fig:MixedStateGlobalMassiveEvolutionTDDetDimensionless},
Fig.\,\ref{fig:MixedStateGlobalMassiveEvolutionTDLargeomega0DetDimensionless},
Fig.\,\ref{fig:MixedStateGlobalMasslessEvolutionTD},
Fig.\,\ref{fig:MixedStateGlobalMassiveEvolutionTD}
and 
Fig.\,\ref{fig:MixedStateGlobalMassiveEvolutionTDLarget}.

A numerical analysis shows that (\ref{stationary complexity})
grows like $\sqrt{L}$ as $L \to \infty$
for fixed values of $\omega$ and $\omega_0$;
hence,  by adapting  (\ref{stationary entropy density}) to the subsystem complexity, 
we expect 
 \be
\label{stationary complexity density}
\lim_{L\to\infty}\frac{\lim_{t\to\infty}\lim_{N\to\infty}\mathcal{C}_{A}}{\sqrt{L}}
=
\lim_{L\to\infty}\frac{\lim_{N\to\infty}\mathcal{C}_{\textrm{\tiny GGE},A}}{\sqrt{L}}=\lim_{N\to\infty}\frac{\mathcal{C}_\textrm{\tiny GGE}}{\sqrt{N}}
\ee
where $\mathcal{C}_\textrm{\tiny GGE}$ is given in (\ref{comp full GGE TD})
and  the order of the limits is important.
Numerical evidences for (\ref{stationary complexity density}) 
are discussed in appendix\;\ref{app:gge}
(see Fig.\,\ref{fig:CAGGEvsCGGE} 
and Fig.\,\ref{fig:MixedStatePBCGlobalMasslessEvolutionTD}).

In the following numerical analysis we show that,
for the harmonic chains that we are exploring,
the asymptotic limit for $t\to \infty$ of the reduced density matrix after the global quench
is the reduced density matrix obtained from the GGE.
This result has been already discussed for a fermionic chain in \cite{Fagotti_2013},
where, considering a global quench of the magnetic field in the transverse-field Ising chain
and the subsystem given by a finite block made by consecutive sites in an infinite chain on the line, 
it has been found that a properly defined distance 
between the reduced density matrix at a generic value of time 
along the evolution and the asymptotic one obtained from the GGE vanishes as $t \to \infty$.

\subsection{Numerical results}
\label{sec:numerics-gge}

In order to test (\ref{comp-initialvsGGE}), infinite harmonic chains must be considered. 
The reference and the target states have been described in Sec.\,\ref{sec:mixedFinSize}.
In this section we study harmonic chains both on the line 
and on the semi-infinite line with DBC imposed at its origin. 
In the latter case, the block $A$ made by $L$ consecutive sites 
is either adjacent to the origin or separated from it. 

The correlators to employ in the numerical analysis
can be obtained from the ones reported in Sec.\,\ref{sec:mixedFinSize}. 
For the infinite harmonic chain on the line, 
we take $N\to\infty$ of (\ref{QPRmat t-dep 1d}), finding 
\be
\label{QPRmat t-dep TD}
\begin{array}{l}
\displaystyle
Q_{i,j}(t) 
=
\frac{1}{\pi} \int_{0}^{\pi} Q_{\theta}(t) \cos\!\big[2\theta\,(i-j)\,\big] \, d\theta
\\
\rule{0pt}{.9cm}
\displaystyle
P_{i,j}(t) 
=
\frac{1}{\pi} \int_{0}^{\pi} P_{\theta}(t) \cos\!\big[2\theta\,(i-j)\,\big] \, d\theta
 \\
 \displaystyle
 \rule{0pt}{.9cm}
 \displaystyle
M_{i,j}(t) 
=
\,\frac{1}{\pi} \int_{0}^{\pi} M_\theta(t) \cos\!\big[2\theta\,(i-j)\,\big] \, d\theta
\end{array}
\ee
where $i,j\in \mathbb{Z}$;
while, for the harmonic chain on the semi-infinite line with DBC, 
the limit $N\to\infty$ of (\ref{QPRmat t-dep 1d DBC}) leads to
\be
\label{QPRmat t-dep DBC TD}
\begin{array}{l}
\displaystyle
Q_{i,j}(t) 
=
\frac{2}{\pi}\int_{0}^{\pi} Q_{\theta}(t)\sin(i\theta) \sin(j\theta)  \, d\theta
\\
\rule{0pt}{.9cm}
\displaystyle
P_{i,j}(t) 
=
\frac{2}{\pi} \int_{0}^{\pi} P_{\theta}(t)\sin(i\theta) \sin(j\theta)  \, d\theta
 \\
 \displaystyle
 \rule{0pt}{.9cm}
 \displaystyle
M_{i,j}(t) 
=
\frac{2}{\pi} \int_{0}^{\pi} M_\theta(t)\sin(i\theta) \sin(j\theta)  \, d\theta
\end{array}
\ee
where $i,j>0$.
The functions $Q_\theta(t)$, $P_\theta(t) $ and $M_\theta(t)$ 
in these integrands
are given by (\ref{QPRmat t-dep-k}) 
where $\Omega_{0,k}$ and $\Omega_k$
are replaced respectively by $\Omega_{0,\theta}$ and $\Omega_\theta$,
which are (\ref{dispersion relations TD}) and (\ref{dispersion DBC TD}) 
for the infinite and for the semi-infinite line respectively.

\begin{figure}[htbp!]
\vspace{-.6cm}
{
\subfigure
{\hspace{-1.05cm}
\includegraphics[width=1.03\textwidth]{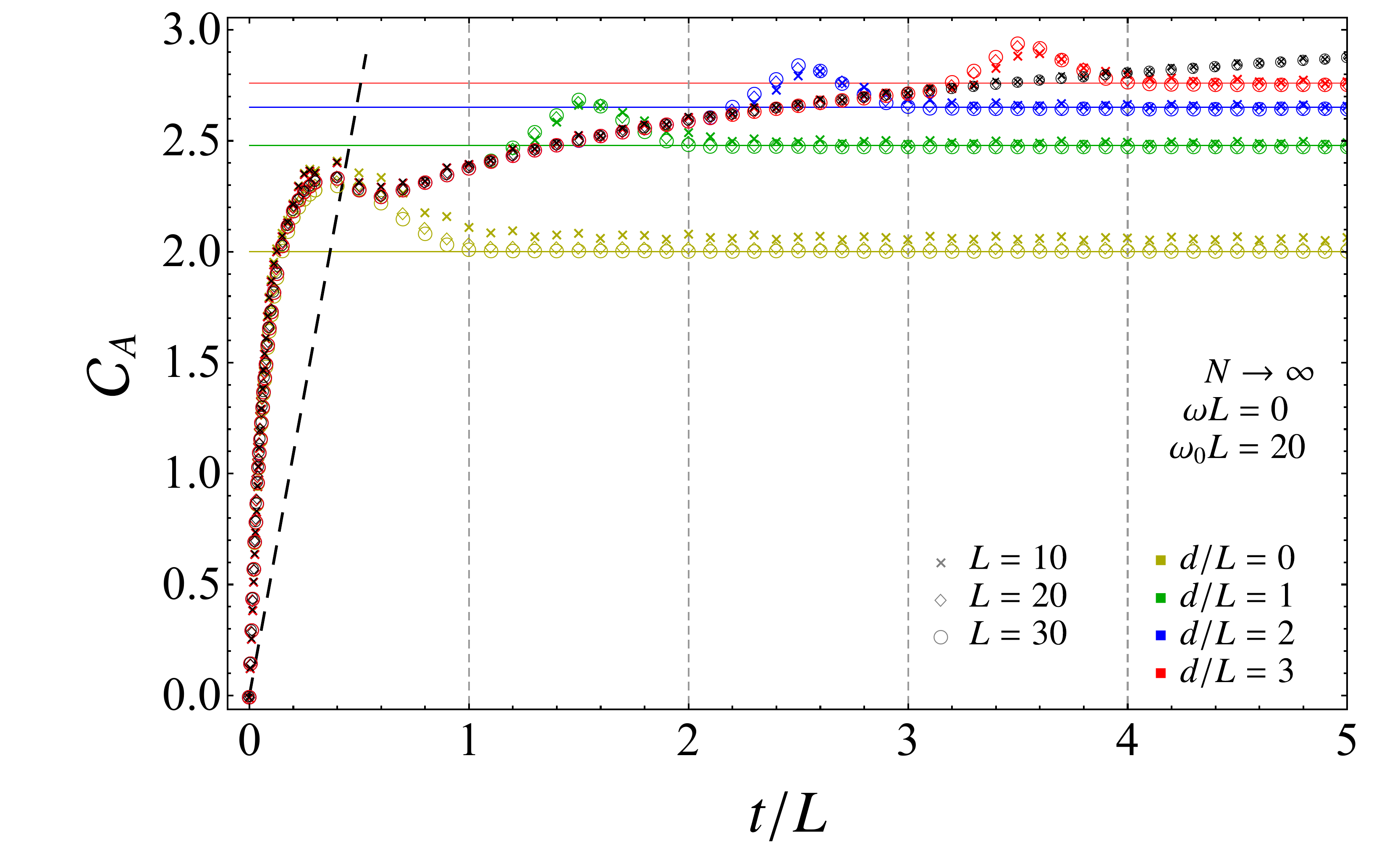}}
\subfigure
{\hspace{-1.05cm}
\includegraphics[width=1.03\textwidth]{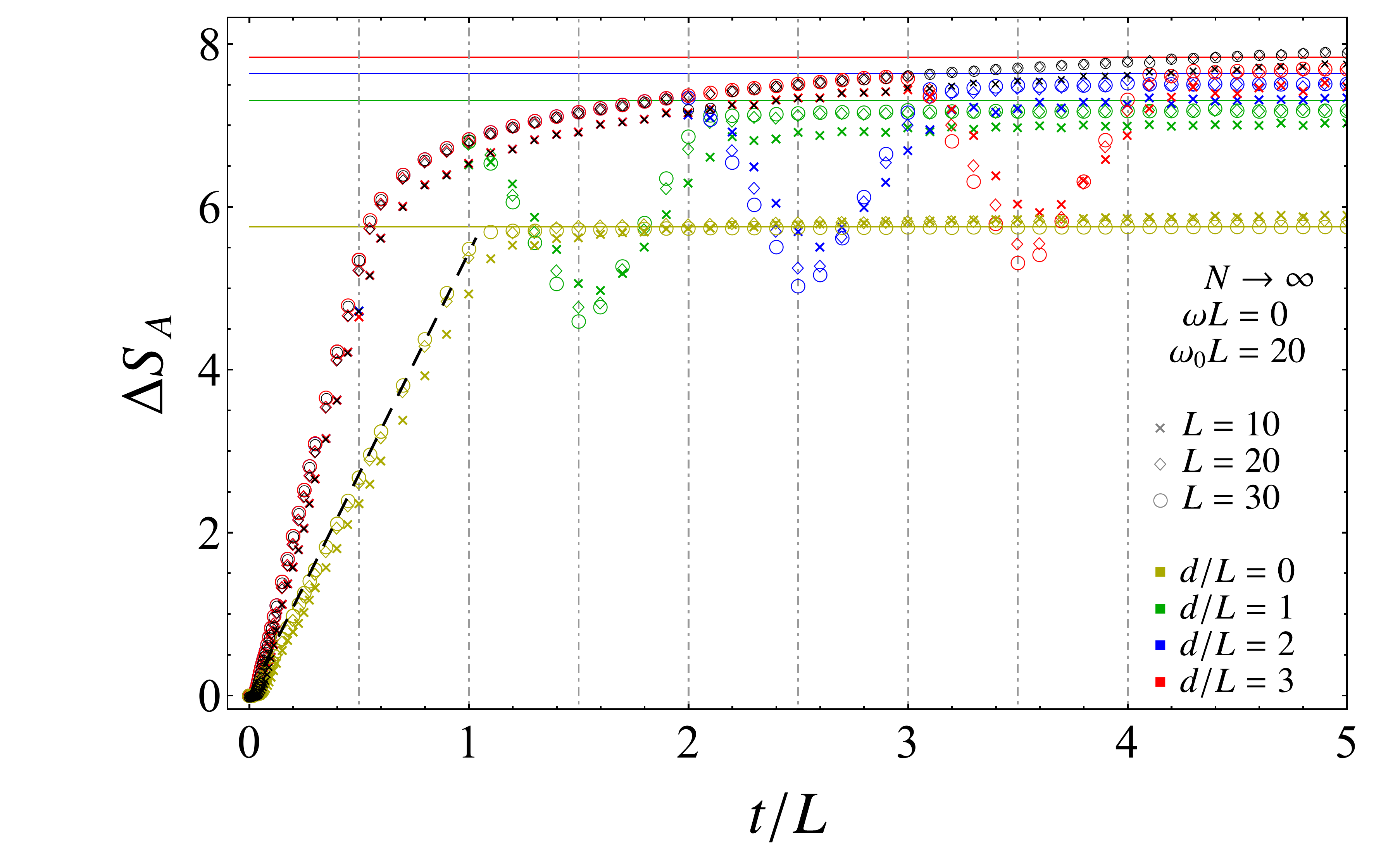}}
}
\caption{
Temporal evolution of $\mathcal{C}_A$ (top panel) and of $\Delta S_A$ (bottom panel) 
after a global quantum quench with a gapless evolution Hamiltonian and $\omega_0 L =20$.
The subsystem is a block $A$ made by $L$ consecutive sites 
either on the infinite line (black data points)
or on the semi-infinite line,
separated by $d$ sites from the origin where DBC hold (coloured data points).
The dashed black straight line is the same in both panels. 
}
\vspace{0.4cm}
\label{fig:MixedStateGlobalMassiveEvolutionTDDetDimensionless}
\end{figure} 

\begin{figure}[t!]
{
\subfigure
{\hspace{-1.05cm}
\includegraphics[width=1.03\textwidth]{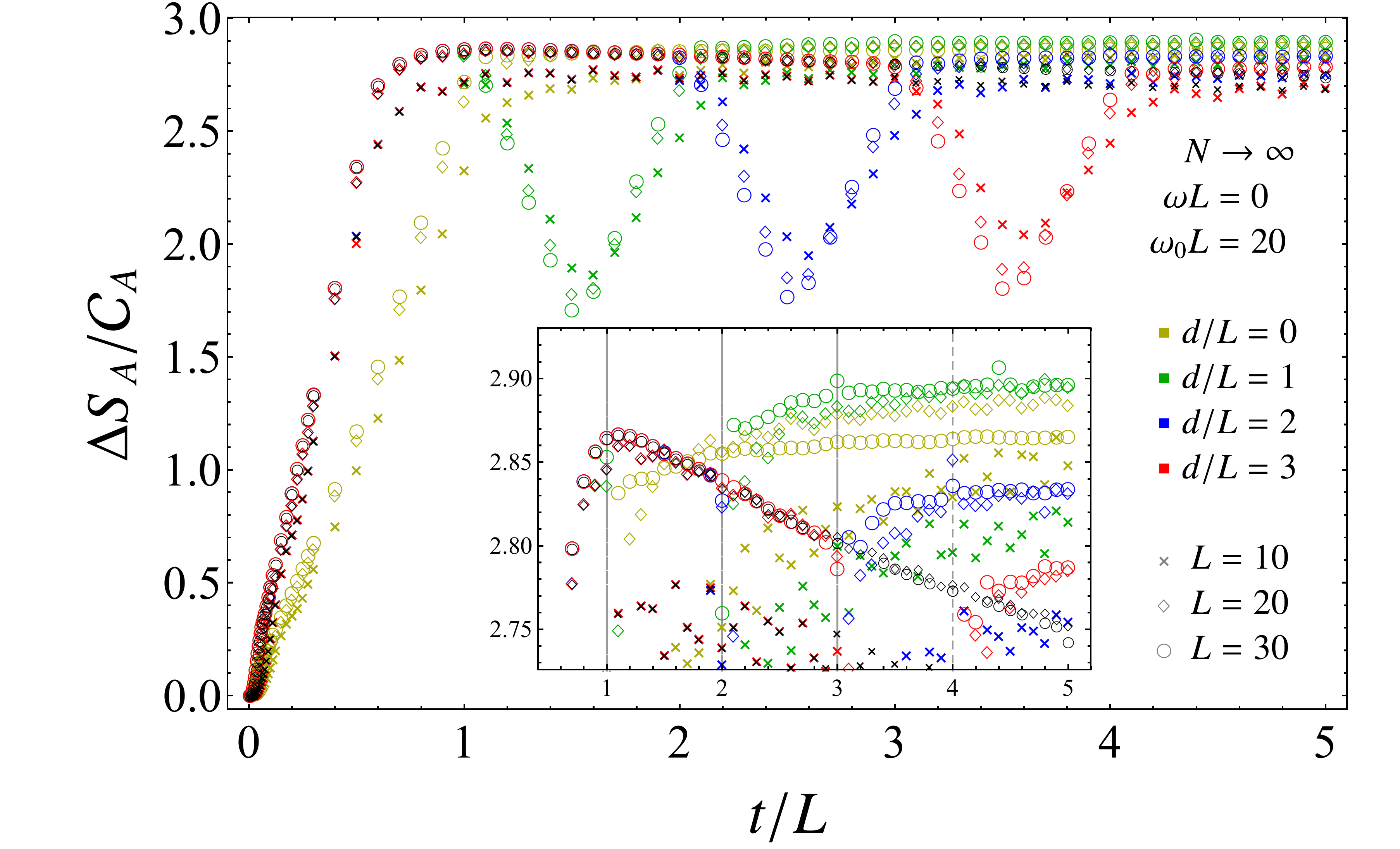}}
}
\caption{Temporal evolution of $\Delta S_A/\mathcal{C}_A$ for the data reported 
in Fig.\,\ref{fig:MixedStateGlobalMassiveEvolutionTDDetDimensionless}.
The inset zooms in to highlight the data points having $t/L >1$.
}
\vspace{0.4cm}
\label{fig:RatioGlobalMassiveEvolutionTDDetDimensionless}
\end{figure}

\begin{figure}[t!]
{
\subfigure
{\hspace{-1.05cm}
\includegraphics[width=1.03\textwidth]{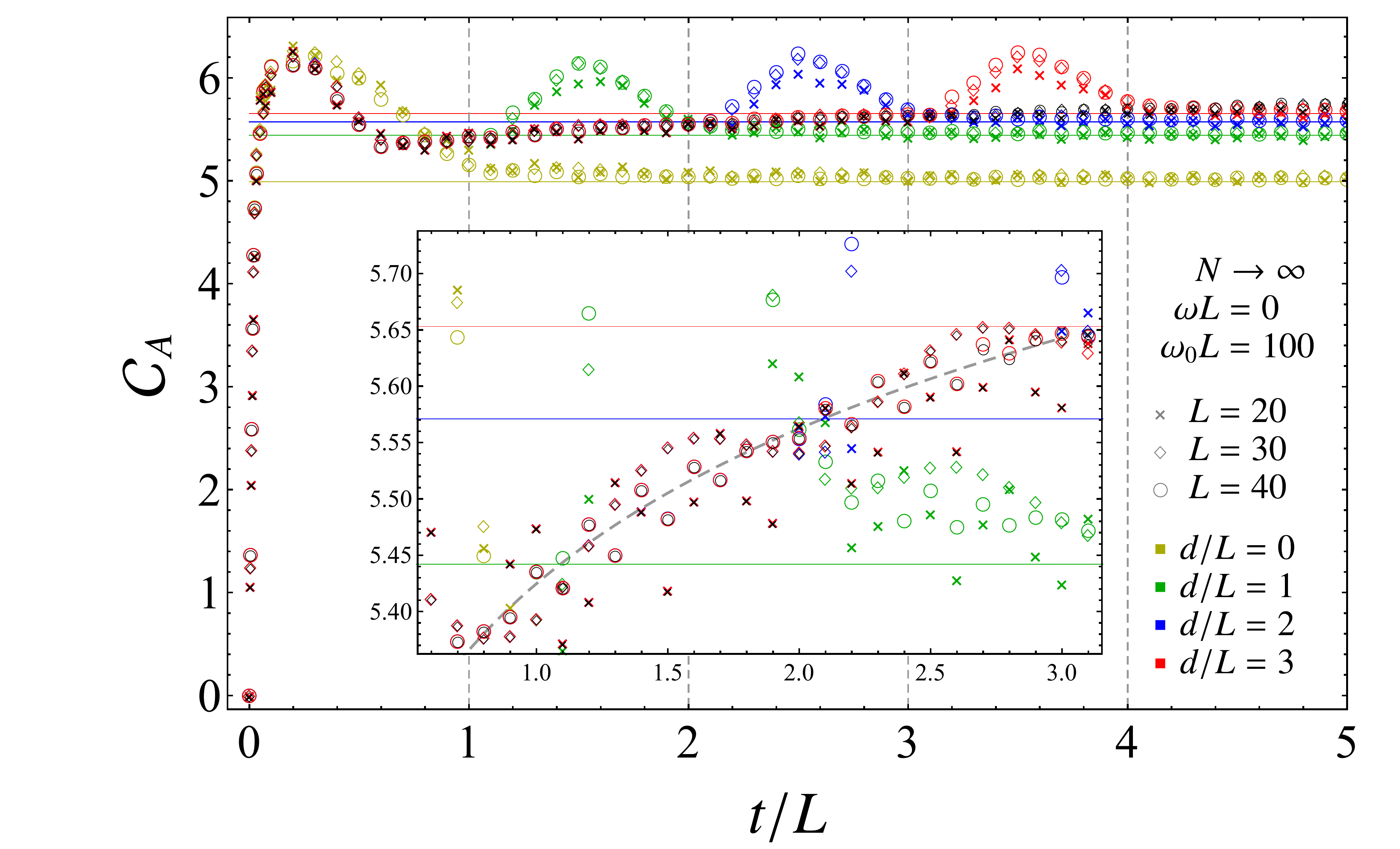}}
}
\caption{Temporal evolution of $\mathcal{C}_A$
after a global quantum quench with a gapless evolution Hamiltonian and $\omega_0 L =100$,
in the same setups of Fig.\,\ref{fig:MixedStateGlobalMassiveEvolutionTDDetDimensionless}.
The inset zooms in on the intermediate temporal regime
between the two local maxima for the data having $d/L =3$.
}
\vspace{0.4cm}
\label{fig:MixedStateGlobalMassiveEvolutionTDLargeomega0DetDimensionless}
\end{figure}


Once the proper correlators on the chain are identified, 
the reduced correlation matrices $Q_A$, $P_A$ and $M_A$ 
are the blocks providing the reduced covariance matrix (\ref{reduced CM}).
These matrices are obtained by restricting the indices of the proper correlators 
to $i,j=1,\dots,L$ when $A$ is on the infinite line 
and to $i,j=1+d,\dots,L+d$ when $A$ is on the semi-infinite line,
where $d$ corresponds to its separation from the origin.


In the following we discuss numerical data sets 
obtained for infinite harmonic chains,
either on the infinite line or on the semi-infinite line, 
where $\omega L$ and $\omega_0 L$ are kept fixed.
In appendix\;\ref{app:gge} we report numerical results characterised
by fixed values of $\omega$ and $\omega_0$.

\begin{figure}[t!]
\subfigure
{\hspace{-1.65cm}
\includegraphics[width=.575\textwidth]{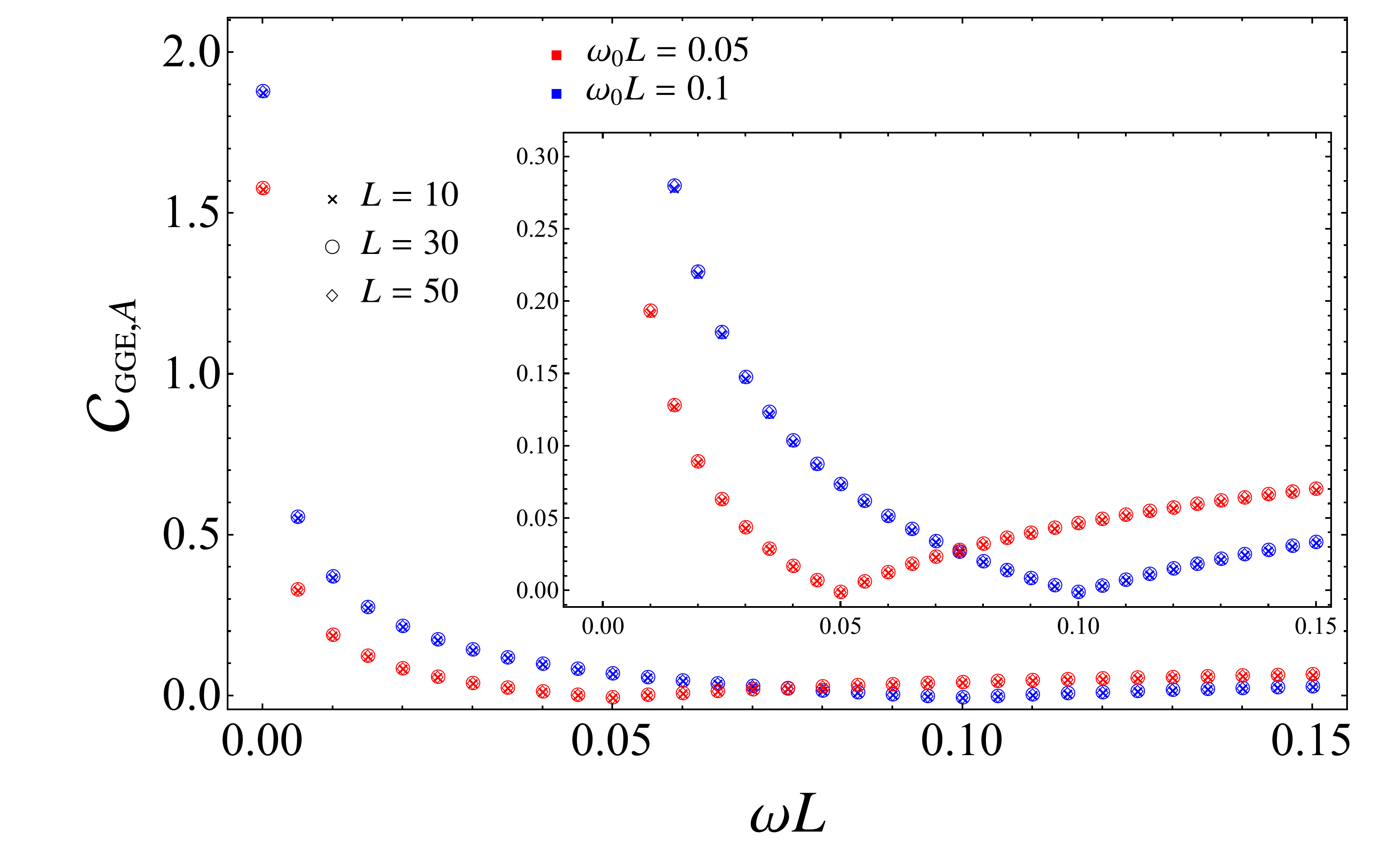}}
\subfigure
{
\hspace{-.3cm}\includegraphics[width=.575\textwidth]{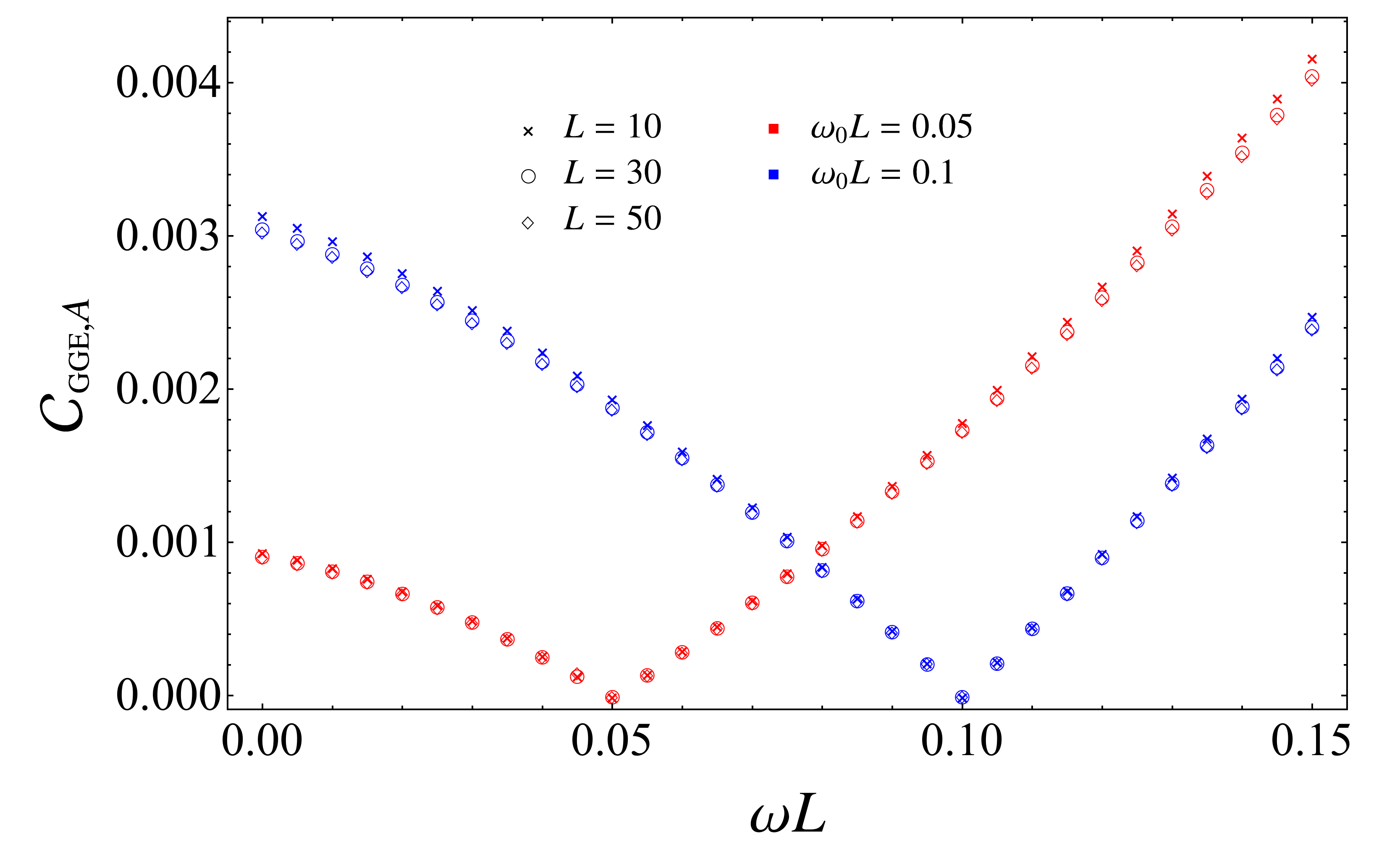}}
\caption{Asymptotic value of $\mathcal{C}_{\textrm{\tiny GGE},A}$ in (\ref{comp-initialvsGGE}) 
for a block $A$ made by $L$ consecutive sites in infinite chains in terms of $\omega L$. 
The block is either in an infinite chain (left panel) 
or adjacent to the origin of the semi-infinite line with DBC (right panel).
}
\vspace{0.4cm}
\label{fig:GGE}
\end{figure}

In Fig.\,\ref{fig:MixedStateGlobalMassiveEvolutionTDDetDimensionless} 
and Fig.\,\ref{fig:RatioGlobalMassiveEvolutionTDDetDimensionless} 
we show the temporal evolutions of $\mathcal{C}_A$,
of $\Delta S_A$ and of $\Delta S_A / \mathcal{C}_A$
after the quench with $\omega_0 L =20$ and $\omega L =0$.
In Fig.\,\ref{fig:MixedStateGlobalMassiveEvolutionTDLargeomega0DetDimensionless}
we display the temporal evolution of $\mathcal{C}_A$ 
with $\omega_0 L =100$ and $\omega L =0$.
The subsystem $A$ is a block made by $L$ consecutive sites 
either on a semi-infinite line,
separated by $d$ sites from the origin where DBC are imposed (coloured symbols),
or on the infinite line (black symbols).
The black and coloured data points  for $\mathcal{C}_A$ 
have been found through (\ref{c2-complexity-rdm-our-case})
with the reduced correlators obtained from either (\ref{QPRmat t-dep TD}) or (\ref{QPRmat t-dep DBC TD}) respectively. 
The coloured horizontal solid lines correspond to
either (\ref{comp-initialvsGGE}) or (\ref{EE-initialvsGGE}),
with the reduced correlators from (\ref{corr-GGE-app-int-dbc}) for the target state
and from (\ref{QPRmat t-dep DBC TD}) at $t=0$ for the reference state,
with $L=50$.
Notice that a black horizontal solid line does not occur 
because the corresponding value is divergent, 
as indicated also by the left panel in Fig.\,\ref{fig:GGE}.

Considering the block on the semi-infinite line, 
in Fig.\,\ref{fig:MixedStateGlobalMassiveEvolutionTDDetDimensionless} 
and Fig.\,\ref{fig:MixedStateGlobalMassiveEvolutionTDLargeomega0DetDimensionless} 
we observe that 
the initial growth of $\mathcal{C}_A$ is the same until the first local maximum, for all the values of $d/L$.
After the first local maximum, the temporal evolution of $\mathcal{C}_A$ depends on whether 
the block is adjacent to the boundary.
If $d/L =0$ the curve decreases until it reaches the saturation value.
Instead, when $d/L > 0$, first $\mathcal{C}_A$ decreases along a different curve 
(see e.g. Fig.\,\ref{fig:MixedStateGlobalMassiveEvolutionTDLargeomega0DetDimensionless})
until a local minimum;
then we observe an intermediate growth,
followed by a second local maximum 
and finally by the saturation regime. 
A fitting procedure shows that the intermediate growth
between the two local maxima is logarithmic
(in the inset of 
Fig.\,\ref{fig:MixedStateGlobalMassiveEvolutionTDLargeomega0DetDimensionless} the grey dashed curve has been found by fitting the data having $L=40$ and $d/L=3$
through a logarithm and a constant).
Its temporal duration is approximatively $d/L-1/2$,
for the three values of non vanishing $d/L$ considered in 
Fig.\,\ref{fig:MixedStateGlobalMassiveEvolutionTDDetDimensionless}
and Fig.\,\ref{fig:MixedStateGlobalMassiveEvolutionTDLargeomega0DetDimensionless}.
Fitting the intermediate growth 
in Fig.\,\ref{fig:MixedStateGlobalMassiveEvolutionTDDetDimensionless} 
and Fig.\,\ref{fig:MixedStateGlobalMassiveEvolutionTDLargeomega0DetDimensionless},
one observes that
the coefficient of the logarithmic growth decreases as $\omega_0 L$ increases.
The first local maximum in the temporal evolution of $\mathcal{C}_A$  occurs for $0 < t/L <1$.
When $d>0$, the second maximum occurs for $d/L < t/L < (d+1)/L$.
Notice that these two local maxima can be seen also in the top panels of  
Fig.\,\ref{fig:MixedStateGlobalMasslessDBCIntDet} for $t/N < 1/2$.

In Fig.\,\ref{fig:MixedStateGlobalMassiveEvolutionTDDetDimensionless},
Fig.\,\ref{fig:RatioGlobalMassiveEvolutionTDDetDimensionless} and
Fig.\,\ref{fig:MixedStateGlobalMassiveEvolutionTDLargeomega0DetDimensionless},
the data points represented through black symbols have been obtained
for a block in the infinite line. 
These data overlap with the ones corresponding to the block on the semi-infinite line with $d>0$ 
until the latter ones display the development of the second local maximum.
For the temporal evolution of $\mathcal{C}_A$ on the infinte line
only one local maximum occurs
and the intermediate logarithmic growth mentioned above does not finish 
within the temporal regime that we have considered. 
This agreement tells us that the second local maximum in 
the temporal evolution of $\mathcal{C}_A$ is due to the presence of the boundary.

  The temporal evolutions of $\Delta S_A$ in the bottom panel of 
 Fig.\,\ref{fig:MixedStateGlobalMassiveEvolutionTDDetDimensionless}
 can be explained  by employing the quasi-particle picture \cite{Calabrese_2005},
 which provides the different temporal regimes 
 and the corresponding qualitative behaviour of $\Delta S_A$
 (for the subsystems where a boundary occurs, 
 the quasi-particle picture has been described e.g. in \cite{Surace:2019mft}).
 The different regimes identified by this analysis correspond to the vertical dot-dashed lines
 in the bottom panel of Fig.\,\ref{fig:MixedStateGlobalMassiveEvolutionTDDetDimensionless}.
 Instead, the vertical dashed grey lines in the top panel of 
 Fig.\,\ref{fig:MixedStateGlobalMassiveEvolutionTDDetDimensionless}
 correspond to $t/L=1+d/L$.
 For $d>0$, when $t/L > 1/2$ 
 we observe a regime of logarithmic growth for $\Delta S_A$
 whose duration depends on $d/L$ according to 
 the quasi-particle picture, until the beginning of a linear decreases.
Considering two sets of data points of  $\Delta S_A$ having different $d/L$,
they collapse  until the first  linear decrease is reached. 

 The initial growths of $\mathcal{C}_A$ and of $\Delta S_A$ 
 in Fig.\,\ref{fig:MixedStateGlobalMassiveEvolutionTDDetDimensionless} 
 are very different.
For instance, 
the growth of $\mathcal{C}_A$ is the same for all the data sets,
 while for $\Delta S_A$ it depends on whether $d$ vanishes.
Moreover, while the growth of $\Delta S_A$ is linear for $t/L < 1$ when $d=0$
and  for $t/L < 1/2$ when $d>0$,
the growth of $\mathcal{C}_A$  is linear only at the very beginning of the temporal evolution
and it clearly deviates from linearity within the regime of $t/L$ where $\Delta S_A$ grows linearly. 
 The dashed black straight line passing through the origin
 in Fig.\,\ref{fig:MixedStateGlobalMassiveEvolutionTDDetDimensionless}
 describes the linear growth of $\Delta S_A$ when $d=0$
 and it is the same in both the panels.
 This straight line intersects the first local maximum of $\mathcal{C}_A$.
 This has been highlighted also for finite systems
 in Fig.\,\ref{fig:MixedStateGlobalMasslessEvolutionDimensionless} 
 and Fig.\,\ref{fig:MixedStateGlobalMasslessDBCIntDet}.

In Fig.\,\ref{fig:RatioGlobalMassiveEvolutionTDDetDimensionless}
we show the ratio $ \Delta S_A/\mathcal{C}_A$ for the data reported in 
Fig.\,\ref{fig:MixedStateGlobalMassiveEvolutionTDDetDimensionless}.
We remark that the two logarithmic growths occurring in 
$ \Delta S_A$ and in $\mathcal{C}_A$ almost cancel in the ratio;
indeed, a mild logarithmic decreasing is observed 
when $t/L > 1$ for the data obtained on the infinite line (black symbols) and 
when $1 < t/L < 3$ for the data obtained on the semi-infinite line 
with $d/L=3$ (red symbols) that are already collapsed.

The curves in 
Fig.\,\ref{fig:MixedStateGlobalMassiveEvolutionTDLargeomega0DetDimensionless}
must be compared with the corresponding ones in top panel in 
Fig.\,\ref{fig:MixedStateGlobalMassiveEvolutionTDDetDimensionless}
in order to explore the effect of $\omega_0 L$.
 The height of the first local maximum in the temporal evolution of $\mathcal{C}_A$
 and also the saturation values for the data obtained on the semi-infinite line
 increase as $\omega_0 L$ increases. 
 Instead, the coefficient of the logarithmic growth after the first local maximum
 decreases as $\omega_0 L$ increases, as already remarked above. 
 Notice that higher values of $L$ are needed to observe data collapse as $\omega_0 L$ increases.

From the numerical results reported in the previous figures,
we conclude that (\ref{comp-initialvsGGE}) 
provides the asymptotic value of the subsystem complexity as $t \to \infty$;
hence it is  worth studying the dependence of this expression on 
the subsystem size and on the parameters of the quench protocol. 

\begin{figure}[t!]
\subfigure
{
\hspace{3.0cm}\includegraphics[width=.57\textwidth]{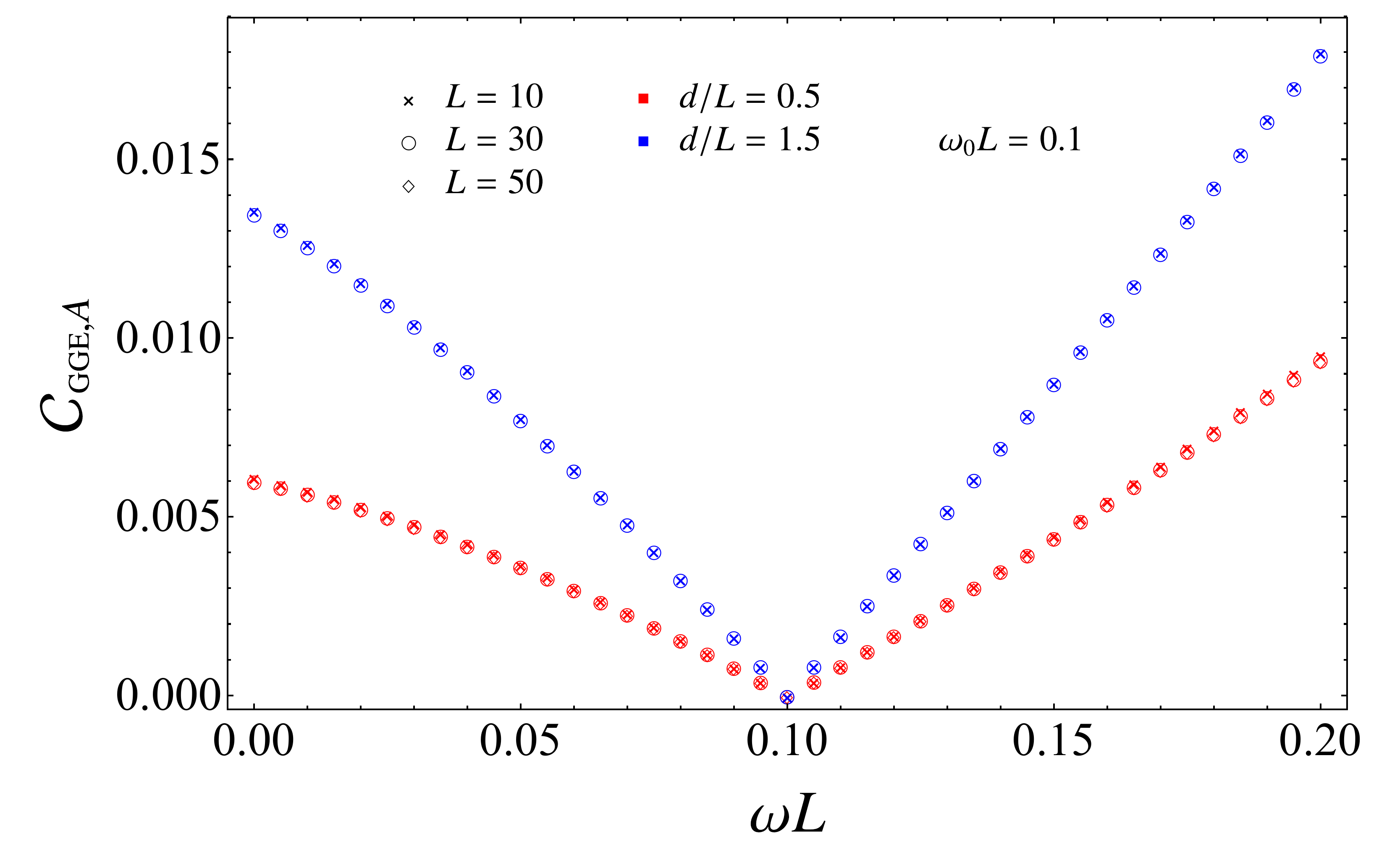}}
\\
\subfigure
{
\hspace{-1.55cm}\includegraphics[width=.57\textwidth]{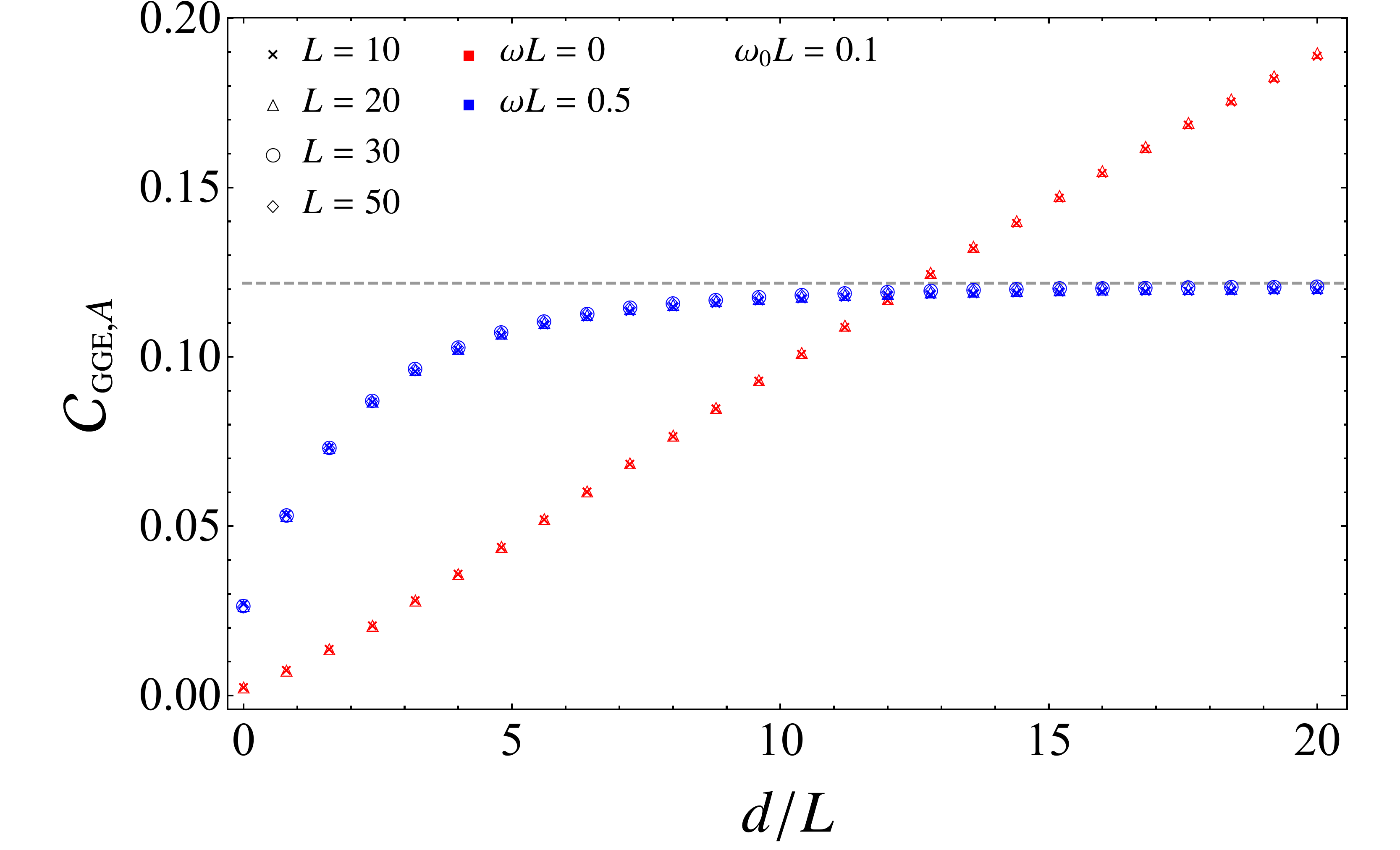}}
\subfigure
{\hspace{-0.45cm}
\includegraphics[width=.57\textwidth]{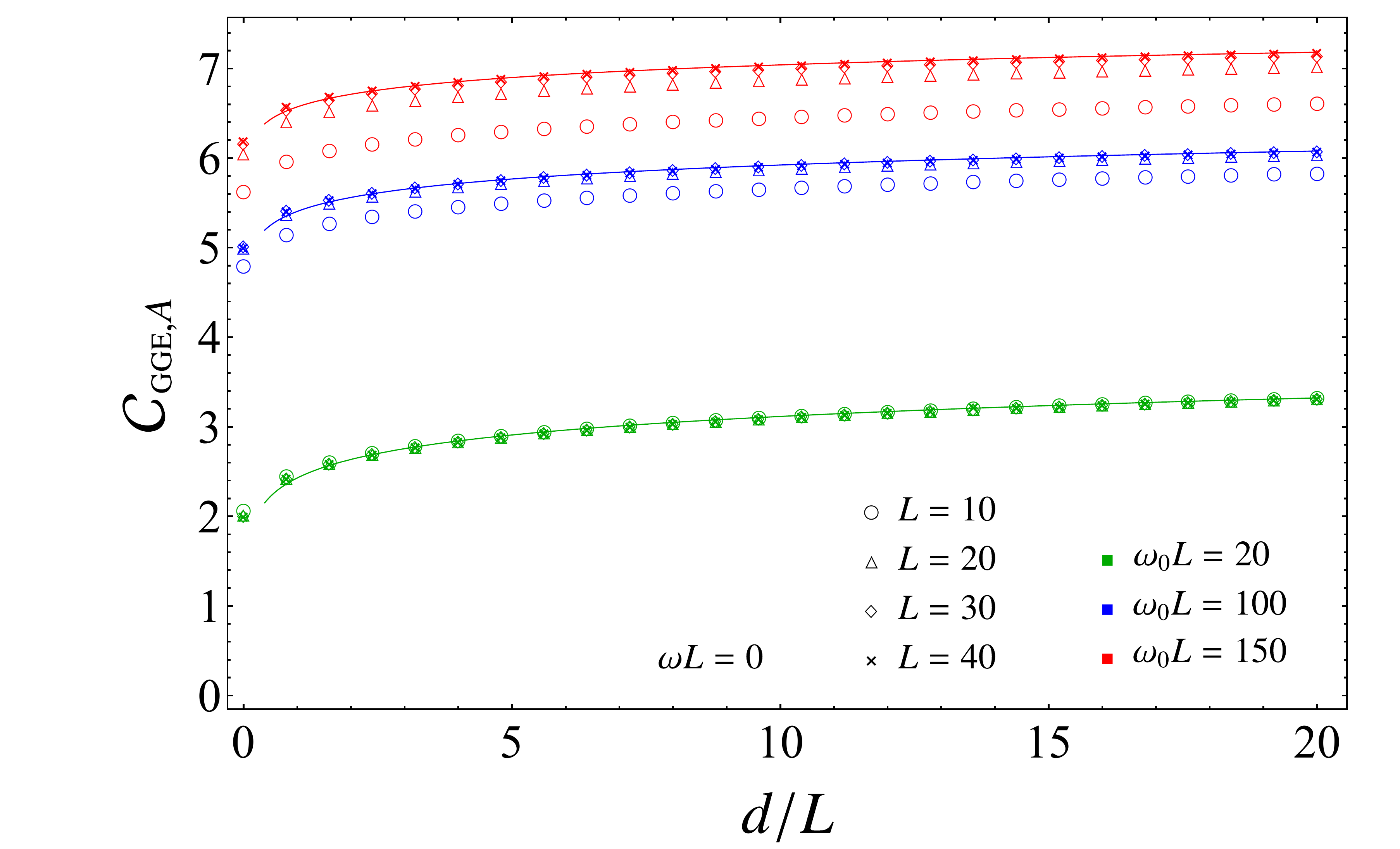}}
\caption{
Asymptotic value of $\mathcal{C}_{\textrm{\tiny GGE},A}$ in (\ref{comp-initialvsGGE}) 
for a block made by $L$ consecutive sites 
and separated by $d$ sites from the origin of a semi-infinite line with DBC,
in terms of $\omega L$ (top panel) and of $d/L$ (bottom panels). 
} 
\vspace{0.4cm}
\label{fig:GGEdet}
\end{figure}

In Fig.\,\ref{fig:GGE} and Fig.\,\ref{fig:GGEdet}  
we show numerical results for (\ref{comp-initialvsGGE}),
obtained by using the reduced correlators 
from (\ref{corr-GGE-app-int-pbc}) and (\ref{corr-GGE-app-int-dbc})
for the target state and the reduced correlators 
from (\ref{QPRmat t-dep TD}) and (\ref{QPRmat t-dep DBC TD}) at $t=0$
for the reference state. 
%

In Fig.\,\ref{fig:GGE} we show  (\ref{comp-initialvsGGE}) as function of $\omega L$ 
when the block is either in the infinite line (left panel) or at the beginning of the semi-infinite line with DBC (right panel).
The main difference between  the two panels of Fig.\,\ref{fig:GGE}
is that the limit $\omega L \to 0$ is finite for the semi-infinite line
while it diverges for the infinite line 
(the correlators (\ref{corr-GGE-app-int-pbc}) are well defined for  $\omega \neq 0$).
This is consistent with the results displayed through the black symbols 
in the top panel of Fig.\,\ref{fig:MixedStateGlobalMassiveEvolutionTDDetDimensionless} 
and in Fig.\,\ref{fig:MixedStateGlobalMassiveEvolutionTDLargeomega0DetDimensionless}.

In Fig.\,\ref{fig:GGEdet} we study (\ref{comp-initialvsGGE}) 
for a block on the semi-infinite line,
separated by $d$ sites from the origin where DBC are imposed. 
For a given value of $\omega_0 L$, we show $\mathcal{C}_{\textrm{\tiny GGE},A}$
as function of $\omega L$ at fixed $d/L$ (top panel) and viceversa (bottom panels).
The qualitative behaviour of the curves in the top panel of Fig.\,\ref{fig:GGEdet}
is similar to the one in the right panel of Fig.\,\ref{fig:GGE}.
In the bottom left panel of Fig.\,\ref{fig:GGEdet}, 
as $d/L \to \infty$,
the data points with $\omega L > 0$
asymptote (horizontal dashed line)
to the value of $\mathcal{C}_{\textrm{\tiny GGE},A}$ 
obtained through (\ref{comp-initialvsGGE})
with the reduced correlators
(\ref{corr-GGE-app-int-pbc})  for the target state
and (\ref{QPRmat t-dep TD}) at $t=0$ for the reference state.
Instead, when $\omega L = 0$ the data in the bottom left panel of Fig.\,\ref{fig:GGEdet}
do not have a limit as $d/L$ increases. 
This is consistent with the divergence of the curves in left panel of Fig.\,\ref{fig:GGE}
as $\omega L \to 0$.
In the bottom right panel of Fig.\,\ref{fig:GGEdet} we consider a critical evolution Hamiltonian
and large values of $\omega_0 L$.
In this regime of parameters, we highlight the logarithmic growth of 
$\mathcal{C}_{\textrm{\tiny GGE},A}$ in terms of $d/L$
(the solid lines are obtained by fitting the data corresponding to $L=40$ 
through the function $a \log(d/L)+b$).

The numerical data sets discussed in this section 
are characterised by fixed values of $\omega L$ and $\omega_0 L$.
In appendix\;\ref{app:gge} we report numerical results
where $\omega$ and $\omega_0$ are kept fixed:
besides supporting further the validity of (\ref{comp-initialvsGGE}),
this analysis provides numerical evidences for (\ref{stationary complexity density}).


Within the context of the gauge/gravity correspondence, 
the temporal evolution of the holographic subsystem complexity 
in the gravitational backgrounds given by Vaidya spacetimes
has been studied numerically through the CV proposal
\cite{Chen:2018mcc,Auzzi:2019mah,Ling:2019ien,Zhou:2019xzc}.

We find it worth remarking that
the qualitative behaviour of  the temporal evolution of $\mathcal{C}_A$
for an interval in the infinite line shown by the black data points 
in Fig.\,\ref{fig:MixedStateGlobalMassiveEvolutionTDDetDimensionless} 
and Fig.\,\ref{fig:MixedStateGlobalMassiveEvolutionTDLargeomega0DetDimensionless}
is in agreement with the results for 
the temporal evolution of the holographic subsystem complexity
reported in \cite{Chen:2018mcc,Auzzi:2019mah}.
The change of regime occurs at $t/L \simeq 1/2$ 
for both these quantities and their qualitative behaviour
in the initial regime given by $0< t/L < 1/2$ is very similar. 

For  $t/L > 1/2$ we observe a logarithmic growth whose coefficient depends on $\omega_0 L$
in Fig.\,\ref{fig:MixedStateGlobalMassiveEvolutionTDDetDimensionless} 
and Fig.\,\ref{fig:MixedStateGlobalMassiveEvolutionTDLargeomega0DetDimensionless},
while the holographic subsystem complexity remains constant. 
However, a similar issue occurs in the corresponding comparison for the entanglement entropy.

\section{Conclusions}
\label{sec:conclusions}

In this manuscript we studied the 
temporal evolution of the subsystem complexity after a global quench of the mass parameter
in harmonic lattices,
focussing our analysis on harmonic chains with either PBC or DBC
and on subsystems given by blocks of consecutive sites.
The initial state is mainly chosen as the reference state of the circuit. 
The circuit complexity of the mixed states described by the reduced density matrices
has been evaluated 
by employing the approach based on the Fisher information geometry \cite{DiGiulio:2020hlz},
which provides also the optimal circuit 
(see (\ref{optimal circuit rdm}) and (\ref{c2-complexity-rdm})).

When the entire system is considered (see Sec.\,\ref{sec-pure-states}, 
Sec.\,\ref{sec:comp-initial-state} and Sec.\,\ref{sec:purestates_HC_glob}),
the optimal circuit is made by pure states \cite{Jefferson:2017sdb,Chapman:2018hou}
and for the temporal evolution of the circuit complexity after the global quench
one obtains the expression given by (\ref{c2-log-lambda-arcosh}) and (\ref{CTR generic}),
which holds in a generic number of dimensions. 
When the reference and the target states are pure states along the time evolution  of a given quench,
we find that the complexity is given by (\ref{c2-log-lambda-arcosh}) and (\ref{CTRomegaReqomegaT}), 
which simplifies to (\ref{comp-pure-global-general})
in the case where the reference state is the initial state.
Specialising the latter result to the harmonic chains where either PBC or DBC are imposed,
one obtains (\ref{comp-pure-global-DBCPBC}), where the contribution of the zero mode for PBC is highlighted.
The occurrence of the zero mode provides 
the logarithmic growth of the complexity when the evolution is critical 
(see (\ref{C-pure-both-eta}) and Fig.\,\ref{fig:PureStateCritical}).
Typical temporal evolutions of the complexity for the entire chain 
when the post-quench Hamiltonian is massive are shown in Fig.\,\ref{fig:TDvsfinite}.

The bounds (\ref{naive-bounds}) and (\ref{new bounds_main})
are obtained for the temporal evolution of the complexity of the entire harmonic lattice.
The former ones are simple but not very accurate
(see Fig.\,\ref{fig:BoundNaive} for harmonic chains with PBC);
instead, the latter ones capture the dynamics of the complexity in a very precise way
but their analytic expressions are more involved. 
In the case of harmonic chains, the bounds (\ref{new bounds_main})
lead to the bounds (\ref{new bounds_main_weaker})
displayed in Fig.\,\ref{fig:BoundsImproved},
which are less constraining but easier to deal with.

The aim of this manuscript is to investigate the temporal evolution of the subsystem complexity
$\mathcal{C}_A$ after a global quench (see Sec.\,\ref{sec:mixedFinSize} and Sec.\,\ref{sec:GGE}).

For a gapless evolution Hamiltonian, our main results are shown in 
Fig.\,\ref{fig:MixedStateGlobalMasslessEvolutionDimensionlessNoentropy}, 
Fig.\,\ref{fig:MixedStateGlobalMasslessEvolutionDimensionless}, 
Fig.\,\ref{fig:CompvsEntStateGlobalMasslessEvolutionDimensionless} 
and Fig.\,\ref{fig:MixedStateGlobalMasslessDBCIntDet} 
for finite chains and in 
Fig.\,\ref{fig:MixedStateGlobalMassiveEvolutionTDDetDimensionless},
Fig.\,\ref{fig:RatioGlobalMassiveEvolutionTDDetDimensionless}, 
and Fig.\,\ref{fig:MixedStateGlobalMassiveEvolutionTDLargeomega0DetDimensionless}
for infinite chains.
In some cases, also the temporal evolutions 
for the corresponding  increment of the entanglement entropy $\Delta S_A$ are reported,
in order to highlight the similar features and the main differences. 
This comparison allows to observe that the initial growths
of $\mathcal{C}_A$ and $\Delta S_A$ are very different,
while the behaviours in the saturation regime are similar,
as highlighted in 
Fig.\,\ref{fig:MixedStateGlobalMasslessEvolutionDimensionless}, 
Fig.\,\ref{fig:CompvsEntStateGlobalMasslessEvolutionDimensionless} 
and Fig.\,\ref{fig:RatioGlobalMassiveEvolutionTDDetDimensionless}, 
where also the temporal evolutions of the ratio $\Delta S_A / \mathcal{C}_A$ 
are shown. 
An important difference between the temporal evolution
of $\mathcal{C}_A$ and of $\Delta S_A$
is that $\mathcal{C}_A$  displays a local maximum 
before the saturation regime (within a revival for finite systems),
as discussed in Sec.\,\ref{sec:mixedFinSize} and Sec.\,\ref{sec:GGE}.
Interestingly, 
within the framework of the gauge/gravity correspondence,
this feature has been observed also 
in the temporal evolution of holographic subsystem complexity
in Vaidya gravitational backgrounds \cite{Chen:2018mcc,Auzzi:2019mah}.

Some temporal evolutions of $\mathcal{C}_A$ 
determined by gapped evolution Hamiltonians have been reported in 
Fig.\,\ref{fig:MixedStateGlobalMassiveEvolutionDimensionlessNoEntropy}
and Fig.\,\ref{fig:MixedStateGlobalMassiveEvolutionDimensionless}.
However, a more systematic analysis is needed to explore their characteristic features.

For the infinite harmonic chains that we have considered
the asymptotic regime is described by a GGE;
hence in Sec.\,\ref{sec:GGE} we have argued that
the asymptotic value of the temporal evolution of $\mathcal{C}_A$
is given by (\ref{comp-initialvsGGE}).
This result has been checked both for $\omega=0$ 
(see Fig.\,\ref{fig:MixedStateGlobalMassiveEvolutionTDDetDimensionless},
Fig.\,\ref{fig:MixedStateGlobalMassiveEvolutionTDLargeomega0DetDimensionless}
and Fig.\,\ref{fig:MixedStateGlobalMasslessEvolutionTD})
and for $\omega>0$
(see Fig.\,\ref{fig:MixedStateGlobalMassiveEvolutionTD}
and Fig.\,\ref{fig:MixedStateGlobalMassiveEvolutionTDLarget}).

In the future research, it would be interesting to investigate the subsystem complexity
and its temporal evolution after a quench in fermionic systems, 
in circuits involving non-Gaussian states and in interacting systems. 
The analysis reported in this manuscript can be extended straightforwardly in various directions. 
For instance, we find it worth exploring 
the dependence of the temporal evolution on the reference state
(e.g. by considering the unentangled product state as the reference state),
the temporal evolution for higher dimensional harmonic lattices 
and
the temporal evolutions of the subsystem complexity 
when the system is driven out of equilibrium through other quench protocols
\cite{Das:2014jna,Alves:2018qfv, Caputa:2017ixa, Camargo:2018eof},
like e.g. local quenches \cite{Eisler_2007, Calabrese:2007mtj, Ageev:2018nye,Ageev:2019fxn}.
In \cite{DiGiulio:2020hlz} 
the subsystem complexity has been studied also by employing the entanglement Hamiltonians 
\cite{Peschel_2009,Casini:2009sr, Eisler:2017cqi,Eisler:2018ugn, 
Tonni:2017jom, Eisler:2019rnr,DiGiulio:2019cxv,Eisler:2020lyn};
hence one can consider the possibility to explore also its temporal evolution 
through these entanglement quantifiers.  


It would be interesting to study the temporal evolutions of the subsystem complexity 
by employing other ways to evaluate the complexity of mixed states,
e.g. through other distances between bosonic Gaussian states
or the approach based on the purification complexity 
\cite{Caceres:2019pgf,Camargo:2020yfv}.
The cost function plays an important role in the evaluation of the circuit complexity \cite{Jefferson:2017sdb};
hence it is worth studying its effect on the temporal evolution of the subsystem complexity.

Finally, it is important to keep exploring the temporal evolutions of the subsystem complexity 
through holographic calculations in order to find qualitative features that are observed
in lattice models. 
They would be crucial tests for quantum field theory methods to evaluate the subsystem complexity.

\vskip 30pt 
\centerline{\bf Acknowledgments} 
\vskip 10pt

We are grateful to Leonardo Banchi, Lucas Hackl, Mihail Mintchev, 
Nadir Samos S\'aenz de Buruaga
and Luca Tagliacozzo
for useful discussions. 
ET's work has been conducted within the framework of the 
Trieste Institute for Theoretical Quantum Technologies.

\newpage
\vskip 30pt 
\appendix

\section{Covariance matrix after a global quantum quench}
\label{app:CMglobalquench}

In this appendix we discuss further the
covariance matrices after the global quench employed in Sec.\,\ref{sec:cov-mat}
and Sec.\,\ref{sec:purestates_HC_glob}.
The explicit expressions for the correlators of the GGE 
that have been used in Sec.\,\ref{sec:GGE} for some numerical computations
are also provided.

\subsection{Covariance matrix}
\label{subapp:covariancematrix}

The matrix $H^{\textrm{\tiny phys}}$ defined in (\ref{HC ham}), 
which characterises the Hamiltonian of the model, reads
\be
\label{Hphys_blocks}
H^{\textrm{\tiny phys}}=Q^{\textrm{\tiny phys}}\oplus P^{\textrm{\tiny phys}}
\ee
where $P^{\textrm{\tiny phys}}=\frac{1}{m}\boldsymbol{1}$ and $Q^{\textrm{\tiny phys}}$ is a $N\times N $ real, symmetric and positive definite matrix 
whose explicit expression is not needed for the subsequent discussion.

Denoting by $\widetilde{V}$ the real orthogonal matrix diagonalising $Q^{\textrm{\tiny phys}}$
(for harmonic chains with PBC the matrix $\widetilde{V}$ is given in (\ref{Vtilde-def-even}) and (\ref{Vtilde-def-odd})),
one notices that (\ref{Hphys_blocks}) can be diagonalised as follows
\be
\label{Hphysgen_diag}
H^{\textrm{\tiny phys}}= \,V\,
\bigg[\, \frac{1}{m}\;\textrm{diag}\big( (m\Omega_1)^2,\dots,(m\Omega_N)^2,1,\dots,1\big) \bigg] \, V^{\textrm{t}}
\;\;\qquad\;\;
V\equiv\widetilde{V}\oplus \widetilde{V}
\ee
where $m \Omega^2_k$ are the real eigenvalues of $ Q^{\textrm{\tiny phys}}$.
Since $\widetilde{V}$ is orthogonal, the $2N\times 2N$ matrix $V$ is symplectic and orthogonal.
The r.h.s. of (\ref{Hphysgen_diag}) can be written as
\be
\label{Hphys-Wdec}
H^{\textrm{\tiny phys}}= \,V\,\mathcal{X}_{\textrm{\tiny phys}}\,
\Big[ \textrm{diag}\big(\Omega_1,\dots,\Omega_N,\Omega_1,\dots,\Omega_N\big) \Big] \, \mathcal{X}_{\textrm{\tiny phys}}\,V^{\textrm{t}}
\ee
where we have introduced the following symplectic and diagonal matrix 
\be
\label{Chiphys-Wdec}
\mathcal{X}_{\textrm{\tiny phys}}=\textrm{diag}\Big( (m\Omega_1)^{1/2},\dots, (m\Omega_N)^{1/2}, (m\Omega_1)^{-1/2},\dots,(m\Omega_N)^{-1/2}\Big)
\equiv
S_{\textrm{\tiny phys}}\oplus S^{-1}_{\textrm{\tiny phys}}\;.
\ee

From (\ref{Hphys-Wdec}), 
the Williamson's decomposition \cite{Williamson36} of the matrix $H^{\textrm{\tiny phys}}$ 
reads
\be
\label{williamson-Hphys-gen}
H^{\textrm{\tiny phys}}
\,= \,
W^{\textrm{t}}_{\textrm{\tiny phys}}  \, \mathcal{D}_{\textrm{\tiny phys}}  \, W_{\textrm{\tiny phys}}
\ee
where
\be
\label{Wphys-mat-def}
\mathcal{D}_{\textrm{\tiny phys}}  =
\textrm{diag}\big(\Omega_1,\dots,\Omega_N,\Omega_1,\dots,\Omega_N\big)
\;\;\qquad\;\;
W_{\textrm{\tiny phys}} = \mathcal{X}_{\textrm{\tiny phys}}\,V^{\textrm{t}} \,.
\ee

The decomposition (\ref{williamson-Hphys-gen}) leads to write the Hamiltonian (\ref{HC ham}) 
in terms of the canonical variables defined through $W_{\textrm{\tiny phys}}$ as follows
\be
\label{Williamson-basis-Hphys}
\widehat{H}=\frac{1}{2}\, \hat{\boldsymbol{s}}^{\textrm t}\, \mathcal{D}_{\textrm{\tiny phys}} \, \hat{\boldsymbol{s}}
\;\;\qquad\;\;
\hat{\boldsymbol{s}} 
\equiv
W_{\textrm{\tiny phys}}\,\hat{\boldsymbol{r}}
 \equiv
\bigg( 
\begin{array}{c}
\hat{\boldsymbol{\mathfrak{q}}} \\  \hat{\boldsymbol{\mathfrak{p}}}
\end{array}  \bigg) \,.
\ee

Following the standard quantisation procedure, 
the annihilation operators $\hat{\mathfrak{b}}_k$ and the creation operators 
$\hat{\mathfrak{b}}_k^\dagger$ are
\be
\label{b_operators def}
\hat{\boldsymbol{b}}
\equiv 
\big(\,
\hat{\mathfrak{b}}_1, \dots ,  \hat{\mathfrak{b}}_N, \, 
\hat{\mathfrak{b}}_1^\dagger,   \dots, \hat{\mathfrak{b}}_N^\dagger\,\big)^{\textrm t}
\equiv
\Theta^{-1} \hat{\boldsymbol{s}}
\;\qquad\;
\hat{\mathfrak{b}}_k \equiv \frac{\hat{\mathfrak{q}}_k +\textrm{i}\,\hat{\mathfrak{p}}_k}{\sqrt{2}}
\;\qquad\;
\Theta \equiv 
\frac{1}{\sqrt{2}}
\bigg(  \begin{array}{cc}
 \boldsymbol{1} &  \boldsymbol{1} \\
 -\textrm{i} \boldsymbol{1} \; & \textrm{i} \boldsymbol{1} \\
\end{array}   \bigg)
\ee
which satisfy $[\hat{\boldsymbol{b}}_i,\hat{\boldsymbol{b}}_j]=J_{ij}$,
where $J$ is the standard symplectic matrix 
\be
\label{Jmat}
J \equiv
\bigg(  \begin{array}{cc}
 \boldsymbol{0} &  \; \boldsymbol{1} \\
 - \boldsymbol{1} &  \; \boldsymbol{0} \\
\end{array}  \, \bigg)
\ee
whose blocks are given by the $N \times N$ identity matrix $\boldsymbol{1}$
and the matrix $ \boldsymbol{0}$ filled by zeros. 
In terms of these operators, the  Hamiltonian (\ref{Williamson-basis-Hphys}) reads
\be
\label{Hphys-op-Omega}
\widehat{H}
=
\sum_{k=1}^N 
\Omega_{k} \! \left( \hat{\mathfrak{b}}_k^\dagger\, \hat{\mathfrak{b}}_k +\frac{1}{2}\, \right) .
\ee
Thus, the symplectic spectrum $\mathcal{D}_{\textrm{\tiny phys}} $ 
in (\ref{Wphys-mat-def}) provides the dispersion relation $\Omega_k$,
that depends both on the dimensionality of the lattice and on the boundary conditions.

By applying the above procedure to the Hamiltonian $\widehat{H}_0$
whose ground state $|\psi_0\rangle$ is the initial state, one finds 
\be
\label{Hphys-op-Omega0}
\widehat{H}_0
=
\sum_{k=1}^N 
\Omega_{0,k} \! \left( \hat{\mathfrak{b}}_{0,k}^\dagger\, \hat{\mathfrak{b}}_{0,k}+\frac{1}{2}\, \right) 
\ee
where $\Omega_{0,k}$ is the dispersion relation of $\widehat{H}_0$.

To evaluate (\ref{heisemberg rps}) and (\ref{time dep corrs}),
from  (\ref{Hphysgen_diag}), (\ref{Chiphys-Wdec}), (\ref{Wphys-mat-def}) and (\ref{Williamson-basis-Hphys}) one obtains
(\ref{time dep corrs trans}), namely
\bea
\label{time dep corrs trans app}
Q(t) &=&
 \widetilde{V}S^{-1}_{\textrm{\tiny phys}} \,
 \langle \psi_0 |e^{\textrm{i} \widehat{H} t}\, \hat{\boldsymbol{\mathfrak{q}}}(0)  \,\hat{\boldsymbol{\mathfrak{q}}}^{\textrm{t}}(0) \,e^{-\textrm{i} \widehat{H} t} | \psi_0 \rangle \,
 S^{-1}_{\textrm{\tiny phys}}\widetilde{V}^{\textrm{t}}
\,\equiv \,
\widetilde{V} \, \mathcal{Q}(t) \, \widetilde{V}^{\textrm{t}}
\\
\rule{0pt}{.6cm}
P(t) &=& 
\widetilde{V}S_{\textrm{\tiny phys}}\,
\langle \psi_0 |e^{\textrm{i} \widehat{H} t} \,\hat{\boldsymbol{\mathfrak{p}}}(0)  \,\hat{\boldsymbol{\mathfrak{p}}}^{\textrm{t}}(0) \,e^{-\textrm{i} \widehat{H} t} | \psi_0 \rangle\,
S_{\textrm{\tiny phys}}\widetilde{V}^{\textrm{t}}
\,\equiv \,
 \widetilde{V} \, \mathcal{P}(t)\, \widetilde{V}^{\textrm{t}}
\\
\rule{0pt}{.6cm}
M(t) &=&
\widetilde{V}S^{-1}_{\textrm{\tiny phys}} \,
\textrm{Re} \big[ \langle \psi_0 |e^{\textrm{i} \widehat{H} t} \,\hat{\boldsymbol{\mathfrak{q}}}(0)  \,\hat{\boldsymbol{\mathfrak{p}}}^{\textrm{t}}(0) e^{-\textrm{i} \widehat{H} t} \,| \psi_0 \rangle \big] \,
S_{\textrm{\tiny phys}}\widetilde{V}^{\textrm{t}}
\,\equiv \,
\widetilde{V} \, \mathcal{M}(t) \, \widetilde{V}^{\textrm{t}}\,.
\eea

In order to find the correlators of the operators $\hat{\boldsymbol{\mathfrak{q}}}(0)$ and $\hat{\boldsymbol{\mathfrak{p}}}(0)$,
one first employs (\ref{b_operators def}) to express 
all the operators in terms of the creation and annihilation operators. 
Then, since the initial state $| \psi_0 \rangle$ is annihilated 
by the operators 
$\hat{\mathfrak{b}}_{0,k}$ and $\hat{\mathfrak{b}}_{0,k}^{\dagger}$
introduced in (\ref{Hphys-op-Omega0}),
we have to express $\hat{\mathfrak{b}}_k$ and $\hat{\mathfrak{b}}_k^{\dagger}$ 
in terms of $\hat{\mathfrak{b}}_{0,k}$ and $\hat{\mathfrak{b}}_{0,k}^{\dagger}$,
as done in \cite{Calabrese:2007rg}.
This leads to write the diagonal matrices
$\mathcal{Q}(t)$, $\mathcal{P}(t)$ and $\mathcal{M}(t)$,
whose non vanishing elements are given by (\ref{QPRmat t-dep-k}).

\subsection{Complexity through the matrix $W_\textrm{\tiny TR}$}
\label{subapp:squeezing}

The Williamson's decomposition 
\cite{Williamson36}
is an important tool to study the circuit complexity of bosonic Gaussian states \cite{Caceres:2019pgf,DiGiulio:2020hlz}.
When the reference and the target states are pure states,
both the optimal circuit and the corresponding complexity can be evaluated through the symplectic matrix
$W_\textrm{\tiny TR} \equiv W_\textrm{\tiny T}\, W_\textrm{\tiny R}^{-1}$,
where $W_\textrm{\tiny R}$ and $W_\textrm{\tiny T}$ occur in the Williamson's decomposition of the
reference and of the target states respectively 
\cite{Chapman:2018hou,Camargo:2018eof,DiGiulio:2020hlz}.

In the following we construct the Williamson's decomposition of the covariance matrix (\ref{covariancematrix_t})
after the global quantum quench, that describes a pure state.

By using (\ref{purity-condition}),
we first observe that  the block matrix in (\ref{covariancematrix_diags_t}) can be decomposed as 
\be
\label{Williamson step1}
\Gamma(t)
=\,
T(t)^{\textrm{t}}
\,
\bigg( 
\begin{array}{cc}
\frac{1}{4}\,\mathcal{P}(t)^{-1} & \boldsymbol{0}  \\
 \boldsymbol{0} & \mathcal{P}(t)  
\end{array}  
\bigg)\,
T(t)
\;\;\qquad\;\;
T(t) \equiv
\bigg( 
\begin{array}{cc}
 \boldsymbol{1} \;& \boldsymbol{0}  
 \\
  \mathcal{P}(t)^{-1} \mathcal{M}(t) \;&  \boldsymbol{1} 
\end{array}  \bigg)
\ee
where the triangular matrix $T(t)$ is symplectic and not orthogonal. 
Then, the symplectic spectrum of the diagonal matrix in (\ref{Williamson step1}) can be obtained 
as discussed e.g. in the appendix D of \cite{DiGiulio:2020hlz}, finding
\be
\label{Williamson step2}
\bigg( 
\begin{array}{cc}
\frac{1}{4}\,\mathcal{P}(t)^{-1} & \!\! \boldsymbol{0}  \\
 \boldsymbol{0} & \mathcal{P}(t)
\end{array}   \bigg)
\,=\,
\frac{1}{2} \, \mathcal{X}^2(t)
\ee
where the symplectic and diagonal matrix $\mathcal{X}(t)$ 
can be defined in terms of $P_k(t)$ in (\ref{QPRmat t-dep-k})
as 
\be
\label{chi mat}
\mathcal{X}(t)
=
\textrm{diag}\bigg( \frac{1}{\sqrt{2 P_1}} \,,\dots, \frac{1}{\sqrt{2 P_N}}\, ,\sqrt{2 P_1} \,,\dots,\sqrt{2 P_N} \, \bigg)\,.
\ee

Plugging (\ref{Williamson step2}) into (\ref{Williamson step1}), 
one finds the Williamson's decomposition of the covariance matrix (\ref{covariancematrix_t})
\be
\label{Williamson final}
\gamma(t)=\frac{1}{2} \,W(t)^{\textrm{t}} \, W(t)
\,\,\qquad\,\,
W(t)= \mathcal{X}(t)\, T(t)\, V^{\textrm{t}}
\ee
which tells us also that all the symplectic eigenvalues of  $\gamma(t)$ 
are equal to $1/2$, as expected for pure states.

By using this decomposition for both the reference and the target states, 
with the same matrix $V$ (see (\ref{VV-diag-hyp})),
we find that
$W_\textrm{\tiny TR} \equiv W_\textrm{\tiny T}\, W_\textrm{\tiny R}^{-1}$
becomes
\be
\label{WTR HC timedep}
W_\textrm{\tiny TR}
=
\mathcal{X}_\textrm{\tiny T} \, T_\textrm{\tiny T} \,T_\textrm{\tiny R}^{-1}\,\mathcal{X}_\textrm{\tiny R}^{-1}\,.
\ee

For the sake of simplicity, let us focus on the complexity w.r.t. the initial state, 
which is also the case mainly explored throughout this manuscript
(hence $t_\textrm{\tiny R}=0$ and $t_\textrm{\tiny T}=t$).

From (\ref{Williamson step1}), it is straightforward to check that
\be 
T(t)^{-1}
=
\bigg( 
\begin{array}{cc}
 \boldsymbol{1} & \!\!\,\,\,\,\boldsymbol{0}  \\
 - \mathcal{P}^{-1}(t)\,\mathcal{M}(t) & \!\!\,\,\,\,  \boldsymbol{1} 
\end{array}  
\bigg)\,.
\ee
Then, since $\mathcal{M}_\textrm{\tiny R}=\boldsymbol{0}$ when $t_\textrm{\tiny R}=0$,
using (\ref{chi mat}) we obtain
\be
W_\textrm{\tiny TR}=
\,
\Bigg( 
\begin{array}{cc}
\sqrt{\mathcal{P}^{-1}_\textrm{\tiny T}\mathcal{P}_\textrm{\tiny R}} & \!\!\,\,\,\, \boldsymbol{0}
 \\
 \rule{0pt}{.5cm}
2 \sqrt{\mathcal{P}^{-1}_\textrm{\tiny T}\mathcal{P}_\textrm{\tiny R}} \;\mathcal{M}_\textrm{\tiny T}
 & \!\!\,\,\,\, \sqrt{\mathcal{P}^{-1}_\textrm{\tiny R}\mathcal{P}_\textrm{\tiny T}}
\\
\end{array}   \Bigg)
\ee
which gives
\be
\label{WTRWTR}
W_\textrm{\tiny TR}^{\textrm{t}}W_\textrm{\tiny TR}
=
\,
\bigg( 
\begin{array}{cc}
\mathcal{P}^{-1}_\textrm{\tiny T}\mathcal{P}_\textrm{\tiny R} \big( \boldsymbol{1}+4\mathcal{M}^2_\textrm{\tiny T} \big) & \!\!\,\,\,\, 
2\mathcal{M}_\textrm{\tiny T} \\
 2\mathcal{M}_\textrm{\tiny T}
 & \!\!\,\,\,\, 
\mathcal{P}^{-1}_\textrm{\tiny R}\mathcal{P}_\textrm{\tiny T} \\
\end{array}  
\bigg)
\ee
whose eigenvalues provide the circuit complexity.
Indeed, by employing the Williamson's decomposition (\ref{Williamson final}) 
for the covariance matrices of the reference and of the target states 
into the expression (\ref{c2 complexity}), 
one finds that it can be written as follows
\be
\label{c2 complexity WTR}
\mathcal{C}
\,=\,
\frac{1}{2\sqrt{2}}\;
\sqrt{\,
\textrm{Tr}\, \Big\{ \big[ \log \big( W_\textrm{\tiny TR}^{\textrm{t}}W_\textrm{\tiny TR} \big) \big]^2 \Big\}
}\;.
\ee

Since the matrix (\ref{WTRWTR}) is a special case of (\ref{det_1}),
its eigenvalues can be found by applying (\ref{eigenvalues block mat diag}).
The resulting spectrum is given by the pairs
$ \big(\chi_{\textrm{\tiny TR}}^2\big)_{k}$ and $\big(\chi_{\textrm{\tiny TR}}^2\big)^{-1}_{k} $, 
labelled by $1 \leqslant k \leqslant N$, with
\bea
\label{squeezing parameters}
\big(\chi_{\textrm{\tiny TR}}^2\big)_{k}
&=&
\frac{P^2_{\textrm{\tiny T},k}+P^2_{\textrm{\tiny R},k}\big( 1+4 M^2_{\textrm{\tiny T},k} \big)
+\sqrt{ \big[ P_{\textrm{\tiny T},k}^2+P_{\textrm{\tiny R},k}^2 \big(1+4 M_{\textrm{\tiny T},k}^2 \big) \big]^2-4\,P_{\textrm{\tiny T},k}^2 P^2_{\textrm{\tiny R},k}}}{2\,P_{\textrm{\tiny T},k} P_{\textrm{\tiny R},k}}
\phantom{xxxx}
\\
\rule{0pt}{.9cm}
\label{squeezing parameters v2}
&=&
\frac{1}{2}\bigg[\frac{Q_{\textrm{\tiny T},k}}{Q_{\textrm{\tiny R},k}}+\frac{P_{\textrm{\tiny T},k}}{P_{\textrm{\tiny R},k}}\bigg]
+
\sqrt{\frac{1}{4}\bigg[\frac{Q_{\textrm{\tiny T},k}}{Q_{\textrm{\tiny R},k}}+\frac{P_{\textrm{\tiny T},k}}{P_{\textrm{\tiny R},k}}\bigg]^2-1}
\eea
where the last expression is obtained by employing the fact that, from (\ref{purity-condition}),
for each $k$ we have
$1+4M_{\textrm{\tiny T},k}^2=4 Q_{\textrm{\tiny T},k} P_{\textrm{\tiny T},k} $ and $1=4Q_{\textrm{\tiny R},k} P_{\textrm{\tiny R},k} $.

Comparing (\ref{squeezing parameters v2}) with (\ref{gpm-from-C}) and (\ref{C-TR-camargo}),
we conclude that, for any $1\leqslant k \leqslant N$, we have 
\be
\label{squeezing-RCM}
\big(\chi_{\textrm{\tiny TR}}^2\big)_{k}=g_{\textrm{\tiny TR},k}^{(+)}\;.
\ee
Thus, the complexity (\ref{c2-log-lambda-arcosh}) can be written in terms of $\big(\chi_{\textrm{\tiny TR}}^2\big)_{k}$.
Since this result is expected for the circuit complexity of bosonic Gaussian pure states, 
(\ref{squeezing-RCM}) provides a non-trivial consistency check of the entire procedure. 
Furthermore, by extending this analysis to the case $t_\textrm{\tiny R}\neq 0$ 
in the straightforward way,
(\ref{squeezing-RCM}) is recovered.

In the space of covariance matrices and after a proper change of basis, 
the optimal circuit made by pure states that connects the reference state to the target state
reads \cite{Chapman:2018hou}
\be
G_s(\gamma_\textrm{\tiny R},\gamma_\textrm{\tiny T})=\frac{1}{2}\big(\mathcal{X}^2_{\textrm{\tiny TR}}\big)^{s}
\ee
where the symplectic diagonal matrix $\mathcal{X}^2_{\textrm{\tiny TR}}$ is defined as follows
\be
\mathcal{X}^2_{\textrm{\tiny TR}}
\,\equiv\,
\textrm{diag}\,\Big\{
\big(\chi_{\textrm{\tiny TR}}^2\big)_{1}\,,\dots\big(\chi_{\textrm{\tiny TR}}^2\big)_{N}\,,
\big(\chi_{\textrm{\tiny TR}}^2\big)^{-1}_{1},\dots,\big(\chi_{\textrm{\tiny TR}}^2\big)^{-1}_{N}
\Big\}
\ee
in terms of the eigenvalues of the matrix (\ref{WTRWTR}), given in (\ref{squeezing parameters v2}).

\subsection{GGE correlators}
\label{app-sec-GGE-corr}

In the following we report the explicit expressions of the correlators 
for the harmonic chains in the GGE state which have been employed to construct the reduced covariance matrix $\gamma_{\textrm{\tiny GGE},A}$ 
from the covariance matrix $\gamma_{\textrm{\tiny GGE}}$ defined in (\ref{gamma-GGE-matrix}). The matrix $\gamma_{\textrm{\tiny GGE},A}$ occurs in the expression (\ref{comp-initialvsGGE}) 
for the complexity $\mathcal{C}_{\textrm{\tiny GGE},A} $.

The harmonic chains where either PBC or DBC are imposed
must be treated separately.

For PBC, by using (\ref{Vtilde-def-even}) or (\ref{Vtilde-def-odd}), 
(\ref{Q-P-GGEdiag}) and (\ref{nk omegak})
into (\ref{QGGE}) and (\ref{PGGE}), we obtain
\bea
\label{QijGGE}
\textrm{Tr}\big(\hat{q}_i \, \hat{q}_j \, \hat{\rho}_{\textrm{\tiny GGE}}\big)
&=&
\frac{1}{N}\sum_{k=1}^N \frac{\Omega_{0,k}^2+\Omega_{k}^2}{4 \,\Omega_{k}^2 \,\Omega_{0,k}}\; \cos\! \big[(i-j)\,2\pi k/N\,\big]
\\
\rule{0pt}{.8cm}
\label{PijGGE}
\textrm{Tr}\big(\hat{p}_i \, \hat{p}_j \, \hat{\rho}_{\textrm{\tiny GGE}}\big)
&=&
\frac{1}{N}\sum_{k=1}^N \frac{\Omega_{0,k}^2+\Omega_{k}^2}{4 \, \Omega_{0,k}} \;\cos\!\big[(i-j)\,2\pi k/N\,\big]
\eea
in terms of the dispersion relations (\ref{dispersion relations}).
Notice that the correlators (\ref{QijGGE}) diverge when  $\omega=0$;
hence for PBC the massless limit must be studied by taking $\omega$ very small, 
but non vanishing. 
In the thermodynamic limit $N\to\infty$, 
these correlators become respectively
\be
\label{corr-GGE-app-int-pbc}
 \int_{0}^{\pi} \frac{\Omega_{0,\theta}^2+\Omega_{\theta}^2}{ \Omega_{\theta}^2 \,\Omega_{0,\theta}} \,\cos\!\big[2\theta\,(i-j)\,\big] \, \frac{d\theta}{4\pi}
\;\;\;\qquad\;\;\;
\int_{0}^{\pi} \frac{\Omega_{0,\theta}^2+\Omega_{\theta}^2}{ \Omega_{0,\theta}}\, \cos\!\big[2\theta\,(i-j)\,\big] \, \frac{d\theta}{4\pi}
\ee
where the dispersion relations are given in (\ref{dispersion relations TD}).

When DBC are imposed, 
by using the matrix $\widetilde{V}$ in (\ref{Vtilde-HC-DBC}), we obtain
\bea
\label{QijGGEDBC}
\textrm{Tr}\big(\hat{q}_i \,\hat{q}_j \, \hat{\rho}_{\textrm{\tiny GGE}}\big)
&=&
\frac{2}{N}\sum_{k=1}^{N-1} \frac{\Omega_{0,k}^2+\Omega_{k}^2}{4\, \Omega_{k}^2 \,\Omega_{0,k}}\,
\sin\big(\pi k i /N\big) \,\sin\big(\pi k j /N\big)
\\
\rule{0pt}{.8cm}
\label{PijGGEDBC}
\textrm{Tr}\big(\hat{p}_i \,\hat{p}_j \, \hat{\rho}_{\textrm{\tiny GGE}}\big)
&=&
\frac{2}{N}\sum_{k=1}^{N-1} \frac{\Omega_{0,k}^2+\Omega_{k}^2}{4 \, \Omega_{0,k}}\,
\sin\big(\pi k i /N\big)\,\sin\big(\pi k j /N\big)
\eea
in terms of the dispersion relations (\ref{dispersion DBC}).
Notice that, in this case, all these correlators are finite when $\omega =0$.
The thermodynamic limit $N\to\infty$ of these correlators gives respectively
\be
\label{corr-GGE-app-int-dbc}
\int_{0}^{\pi} \frac{\Omega_{0,\theta}^2+\Omega_{\theta}^2}{ \Omega_{\theta}^2 \,\Omega_{0,\theta}} \sin(i \theta) \sin(j \theta) \, \frac{d\theta}{2\pi}
\;\;\;\qquad\;\;\;
\int_{0}^{\pi} \frac{\Omega_{0,\theta}^2+\Omega_{\theta}^2}{ \Omega_{0,\theta}} \sin(i \theta) \sin(j \theta) \, \frac{d\theta}{2\pi}
\ee
where (\ref{dispersion DBC TD}) must be employed.

These correlators have been used to construct $\gamma_{\textrm{\tiny GGE},A}$, 
that occurs in $\mathcal{C}_{\textrm{\tiny GGE},A}$ defined in (\ref{comp-initialvsGGE}).
In particular, 
the expressions (\ref{corr-GGE-app-int-pbc}) have been exploited to draw
the horizontal lines in the left panels of 
Fig.\,\ref{fig:MixedStateGlobalMasslessEvolutionTD}, 
Fig.\,\ref{fig:MixedStateGlobalMassiveEvolutionTD}
and Fig.\,\ref{fig:MixedStateGlobalMassiveEvolutionTDLarget}
and in the bottom left panel of Fig.\,\ref{fig:GGEdet}.
They have also provided 
the data points in the left panels of Fig.\,\ref{fig:GGE} and Fig.\,\ref{fig:CAGGEvsCGGE}.
Instead, the horizontal lines in 
Fig.\,\ref{fig:MixedStateGlobalMassiveEvolutionTDDetDimensionless}, 
Fig.\,\ref{fig:MixedStateGlobalMassiveEvolutionTDLargeomega0DetDimensionless}
and in the right panels of 
Fig.\,\ref{fig:MixedStateGlobalMasslessEvolutionTD}, 
Fig.\,\ref{fig:MixedStateGlobalMassiveEvolutionTD}
and Fig.\,\ref{fig:MixedStateGlobalMassiveEvolutionTDLarget} 
have been obtained through the correlators (\ref{corr-GGE-app-int-dbc}),
which have provided also
the data points in Fig.\,\ref{fig:GGEdet} and in the right panels 
of Fig.\,\ref{fig:GGE} and of Fig.\,\ref{fig:CAGGEvsCGGE}.

As consistency check of the fact that the GGE describes the limit $t\to \infty$ after the global quench,
one observes that
the correlators in (\ref{corr-GGE-app-int-pbc}) and (\ref{corr-GGE-app-int-dbc}) are recovered
by taking (\ref{QPRmat t-dep TD}) and (\ref{QPRmat t-dep DBC TD}) respectively  
and replacing all the oscillatory functions with their averages
(i.e. $[\sin(\Omega_\theta t)]^2$ and $[\cos(\Omega_\theta t)]^2$ by $1/2$ and $\sin(\Omega_\theta t)$ by $0$).

\section{Complexity w.r.t. the unentangled product state}
\label{app:unentangled}

In this appendix we consider the temporal evolution of the complexity between the target state defined as the state 
at time $t$ after the quench, characterised by the parameters 
$(\kappa_\textrm{\tiny T},m_\textrm{\tiny T},\omega_\textrm{\tiny T},\omega_{0,\textrm{\tiny T}})\equiv (\kappa,m,\omega,\omega_{0})$, 
and the reference state defined as the state at $t=0$, 
when the system is prepared in the unentangled product state, 
characterised by the parameters 
$(\kappa_\textrm{\tiny R},m_\textrm{\tiny R},\omega_\textrm{\tiny R}) \equiv ( 0,m,\mu)$.
This unentangled product state has been largely employed as reference state to explore the circuit complexity
\cite{Jefferson:2017sdb,Chapman:2018hou,Guo:2018kzl,Caceres:2019pgf}, 
also in time-dependent settings \cite{Alves:2018qfv},
hence we find it worth providing a brief discussion for the complexity 
when this state is chosen as reference state.

For circuits made by pure states,
the complexity is (\ref{c2-log-lambda-arcosh}) with $C_{\textrm{\tiny TR},k}$ given by (\ref{complexity_equalm}). 
When the reference state is the unentangled product state,
from (\ref{dispersion relations}) and (\ref{dispersion DBC})
one observes that $\Omega_{0,\textrm{\tiny R},k}=\mu$ for any $k$,
independently of whether PBC or BDC are imposed. 
Thus, the expression of $C_{\textrm{\tiny TR},k}$ simplifies to
\be
\label{CTR_equalm_prod ref state}
C_{\textrm{\tiny TR},k}
=
\frac{1}{2\mu\, \Omega_{0,k}} 
\left(\,
\Omega^2_{0,k}+\mu^2
+
\frac{(\Omega_k^2-\Omega_{0,k}^2)(\Omega_k^2-\mu^2)}{\Omega_{k}^2} \, [ \sin(\Omega_{k}t)]^2
\right)
\ee
where we have defined 
$\Omega_{\textrm{\tiny T},k}\equiv \Omega_{k} $ 
and $\Omega_{0,\textrm{\tiny T},k}\equiv \Omega_{0,k}$
to enlighten the expression. 
Isolating the zero mode contribution, as done also in Sec.\,\ref{subsec:zeromodesboundsgeneral}, 
the complexity (\ref{c2-log-lambda-arcosh}) reads
\bea
\label{complexity_equalm_prod ref state}
\mathcal{C}^2
&=&
\frac{\eta}{4}\,
\bigg\{
\textrm{arccosh}\bigg[\,\frac{\omega_0^2+\mu^2}{2\mu \, \omega_0}
+
\frac{(\omega^2-\omega_0^2)(\omega^2-\mu^2)}{2\mu \, \omega_0\,\omega^2} \, [ \sin(\omega t) ]^2\, \bigg]
\bigg\}^2
\\
&&
+ \,\frac{1}{4}\sum_{k=1}^{N-1}\bigg\{
\textrm{arccosh}\bigg[\,\frac{\Omega^2_{0,k}+\mu^2}{2\mu \, \Omega_{0,k}}
+
\frac{(\omega^2-\omega_0^2)(\Omega_k^2-\mu^2)}{2\mu \, \Omega_{0,k} \, \Omega_{k}^2} \,[\sin(\Omega_{k}t)]^2 \bigg]
\bigg\}^2
\nonumber
\eea
where $\eta$ has been introduced in (\ref{comp-pure-global-DBCPBC}) 
and the dispersion relations (\ref{dispersion relations}) or (\ref{dispersion DBC}) must be employed,
depending respectively on whether PBC or DBC hold.

When $\omega=0$, the zero mode term having $k=N$ in (\ref{CTR_equalm_prod ref state}) simplifies to
\be
C_{\textrm{\tiny TR},N}
=
 \frac{\omega_0^2+\mu^2}{2\mu\,\omega_0}+\frac{\mu\,\omega_0}{2}\, t^2
\ee
which is divergent when $t\to\infty$, 
while $C_{\textrm{\tiny TR},k}$ with $k\neq N$ is bounded for any value of $t$.
Thus, for the critical evolution the complexity (\ref{complexity_equalm_prod ref state}) becomes
\bea
\label{complexity_equalm_prod ref state massless}
\mathcal{C}^2
&=&
\frac{\eta}{4}\bigg[\textrm{arccosh}\bigg(\frac{\omega_0^2+\mu^2}{2\mu\,\omega_0}
+\frac{\mu\,\omega_0}{2} \, t^2 \bigg)\bigg]^2
\\
& &
+\,
\frac{1}{4}\sum_{k=1}^{N-1}
\bigg\{\textrm{arccosh}\bigg[\,\frac{\Omega^2_{0,k}+\mu^2}{2\mu\,\Omega_{0,k}}
+\frac{\omega_0^2(\mu^2-\Omega_k^2)}{2\mu\,\Omega_{0,k}\,\Omega_{k}^2}
\,[ \sin(\Omega_{k}t)]^2 \bigg]
\bigg\}^2\,.
\nonumber
\eea
We remark that, when PBC are imposed (hence $\eta=1$),
the complexity (\ref{complexity_equalm_prod ref state massless}) 
diverges logarithmically as $t\to\infty$
because of the occurrence of the zero mode contribution.
This feature is observed also when the reference state is the initial state 
(see Sec.\,\ref{subsec:zeromodesboundsgeneral}).

When the reference state is the unentangled product state, 
the complexity is non vanishing at $t=0$.
In particular, from (\ref{complexity_equalm_prod ref state}) we obtain
\be
\label{complexity_equalm_prod ref state t equal zero}
\mathcal{C}^2\big|_{t=0}
\,=\,
\frac{\eta}{4} \bigg[\textrm{arccosh}\bigg(\frac{\omega_0^2+\mu^2}{2\mu \, \omega_0}\bigg)\bigg]^2
+\frac{1}{4}\sum_{k=1}^{N-1}\bigg[\textrm{arccosh}\bigg(\frac{\Omega^2_{0,k}+\mu^2}{2\mu\,\Omega_{0,k}}\bigg)\bigg]^2
\ee
Specialising this expression to PBC and DBC, 
one recovers the results found in \cite{Jefferson:2017sdb}
and \cite{Braccia:2019xxi} respectively. 
The expression (\ref{complexity_equalm_prod ref state t equal zero}) 
provides the leading term in the expansion of (\ref{complexity_equalm_prod ref state}) 
as $t \to 0$, that reads
\be
\label{complexity_equalm_prod ref state small t}
\mathcal{C}^2
\,=\,
\big(\mathcal{C}|_{t=0} \big)^2
+
\frac{(\omega^2-\omega_0^2)}{2}\,
\Bigg[
\sum_{k\,=\,1}^{N-1+\eta}\frac{\Omega_k^2-\mu^2 }{\big| \Omega_{0,k}^2-\mu^2 \big|}
\; \textrm{arccosh}\!
\left(\frac{\Omega^2_{0,k}+\mu^2}{2\mu\, \Omega_{0,k}}\right)
\!\Bigg]\,
t^2
+
O(t^4)\,.
\ee
From this expansion it is straightforward to realise that $\mathcal{C} - \mathcal{C}|_{t=0} = O(t^2)$ as $t\to 0$, where the sign of the r.h.s. is not well defined. 
This quadratic behaviour in $t$ as $t \to 0$ 
represents an interesting difference w.r.t. the behaviour of the complexity w.r.t. the initial state in the same temporal regime
(indeed, the latter one grows linearly, as highlighted in (\ref{comp-pure-initialgrowth})).
This difference is due to the fact that $ \mathcal{C}|_{t=0}$ is non vanishing. 
Furthermore, the sign of the $O(t^2)$ term in (\ref{complexity_equalm_prod ref state small t}) 
determines whether the complexity increases or decreases with respect to its initial value during the early time regime.
A similar feature has been observed also in the temporal evolution of the complexity considered in \citep{Alves:2018qfv}.

\section{Technical details about some limiting regimes}
\label{app:large-N}

In this appendix we report some technical details about the large $N$ regimes 
discussed in Sec.\,\ref{subsec:TD limit HC} 
for the temporal evolution of the complexity of the entire harmonic chain.

\subsection{Approximation for small $\frac{k}{N}$ at finite $N$}
\label{subapp:smallkapprox}

In the following we provide some details about the derivation of 
the expressions given by 
(\ref{smallkapprox_PBC_main}) and (\ref{smallkapprox_DBC_main}) 
for the complexity
and by (\ref{lineargrowth_app_PBC_text}) and (\ref{lineargrowth_app_DBC_text}) 
for the slope of its linear initial growth,
obtained in the approximation introduced at the beginning of Sec.\,\ref{subsec:TD limit HC}.

When DBC hold and therefore the dispersion relations (\ref{dispersion DBC}) are employed, 
the argument of the sum in (\ref{comp-pure-global-DBCPBC}) 
is a function of $\frac{k}{N}$ 
whose main contribution comes from the regime where $\frac{k}{N}\ll 1$.
This suggests to introduce the approximation 
$[ \sin\big(\frac{\pi k}{2N}\big) ]^2 \simeq \big(\frac{\pi k}{2N}\big)^2$ in (\ref{dispersion DBC}),
which leads to the approximate expression for 
(\ref{comp-pure-global-DBCPBC})
given in (\ref{smallkapprox_DBC_main}),
which depends only on $\omega N$, $\omega_0 N$ and $t/N$. 
The argument of the sum in (\ref{smallkapprox_DBC_main}) 
decreases very rapidly as $k$ increases;
hence increasing $N$ does not change significantly
the value of $\mathcal{C}_{\textrm{\tiny approx}}$.

When PBC hold, in the expression (\ref{comp-pure-global-zm}) for the complexity let us observe that
$c_0$ defined in (\ref{low-bound}) can be written as the following function of $\omega N$, $\omega_0 N$ and $t/N$
\be
\label{zm-scaling}
c_{0}(t) 
=
\left[ \textrm{arcsinh}\! 
\left( \,
 \frac{(\omega N)^2 - (\omega_0 N)^2}{2\,(\omega N)\, (\omega_{0} N)} \,
 \sin \bigg(\omega N   \frac{t}{N} \bigg) 
 \right)
 \right]^2
\ee
without any approximation.
If we restrict $1\leqslant k \leqslant N/2$, the argument of the sum in (\ref{comp-pure-global-zm}) is non vanishing when $k/N \ll 1$. 
This suggests to approximate $[ \sin\big(\frac{\pi k}{N}\big)]^2 \simeq \big(\frac{\pi k}{N}\big)^2$ in (\ref{dispersion relations})
which leads  (\ref{comp-pure-global-zm}) to become the following function of $\omega N$, $\omega_0 N$ and $t/N$
\be
\label{compPBCapproxim}
\mathcal{C}_{\textrm{\tiny approx}}
\,=\,
\sqrt{c_{0}(t) 
+
2 \sum_{k=1}^{[\frac{N-1}{2}]}\! 
\left[ \textrm{arcsinh}\! 
\left( \,
 \frac{(\omega N)^2 - (\omega_0 N)^2}{2\,\widetilde{\Omega}^\textrm{\tiny (P)}_{k}\, \widetilde{\Omega}^\textrm{\tiny (P)}_{0,k}} \,
 \sin \bigg(\widetilde{\Omega}^\textrm{\tiny (P)}_{k} \frac{t}{N} \bigg) 
 \right)
 \right]^2
 + c^{\textrm{\tiny approx}}_{N/2}(t)
 }
\ee
where 
\be
c^{\textrm{\tiny approx}}_{N/2}(t) \equiv
\left\{ \begin{array}{l l}
\displaystyle 
\left[ \textrm{arcsinh}\! 
\left( \,
 \frac{(\omega N)^2 - (\omega_0 N)^2}{2\,\widetilde{\Omega}^\textrm{\tiny (P)}_{N/2}\, \widetilde{\Omega}^\textrm{\tiny (P)}_{0,N/2}} \,
 \sin \bigg(\widetilde{\Omega}^\textrm{\tiny (P)}_{N/2} \frac{t}{N} \bigg)
 \right)
 \right]^2
\hspace{1cm}&
\textrm{even $N$}
\\
\rule{0pt}{.7cm}
0
&
\textrm{odd $N$}
\end{array}
\right.
\ee
and $\widetilde{\Omega}^\textrm{\tiny (P)}$ is defined in (\ref{smallkapprox_dispersion}).
The expression (\ref{compPBCapproxim})
does not grow with $N$ because the terms of the sum in (\ref{compPBCapproxim}) become negligible from a certain value of $k$.
Let us observe that, consistently with this approximation, 
the term $c^{\textrm{\tiny approx}}_{N/2}(t)$ in (\ref{compPBCapproxim}) can be neglected and (\ref{smallkapprox_PBC_main}) is obtained. 

In this approximation $N$ is kept finite, both for PBC and DBC, as long as it is large enough. 

 The initial growth within this approximation can be obtained by applying the steps discussed above 
 (\ref{initial growth HC 1d}), finding
\bea
\label{compDBCapproximinitialt}
& & \hspace{-1.2cm}
\mathcal{C}_{\textrm{\tiny approx}}
=
\frac{t}{2 N}\sqrt{\sum_{k=1}^{N-1}\frac{\big[(\omega N)^2 - (\omega_0 N)^2\big]^2}{(\omega_0 N)^2+\pi^2 k^2\kappa/m}}+O(t^3)
\hspace{5.1cm} 
\textrm{DBC}
\\
\rule{0pt}{1.4cm}
\label{compPBCapproximinitialt}
& & \hspace{-1.2cm}
\mathcal{C}_{\textrm{\tiny approx}}
=
\frac{t}{2N}\sqrt{\frac{\big[(\omega N)^2 - (\omega_0 N)^2\big]^2}{(\omega_0 N)^2} 
+
2 \sum_{k=1}^{[\frac{N-1}{2}]}\! 
\frac{\big[(\omega N)^2 - (\omega_0 N)^2\big]^2}{(\omega_0 N)^2+4\pi^2 k^2\kappa/m}
 }+O(t^3)
 \hspace{.8cm} 
\textrm{PBC\,.}
\eea
Since the arguments of the sums in (\ref{compDBCapproximinitialt}) and (\ref{compPBCapproximinitialt}) are negligible from a certain value of $k$, 
we are allowed to extend the sums up to infinite. 
Then, using $\sum_{k=1}^{\infty} \tfrac{1}{k^2+a^2}= \tfrac{a\pi \coth(a \pi)-1}{2 a^2}$,
we finally get $\mathcal{C}_{\textrm{\tiny approx}} = a_{\textrm{\tiny (B)}} \, t/N + \dots$
for (\ref{compDBCapproximinitialt}) and (\ref{compPBCapproximinitialt}) 
with $\textrm{B} \in \{ \textrm{P}, \textrm{D}\}$,
where the dots represent higher orders in $t/N$ and the slopes $a_{\textrm{\tiny (P)}} $ and $a_{\textrm{\tiny (D)}} $ are given in (\ref{lineargrowth_app_PBC_text}) and (\ref{lineargrowth_app_DBC_text}) respectively.

\subsection{Thermodynamic limit}
\label{app:EulerMcLaurin}

In order to study the thermodynamic limit of the complexity discussed in Sec.\,\ref{sec:purestates_HC_glob}, 
let us recall some basic facts about the Euler-Maclaurin formula.

The Euler-Maclaurin formula quantifies 
the discrepancy between the sum $S=\sum_{n=a+1}^b f(n)$ 
and the integral $I=\int_a^b f(x)dx$.
It reads \cite{Kac02bookCalculus} 
\be
\label{Euler-MacLaurin}
S-I
\,=\,
\frac{f(b)-f(a)}{2}+\sum_{j=1}^p \frac{B_{2j}}{(2j)!}\big[f^{(2j-1)}(b)-f^{(2j-1)}(a)\big]+R_{2p+1}
\ee
where $B_j$ are the Bernoulli numbers, $f^{(j)} \equiv \partial^j_x f$ and the remainder
\be 
\label{remainder}
R_{2p+1}\,\equiv\,- \frac{1}{(2p+1)!} \int_a^b P_{2p+1}(x) \, f^{(2p+1)}(x)\,dx
\ee 
where $P_k(x)=B_k(x-\lfloor x \rfloor)$ are expressed in terms of the Bernoulli polynomials $B_k(x)$.
The reminder $R_{2p+1}$ is bounded as follows 
\be
\label{bound rest}
|R_{2p+1}|< \frac{4 e^{2\pi}}{(2\pi)^{2p+1}}\int_a^b|f^{(2p+1)}(x)|dx\,.
\ee

Let us consider the cases where $p=0$ in (\ref{Euler-MacLaurin}) and  (\ref{bound rest}), 
which leads to
\be
\label{Euler-MacLaurin-p0}
S-I=\frac{f(b)-f(a)}{2}+R_{1}
\;\;\;\qquad\;\;\;
|R_{1}|< \frac{2 e^{2\pi}}{\pi}\int_a^b |f'(x)| \, dx\,.
\ee

By applying (\ref{Euler-MacLaurin-p0}) for $S=\mathcal{C}^2$ and the extrema $a=0$ and $b=N-1+\eta$
(where $\eta=1$ for PBC and $\eta =0$ for DBC),
for the complexity  (\ref{comp-pure-global-DBCPBC}) we find
\be
\label{EmL_leading}
I=\int_0^{N-1+\eta} \! f_{\textrm{\tiny B}}(k) \,dk
\equiv I_N^{(\textrm{\tiny B})}
\;\;\qquad\;\;
\textrm{B} \in \big\{ \textrm{P}, \textrm{D} \big\}
\ee
where, for PBC and DBC, we have respectively
\be
\label{f_def}
f_{\textrm{\tiny P}}(k)
\equiv
 \left[ \textrm{arcsinh}\! 
\left( \,
 \frac{\omega^2 - \omega_0^2}{2\,\Omega_{k}\, \Omega_{0,k}} \,
 \sin (\Omega_{k} t ) 
 \right)
 \right]^2
 \;\;\qquad\;\;
 f_{\textrm{\tiny D}}(k)
 \equiv
 f_{\textrm{\tiny P}}(k/2)\,.
\ee
Since in $f_{\textrm{\tiny P}}(k)$ the dependence on $k$ occurs 
only through $\sqrt{4\kappa/m}\,\sin(\pi k/N)$, we find it convenient to introduce 
\be
\label{Fdef}
F_t(y)
\equiv
\bigg[\textrm{arcsinh} \bigg( \,
 \frac{\omega^2 - \omega_0^2}{2\,\sqrt{\omega^2 +y^2} \, \sqrt{\omega_0^2 + y^2} } \,
 \sin\! \big( \sqrt{\omega^2 + y^2}\; t \, \big) 
 \bigg)\bigg]^2 
\ee
which leads to write (\ref{f_def}) as
\be
\label{fandF}
f_{\textrm{\tiny P}}(k)
=
F_t(s_k)
\;\;\qquad\;\;
s_k \equiv \sqrt{4\kappa/m}\,\sin(\pi k/N)\,.
\ee

By introducing the integration variable  $\theta=\pi k /N$ in (\ref{EmL_leading})
and taking $N\to\infty$, the expression (\ref{comp TD limit}) is obtained
independently of the boundary conditions.

The remainder (\ref{remainder}) for $p=0$, which depends on the boundary conditions, 
is denoted by $R_{1,N}^{\textrm{\tiny (P)}}$  for PBC 
and by $R_{1,N}^{\textrm{\tiny(D)}}$ for DBC, where we have indicated explicitely the dependence on the size $N$ of the chain. 
In order to investigate the behaviour of $R_{1,N}^{\textrm{\tiny(B)}}$,
with $\textrm{B} \in \big\{ \textrm{P}, \textrm{D} \big\}$,
at large $N$,
we approximate its expression through its bound given in (\ref{Euler-MacLaurin-p0}).
Thus, for $S=\mathcal{C}^2$, we find that (\ref{Euler-MacLaurin-p0}) becomes
\be 
\label{comp-EmL}
S-I_N^{(\textrm{\tiny B})}
\,=\,
\mathcal{C}^2-I_N^{(\textrm{\tiny B})}
\,=\,
\frac{f_{\textrm{\tiny B}}(N-1+\eta)-f_{\textrm{\tiny B}}(0)}{2}
+R^{_{\textrm{\tiny (B)}}}_{1,N}
\;\;\qquad\;\;
\textrm{B} \in \big\{ \textrm{P} , \textrm{D}  \big\}\,.
\ee
When PBC are imposed, the expression (\ref{comp-EmL}) becomes
\be
\label{EMacL_v2_PBC}
\mathcal{C}^2-I_N^{(\textrm{\tiny P})}
=
\frac{f_{\textrm{\tiny P}}(N)-f_{\textrm{\tiny P}}(0)}{2}
+R_{1,N}^{\textrm{\tiny (P)}}
=
R_{1,N}^{\textrm{\tiny (P)}}\,.
\ee

In order to estimate $R_{1,N}^{\textrm{\tiny (P)}}$ we consider its bound in (\ref{Euler-MacLaurin-p0}) and therefore we have
\be
\label{bound rest_ p0PBC}
R_{1,N}^{\textrm{\tiny (P)}}
=
\frac{2 e^{2\pi}}{\pi}\int_0^N \bigg| \frac{d f_{\textrm{\tiny P}}(k)}{dk} \bigg| \,dk
\,=\,
\frac{2 e^{2\pi}}{N}\,\sqrt{\frac{4\kappa}{m}}
\int_0^N
\big|
F_t'(s_k)\, \cos(\pi k/N)
\big| \,dk
\ee
where (\ref{fandF}) has been used. 
By introducing $\theta= \pi k /N$, one can take $N\to\infty$, obtaining
\be
\label{bound rest_ p0PBC v2}
R_{1,\infty}^{\textrm{\tiny (P)}}
\equiv
\lim_{N\to\infty}
R_{1,N}^{\textrm{\tiny (P)}}
\,=\,
\frac{2 e^{2\pi}}{\pi} \, \sqrt{\frac{4\kappa}{m}}
\int_0^\pi \big| F_t' \big(\sqrt{4\kappa/m}\,\sin\theta \big) \,\cos\theta\,\big| \,d\theta
\ee
where $F_t'$ can be computed from (\ref{Fdef}).
This calculation provides a complicated expression in the integrand of (\ref{bound rest_ p0PBC v2}), 
hence we evaluate $R_{1,\infty}^{\textrm{\tiny (P)}}$ numerically. 
Since $I_N^{(\textrm{\tiny P})}\to \mathcal{C}^2_{\textrm{\tiny TD}}$ when $N\to\infty$, in this limit (\ref{EMacL_v2_PBC}) 
gives (\ref{EmcL maintext}) with $\textrm{B}=\textrm{P}$.

In the case of DBC, the expression in (\ref{comp-EmL}) becomes
\be
\label{EMacL_v1_DBC}
\mathcal{C}^2-I_N^{(\textrm{\tiny D})}
=
\frac{f_{\textrm{\tiny D}}(N-1)-f_{\textrm{\tiny D}}(0)}{2} + R_{1,N}^{\textrm{\tiny (D)}}
=
\frac{f_{\textrm{\tiny P}}((N-1)/2)-f_{\textrm{\tiny P}}(0)}{2}+R_{1,N}^{\textrm{\tiny (D)}}
\ee
where in the last step we have emplyed 
the relation between $f_{\textrm{\tiny D}}$ and $f_{\textrm{\tiny P}}$. 
Using (\ref{f_def}), the limit $N\to\infty$ of the first term in the r.h.s. of (\ref{EMacL_v1_DBC})
gives 
\bea
\label{corr_Csq Dir}
\zeta
&\equiv&
\lim_{N\to\infty}
\frac{f_{\textrm{\tiny P}}((N-1)/2)-f_{\textrm{\tiny P}}(0)}{2}
\\
&=&
\frac{1}{2}\,\bigg\{
\bigg[\textrm{arcsinh} \bigg( \,
 \frac{(\omega^2 - \omega_0^2) \,  \sin\! \big(\sqrt{\omega^2 + 4\kappa/m } \;t \,\big) 
 }{2 \sqrt{\omega^2 + 4\kappa/m} \; \sqrt{\omega_0^2 + 4\kappa/m} } 
 \bigg)\bigg]^2 
 -
 \bigg[\textrm{arcsinh} \bigg(
 \frac{\omega^2 - \omega_0^2}{2\,\omega \, \omega_0 } \,
 \sin ( \omega t ) 
 \bigg)\bigg]^2
 \bigg\}\,.
 \nonumber
\eea

Now we estimate $R_{1,N}^{\textrm{\tiny (D)}}$ by approximating it 
through its bound in (\ref{Euler-MacLaurin-p0}).
From  the relation between $f_{\textrm{\tiny D}} $ and $f_{\textrm{\tiny P}} $ 
and (\ref{fandF}), we get
\be
\label{bound rest_ p0DBC}
R_{1,N}^{\textrm{\tiny (D)}}=\frac{2 e^{2\pi}}{\pi}\int_0^{N-1}|f_{\textrm{\tiny D}}'(k)|dk
=
\frac{2 e^{2\pi}}{2N}\sqrt{\frac{4\kappa}{m}}\int_0^{N-1}\bigg|\cos\bigg(\frac{\pi k}{2N}\bigg)
\,
F_t'\bigg(\sqrt{\frac{4\kappa}{m}}\sin\bigg(\frac{\pi k}{2N}\bigg)\bigg)\bigg|dk\,.
\ee 
Changing the integration variable to $\theta=\frac{\pi k}{2N}$ and taking the limit $N\to\infty$, we get
\be
\label{bound rest_ p0DBC v2}
R_{1,\infty}^{\textrm{\tiny (D)}}\equiv
\lim_{N\to\infty}R_{1,N}^{\textrm{\tiny (D)}}=\frac{2 e^{2\pi}}{\pi}\sqrt{\frac{4\kappa}{m}}
\int_0^{\pi/2} \bigg| F_t'\bigg(\sqrt{\frac{4\kappa}{m}}\sin\theta\bigg) \cos\theta\bigg| d\theta
=
\frac{R_{1,\infty}^{\textrm{\tiny (P)}}}{2}
\ee
where in the last step we used that
the integrand is symmetric under $\theta\to\pi-\theta$.

Thus, (\ref{EMacL_v1_DBC}) becomes (\ref{EmcL maintext}) 
with  $\textrm{B}=\textrm{D}$ as $N\to\infty$,
given that $I_N^{(\textrm{\tiny D})}\to \mathcal{C}^2_{\textrm{\tiny TD}}$.
When $\omega=0$, from (\ref{corr_Csq Dir}) we get $\zeta\to-(\log t)^2$ as $t\to\infty$. 
Since $\mathcal{C}^2$ is finite for any value of time when $\omega=0$ and DBC are imposed, 
from (\ref{EMacL_v1_DBC}) we have $\mathcal{C}_{\textrm{\tiny TD}}^2+R_{1,\infty}^{\textrm{\tiny (D)}}\to(\log t)^2 $ for $t\to\infty$.
We are not able to identify the asymptotic behaviour of 
$\mathcal{C}^2_{\textrm{\tiny TD}}$ and $R_{1,\infty}^{\textrm{\tiny (D)}}$ separately.

As for the initial growth, transforming the sum in (\ref{initial growth HC 1d}) 
into an integral as shown in Sec.\,\ref{subsec:TD limit HC}, 
we can write an explicit expression for the slope of the initial growth. 
The same limit can be done for the higher order terms in (\ref{comp-pure-initialgrowth}), obtaining
\be
\label{comp-pure-initialgrowth TD}
\frac{\mathcal{C}_{\textrm{\tiny TD}}}{\sqrt{N}} 
=
\frac{t|\omega^2-\omega_0^2|}{2\sqrt[4]{\omega_0^2\big(\frac{4\kappa}{m}+\omega_0^2\big)}} \;
 \bigg[
 1-
 \frac{t^2}{24}\, B_2
+
O(t^4)\bigg]
\ee
where
\be
B_2 \equiv
 \frac{\sqrt{\omega_0^2\big(\frac{4\kappa}{m}+\omega_0^2\big)}\big(16\omega_0^2\frac{\kappa}{m}+4\omega_0^4\big)-(\omega_0^2-\omega^2)\big(\omega_0^2(3\omega_0^2+\omega^2)+2\frac{\kappa}{m}(7\omega_0^2+\omega^2)\big)}{\omega_0^2\big(\frac{4\kappa}{m}+\omega_0^2\big)}  \,.
\ee

Let us stress that the formula (\ref{comp-pure-initialgrowth TD}) for the initial growth does not distinguish between PBC or DBC, differently from
(\ref{compDBCapproximinitialt}) and (\ref{compPBCapproximinitialt}).

The thermodynamic limit discussed above can be easily applied also to the case discussed in appendix \ref{app:unentangled}, where the reference state is
the unentangled product state. 
From (\ref{complexity_equalm_prod ref state}), at leading order in $N$ we obtain
\be
\label{complexity_equalm_prod ref state TD}
\mathcal{C}^2_{\textrm{\tiny TD}}
=
\frac{N}{4\pi}\int_{0}^\pi
\bigg\{\textrm{arccosh}\bigg[\,
\frac{\Omega^2_{0,\theta}+\mu^2}{2\mu\,\Omega_{0,\theta}}
+
\frac{(\omega^2-\omega_0^2)(\Omega_\theta^2-\mu^2)}{2\mu\,\Omega_{0,\theta}\,\Omega_{\theta}^2}\,
[\sin(\Omega_{\theta}t)]^2
\bigg]\bigg\}^2
d\theta
\ee
where the dispersion relations are given by (\ref{dispersion relations TD}).

\subsection{Continuum limit}
\label{app:continuum}

In this appendix we report some details on the continuum limit procedure that leads to (\ref{comp HC cont}) which is valid for both PBC and DBC.

Starting from PBC, we can exploit the identity $\sin(x) = \sin (\pi -x)$ 
to rewrite the complexity (\ref{comp-pure-global-DBCPBC}) as follows
\be
\label{C2 sq even}
\mathcal{C}^2=
\sum_{k=-N/2+1}^{N/2}\left[ \textrm{arcsinh}\! 
\left( \,
 \frac{\omega^2 - \omega_0^2}{2\,\Omega_{k}\, \Omega_{0,k}} \,
 \sin (\Omega_{k} t ) 
 \right)
 \right]^2
\,\,\qquad\,\,
\textrm{even $N$}
\ee
and
\be
\label{C2 sq odd}
\mathcal{C}^2=
\sum_{k=-(N-1)/2}^{(N-1)/2}\left[ \textrm{arcsinh}\! 
\left( \,
 \frac{\omega^2 - \omega_0^2}{2\,\Omega_{k}\, \Omega_{0,k}} \,
 \sin (\Omega_{k} t ) 
 \right)
 \right]^2
\,\,\qquad\,\,
\textrm{odd $N$.}
\ee
In the continuum limit $N\to\infty$ and the lattice spacing $a \equiv \sqrt{m/\kappa} \to 0$ 
while $N a\equiv \ell$ is kept fixed. 
In this limit the dispersion relation (\ref{dispersion relations}) becomes
$\sqrt{\omega^2 + (2\pi k/\ell)^2 }  =\sqrt{\omega^2 + p^2} =\Omega_p$, 
where $\Omega_p$ has been defined in (\ref{dispersion continuum HC})
and  $p\equiv\frac{2k\pi}{a N} \in \mathbb{R}$,
because of the range of $k$ in (\ref{C2 sq even}) and (\ref{C2 sq odd}).
The resulting dispersion relation  identifies the frequency of the harmonic chain with the mass of the underlying continuum field theory, which is the Klein-Gordon field theory.
Replacing the sum over the integers $k$ with the integral $\ell\int_{-\infty}^{\infty}\frac{dp}{2\pi}$ over the momenta $p$ we obtain (\ref{comp HC cont}) at leading order in $\ell$.

When DBC are imposed we cannot exploit the identity for $\sin x$ mentioned above.
In this case, we first observe 
that the dispersion relation (\ref{dispersion DBC}) in this limit becomes
$\sqrt{\omega^2 + (\pi k/\ell)^2 }  =\sqrt{\omega^2 + p^2} =\Omega_p$, 
with $\Omega_p$ given by (\ref{dispersion continuum HC})
and $p=\frac{\pi k}{a N} \in (\pi/\ell ,\infty)$ because $k \geqslant 1$ in (\ref{dispersion DBC}).
Similarly, the dispersion relation of the pre-quench hamiltonian becomes 
$\Omega_{0,p}=\sqrt{\omega_0^2 + p^2}$.
At  leading order in $\ell$, we have that $\pi/\ell$ vanishes
and therefore $p \in [0,\infty)$.
By substituting the sum over $k$ in (\ref{comp-pure-global-DBCPBC}) 
with $\ell\int_{0}^{\infty}\frac{dp}{\pi}$, we get
\be
\label{C_cont_DBC}
\mathcal{C}_{\textrm{\tiny cont}}
=
\sqrt{\frac{\ell}{\pi}}\;
\sqrt{\int_{0}^{\infty}
 \left[ \textrm{arcsinh}\! 
\left( \,
 \frac{\omega^2 - \omega_0^2}{2\,\Omega_p\, \Omega_{0,p}} \,
 \sin (\Omega_p t ) 
 \right)
 \right]^2
 d p
}\;.
\ee
The expression (\ref{comp HC cont}) is easily recovered
by using that the integrand in (\ref{C_cont_DBC}) is even in $p$.

This procedure can be applied also to study 
the continuum limit of (\ref{complexity_equalm_prod ref state})
where the reference state is the unentangled product state. 
For both PBC and DBC, for the leading term we find
\be
\label{complexity_equalm_prod ref state TD}
\mathcal{C}_{\textrm{\tiny cont}}^2
=
\frac{\ell}{8\pi}
\int_{-\infty}^\infty
\bigg\{\textrm{arccosh}\bigg[\,
\frac{\Omega^2_{0,p}+\mu^2}{2\mu\,\Omega_{0,p}}
+
\frac{(\omega^2-\omega_0^2)(\Omega_p^2-\mu^2)}{2\mu\,\Omega_{0,p}\,\Omega_{p}^2}\,
[\sin(\Omega_{p}t)]^2\bigg]\bigg\}^2
dp
\ee
where the dispersion relations are given by (\ref{dispersion continuum HC}).
This result does not coincide with
the one reported in \cite{Alves:2018qfv} for the temporal evolution of the complexity
because of the different choice of gates. 
The role of the set of allowed gates in the determination of the 
temporal evolution of the complexity deserves further future analyses.

Let us remark that, while the expression (\ref{comp HC cont}),
obtained by choosing the initial state as reference state, is UV finite, 
(\ref{complexity_equalm_prod ref state TD}) is UV divergent.
This UV divergence 
can be regularised by introducing a cutoff $|p|\leqslant \Lambda$ on the momenta.
Alternatively, 
since the UV divergence comes from $\mathcal{C}_{\textrm{\tiny cont}}^2|_{t=0}$, 
it is natural to introduce the following UV finite quantity
\be
\label{uv-reg-C-cont}
\Delta\mathcal{C}_{\textrm{\tiny cont}}=
\mathcal{C}_{\textrm{\tiny cont}}^2-\mathcal{C}_{\textrm{\tiny cont}}^2\big|_{t=0}
\ee
whose sign is not definite for $t>0$.

A similar analysis has been carried out also in \cite{Alves:2018qfv},
whose result in the sudden quench limit can be compared against 
(\ref{complexity_equalm_prod ref state TD}).
These two expressions coincide at $t=0$,
when the result of \cite{Jefferson:2017sdb} is recovered,
and they both display a UV divergence that can be regularised 
as done in (\ref{uv-reg-C-cont}).
For $t>0$, after an initial growth both the expressions 
show persistent oscillations but they do not coincide.
For instance, 
while the initial growth of the result of \cite{Alves:2018qfv} is linear, 
the next term after the constant 
in the expansion of (\ref{complexity_equalm_prod ref state TD}) 
as $t \to 0$ is quadratic
(see also (\ref{complexity_equalm_prod ref state small t})).

\section{Further numerical results on the relaxation to the GGE}
\label{app:gge}

\begin{figure}[t!]
\subfigure
{\hspace{-1.55cm}
\includegraphics[width=.58\textwidth]{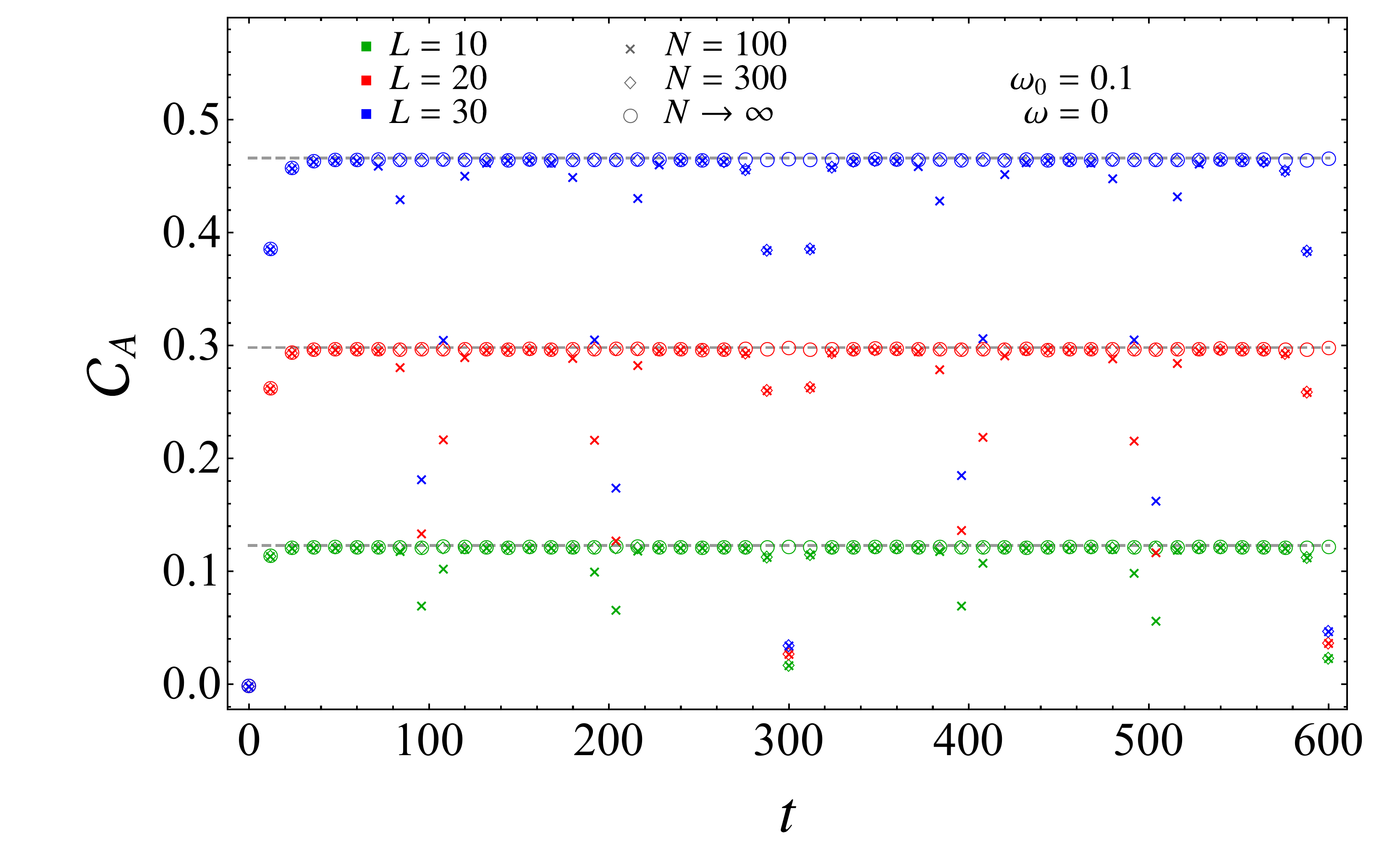}}
\subfigure
{
\hspace{-.65cm}\includegraphics[width=.58\textwidth]{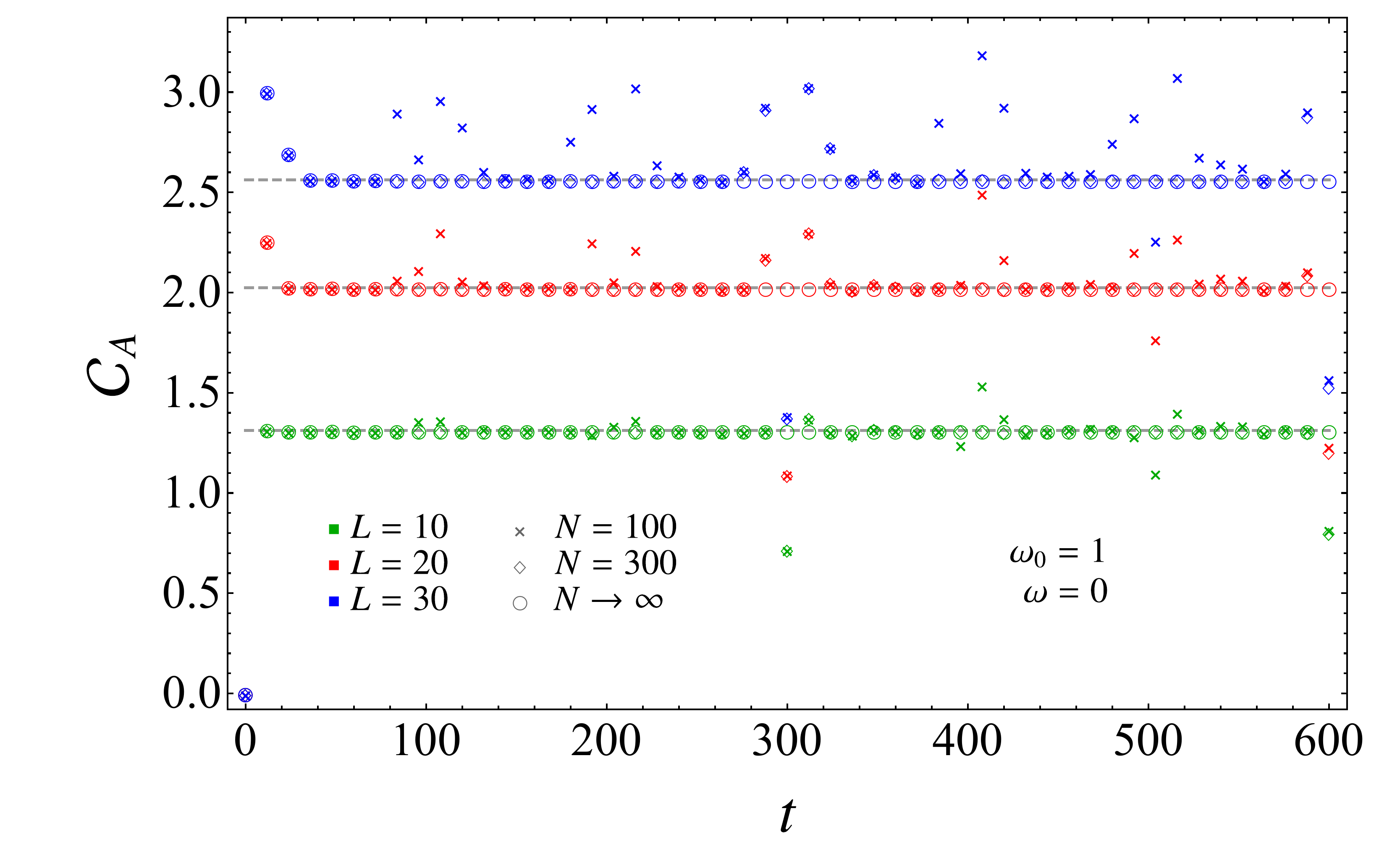}}
\caption{Temporal evolution of $\mathcal{C}_A$ 
after a global quantum quench with a gapless evolution Hamiltonian 
for a block $A$ made by $L$ consecutive sites 
adjacent to a boundary of harmonic chains with DBC made by $N$ sites.
The data corresponding to $N \to \infty$ are obtained through a chain on the semi-infinite line.
The horizontal dashed grey lines correspond to (\ref{comp-initialvsGGE}).
}
\vspace{0.4cm}
\label{fig:MixedStateGlobalMasslessEvolutionTD}
\end{figure}

\begin{figure}[t!]
\subfigure
{
\hspace{-1.55cm}\includegraphics[width=.58\textwidth]{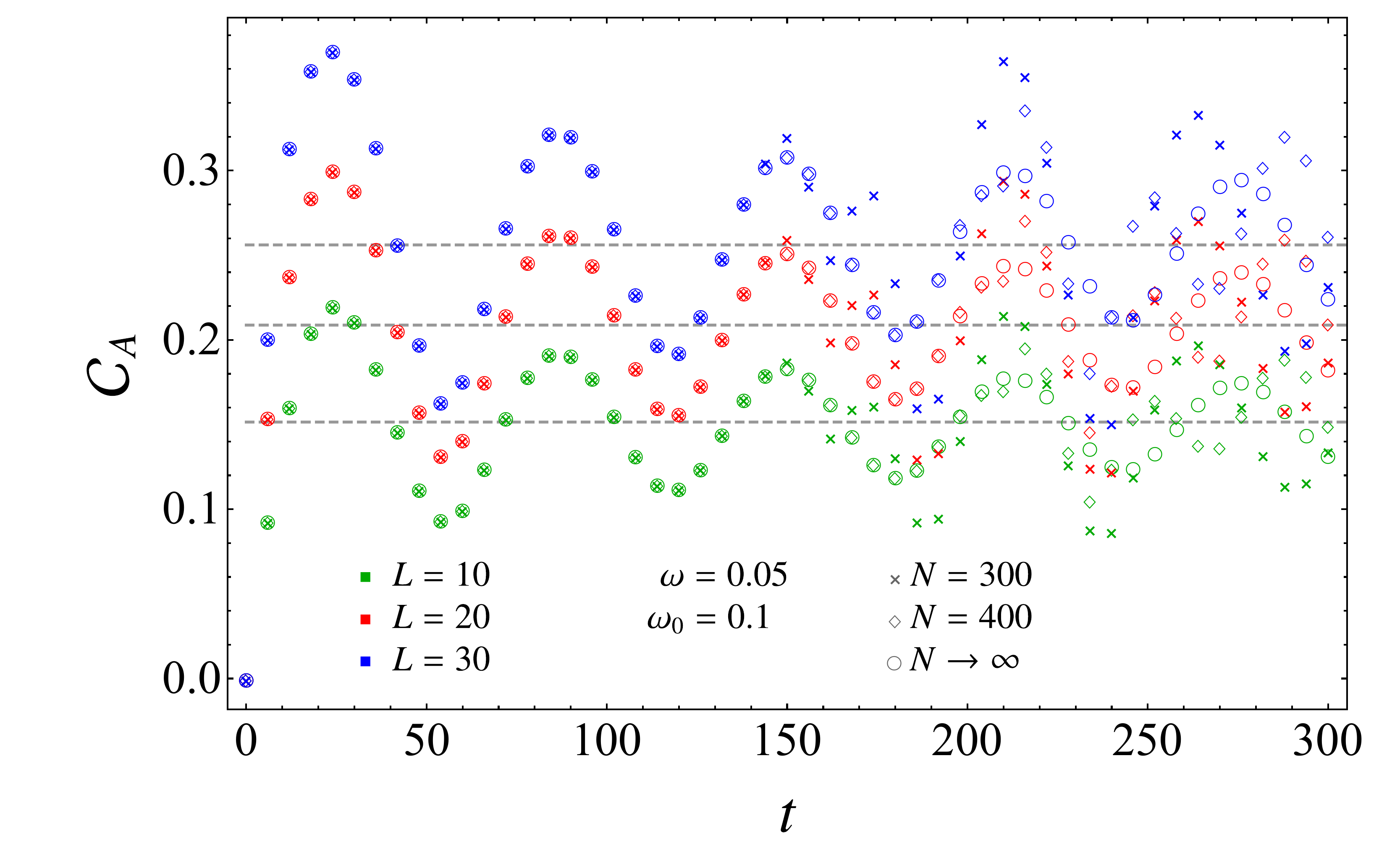}}
\subfigure
{
\hspace{-.65cm}\includegraphics[width=.58\textwidth]{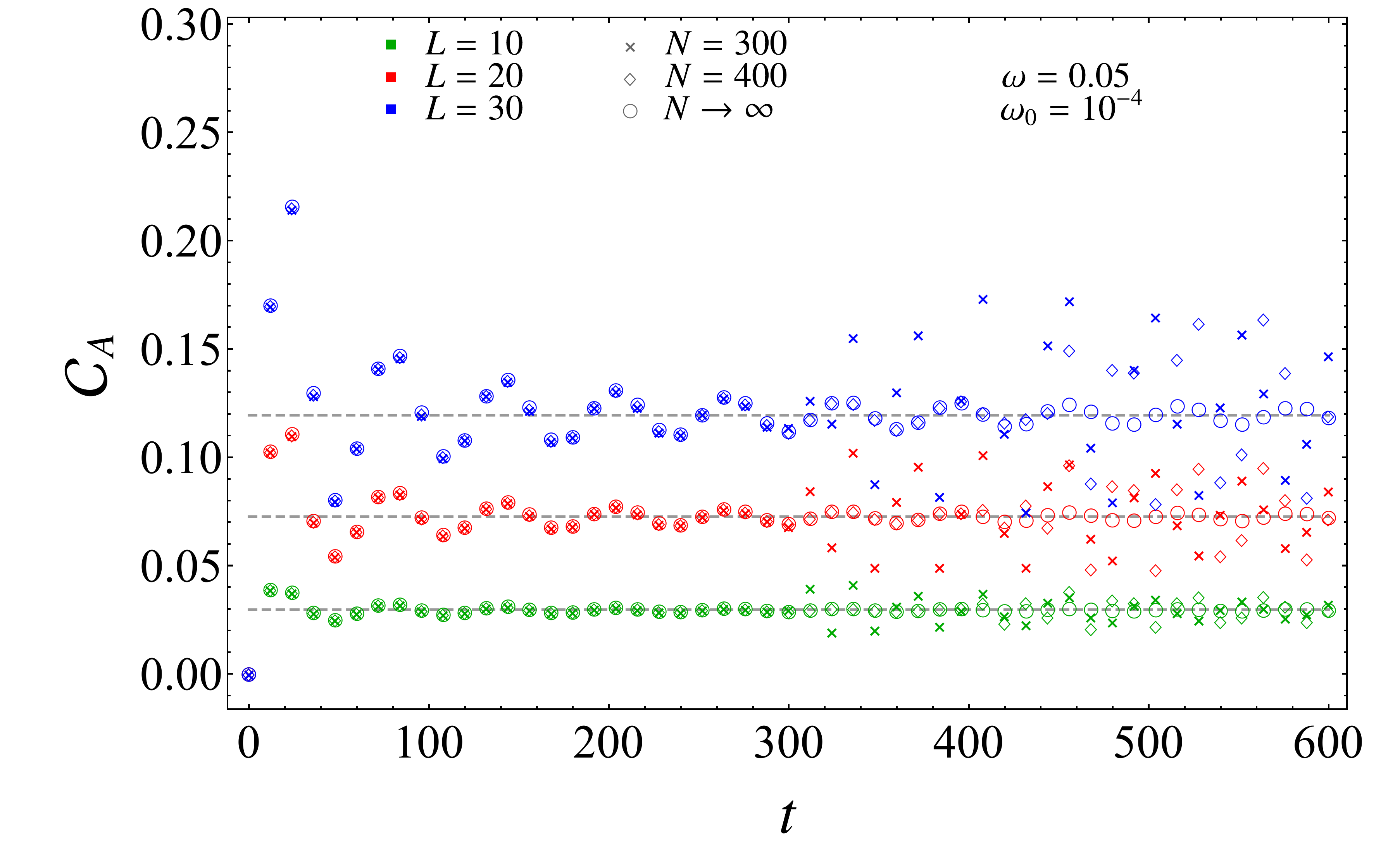}}
\subfigure
{
\hspace{-1.55cm}\includegraphics[width=.58\textwidth]{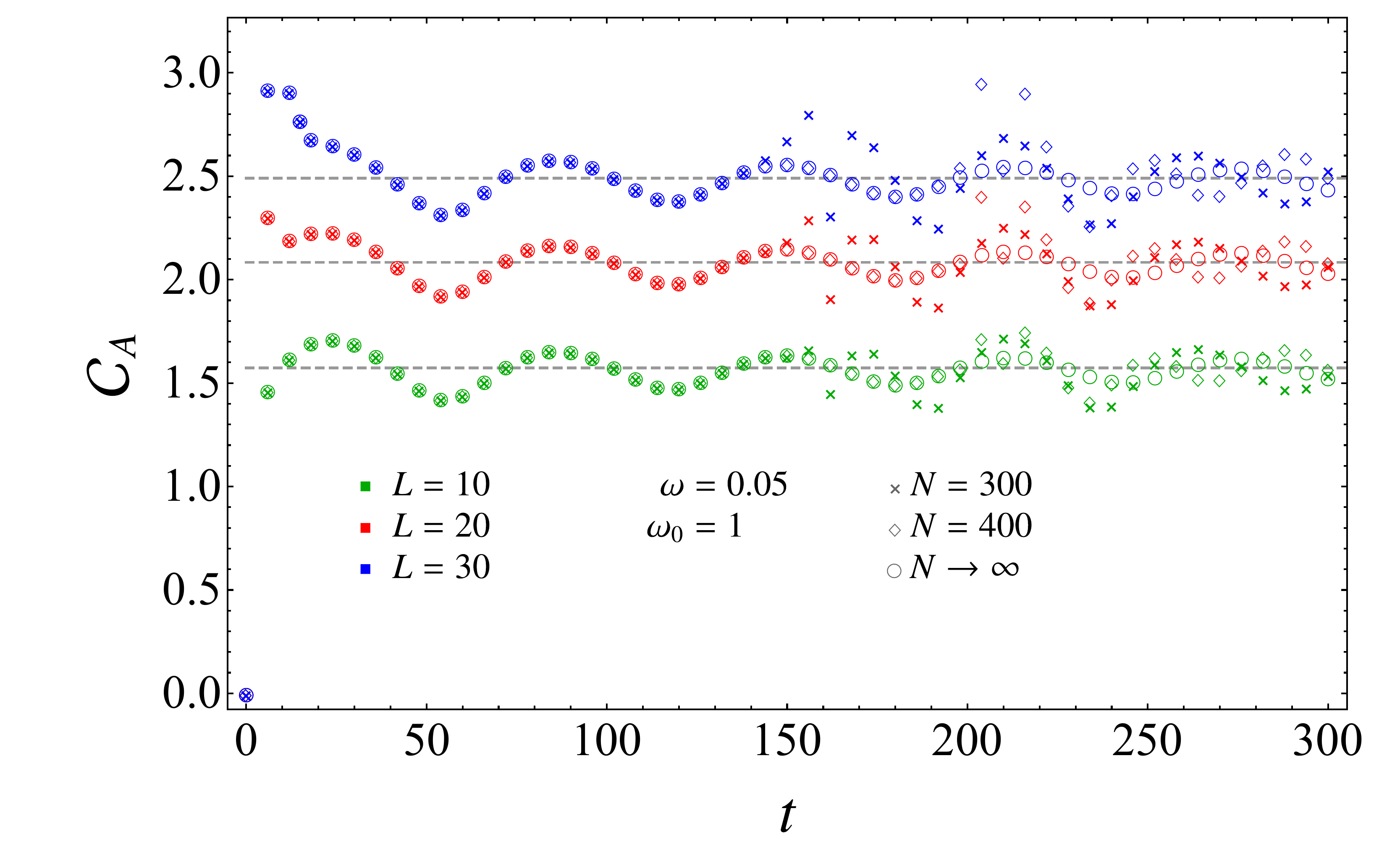}}
\subfigure
{\hspace{-.65cm}
\includegraphics[width=.58\textwidth]{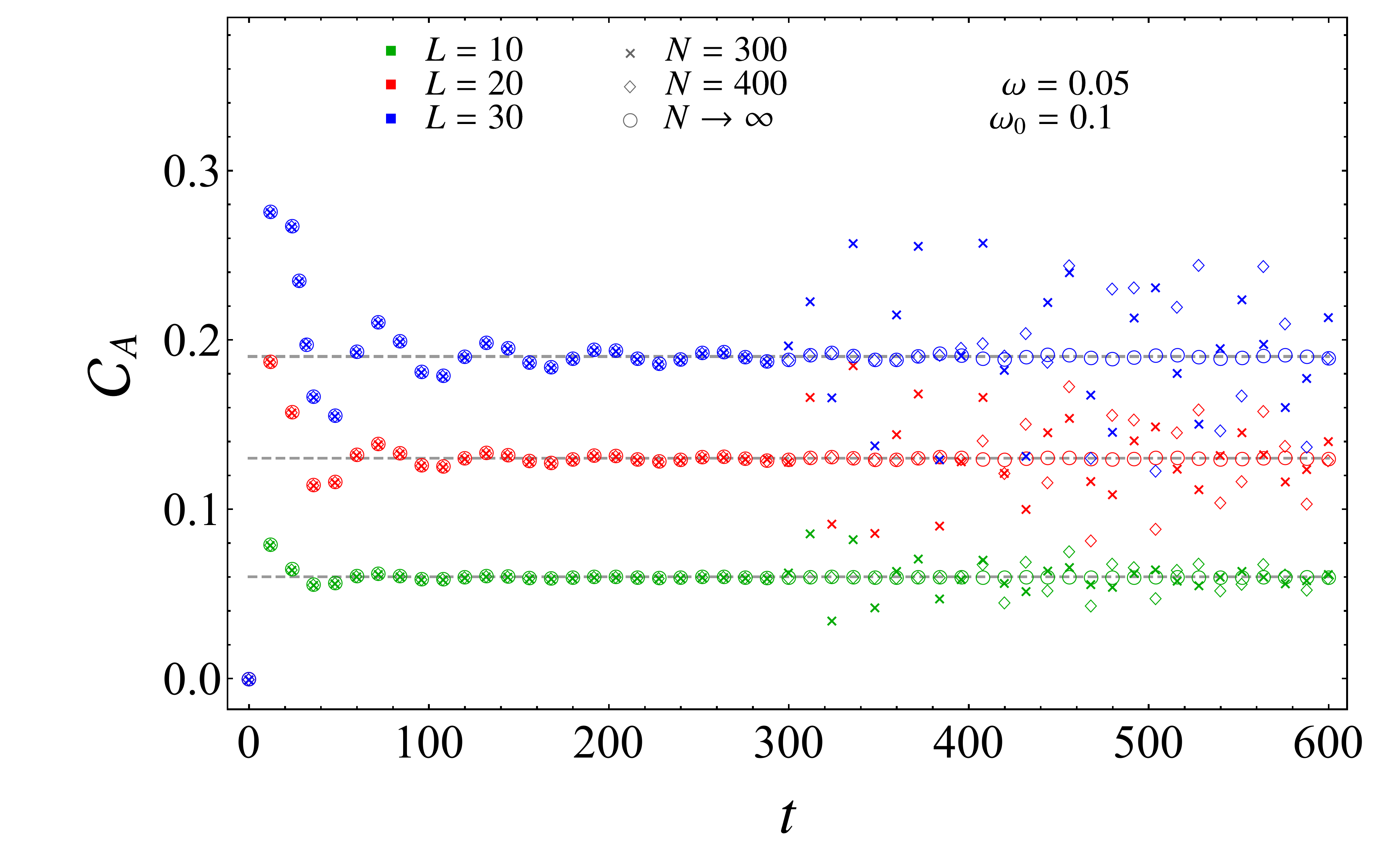}}
\caption{
Temporal evolution of $\mathcal{C}_A$ 
after a global quantum quench with a gapped evolution Hamiltonian.
In the left panels the chains are either on the circle or on the infinite line,
while in the right panels the chains are either on the segment or on the
semi-infinite line with DBC and $A$ is adjacent to a boundary.
The dashed grey lines correspond to (\ref{comp-initialvsGGE}).
}
\vspace{0.4cm}
\label{fig:MixedStateGlobalMassiveEvolutionTD}
\end{figure}

In this appendix we report further numerical results supporting 
(\ref{comp-initialvsGGE}) and (\ref{stationary complexity density}).

\begin{figure}[t!]
\subfigure
{\hspace{-1.55cm}
\includegraphics[width=.58\textwidth]{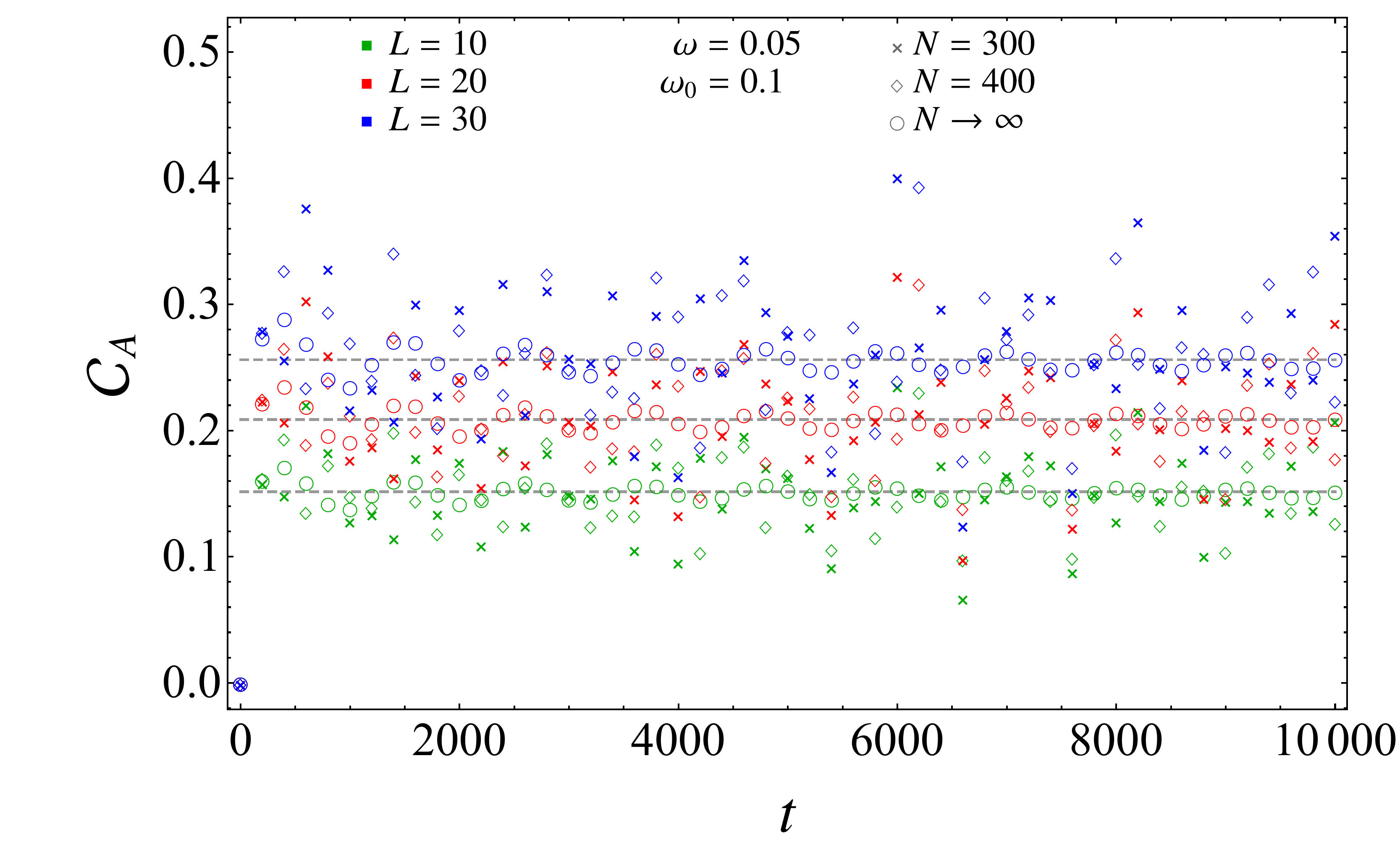}}
\subfigure
{
\hspace{-.65cm}\includegraphics[width=.58\textwidth]{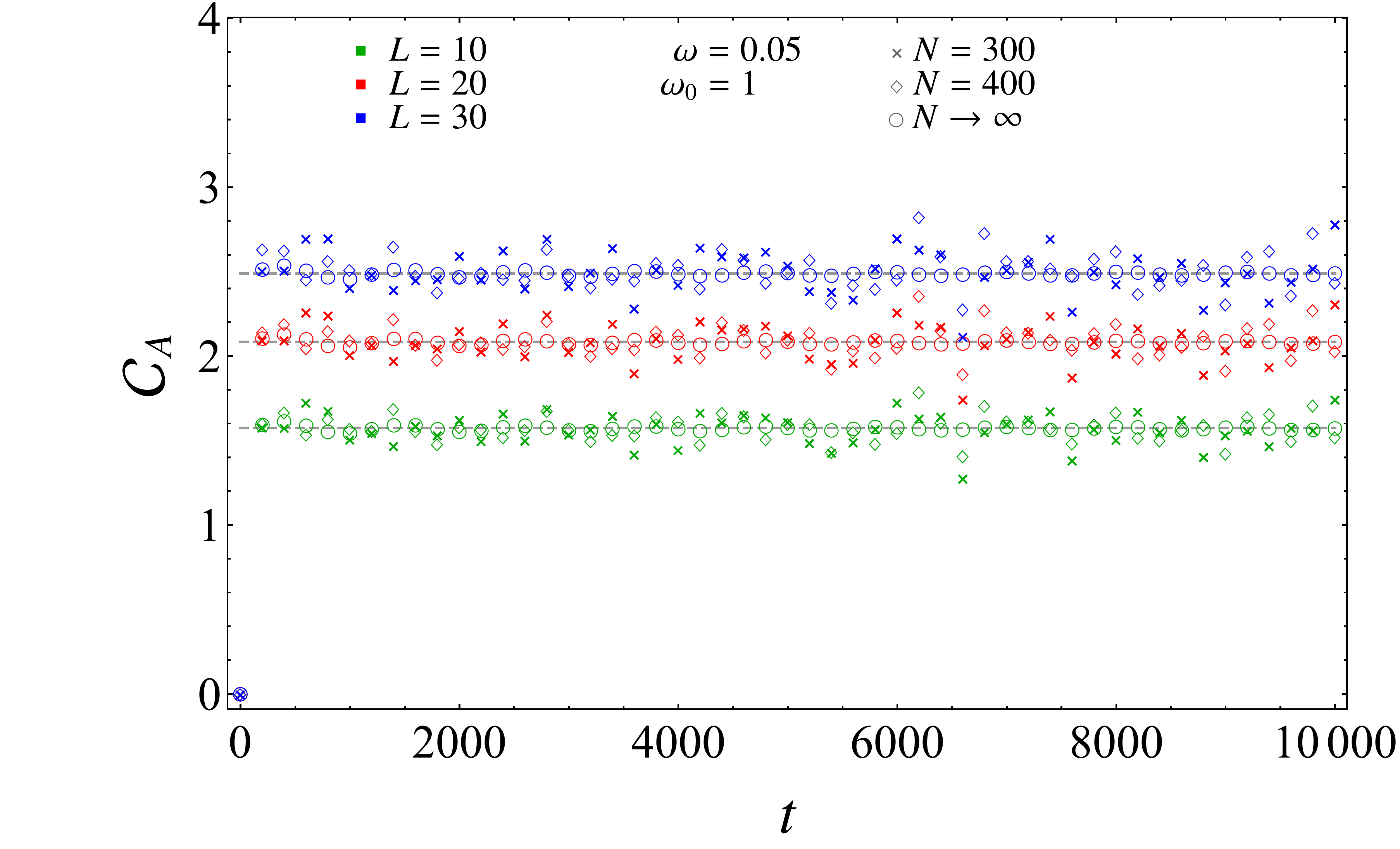}}
\caption{
Temporal evolution of $\mathcal{C}_A$ 
after a global quantum quench with a gapped evolution Hamiltonian
for a block made by $L$ consecutive sites in harmonic chains 
either with PBC or on the infinite line. 
The dashed grey lines correspond to (\ref{comp-initialvsGGE}).}
\vspace{0.4cm}
\label{fig:MixedStateGlobalMassiveEvolutionTDLarget}
\end{figure}

\begin{figure}[htbp!]
\vspace{.1cm}
\subfigure
{\hspace{-1.55cm}
\includegraphics[width=.58\textwidth]{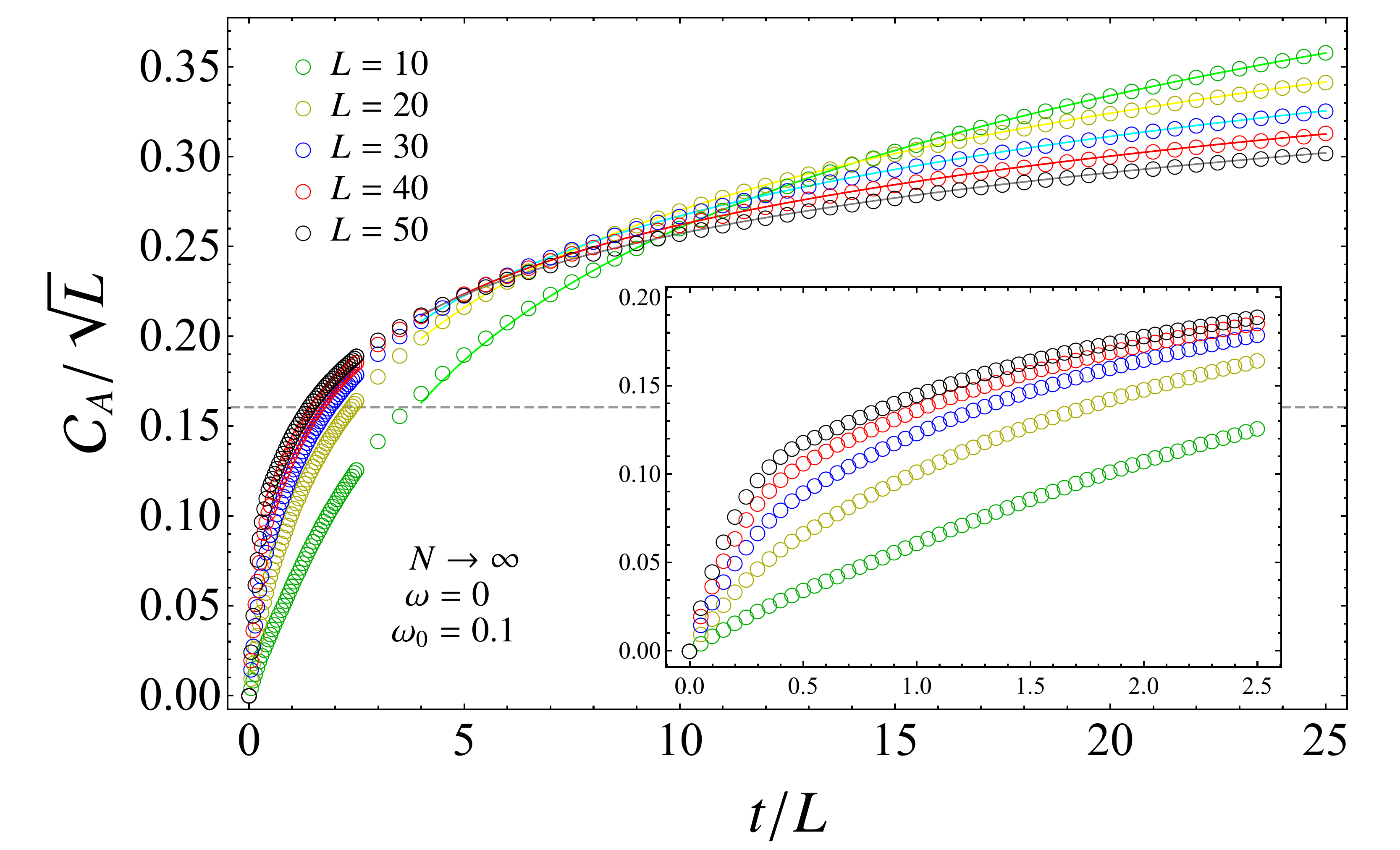}}
\subfigure
{
\hspace{-.65cm}\includegraphics[width=.58\textwidth]{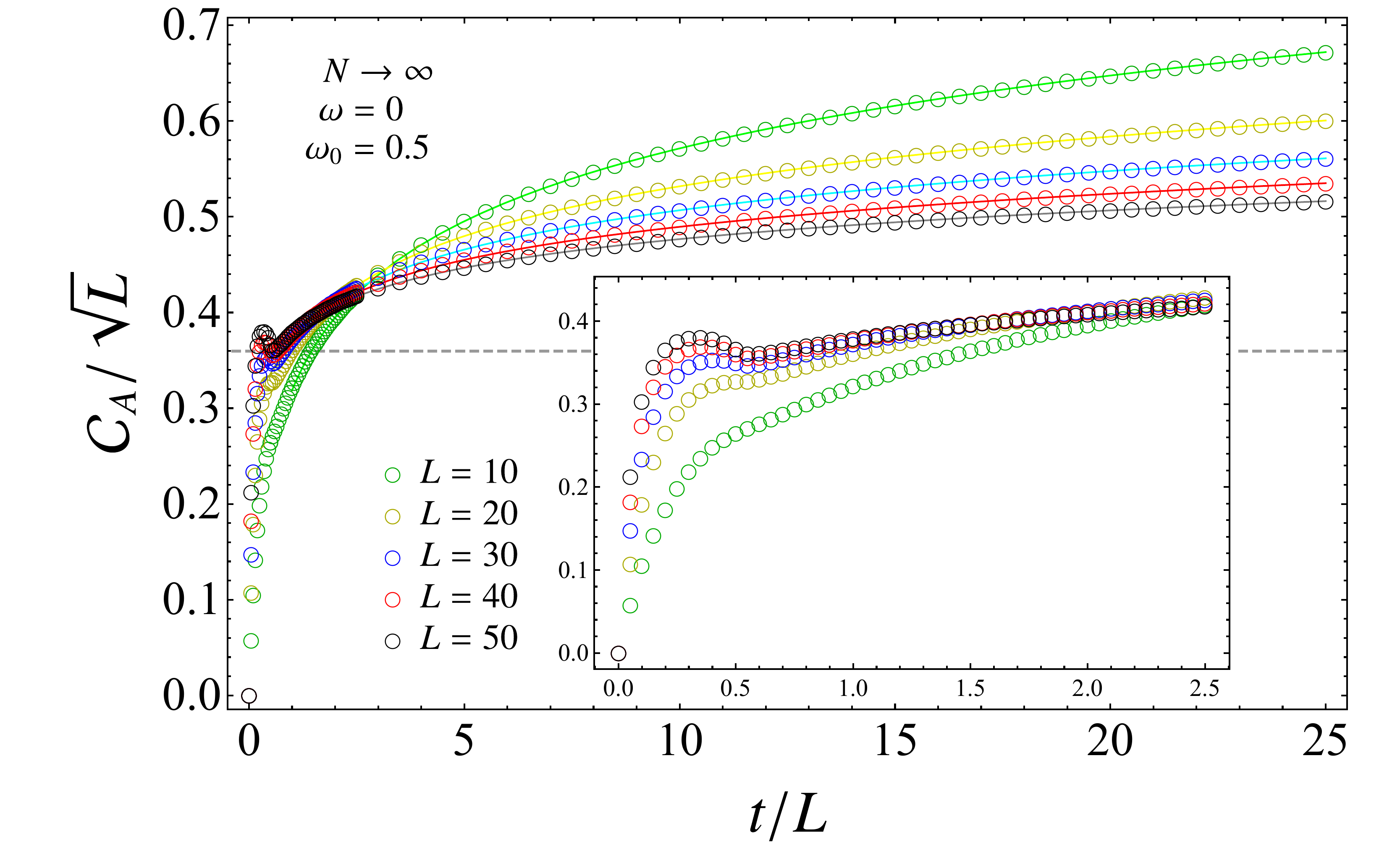}}
\subfigure
{\hspace{-1.55cm}
\includegraphics[width=.58\textwidth]{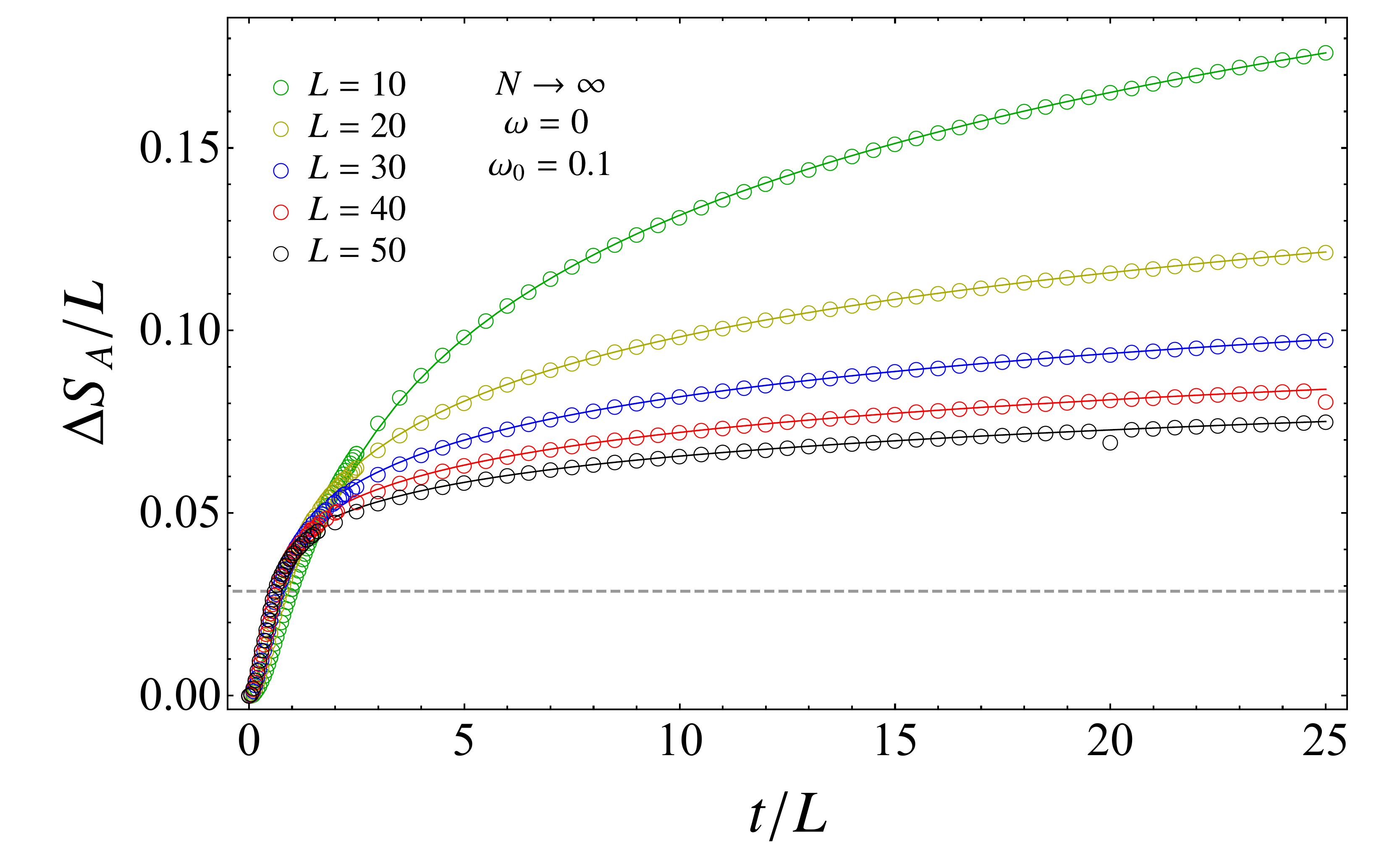}}
\subfigure
{
\hspace{-.65cm}\includegraphics[width=.58\textwidth]{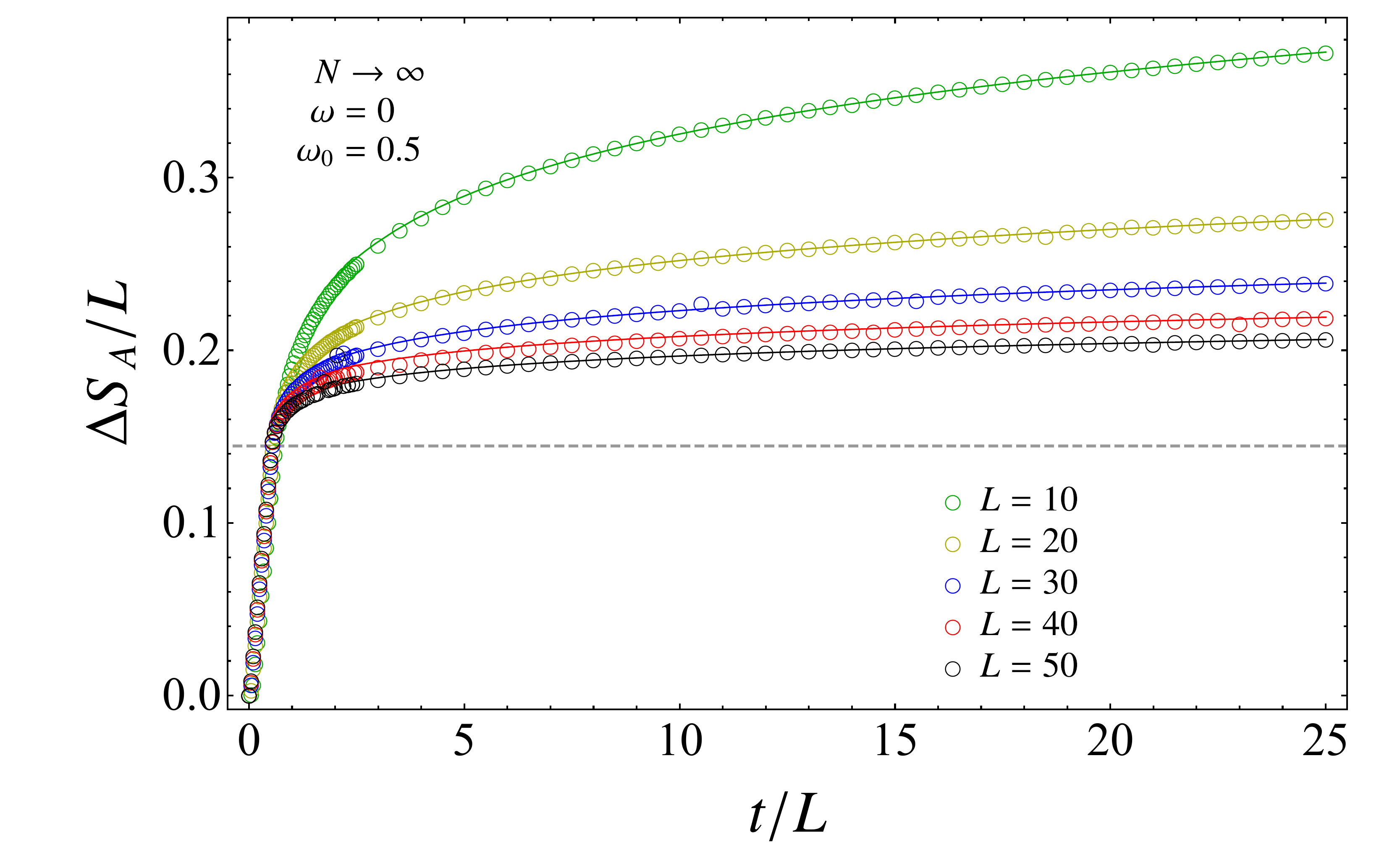}}
\subfigure
{\hspace{-1.55cm}
\includegraphics[width=.58\textwidth]{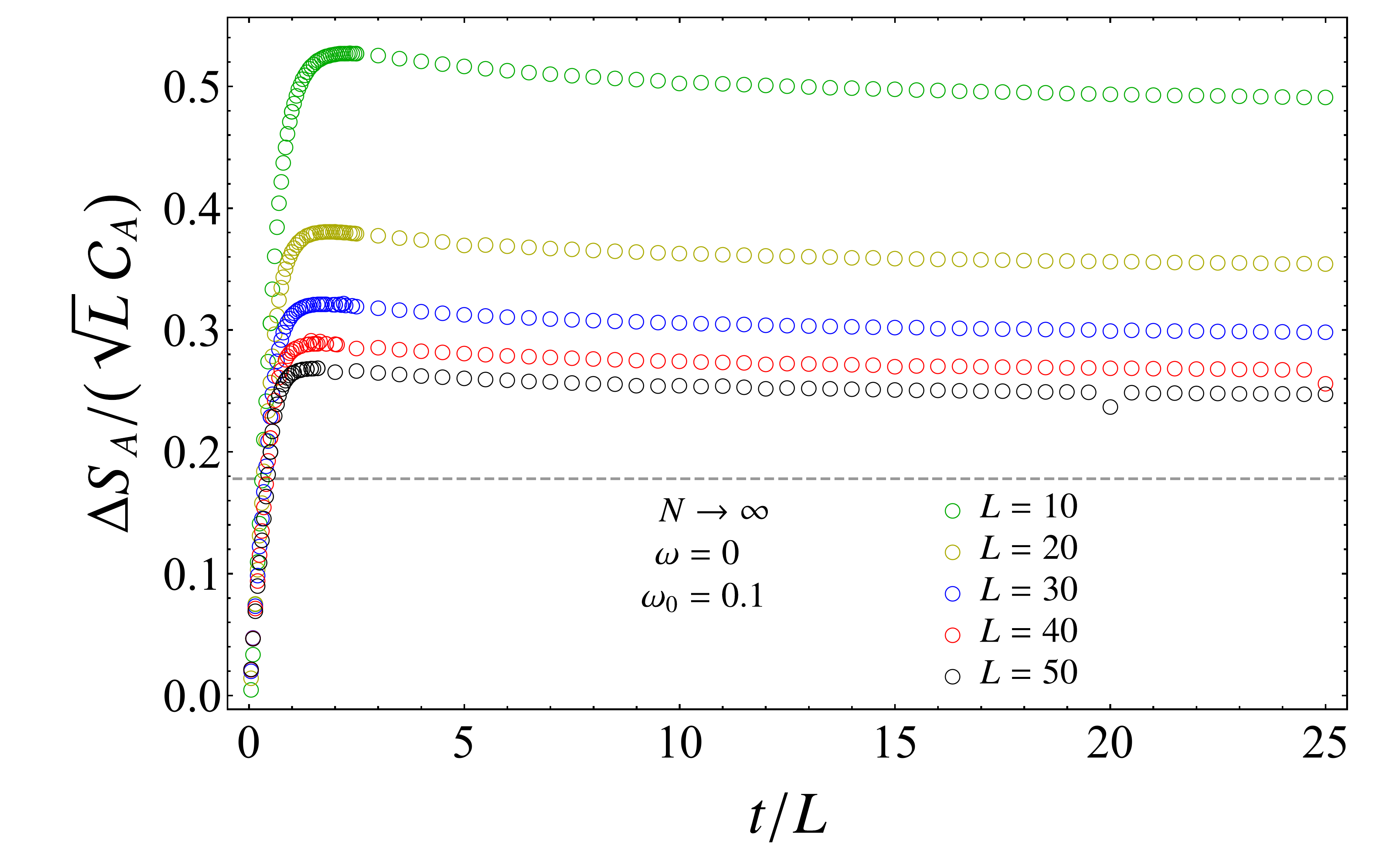}}
\subfigure
{
\hspace{-.65cm}\includegraphics[width=.58\textwidth]{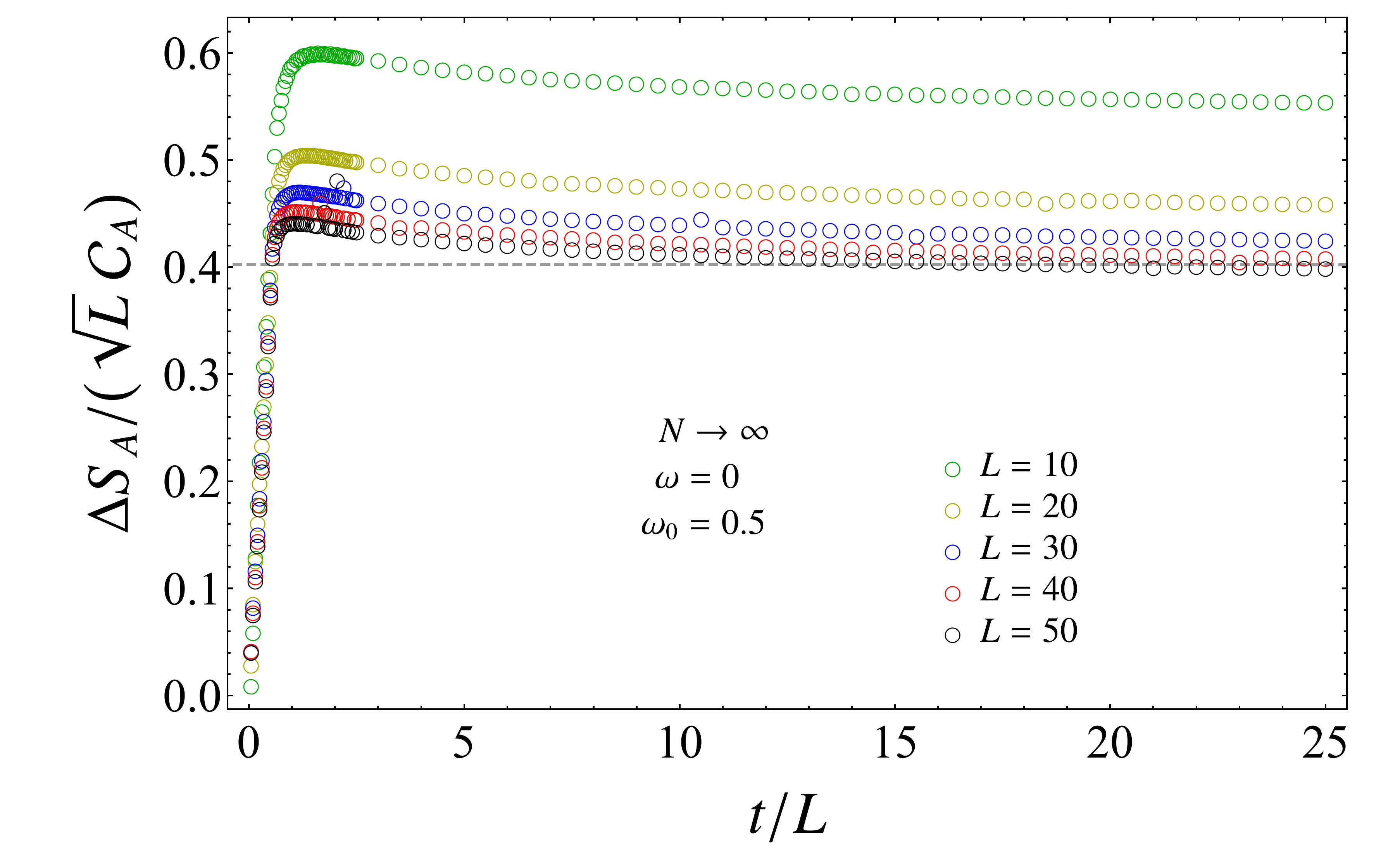}}
\caption{Temporal evolutions 
after a global quantum quench with a gapless evolution Hamiltonian
and either $\omega_0 = 0.1$ (left panels) or $\omega_0 = 0.5$ (right panels)
of $\mathcal{C}_A$  (top panels), of $\Delta S_A$ (middle panels) 
and of $ \Delta S_A/\mathcal{C}_A $ (bottom panels)
for a block made by $L$ consecutive sites in an infinite harmonic chain.
}
\vspace{0.4cm}
\label{fig:MixedStatePBCGlobalMasslessEvolutionTD}
\end{figure}

In Fig.\,\ref{fig:MixedStateGlobalMasslessEvolutionTD}, 
Fig.\,\ref{fig:MixedStateGlobalMassiveEvolutionTD}. 
Fig.\,\ref{fig:MixedStateGlobalMassiveEvolutionTDLarget} 
and Fig.\,\ref{fig:MixedStatePBCGlobalMasslessEvolutionTD} 
we show some temporal evolutions of $\mathcal{C}_A$
for a block $A$ made by $L$ consecutive  sites in harmonic chains with $N$ sites
where $N$ is either finite or infinite,
with the aim to check that (\ref{comp-initialvsGGE}) 
provides the correct asymptotic value as $t \to \infty$.

Each set of data corresponds to a choice of 
$N$, $L$, $\omega_0$ and $\omega$.
The data represented by coloured markers have been found through (\ref{c2-complexity-rdm-our-case}),
with the reduced correlators 
obtained 
either from (\ref{QPRmat t-dep 1d}) 
or from (\ref{QPRmat t-dep 1d DBC}) 
when $N$ is finite
(for PBC and DBC respectively)
and 
either from (\ref{QPRmat t-dep TD}) 
or from (\ref{QPRmat t-dep DBC TD})
when $N \to \infty$
(on the infinite line and on the semi-infinite line respectively).
The horizontal dashed lines show the subregion complexity between the initial state and the GGE 
given by (\ref{comp-initialvsGGE}),
obtained by reducing 
the correlators (\ref{corr-GGE-app-int-pbc}) and (\ref{corr-GGE-app-int-dbc})
for the target state
and the correlators (\ref{QPRmat t-dep TD}) and (\ref{QPRmat t-dep DBC TD}) at $t=0$
for the reference state
(for the infinite line and the semi-infinite line respectively).

In Fig.\,\ref{fig:MixedStateGlobalMasslessEvolutionTD}
we consider the temporal evolution of $\mathcal{C}_A$ when $\omega = 0$,
DBC are imposed and the block is adjacent to a boundary.
The data obtained for finite $N$ are compared against the ones for $N \to \infty$,
found for a block at the beginning of the semi-infinite chain. 
The main feature to highlight in these temporal evolutions are the plateaux
occurring both for finite $N$ and for $N \to \infty$.
The horizontal dashed lines in Fig.\,\ref{fig:MixedStateGlobalMasslessEvolutionTD}
correspond to the subregion complexity (\ref{comp-initialvsGGE}).
These agreements support the assumption that
the target state relaxes to an asymptotic state locally described by the GGE 
in (\ref{rho_gge}) as $t \to \infty$.
For a given set of parameters,
the height of the plateaux is independent of $N$,
while it increases as either $L$ or $\omega_0$ increases. 
Comparing the two panels of Fig.\,\ref{fig:MixedStateGlobalMasslessEvolutionTD},
where different values of $\omega_0$ are considered, 
one observes that the local maxima occur (when $N \to \infty$ there is only the first one)
for large enough $\omega_0$.
We also highlight the absence of oscillations in the formation of the plateaux 
when the evolution Hamiltonian is gapless. 
%

In Fig.\,\ref{fig:MixedStateGlobalMassiveEvolutionTD}
and Fig.\,\ref{fig:MixedStateGlobalMassiveEvolutionTDLarget}
the evolution Hamiltonians are gapped with $\omega = 0.05$.
We show data obtained
for harmonic chains either with PBC or on the infinite line
in the left panels and
for harmonic chains either with DBC or on the semi-infinite line
(with the block adjacent to a boundary)
in the right panels.
In these evolutions, data corresponding to the same $\omega$ and $\omega_0$ 
collapse for $t<N/2$ in the left panels and for  $t<N$ in the right panels. 
The main difference with respect to the gapless evolutions 
in Fig.\,\ref{fig:MixedStateGlobalMasslessEvolutionTD} 
are the oscillations after the initial growth around the asymptotic value,
which is evaluated through (\ref{comp-initialvsGGE})
and corresponds to the horizontal dashed grey lines,
whose height depends on $\omega$ and $\omega_0$.
Fig.\,\ref{fig:MixedStateGlobalMassiveEvolutionTDLarget}
highlights the fact that, for harmonic chains with PBC or on the infinite line, 
very long time is needed to reach the asymptotic value
given by (\ref{comp-initialvsGGE}).
Comparing the two panels in Fig.\,\ref{fig:MixedStateGlobalMassiveEvolutionTDLarget},
one notices that the amplitude of the oscillations 
decreases as $|\omega - \omega_0|$ increases.

\begin{figure}[t!]
\subfigure
{\hspace{-1.55cm}
\includegraphics[width=.58\textwidth]{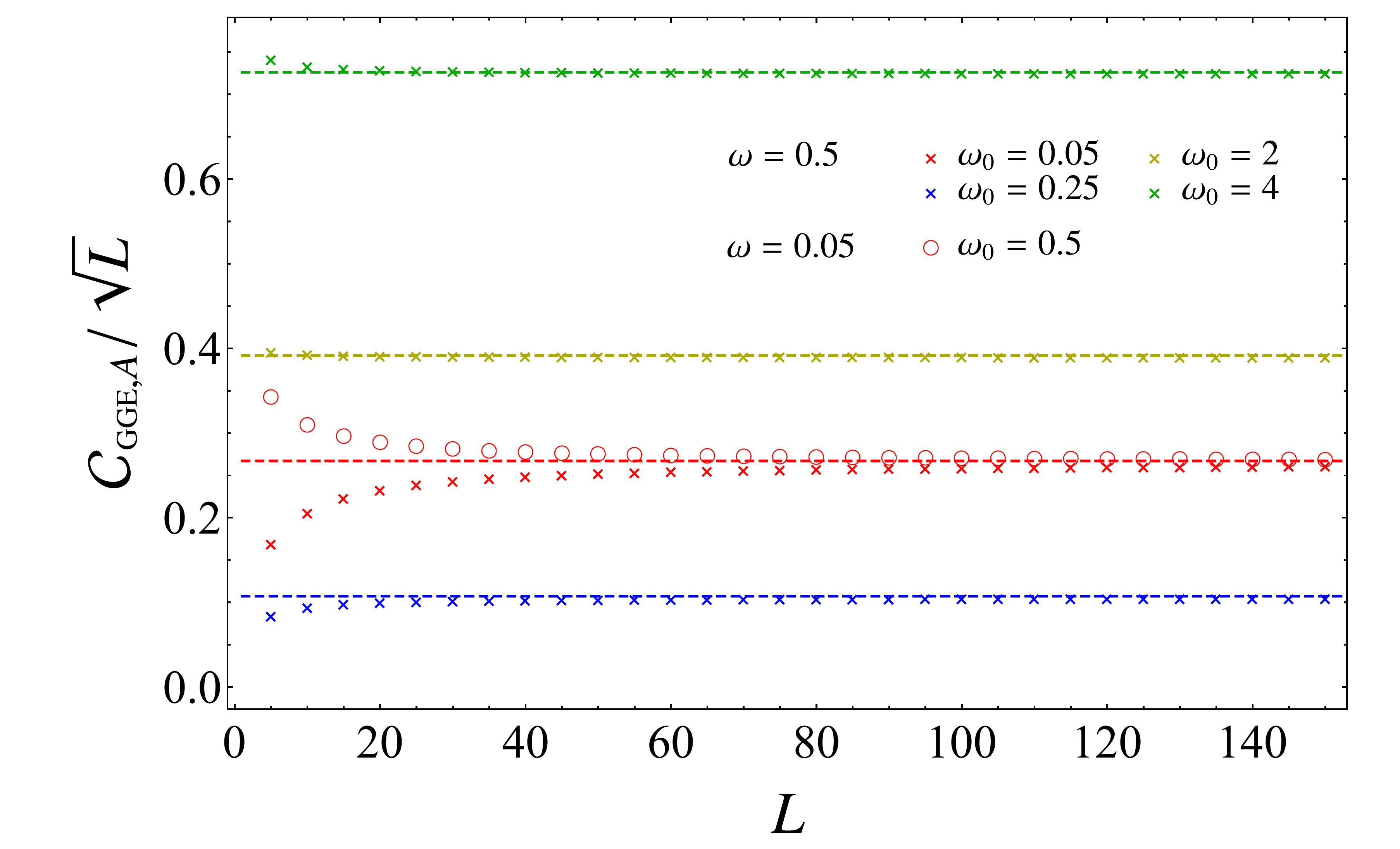}}
\subfigure
{
\hspace{-.65cm}\includegraphics[width=.58\textwidth]{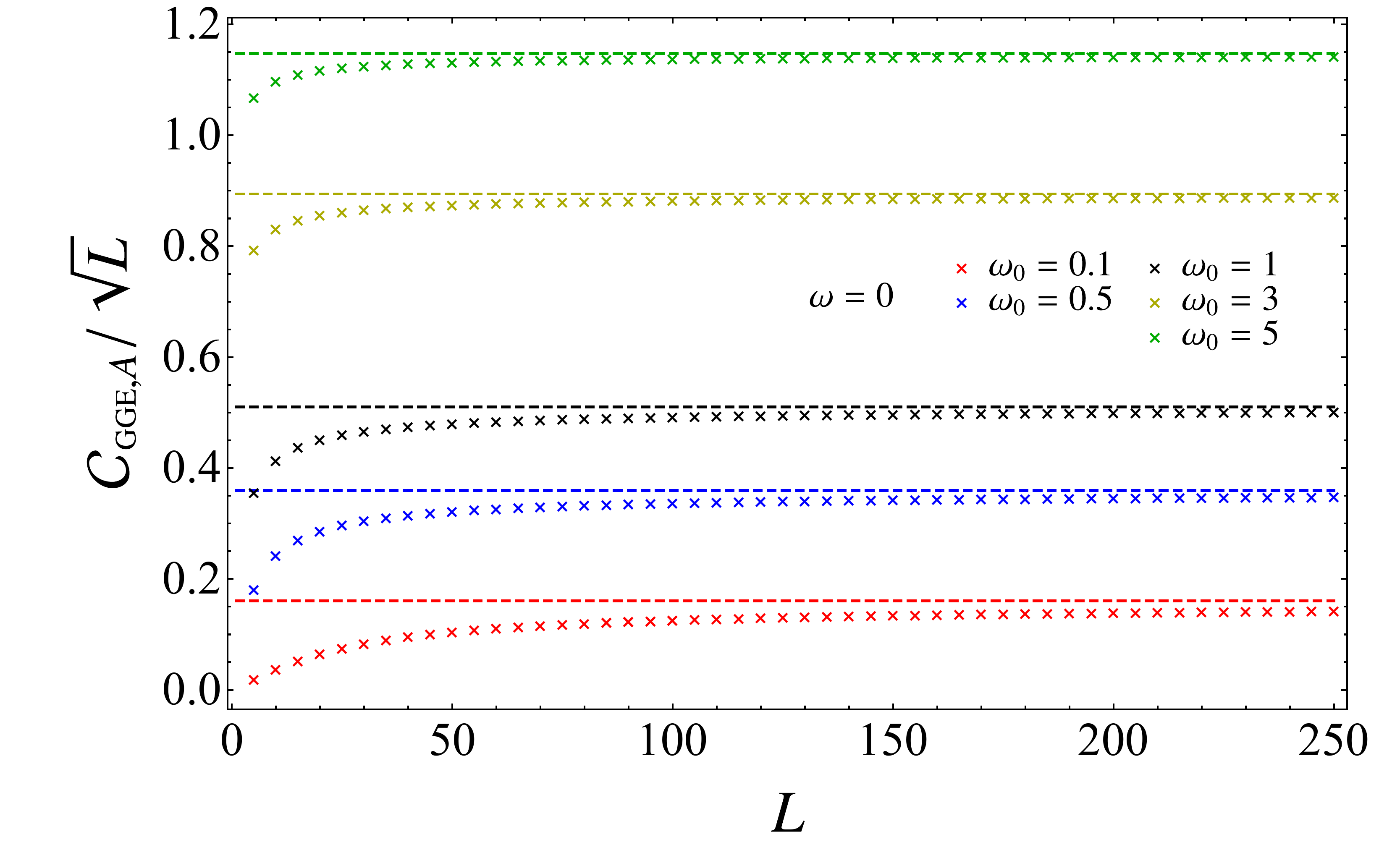}}
\caption{
Asymptotic value of $\mathcal{C}_{\textrm{\tiny GGE},A}$ in (\ref{comp-initialvsGGE}) 
for a block made by $L$ consecutive sites 
which is either in the infinite chain (left panel) or 
adjacent to the origin of a semi-infinite chain with DBC (right panel).
The horizontal dashed lines show $\mathcal{C}_{\textrm{\tiny GGE}} / \sqrt{N}$ as $N \to \infty$,
from (\ref{comp full GGE TD}).
These results support the last equality in (\ref{stationary complexity density}).
}
\vspace{0.4cm}
\label{fig:CAGGEvsCGGE}
\end{figure}

The numerical results in Fig.\,\ref{fig:CAGGEvsCGGE}
provide the main outcome of this appendix. 
Since (\ref{comp full GGE TD}) tells us that $\mathcal{C}_{\textrm{\tiny GGE}}/\sqrt{N}$ 
is finite as $N \to \infty$ 
(see the left panel of Fig.\,\ref{fig:CGGEanalytic})
let us consider $\mathcal{C}_{\textrm{\tiny GGE},A} / \sqrt{L}$ in the limit $L \to \infty$,
where $\mathcal{C}_{\textrm{\tiny GGE},A}$ is given by  (\ref{comp-initialvsGGE}).
The data reported in Fig.\,\ref{fig:CAGGEvsCGGE} for this quantity 
support the validity of the last equality in (\ref{stationary complexity density}).
The coloured data points have been obtained from (\ref{comp-initialvsGGE}),
by employing the reduced correlators from (\ref{corr-GGE-app-int-pbc}) and (\ref{corr-GGE-app-int-dbc})
for the target state
and the reduced correlators from (\ref{QPRmat t-dep TD}) e (\ref{QPRmat t-dep DBC TD}) at $t=0$
for the reference state (like in Fig.\,\ref{fig:GGE} and Fig.\,\ref{fig:GGEdet}). 
The horizontal dashed lines represent the asymptotic values obtained from 
(\ref{comp full GGE TD}), which depend only on $\omega_0$ and $\omega$. 
Comparing the two panels in Fig.\,\ref{fig:CAGGEvsCGGE},
one realises that larger $L$'s are needed to reach the asymptotic value when 
the evolution Hamiltonian is gapless. 
Considering the red data points in the left panel of Fig.\,\ref{fig:CAGGEvsCGGE},
notice that the asymptotic value (\ref{comp full GGE TD})
is symmetric under the exchange $\omega \leftrightarrow \omega_0$,
as already remarked in the text above (\ref{comp full GGE TD}),
while the sets of data points converging to it do not display this symmetry.

In Fig.\,\ref{fig:MixedStatePBCGlobalMasslessEvolutionTD}
we consider harmonic chains on the infinite line and gapless evolution Hamiltonians. 
In particular, we study the temporal evolutions of $\mathcal{C}_{A}/\sqrt{L}$
(from (\ref{c2-complexity-rdm-our-case})), 
of $\Delta S_{A}/L$ and of the ratio $\Delta S_{A} / ( \sqrt{L}\, \mathcal{C}_{A} )$
in terms of $t/L$, for various $L$'s and two values of $\omega_0$.
The reduced covariance matrices have been obtained 
from the correlators (\ref{QPRmat t-dep TD}).
The growths of $\mathcal{C}_{A}$ and of $\Delta S_{A}$
from $t/L\simeq 7$ to $t/L\simeq 25$
have been fitted through the function $ a\log (t/L)+b$
(coloured solid lines in Fig.\,\ref{fig:MixedStatePBCGlobalMasslessEvolutionTD}),
finding that the coefficient of the logarithmic term 
is positive and decreases as $L$ increases. 
In the top panels of Fig.\,\ref{fig:MixedStatePBCGlobalMasslessEvolutionTD},
after the initial growth, $\mathcal{C}_{A}$ reaches a local maximum, 
then it decreases until $t/L\simeq 1/2$ 
and finally the curves follow the logarithmic growth mentioned above. 
This behaviour, which is more evident as $\omega_0$ increases,
 is highlighted in the insets.
 It would be interesting to explore higher value of $L$ 
 in order to check whether, in the limit of $L\to\infty$,
 a saturation is observed to the value given by (\ref{comp full GGE TD}),
 which provides the horizontal dashed lines in the top panels of
 Fig.\,\ref{fig:MixedStatePBCGlobalMasslessEvolutionTD}.
 The horizontal dashed lines in the  middle panels 
 are obtained from (\ref{SGGE}).
In the bottom panels of Fig.\,\ref{fig:MixedStatePBCGlobalMasslessEvolutionTD},
we show the temporal evolutions of the ratio $\Delta S_{A} / ( \sqrt{L}\, \mathcal{C}_{A} )$,
which exhibit a mild logarithmic decreasing for large values of $t/L$.
This tells us that the logarithmic growths of $\mathcal{C}_{A}$ and $\Delta S_{A}$
are very similar. 
However, the values of $L$ are not large enough to determine whether the numerical data points 
asymptote to a constant value as $t/L \to \infty$.

\newpage

\bibliographystyle{nb}

\bibliography{refsMSC}

\end{document}